\documentclass[final]{IEEEtran}
\usepackage{tcolorbox}
\usepackage{adjustbox}

\usepackage{graphicx}
\usepackage{epstopdf}
\usepackage{epsfig}
\usepackage{pdfpages}
\usepackage{psfrag}
\usepackage{ragged2e}
\usepackage{caption}
\usepackage{multirow}
\usepackage{array}
\usepackage{tabularx}
\usepackage{makecell}
\usepackage{threeparttable}
\usepackage{multicol}

\usepackage{subfigure} 
\usepackage[justification=centering]{caption}

\usepackage{tikz}
\usepackage{xcolor}
\usepackage{color}
\usepackage{soul}
\usepackage{framed}

\usepackage{amsmath,amssymb,amsfonts,amsthm}
\usepackage{mathtools}
\usepackage{mathrsfs}
\usepackage{bm}
\usepackage{bbm}
\usepackage{dsfont}
\usepackage{tensor}
\usepackage{mhchem}

\usepackage{algorithm}
\usepackage{algorithmic}

\usepackage{url}
\usepackage{cite}
\usepackage[nolist]{acronym}
\usepackage{textcomp}
\usepackage{setspace}
\usepackage{cuted}
\usepackage{fancyhdr}
\usepackage{lipsum}
\usepackage{xpatch}
\def\AF{{\rm AF}}
\def\FDA{{\rm FDA}}
\usepackage[commandnameprefix=ifneeded]{changes}

\theoremstyle{remark}

\def\nb0{{\mathbf{0}}}
\def\nb1{{\mathbf{1}}}

\newtheorem{definition}{Definition}

\begin{document}
\graphicspath{{./Figures/}}
	\begin{acronym}

\acro{5G-NR}{5G New Radio}
\acro{3GPP}{3rd Generation Partnership Project}
\acro{ABS}{aerial base station}
\acro{AC}{address coding}
\acro{ACF}{autocorrelation function}
\acro{ACR}{autocorrelation receiver}
\acro{ADC}{analog-to-digital converter}
\acrodef{aic}[AIC]{Analog-to-Information Converter}     
\acro{AIC}[AIC]{Akaike information criterion}
\acro{aric}[ARIC]{asymmetric restricted isometry constant}
\acro{arip}[ARIP]{asymmetric restricted isometry property}

\acro{ARQ}{Automatic Repeat Request}
\acro{AUB}{asymptotic union bound}
\acrodef{awgn}[AWGN]{Additive White Gaussian Noise}     
\acro{AWGN}{additive white Gaussian noise}

\acro{APSK}[PSK]{asymmetric PSK} 

\acro{waric}[AWRICs]{asymmetric weak restricted isometry constants}
\acro{warip}[AWRIP]{asymmetric weak restricted isometry property}
\acro{BCH}{Bose, Chaudhuri, and Hocquenghem}        
\acro{BCHC}[BCHSC]{BCH based source coding}
\acro{BEP}{bit error probability}
\acro{BFC}{block fading channel}
\acro{BG}[BG]{Bernoulli-Gaussian}
\acro{BGG}{Bernoulli-Generalized Gaussian}
\acro{BPAM}{binary pulse amplitude modulation}
\acro{BPDN}{Basis Pursuit Denoising}
\acro{BPPM}{binary pulse position modulation}
\acro{BPSK}{Binary Phase Shift Keying}
\acro{BPZF}{bandpass zonal filter}
\acro{BSC}{binary symmetric channels}              
\acro{BU}[BU]{Bernoulli-uniform}
\acro{BER}{bit error rate}
\acro{BS}{base station}
\acro{BW}{BandWidth}
\acro{BLLL}{ binary log-linear learning }

\acro{CP}{Cyclic Prefix}
\acrodef{cdf}[CDF]{cumulative distribution function}   
\acro{CDF}{Cumulative Distribution Function}
\acrodef{c.d.f.}[CDF]{cumulative distribution function}
\acro{CCDF}{complementary cumulative distribution function}
\acrodef{ccdf}[CCDF]{complementary CDF}               
\acrodef{c.c.d.f.}[CCDF]{complementary cumulative distribution function}
\acro{CD}{cooperative diversity}

\acro{CDMA}{Code Division Multiple Access}
\acro{ch.f.}{characteristic function}
\acro{CIR}{channel impulse response}
\acro{cosamp}[CoSaMP]{compressive sampling matching pursuit}
\acro{CR}{cognitive radio}
\acro{cs}[CS]{compressed sensing}                   
\acrodef{cscapital}[CS]{Compressed sensing} 
\acrodef{CS}[CS]{compressed sensing}
\acro{CSI}{channel state information}
\acro{CCSDS}{consultative committee for space data systems}
\acro{CC}{convolutional coding}
\acro{Covid19}[COVID-19]{Coronavirus disease}

\acro{DAA}{detect and avoid}
\acro{DAB}{digital audio broadcasting}
\acro{DCT}{discrete cosine transform}
\acro{dft}[DFT]{discrete Fourier transform}
\acro{DR}{distortion-rate}
\acro{DS}{direct sequence}
\acro{DS-SS}{direct-sequence spread-spectrum}
\acro{DTR}{differential transmitted-reference}
\acro{DVB-H}{digital video broadcasting\,--\,handheld}
\acro{DVB-T}{digital video broadcasting\,--\,terrestrial}
\acro{DL}{DownLink}
\acro{DSSS}{Direct Sequence Spread Spectrum}
\acro{DFT-s-OFDM}{Discrete Fourier Transform-spread-Orthogonal Frequency Division Multiplexing}
\acro{DAS}{Distributed Antenna System}
\acro{DNA}{DeoxyriboNucleic Acid}

\acro{EC}{European Commission}
\acro{EED}[EED]{exact eigenvalues distribution}
\acro{EIRP}{Equivalent Isotropically Radiated Power}
\acro{ELP}{equivalent low-pass}
\acro{eMBB}{Enhanced Mobile Broadband}
\acro{EMF}{ElectroMagnetic Field}
\acro{EU}{European union}
\acro{EI}{Exposure Index}
\acro{eICIC}{enhanced Inter-Cell Interference Coordination}

\acro{FC}[FC]{fusion center}
\acro{FCC}{Federal Communications Commission}
\acro{FEC}{forward error correction}
\acro{FFT}{fast Fourier transform}
\acro{FH}{frequency-hopping}
\acro{FH-SS}{frequency-hopping spread-spectrum}
\acrodef{FS}{Frame synchronization}
\acro{FSsmall}[FS]{frame synchronization}  
\acro{FDMA}{Frequency Division Multiple Access}

\acro{GA}{Gaussian approximation}
\acro{GF}{Galois field }
\acro{GG}{Generalized-Gaussian}
\acro{GIC}[GIC]{generalized information criterion}
\acro{GLRT}{generalized likelihood ratio test}
\acro{GPS}{Global Positioning System}
\acro{GMSK}{Gaussian Minimum Shift Keying}
\acro{GSMA}{Global System for Mobile communications Association}
\acro{GS}{ground station}
\acro{GMG}{ Grid-connected MicroGeneration}

\acro{HAP}{high altitude platform}
\acro{HetNet}{Heterogeneous network}

\acro{IDR}{information distortion-rate}
\acro{IFFT}{inverse fast Fourier transform}
\acro{iht}[IHT]{iterative hard thresholding}
\acro{i.i.d.}{independent, identically distributed}
\acro{IoT}{Internet of Things}                      
\acro{IR}{impulse radio}
\acro{lric}[LRIC]{lower restricted isometry constant}
\acro{lrict}[LRICt]{lower restricted isometry constant threshold}
\acro{ISI}{intersymbol interference}
\acro{ITU}{International Telecommunication Union}
\acro{ICNIRP}{International Commission on Non-Ionizing Radiation Protection}
\acro{IEEE}{Institute of Electrical and Electronics Engineers}
\acro{ICES}{IEEE international committee on electromagnetic safety}
\acro{IEC}{International Electrotechnical Commission}
\acro{IARC}{International Agency on Research on Cancer}
\acro{IS-95}{Interim Standard 95}

\acro{KPI}{Key Performance Indicator}

\acro{LEO}{low earth orbit}
\acro{LF}{likelihood function}
\acro{LLF}{log-likelihood function}
\acro{LLR}{log-likelihood ratio}
\acro{LLRT}{log-likelihood ratio test}
\acro{LoS}{Line-of-Sight}
\acro{LRT}{likelihood ratio test}
\acro{wlric}[LWRIC]{lower weak restricted isometry constant}
\acro{wlrict}[LWRICt]{LWRIC threshold}
\acro{LPWAN}{Low Power Wide Area Network}
\acro{LoRaWAN}{Low power long Range Wide Area Network}
\acro{NLoS}{Non-Line-of-Sight}
\acro{LiFi}[Li-Fi]{light-fidelity}
 \acro{LED}{light emitting diode}
 \acro{LABS}{LoS transmission with each ABS}
 \acro{NLABS}{NLoS transmission with each ABS}

\acro{MB}{multiband}
\acro{MC}{macro cell}
\acro{MDS}{mixed distributed source}
\acro{MF}{matched filter}
\acro{m.g.f.}{moment generating function}
\acro{MI}{mutual information}
\acro{MIMO}{Multiple-Input Multiple-Output}
\acro{MISO}{multiple-input single-output}
\acrodef{maxs}[MJSO]{maximum joint support cardinality}                       
\acro{ML}[ML]{maximum likelihood}
\acro{MMSE}{minimum mean-square error}
\acro{MMV}{multiple measurement vectors}
\acrodef{MOS}{model order selection}
\acro{M-PSK}[${M}$-PSK]{$M$-ary phase shift keying}                       
\acro{M-APSK}[${M}$-PSK]{$M$-ary asymmetric PSK} 
\acro{MP}{ multi-period}
\acro{MINLP}{mixed integer non-linear programming}

\acro{M-QAM}[$M$-QAM]{$M$-ary quadrature amplitude modulation}
\acro{MRC}{maximal ratio combiner}                  
\acro{maxs}[MSO]{maximum sparsity order}                                      
\acro{M2M}{Machine-to-Machine}                                                
\acro{MUI}{multi-user interference}
\acro{mMTC}{massive Machine Type Communications}      
\acro{mm-Wave}{millimeter-wave}
\acro{MP}{mobile phone}
\acro{MPE}{maximum permissible exposure}
\acro{MAC}{media access control}
\acro{NB}{narrowband}
\acro{NBI}{narrowband interference}
\acro{NLA}{nonlinear sparse approximation}
\acro{NLOS}{Non-Line of Sight}
\acro{NTIA}{National Telecommunications and Information Administration}
\acro{NTP}{National Toxicology Program}
\acro{NHS}{National Health Service}

\acro{LOS}{Line of Sight}

\acro{OC}{optimum combining}                             
\acro{OC}{optimum combining}
\acro{ODE}{operational distortion-energy}
\acro{ODR}{operational distortion-rate}
\acro{OFDM}{Orthogonal Frequency-Division Multiplexing}
\acro{omp}[OMP]{orthogonal matching pursuit}
\acro{OSMP}[OSMP]{orthogonal subspace matching pursuit}
\acro{OQAM}{offset quadrature amplitude modulation}
\acro{OQPSK}{offset QPSK}
\acro{OFDMA}{Orthogonal Frequency-division Multiple Access}
\acro{OPEX}{Operating Expenditures}
\acro{OQPSK/PM}{OQPSK with phase modulation}

\acro{PAM}{pulse amplitude modulation}
\acro{PAR}{peak-to-average ratio}
\acrodef{pdf}[PDF]{probability density function}                      
\acro{PDF}{probability density function}
\acrodef{p.d.f.}[PDF]{probability distribution function}
\acro{PDP}{power dispersion profile}
\acro{PMF}{probability mass function}                             
\acrodef{p.m.f.}[PMF]{probability mass function}
\acro{PN}{pseudo-noise}
\acro{PPM}{pulse position modulation}
\acro{PRake}{Partial Rake}
\acro{PSD}{power spectral density}
\acro{PSEP}{pairwise synchronization error probability}
\acro{PSK}{phase shift keying}
\acro{PD}{power density}
\acro{8-PSK}[$8$-PSK]{$8$-phase shift keying}
\acro{PPP}{Poisson point process}
\acro{PCP}{Poisson cluster process}
 
\acro{FSK}{Frequency Shift Keying}

\acro{QAM}{Quadrature Amplitude Modulation}
\acro{QPSK}{Quadrature Phase Shift Keying}
\acro{OQPSK/PM}{OQPSK with phase modulator }

\acro{RD}[RD]{raw data}
\acro{RDL}{"random data limit"}
\acro{ric}[RIC]{restricted isometry constant}
\acro{rict}[RICt]{restricted isometry constant threshold}
\acro{rip}[RIP]{restricted isometry property}
\acro{ROC}{receiver operating characteristic}
\acro{rq}[RQ]{Raleigh quotient}
\acro{RS}[RS]{Reed-Solomon}
\acro{RSC}[RSSC]{RS based source coding}
\acro{r.v.}{random variable}                               
\acro{R.V.}{random vector}
\acro{RMS}{root mean square}
\acro{RFR}{radiofrequency radiation}
\acro{RIS}{Reconfigurable Intelligent Surface}
\acro{RNA}{RiboNucleic Acid}
\acro{RRM}{Radio Resource Management}
\acro{RUE}{reference user equipments}
\acro{RAT}{radio access technology}
\acro{RB}{resource block}

\acro{SA}[SA-Music]{subspace-augmented MUSIC with OSMP}
\acro{SC}{small cell}
\acro{SCBSES}[SCBSES]{Source Compression Based Syndrome Encoding Scheme}
\acro{SCM}{sample covariance matrix}
\acro{SEP}{symbol error probability}
\acro{SG}[SG]{sparse-land Gaussian model}
\acro{SIMO}{single-input multiple-output}
\acro{SINR}{signal-to-interference plus noise ratio}
\acro{SIR}{signal-to-interference ratio}
\acro{SISO}{Single-Input Single-Output}
\acro{SMV}{single measurement vector}
\acro{SNR}[\textrm{SNR}]{signal-to-noise ratio} 
\acro{sp}[SP]{subspace pursuit}
\acro{SS}{spread spectrum}
\acro{SW}{sync word}
\acro{SAR}{specific absorption rate}
\acro{SSB}{synchronization signal block}
\acro{SR}{shrink and realign}

\acro{tUAV}{tethered Unmanned Aerial Vehicle}
\acro{TBS}{terrestrial base station}

\acro{uUAV}{untethered Unmanned Aerial Vehicle}
\acro{PDF}{probability density functions}

\acro{PL}{path-loss}

\acro{TH}{time-hopping}
\acro{ToA}{time-of-arrival}
\acro{TR}{transmitted-reference}
\acro{TW}{Tracy-Widom}
\acro{TWDT}{TW Distribution Tail}
\acro{TCM}{trellis coded modulation}
\acro{TDD}{Time-Division Duplexing}
\acro{TDMA}{Time Division Multiple Access}
\acro{Tx}{average transmit}

\acro{UAV}{Unmanned Aerial Vehicle}
\acro{uric}[URIC]{upper restricted isometry constant}
\acro{urict}[URICt]{upper restricted isometry constant threshold}
\acro{UWB}{ultrawide band}
\acro{UWBcap}[UWB]{Ultrawide band}   
\acro{URLLC}{Ultra Reliable Low Latency Communications}
         
\acro{wuric}[UWRIC]{upper weak restricted isometry constant}
\acro{wurict}[UWRICt]{UWRIC threshold}                
\acro{UE}{User Equipment}
\acro{UL}{UpLink}

\acro{WiM}[WiM]{weigh-in-motion}
\acro{WLAN}{wireless local area network}
\acro{wm}[WM]{Wishart matrix}                               
\acroplural{wm}[WM]{Wishart matrices}
\acro{WMAN}{wireless metropolitan area network}
\acro{WPAN}{wireless personal area network}
\acro{wric}[WRIC]{weak restricted isometry constant}
\acro{wrict}[WRICt]{weak restricted isometry constant thresholds}
\acro{wrip}[WRIP]{weak restricted isometry property}
\acro{WSN}{wireless sensor network}                        
\acro{WSS}{Wide-Sense Stationary}
\acro{WHO}{World Health Organization}
\acro{Wi-Fi}{Wireless Fidelity}

\acro{sss}[SpaSoSEnc]{sparse source syndrome encoding}

\acro{VLC}{Visible Light Communication}
\acro{VPN}{Virtual Private Network} 
\acro{RF}{Radio Frequency}
\acro{FSO}{Free Space Optics}
\acro{IoST}{Internet of Space Things}

\acro{GSM}{Global System for Mobile Communications}
\acro{2G}{Second-generation cellular network}
\acro{3G}{Third-generation cellular network}
\acro{4G}{Fourth-generation cellular network}
\acro{5G}{Fifth-generation cellular network}	
\acro{gNB}{next-generation Node-B Base Station}
\acro{NR}{New Radio}
\acro{UMTS}{Universal Mobile Telecommunications Service}
\acro{LTE}{Long Term Evolution}

\acro{QoS}{Quality of Service}
\end{acronym}
	
\newcommand{\SAR} {\mathrm{SAR}}
\newcommand{\WBSAR} {\mathrm{SAR}_{\mathsf{WB}}}
\newcommand{\gSAR} {\mathrm{SAR}_{10\si{\gram}}}
\newcommand{\Sab} {S_{\mathsf{ab}}}
\newcommand{\Eavg} {E_{\mathsf{avg}}}
\newcommand{\ft}{f_{\textsf{th}}}
\newcommand{\alphatf}{\alpha_{24}}

\title{
Frequency Diverse Arrays: Fundamentals, Key Insights, and Future Directions
}
\author{Bang Huang, {\em Member, IEEE}, Sajid Ahmed, {\em Senior Member, IEEE}, Wenkai Jia,  Mohamed-Slim Alouini, {\em Fellow, IEEE}, and Wen-Qin~Wang {\em Senior Member, IEEE}

\thanks{Bang Huang is a  postdoctoral fellow 
with King Abdullah University of Science and Technology (KAUST), CEMSE division , Thuwal 23955-6900, Saudi Arabia. Besides, he was a PhD student from School of Information and Communication Engineering, University of Electronic Science and Technology of China, Chengdu, 611731, P. R. China.(e-mail:  bang.huang@kaust.edu.sa) (Corresponding author: Bang Huang).}

\thanks{
Sajid Ahmed and Mohamed-Slim Alouini are with King Abdullah University of Science and Technology (KAUST), CEMSE division, Thuwal 23955-6900, Saudi Arabia (e-mails:  sajid.ahmed@kaust.edu.sa; slim.alouini@kaust.edu.sa).}
\thanks{Wenkai Jia and Wen-Qin Wang are with the School of Information and Communication Engineering, University of Electronic Science and Technology of China, Chengdu, 611731, P. R. China. (e-mails:  mrwenkaij@126.com; wqwang@uestc.edu.cn).}
\vspace{-6mm}
}
\maketitle
\thispagestyle{empty}

\begin{abstract}
Frequency diverse arrays (FDA) have attracted sustained interest as a promising architecture for introducing range-dependent responses into array systems. Unlike conventional phased arrays (PA), whose transmit behavior is primarily angle-dependent, FDA employs inter-element frequency offsets to generate time- and range-dependent phase structures, thereby producing a joint time--range--angle array response. Despite extensive research, the physical meaning of FDA-induced degrees of freedom remains debated, particularly in relation to range--angle coupling, the feasibility of time-invariant focusing, and the distinction between frequency-driven and waveform-driven range selectivity.
This paper reexamines FDA from a structural and manifold-based perspective. A central contribution is the introduction of an irreducibility criterion, which distinguishes genuine range-domain physical degrees of freedom from effects that can be reproduced by equivalent signal-processing transformations. Based on this perspective, PA, multiple-input multiple-output (MIMO), FDA, and FDA--MIMO are comparatively interpreted according to the physical origin of their effective degrees of freedom, including spatial phase, waveform orthogonality, frequency gradients, and their interaction. The paper further clarifies the role of frequency across different array paradigms, contrasts FDA with time-coding-based architectures, and explains how key FDA properties such as manifold expansion, range--angle coupling, time variation, and multi-frequency diversity translate into system capabilities.
Building on these structural insights, the paper connects FDA to a broad range of radar and communication functionalities, including parameter estimation, target detection, imaging, physical-layer security, and integrated sensing and communication. It also discusses major limitations and open issues, including hardware realizability, wideband constraints, experimental reproducibility, near-field modeling, and the physical boundary of time-invariant focusing. Overall, this survey aims to clarify what is fundamentally distinctive about FDA, identify the conditions under which its gains are physically meaningful, and outline promising directions for future research in next-generation sensing and communication systems.
\end{abstract}
\begin{IEEEkeywords}
Frequency diverse array (FDA), multiple-input multiple-output (MIMO), phased arrays (PA), radar applications, communication applications, range-angle dependence.
\end{IEEEkeywords}
\section{Introduction}
\label{sec_introduction}

\subsection{Background and Motivation}

Modern sensing and communication systems have evolved through a progressive expansion
of exploitable signal dimensions \cite{rahman2019fundamental,rappaport2002wireless}. 
In radar systems, early designs primarily relied on echo detection and propagation delay estimation,
with signal processing largely confined to the time domain \cite{lutkepohl1997handbook}.
The introduction of pulse compression enabled energy focusing via matched filtering,
improving range resolution and partially suppressing uncorrelated interference \cite{eaves2012principles}.
Subsequently, pulse-Doppler radar incorporated Doppler frequency shifts,
introducing a frequency-domain discrimination dimension and enabling joint time–frequency processing \cite{Skolnik2001IntroductionRadarbook}.
Further advances were achieved through phased-array (PA) radar,
which introduced spatial-domain processing via beamforming and directional selectivity \cite{mailloux2017phased}.
Combined with space–time adaptive processing (STAP),
modern radar systems now operate in a joint space–time domain,
significantly enhancing detection performance in complex environments \cite{stoica2005spectral,barton1988modern}.

A similar evolution can be observed in communication systems, whose development has also been marked by the progressive exploitation of new transmission dimensions.
Early communication systems relied primarily on time-domain signaling, where information delivery and user access were largely organized over time.
Subsequent advances extended communication design into the code domain, exemplified by code-division multiple access (CDMA) \cite{buehrer2022code,zigangirov2004theory}, which enabled multiuser separation through spreading codes and provided enhanced robustness against interference.
The frequency domain was progressively incorporated into communication system design, initially through frequency-division multiplexing and multiple-access schemes such as frequency-division multiple access (FDMA) \cite{faruque2018frequency,saadia2020single}, and later in a more structured manner through multicarrier techniques such as orthogonal frequency division multiplexing (OFDM) \cite{li2006orthogonal,weinstein2009history}, which improved spectral efficiency and facilitated reliable transmission over frequency-selective channels.
More recently, multiple-input multiple-output (MIMO) \cite{hampton2013introduction,gershman2005space,duman2008coding,de2008multiantenna} technology has elevated the spatial domain into a fundamental communication resource, enabling spatial multiplexing, beamforming, and diversity gains.
Taken together, these advances reflect a broader evolution from one-dimensional signaling toward the joint exploitation of time, code, frequency, and space.

Despite these advances, a common limitation persists across both radar and communication systems:
while time, frequency, code, and spatial dimensions have been extensively exploited,
the {range dimension} remains largely passive \cite{basit2018development,Wang2015FrequencyDiverseArray}.
In conventional array systems, the array response is primarily a function of angle,
whereas range information is only implicitly embedded in propagation delay or path loss,
rather than being explicitly controllable as a design variable.
Existing approaches attempt to utilize range information through waveform design or receiver processing,
such as matched filtering or delay estimation.
However, these mechanisms operate mainly in the signal processing stage,
without fundamentally altering the propagation structure of the transmitted field.
As a result, range cannot be directly incorporated into the array response
in the same manner as angle in phased arrays or space in MIMO systems.
This limitation becomes particularly critical in emerging scenarios,
including range-dependent interference suppression,
ambiguity resolution, near-field communications, holographic MIMO,
physical-layer security, and integrated sensing and communication (ISAC),
where fine-grained control over the range dimension is highly desirable.
These challenges motivate the need for array architectures
that can explicitly incorporate range into the design space.

\subsection{The Concept of Frequency Diverse Array}

Motivated by the aforementioned limitations, particularly the lack of controllable range-domain response in conventional array systems, 
the frequency diverse array (FDA) emerges as a natural extension that enables the explicit incorporation of the range dimension into array design.
The concept of FDA was first introduced by Antonik in 2006 at the IEEE Radar Conference \cite{Antonik2006Frequencydiverse}. 
Unlike conventional phased arrays (PAs), where all elements operate at a common carrier frequency, 
FDA assigns slightly different carrier frequencies to different array elements, i.e.,
$f_m = f_c + \Delta f_m.$
Although seemingly minor, this modification fundamentally alters the propagation-induced phase structure \cite{antonik2009investigation,XuLiao2018Anovervoew,wang2016overview}. 
Since the radiated field from each element accumulates phase according to its own carrier frequency over the propagation delay, the inter-element phase difference becomes a function not only of angle, but also of range and time.

For a conventional far-field phased array, the transmit array response at an instantaneous snapshot is determined primarily by angular steering, while range enters only through common propagation delay and attenuation rather than as an independently controllable array-response dimension \cite{fulton2016digital,visser2006array}. 
In contrast, the element-dependent frequency offsets in FDA introduce additional time- and range-dependent phase terms, resulting in a transmit response that depends jointly on time, range, and angle.
Consequently, the array manifold is extended from $\mathbf{a}(\theta)$ to a higher-dimensional representation $\mathbf{a}(t,r,\theta)$.
This structural extension gives rise to several distinctive phenomena, including range--angle coupling, time-varying beam scanning, and dynamic focusing behavior.
Unlike conventional arrays that form essentially static beams, FDA produces a spatiotemporal energy distribution that evolves continuously over time, often visualized as a trajectory in the range--angle plane.
These properties create new opportunities for range-aware beam synthesis, ambiguity shaping, interference suppression, and ISAC design.
At the same time, the resulting range dependence is intrinsically coupled with temporal evolution, which fundamentally distinguishes FDA from conventional static-beam array architectures.


Importantly, FDA should not be understood as merely introducing frequency diversity into an array.
Its real significance lies in altering the dimensionality of the array manifold itself, by embedding range and time into the propagation-induced response.
In this sense, FDA extends the conventional angle-dependent array model toward a joint time--range--angle representation.
The canonical FDA usually employs linear frequency increments across the aperture.
However, this simplest form also reveals intrinsic range--angle coupling and time-varying focusing behavior, which soon motivated a broad family of generalized FDA architectures, including nonlinear-offset, random-offset, time-modulated, subarray, and geometry-modified designs.
These developments mark an important transition from the original FDA concept toward a richer research landscape centered on controllable propagation-structure design \cite{Khan2015FrequencyDiverse,wang2020dot,liao2019frequency,gao2016decoupled,Shao2021FrequencyDiverse,basit2017beam,Liao2020FrequencyDiverse,Liao202AntennaBeampattern,saeed2016tangent,wang2016range,Choi2023Analysisof,liu2016random,huang2020frequency,Xiong2017FrequencyDiverse,yaw2020frequency,Ge2022FuzzyEntropy}.
This evolutionary trajectory, together with the subsequent emergence of FDA--MIMO and other related range--angle responsive architectures, is reviewed next.

\subsection{Research History for FDA }

\begin{figure}[htp]
\centering
{\includegraphics[width=0.48\textwidth]{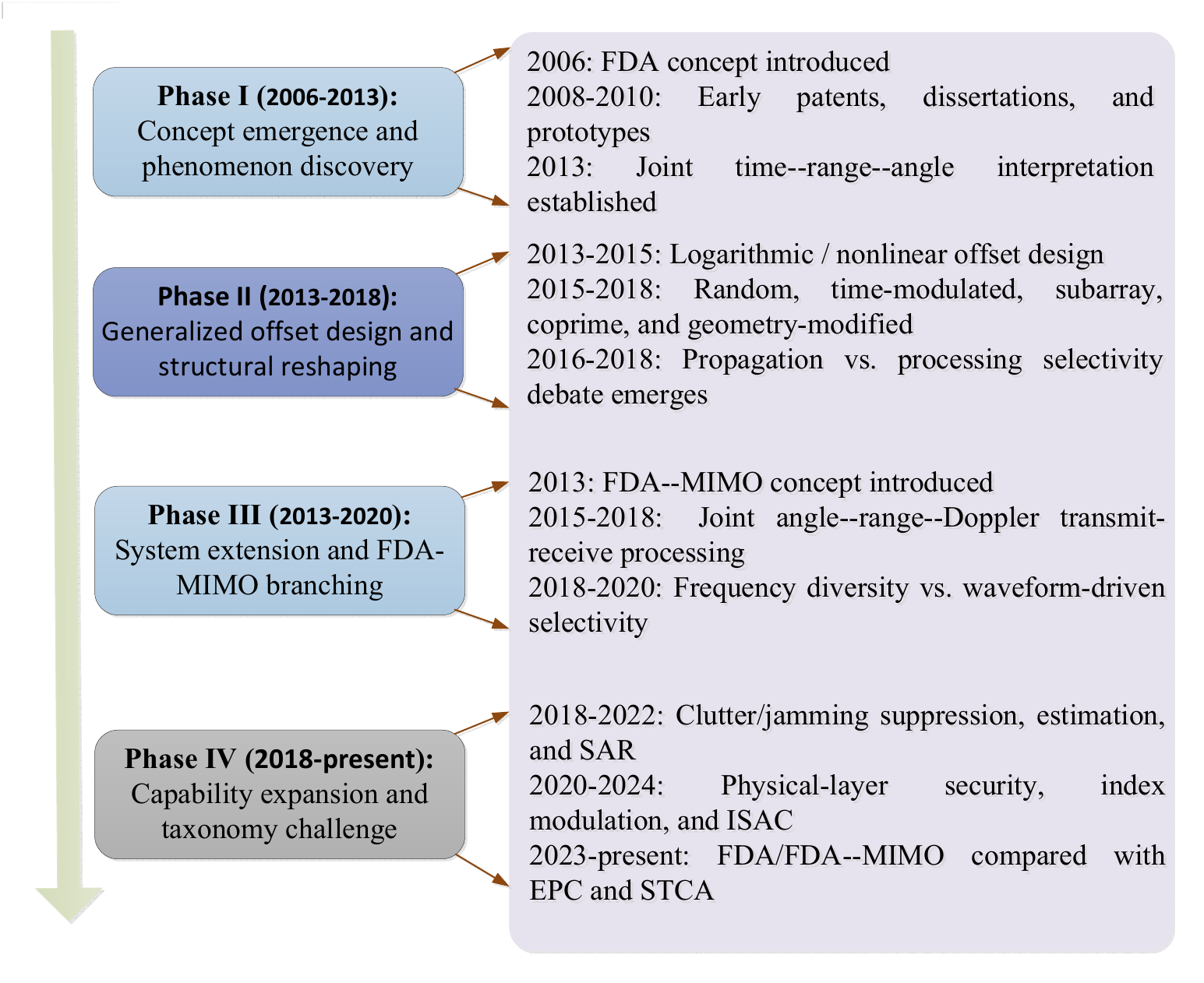}}
\caption{Representative development trajectory of FDA research, from concept emergence to generalized structural redesign, FDA--MIMO branching, capability expansion, and taxonomy-oriented comparison.}
\label{Fig:his}	
\end{figure}

Since its introduction by Antonik in 2006 \cite{Antonik2006Frequencydiverse}, FDA research has evolved from a concept centered on unusual transmit beampattern phenomena into a broader framework for propagation-structure engineering.
Early patents, doctoral-level investigations, and later prototype studies further indicate that FDA quickly attracted attention beyond purely conceptual analysis \cite{wicks2008frequency,wicks2009method,antonik2009investigation,aytun2010frequency}.
As also reflected in later overview and survey articles \cite{wang2016overview}, the development of FDA has not proceeded merely as a chronological accumulation of applications, but rather through a sequence of conceptual reinterpretations, structural extensions, and task-driven generalizations.
From this perspective, the evolution of FDA research can be broadly understood through four interconnected stages, as shown in Fig.\ref{Fig:his}.

\textbf{(i) Concept emergence and phenomenon discovery (2006--2013):}
The earliest studies were primarily devoted to understanding the unconventional transmit behavior induced by element-dependent frequency offsets \cite{Antonik2006Frequencydiverse,antonik2009investigation}.
Relative to conventional phased arrays, FDA exhibited range--angle coupling, time-varying beam scanning, and dynamic focusing effects, thereby extending the array response from an angle-only description to a joint time--range--angle structure \cite{wang2013range,wang2014linear,liao2018generalized}.
During this stage, the main research effort focused on revealing these phenomena, clarifying their physical origin, and assessing whether the resulting range dependence could be interpreted as a genuine new structural degree of freedom \cite{antonik2009investigation,wang2016overview,XuLiao2018Anovervoew}.
At the same time, the canonical FDA with linear frequency increments quickly exposed its intrinsic limitations: the transmit response was strongly range--angle coupled, and the focusing behavior was generally nonstationary rather than permanently fixed at a prescribed location \cite{chen2017transmit,chen2019accurate,liao2023time}.
These observations triggered extensive discussions on the physical interpretability of FDA focusing and the feasibility of time-invariant beamforming.

\textbf{(ii) Generalized offset design and structural reshaping (2013--2018):}
Once the basic FDA mechanism became better understood, research rapidly moved beyond the original linear-offset configuration.
A wide range of generalized FDA architectures were proposed, including logarithmic-offset FDA \cite{Khan2015FrequencyDiverse}, nonlinear polynomial-offset FDA \cite{gao2016decoupled}, random-offset and logarithmic-random-offset FDA \cite{liu2016random,huang2020frequency}, time-modulated and sparse FDA \cite{yang2018optimization,gong2018time}, subarray-based FDA \cite{wang2014transmit,wang2014subarray,xu2015flat,xu2017range}, coprime FDA \cite{qin2016frequency,Ni2021RangeDependent}, and geometry-modified FDA such as semicircular or nonuniform-spacing structures \cite{mahmood2018frequency,Xu2022RangeAngle,ulrich2018wavelength}.
Related efforts also explored beampattern synthesis using windowing, weighting, Costas-type offsets, and intelligent optimization methods in order to improve focusing, reduce sidelobes, or achieve range--angle decoupling \cite{basit2017beam,Liao2020FrequencyDiverse,Liao202AntennaBeampattern,saeed2016tangent,Xiong2017FrequencyDiverse,yaw2020frequency,Ge2022FuzzyEntropy}.
Although these variants differ in implementation, they were largely motivated by a common objective: to reshape the propagation-induced phase structure so as to mitigate range--angle coupling, improve focusing behavior, suppress sidelobes, or enhance controllability in the range--angle plane.
In this sense, the evolution from linear FDA to generalized FDA should not be understood as a simple accumulation of design variants, but as a deeper transition from {phenomenon observation} to {manifold engineering}.

A particularly important thread in this stage concerned attempts to realize quasi-static or time-invariant focusing.
Pulse-based FDA, optimized offset design, and receive-assisted equivalent focusing schemes were all proposed to approximate or synthesize more stable range--angle responses.
However, these efforts also made clear that one must carefully distinguish between genuinely propagation-induced range dependence and range selectivity that is produced only after matched filtering, coding, or transmit--receive synthesis \cite{chen2017transmit,chen2019accurate,TanWang2021CorrectionAnalysis,liao2023time}.
This distinction later became central to the interpretation of FDA-like architectures more broadly.

\textbf{(iii) System extension and architectural branching (2013--2020):}
As FDA matured, the research frontier expanded from single-aperture coherent FDA toward multi-channel and hybrid architectures, most notably FDA--MIMO \cite{sammartino2013frequency,xu2015joint,khan2015double,Xiong2018FDAMIMO}.
By combining frequency offsets with waveform orthogonality, FDA--MIMO enlarged the design space and enabled joint transmit--receive processing in angle, range, and Doppler.
This significantly accelerated research in range-ambiguous clutter suppression, deceptive jamming mitigation, target localization, and adaptive detection \cite{xu2015range,xu2016space,Xu2017AnAdaptive,xu2015deceptive,Xu2015SpaceTimeRange,Lan2018Suppression,Wen2018Enhanced,Gui2020Low,HuangBasit2022AdaptiveMoving,Lan2020GLRTbasedAdaptive}.
At the same time, FDA--MIMO also introduced an important structural ambiguity: once waveform orthogonality and receiver-side matched filtering are incorporated, the effective system degrees of freedom may arise jointly from propagation-level frequency diversity and signal-processing-level waveform diversity.
As a result, the relationship between coherent FDA and FDA--MIMO is not merely a matter of implementation, but a question of mechanism origin.
This is also why the evolution of FDA research cannot be fully understood without explicitly distinguishing propagation-driven array effects from processing-driven virtual responses.

\textbf{(iv) Capability expansion, cross-domain applications, and taxonomy challenge (since 2018):}
With the proliferation of FDA variants and FDA--MIMO architectures, the field increasingly moved toward application-oriented development.
FDA-related techniques have been applied to a broad range of tasks, including range-ambiguous clutter suppression \cite{WangZhu2021RangeAmbiguousClutter,Wang2021RangeAmbiguous,Liu2023RangeAmbiguous,Qiu2023RangeAmbiguous}, deceptive jamming suppression \cite{Wang2020MainBeamRange,Lan2020SuppressionofMainbeam}, target detection and parameter estimation \cite{GuiWang2018CoherentPulsedFDA,HuangJian2022AdaptiveDistributed,HuangBasit2022AdaptiveDetection,Huang2023Adaptivemultiple,chen2023monopulse}, synthetic aperture radar (SAR) imaging \cite{farooq2007application,farooq2008exploiting,Wang2017ARangeAmbiguity,lin2017unambiguous,Zhou2021HighResolutionandWideSwath,Zhang2023HighResolution}, low-probability-of-intercept (LPI) radar \cite{Wang2021LPI,Gong2022JointDesign}, physical-layer security \cite{Ji2018SecrecyCapacityAnalysis,Qiu2018ArtificialNoiseAided,Hu2017ArtificialNoiseAided,ChengWang2021PhysicalLayerSecurity,WangYan2021SecrecyZone,Qiu2019MultiBeamDirectional,Jian2022physical,Jian2023PhysicalLayer}, and ISAC \cite{Nusenu2018Dualfunction,Li2023JointRadarCommunication,Gong2023Optimizationof}.
During this stage, FDA was no longer viewed merely as a source of unusual beampatterns, but increasingly as a flexible mechanism for shaping propagation structures to satisfy task-specific objectives.
Accordingly, design criteria also evolved from qualitative beampattern visualization toward performance-oriented metrics such as ambiguity suppression, estimation accuracy, interference rejection, identifiability, secrecy enhancement, and sensing--communication integration.

In parallel, the literature also began to discuss FDA and FDA--MIMO together with other array paradigms capable of generating range--angle dependent responses, such as element--pulse coding (EPC) arrays and space--time coding arrays (STCA) \cite{xu2020resolving,liu2023resolving,liu2025range,lan2024mainlobe,zhu2024simultaneous,lan2020mainlobe,lan2018subarray,liu2023signal,wang2021transmit,wang2024sidelobe,wang2025range,wang2025mainlobe,wang2025monopulse}.
This broader comparison is highly significant, because it reveals that not all range--angle dependent responses arise from the same mechanism.
For some architectures, the additional selectivity is rooted in transmit-side propagation physics; for others, it is produced through coding, pulse diversity, or receiver-side synthesis.
Consequently, a key issue in the modern FDA literature is no longer only how to generate range-dependent beampatterns, but how to interpret their physical origin, distinguish genuine and synthesized degrees of freedom, and classify related architectures within a unified structural taxonomy.

It is also worth noting that the historical development of FDA research has included not only theoretical analysis and numerical design, but also patents, prototype exploration, and experimental validation \cite{wicks2008frequency,wicks2009method,antonik2009investigation,eker2013exploitation}.
The literature has reported early proof-of-concept platforms and later prototype demonstrations, indicating that FDA has evolved beyond a purely conceptual topic \cite{antonik2009investigation,eker2013exploitation}.
Nevertheless, compared with the diversity of theoretical models, experimental validation remains relatively limited, and many important issues---including hardware complexity, coherent implementation, waveform calibration, and practical robustness---are still far from fully resolved.

Overall, the historical evolution of FDA suggests that the field has progressed through a clear internal logic: from discovering unconventional range--angle--time behavior, to redesigning the array structure for better controllability, to branching into FDA--MIMO and related architectures, and finally to expanding toward diverse sensing and communication tasks.
Yet despite this rich development, several foundational questions remain open: whether truly time-invariant focusing is physically achievable, whether range--angle coupling should be viewed as a useful degree of freedom or a design burden, and how FDA, FDA--MIMO, EPC, STCA, and related architectures should be positioned within a unified modeling framework.
These unresolved issues strongly motivate the need for a structural re-examination of FDA research.

Motivated by these observations, this paper revisits FDA and its related architectures from a unified signal modeling and taxonomy perspective.
Rather than organizing the literature solely by application domain, we aim to establish a principled mapping from design mechanism to manifold structure, from manifold structure to exploitable degrees of freedom, and from these degrees of freedom to sensing and communication capabilities.

\subsection{Core Innovations and Key Debates of FDA}

Although FDA has been extensively studied, its essential innovation and the boundary of its claimed capabilities remain subjects of ongoing debate.
Much of the existing literature interprets FDA from application-oriented, beampattern-oriented, or architecture-oriented viewpoints.
In contrast, we revisit FDA from a structural perspective and argue that its significance lies not merely in generating range-dependent responses, but in altering how propagation variables enter the array manifold itself.
From this viewpoint, FDA exhibits several core innovations, while also raising a number of unresolved debates concerning the physical meaning, structural origin, and practical validity of its claimed degrees of freedom.

\subsubsection{From Angle-Only Response to a Joint Time--Range--Angle Manifold}

The most fundamental distinction of FDA lies in introducing inter-element frequency gradients across the array aperture.
Unlike conventional phased arrays, whose manifold is essentially angle-dependent at an instantaneous snapshot, FDA embeds range dependence directly into the propagation process through element-dependent carrier frequencies.
As a result, the array response is extended from an angle-only representation $\mathbf{a}(\theta)$ to a higher-dimensional manifold $\mathbf{a}(t,R_0,\theta)$.
This transformation constitutes a genuine structural extension of the array model, enabling joint dependence on time, range, and angle.
At the same time, it also introduces intrinsic time variation and range--angle coupling, which fundamentally distinguish FDA from conventional static-beam array architectures.

\subsubsection{Frequency as a Propagation-Domain Design Variable}

A second conceptual innovation of FDA is that frequency is no longer used merely as a signal-domain resource for multiplexing, waveform construction, or spectral allocation.
Instead, FDA employs frequency offsets to generate phase gradients across the array aperture, thereby making frequency a propagation-domain design variable.
This shift is conceptually important: the role of frequency is moved from shaping the transmitted signal alone to shaping how the signal propagates and interferes across space.
In other words, FDA does not simply use different frequencies; it uses frequency differences to alter the spatial-temporal structure of the transmitted field.
This naturally raises a fundamental question: do such frequency gradients create genuinely new physical degrees of freedom, or do they merely re-parameterize effects that can already be synthesized through other means?

\subsubsection{From Beamforming to Propagation-Structure Shaping}

The above viewpoint leads to a broader reinterpretation of what FDA contributes to array design.
In conventional arrays, beamforming is typically performed over a fixed manifold, where the main task is to optimize weights, phases, or waveforms under a given propagation structure.
FDA goes one step further: rather than operating solely on top of a fixed angular manifold, it modifies the mechanism by which the manifold itself is generated.
Hence, FDA should not be understood merely as another beam synthesis technique, but as a framework for propagation-structure shaping.
Its novelty lies in altering how time, range, and angle jointly enter the array response, thereby transforming the design problem from beam control over a given manifold to structural control over the manifold-forming process itself.

\subsubsection{Irreducibility as a Criterion for Genuine Physical Degrees of Freedom}

From this structural perspective, the mere existence of frequency offsets is not sufficient to claim new physical degrees of freedom.
We argue that a more meaningful criterion is  {irreducibility}.
Specifically, a range-dependent phase term should be regarded as a genuine physical degree of freedom only if it cannot be removed, absorbed, or equivalently transformed away without collapsing the underlying propagation structure into an angle-only or waveform-synthesized form.
If the apparent range dependence can be eliminated through equivalent linear compensation, matched filtering, coding synthesis, or other receiver-side transformations, then the resulting selectivity does not necessarily correspond to an independent propagation-domain degree of freedom.
Under this viewpoint, the key issue is not whether range-dependent responses can be observed, but whether they are intrinsically embedded in the propagation manifold itself.
Only when the range-dependent phase structure is irreducible in this sense does FDA introduce a genuinely new physical design dimension. This criterion also has an important implication for FDA--MIMO.
If range is not merely a delay-domain quantity but an irreducible parameter embedded in the transmit--receive manifold, then distance estimation may no longer need to be interpreted exclusively within the classical pulse-compression framework.
Under such a viewpoint, range could in principle be treated as a manifold parameter, much like angle in phased arrays, thereby opening the possibility of a manifold-based ranging paradigm.
This, in turn, raises the further question of whether range resolution should still be defined solely by waveform bandwidth, or instead by the distinguishability and identifiability of range-dependent manifold responses.

\subsubsection{Key Debates in FDA Research}

The above structural insights, especially the criterion of irreducibility for genuine physical degrees of freedom, lead to several fundamental debates that continue to shape the interpretation of FDA and FDA-related architectures:

\begin{itemize}
\item \textbf{Time-invariant focusing:} whether strictly time-invariant spatial focusing can be achieved under nonzero frequency gradients, or whether such focusing is inherently dynamic in FDA-like architectures;

\item \textbf{Range--angle coupling:} whether the coupling observed in FDA should be interpreted as a useful structural degree of freedom or as an additional source of design complexity and signal-processing burden;

\item \textbf{Origin of selectivity:} whether the observed range-dependent response is induced directly by propagation physics or synthesized through coding, matched filtering, or receiver-side processing;

\item \textbf{Ranging paradigm:} whether range should still be interpreted exclusively under the classical delay-resolution framework, or whether the embedding of range into the transmit--receive manifold opens the possibility of manifold-based ranging and a corresponding redefinition of range resolution;

\item \textbf{Paradigm interpretation:} whether FDA--MIMO should be understood primarily as a frequency-driven extension of FDA or as a waveform-driven hybrid architecture whose additional selectivity is not purely propagation-originated;

\item \textbf{Practical realizability:} how hardware nonidealities, bandwidth limitations, calibration errors, coherence requirements, and implementation complexity affect the existence and usability of the theoretically claimed degrees of freedom.

\end{itemize}

Among the above issues, this paper places particular emphasis on three major structural debates, namely the physical feasibility of time-invariant focusing, the interpretation of range--angle coupling as either redundancy or exploitable DoF, and the taxonomy of FDA--MIMO with respect to propagation-driven versus processing-driven selectivity.
The remaining debates, especially those related to manifold-based ranging and practical realizability, are also discussed throughout the paper, but are treated primarily as forward-looking perspective questions and implementation-level challenges rather than as fully resolved theoretical conclusions.

Among these open perspective questions, one particularly intriguing issue arises in the context of FDA.
If range is explicitly embedded into the transmit--receive manifold, then distance information may no longer need to be interpreted exclusively through the conventional delay-domain framework.
This raises the possibility that range could be estimated, at least in principle, as a structural parameter of the array manifold, in a manner conceptually analogous to angle estimation in phased arrays, rather than solely through pulse compression and matched filtering.
Under such a viewpoint, the classical definition of range resolution based only on waveform bandwidth may no longer be sufficient.
Instead, range resolution may need to be reconsidered from a manifold-discrimination perspective, for example through distinguishability, identifiability, or Fisher-information-based metrics of range-dependent steering responses.

Taken together, these debates suggest that FDA should be interpreted not simply as a special waveform configuration or a modified array structure, but as a structural extension of the propagation manifold.
For this reason, FDA is more appropriately understood through a framework that links design mechanism, manifold transformation, and exploitable capability, rather than through application categories alone.
This perspective also motivates the structural taxonomy adopted in the remainder of this paper.

\subsection{Contributions and Organization}

To address the above challenges and establish a unified understanding of FDA, this paper develops a structure-oriented framework that connects array design variables, propagation-induced manifold structure, physical degrees of freedom, and ultimately system capabilities.
Rather than treating FDA as a collection of application-specific techniques, we reinterpret it as a propagation-structure-driven array paradigm and revisit its related architectures from a unified taxonomic perspective.

The main contributions of this paper are summarized as follows:

\begin{itemize}

\item \textbf{Unified structural taxonomy of array paradigms:}
We establish a unified signal and array-level framework covering PA, FDA, MIMO, and FDA--MIMO, and show that these paradigms can be distinguished according to the physical source of their degrees of freedom.
In particular, PA is interpreted as a {spatial-phase-driven} paradigm, MIMO as a {waveform-orthogonality-driven} paradigm, FDA as a {frequency-gradient-driven} paradigm, and FDA--MIMO as a {frequency--waveform jointly driven} paradigm.
This taxonomy provides a clearer physical interpretation of how different array architectures generate their effective capabilities.

\item \textbf{Clarification of the role of frequency across different systems:}
We systematically analyze the fundamentally different roles of frequency in FDMA/OFDM-type systems and FDA-type systems.
In contrast to FDMA and OFDM, where frequency mainly serves as a multiplexing, modulation, or indexing resource, FDA employs inter-element frequency offsets to reshape the propagation kernel itself.
This distinction clarifies why FDA should not be reduced to a conventional frequency-domain communication or signal-processing mechanism.

\item \textbf{Irreducibility criterion for genuine range-domain DoFs:}
We introduce an irreducibility criterion to distinguish genuinely new physical range-domain DoFs from equivalent processing-level reformulations.
Under this viewpoint, frequency offsets alone do not automatically imply a new physical DoF; a genuine range DoF arises only when the resulting range-phase-gradient structure is irreducible and cannot be removed without collapsing the system into an angle-only or waveform-synthesized form.
This provides a physics-grounded answer to the long-standing question of whether FDA truly introduces a new distance-related design dimension.

\item \textbf{The answers to three central structural debates:}
Based on the proposed framework, we provide unified answers to three long-standing structural debates in FDA research, namely the physical feasibility of time-invariant focusing, the interpretation of range--angle coupling as redundancy or exploitable DoF, and the taxonomy of FDA--MIMO in terms of propagation-driven versus processing-driven selectivity.
We further discuss manifold-based ranging and practical realizability as two forward-looking debates that remain open, thereby distinguishing between questions that can be structurally clarified within the present framework and those that still require future theoretical and experimental investigation.

\item \textbf{Systematic classification of FDA--MIMO realizations:}
We classify FDA--MIMO architectures according to their orthogonality mechanisms, namely waveform-orthogonal and frequency-orthogonal realizations, and clarify how these mechanisms affect the preservation, suppression, or reconstruction of propagation-induced range dependence.
This classification explains why FDA--MIMO may exhibit different range--angle behaviors at the propagation level and the equivalent processing level, and provides a more precise answer to the question of what should and should not be regarded as FDA-originated.

\item \textbf{Manifold-based explanation of core FDA phenomena:}
From the perspective of manifold expansion, we systematically analyze the intrinsic time variability, frequency-gradient effect, beam scanning behavior, range--angle coupling, integrated transmit beampattern, ambiguity-function structure, and scattering-response characteristics of FDA.
This provides a coherent physical interpretation of how element-dependent carrier frequencies reshape the propagation manifold from $\mathbf{a}(\theta)$ to $\mathbf{a}(t,r,\theta)$.

\item \textbf{Unified comparison with time-coding array paradigms:}
We establish a structural comparison between frequency-gradient-driven FDA/FDA--MIMO and time-coding-driven architectures such as EPC arrays and STCA.
Although these systems may produce similar range--angle dependent responses at the output level, we show that their underlying mechanisms are fundamentally different:
FDA creates physical range dependence through propagation-phase gradients, whereas EPC/STCA mainly generate processing-induced range selectivity through temporal coding and receiver-side synthesis.

\item \textbf{Capability mapping from mechanism to function:}
We build a capability-oriented mapping framework that connects FDA-induced structural properties to system-level functionalities.
In particular, we identify three representative FDA capabilities, including range--angle selectivity, time-evolving beam control, and frequency-domain diversity, and show how they arise from the underlying propagation structure.
This framework bridges the gap between abstract DoF analysis and practical sensing/communication functions.

\item \textbf{Task-oriented review of radar and communication applications:}
Using the above framework, we systematically review how FDA capabilities translate into practical advantages across radar and communication applications.
On the radar side, we analyze parameter estimation and target localization, target detection under range-ambiguous clutter, deceptive jamming and high-speed-target scenarios, anti-jamming and anti-deception processing, LPI radar, high-resolution wide-swath imaging, and target tracking.
On the communication side, we further review FDA-enabled physical-layer security, index modulation, and FDA-based ISAC, thereby showing how the same propagation-structure mechanism manifests across sensing and communication tasks.

\item \textbf{Critical discussion of limitations and future directions:}
Finally, we discuss the current limitations of FDA research and identify future directions, including near-field modeling, wideband applicability, hardware coherence constraints, experimental validation, the feasibility of manifold-based ranging, and the integration of FDA with emerging near-field/holographic MIMO and next-generation ISAC systems.
\end{itemize}

\begin{figure}[htp]
\centering
{\includegraphics[width=0.48\textwidth]{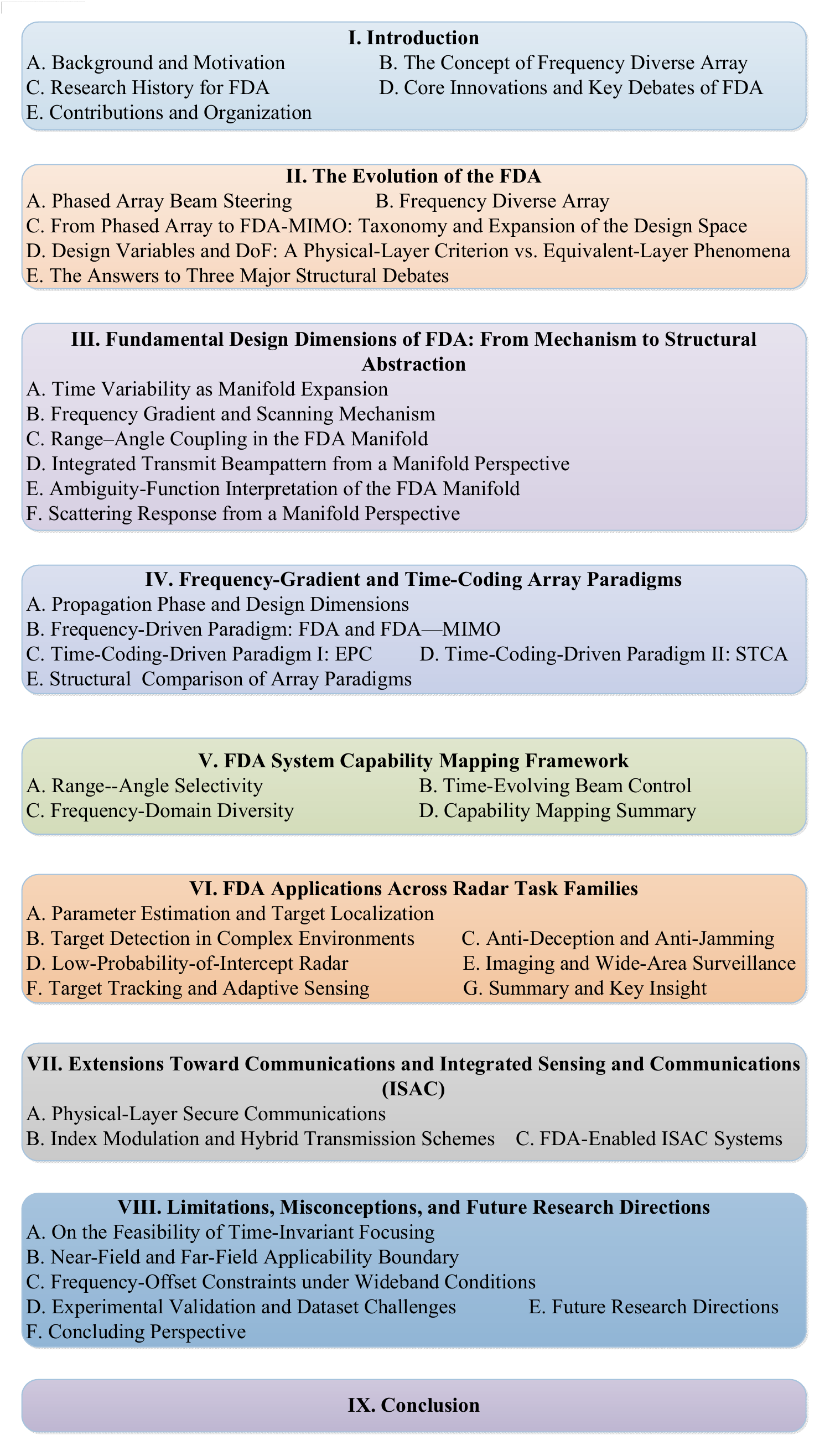}}
\caption{\justifying Organization of this paper.}\label{Fig:org}	
\end{figure}

\noindent

The organization of this paper is illustrated in Fig.~\ref{Fig:org}.
Specifically, Section~\ref{sec_introduction} introduces the background and motivation of FDA, together with its basic concept, research history, core innovations, and key debates.
Section~\ref{sec:unified-model-taxonomy} presents the evolution from phased arrays to FDA and FDA--MIMO, and develops a unified array taxonomy, a physical-layer DoF criterion, and unified answers to three major structural debates.
Section~\ref{sec:fundamental-dimensions} analyzes the fundamental design dimensions of FDA from a manifold perspective, including time variability, frequency-gradient-induced beam scanning, range--angle coupling, integrated transmit beampatterns, ambiguity-function characteristics, and scattering responses.
Section~\ref{sec:paradigm-dof-comparison} compares frequency-gradient-driven FDA paradigms with time-coding-based array architectures, including EPC and STCA, from a unified structural viewpoint.
Section~\ref{sec:capability-mapping} establishes an FDA capability mapping framework that links structural mechanisms to representative system capabilities.
Section~\ref{sec:task-family} reviews FDA applications across major radar task families, including parameter estimation, target detection, anti-jamming, LPI radar, imaging, and target tracking.
Section~\ref{sec_communication} extends the discussion to communication and integrated sensing and communication (ISAC) systems, with particular emphasis on physical-layer security, index modulation, and FDA-enabled ISAC architectures.
Section~\ref{sec:limitations_future} discusses current limitations, common misconceptions, and future research directions.
Finally, Section~\ref{sec:conclusion} concludes the paper.

\begin{figure}[t]
\centering
{\includegraphics[width=0.5\textwidth]{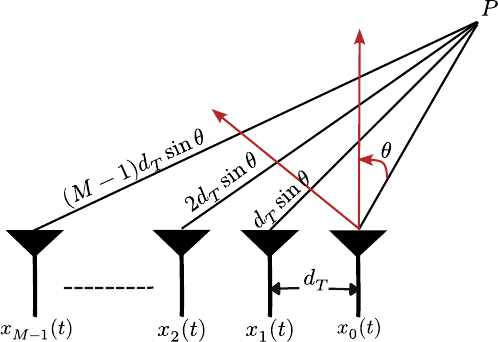}}
\caption{\justifying Geometry of a uniform linear transmit array with $M$ antennae and inter-element-spacing of $d_T$ showing the angle-dependent path differences toward a far-field point $P$.}\label{Fig:ULA}	
\end{figure}
%
\section{The Evolution of the FDA}\label{sec:unified-model-taxonomy}

This section aims to clarify how the FDA evolves from conventional array architectures and why it should be regarded as an independent array paradigm rather than a minor variation of phased arrays or MIMO radar. 
To this end, we first establish a unified modeling starting point based on a uniform linear array (ULA), and then compare the array-factor structures and effective degrees of freedom of PA, MIMO, FDA, and FDA--MIMO. 
Building on this framework, we further clarify the distinct role of frequency in FDA as compared with conventional frequency-domain mechanisms such as FDMA and OFDM, and discuss two representative implementation routes for FDA--MIMO. 
Finally, this unified perspective naturally leads to an open structural question: once range has entered the array manifold through frequency-induced phase gradients, is the conventional pulse-compression-based treatment of the range dimension still indispensable, or should FDA motivate a different view of range processing?

We begin with the standard transmit model of a ULA, which serves as the common reference for all subsequent developments. 
Consider a ULA with $M$ transmit antenna elements and $d_T=\lambda/2$ inter-element spacing, as shown in Fig.~\ref{Fig:ULA}. 
The signal transmitted by the $m$-th antenna is written as
\begin{align}
    x_m(t) = A_m e^{j(2\pi f_ct + \phi_o)},
\end{align}
where $A_m$ and $\phi_m$ are the amplitude and phase applied to the signal transmitted by the $m$-th element, respectively, while $f_c$ is the carrier frequency. The position of the $m$-th antenna relative to the reference antenna located at the origin can be found as 
\begin{equation}
d_m = m d_T,\quad m=0,1,\ldots,M-1. 
\end{equation}
Under the far-field approximation, the wavefront observed at a distant point can be approximated as planar. Accordingly, for a target located at range $R_o$ and direction $\theta$ with respect to the reference antenna (the rightmost element), the propagation distance associated with the $m$-th element is given by
\begin{equation}
R_m = R_o + m d_T\sin\theta.
\end{equation}  
The propagation delay associated with the $m$-th element is given by
\begin{align}
    \tau_m(\theta) = \frac{R_o + m d_T\sin\theta}{c},
\end{align}
where $c$ denotes the speed of light. The contribution of the $m$-th antenna element at the target is given by 
\begin{align}
r_m(t,\theta) &= A_m e^{j\left(2\pi f_c \left(t - \tau_m(\theta)\right) + \phi_o\right)}, \notag 
\end{align}
Aggregating the contributions from all transmit antenna elements, the received signal can be expressed as
\begin{align}
    r(t,\theta) &= \sum_{m=0}^{M-1} A_m e^{j\left(2\pi f_c \left(t - \tau_m(\theta)\right) + \phi_o\right)},\notag\\
                &= \sum_{m=0}^{M-1} A_me^{j\left(2\pi f_c t - 2\pi\frac{R_0}{\lambda} + \phi_o\right)} e^{-j2\pi \frac{md_T\sin\theta}{c}}.
\end{align}
For the given value of $R_o$, the amplitude of the overall received signal depends on the angular location $\theta$. This angular dependence forms the basis for directional transmission using controlled phase excitation across the array elements.
\begin{figure}[t]
\centering
{\includegraphics[width=0.5\textwidth]{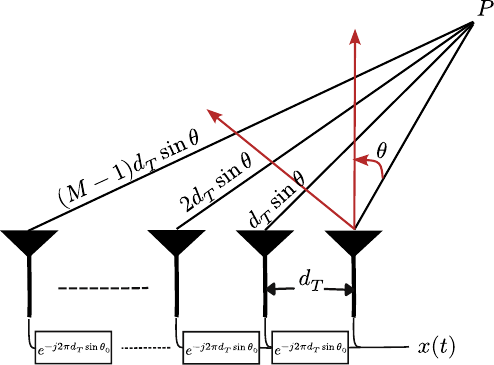}}
\caption{\justifying In phased-array, the progressive phase shifts across antenna elements steer the beam in the direction $\theta_0$.} \label{Fig:PARadar}
\end{figure}.
\subsection{Phased Array Beam Steering}
A uniform linear array can operate as a phased array when a controlled phase shift is applied across its antenna elements as shown in Fig. \ref{Fig:PARadar}. By properly selecting the excitation phases, the radiated signals from different elements combine constructively in a desired direction and destructively in other directions. This enables electronic steering of the main beam without physically rotating the array. With these pre-compensated phases, the transmitted signal from the $m$-th antenna element can be expressed as
\begin{align}
x_m(t) = A_me^{j\left(2\pi f_c t + 2\pi f_c\frac{d_m\sin\theta_o}{c} + \phi_o\right)}. \notag
\end{align}
The relative phase of the transmit signal from the $m$-th antenna element is then
\begin{equation}
\phi_m(t,\theta) = 2\pi f_c\frac{d_m\sin\theta_o}{c}. \notag
\end{equation}
Accordingly, the combined received signal from all transmit antenna elements at point $P$ is given by
\begin{align}
r(t,\theta) = e^{j2\pi\left(f_c t - \frac{R_0}{\lambda}\right)}
    \sum_{m=0}^{M-1} \tilde{A}_m e^{j2\pi f_c\left(\frac{d_m\sin\theta_o}{c} - \frac{d_m\sin\theta}{c}\right)},
\end{align}
where $\tilde{A}_m=A_me^{j\phi_o}$. The term inside the summation defines the array factor, which governs the directional radiation pattern of the phased array:
\begin{align}
    \AF(\theta) = \sum_{m=0}^{M-1} \tilde{A}_m e^{j 2\pi f_c \left(\frac{d_m\sin\theta_o}{c} - \frac{d_m\sin\theta}{c}\right)}. \label{eq:AF_PA}
\end{align}
Therefore, by controlling the excitation phases $\theta_o$, the array radiation can be focused toward the intended direction $\theta_0$. 
However, despite this beam-steering capability, a conventional phased array exhibits a radiation pattern that is solely angle-dependent in the far field and does not provide range-dependent focusing. This angle-only dependence limits the capability of phased arrays in applications requiring joint range and angle localization.
\subsection{Frequency Diverse Array}
To overcome the angle-only limitation of the phased array, the FDA extends the phased-array concept by introducing small frequency increments across the transmitting elements, resulting in a beampattern that depends on both range and angle. In an FDA system, the carrier frequency of the $m$-th antenna element is defined as \cite{WangSo2014TransmitSubaperturing}
\begin{equation}
f_m=f_c+\Delta f_m,\quad m=0,1,\ldots,M-1,
\end{equation}
where $\Delta f_m$ represent the frequency offset of the $m$-th antenna. The offset frequency can vary linearly, for example $\Delta f_m=mf_a$, or it can be assigned independently for each antenna element. Assuming $\Delta f_m \ll f_c$, the phase term of the received signal associated with the signal transmitted from the $m$th antenna can be written as \cite{LiaoWang2019FrequencyDiverseArray,Liao2020FrequencyDiverseArray,GuiWang2018CoherentPulsedFDA}
\begin{equation}
\phi_m^{\FDA}(t,R_o,\theta) = 2\pi\Big(\Delta f_m t-\Delta f_m\frac{R_o}{c} - f_c\frac{d_m\sin\theta}{c}\Big). \notag
\end{equation}
This leads to the array factor (AF)
\begin{equation}\label{eq:AF_unified}
\AF(t,R_o,\theta)= \sum_{m=0}^{M-1}e^{j\phi_m(t,R_o,\theta)}. 
\end{equation}
In the PA case, we only have the spatial phase term, whereas FDA additionally introduces two phase components that are linear in both time and range. In particular, the terms $\Delta f_m t$ and $-\Delta f_m\frac{R_0}{c}$ introduce phase variations with respect to $t$ and $R_o$, respectively. These terms induce controllable phase gradients whose slopes are directly governed by the frequency offset $\Delta f_m$. Consequently, unlike phased arrays, FDA systems inherently couple time and range through a common parameter, enabling joint control of the spatio--temporal--range response within the array manifold.  
%
\begin{figure}
    \centering
    \includegraphics[width=1\linewidth]{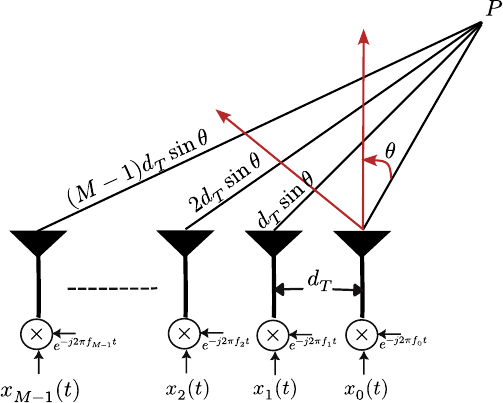}
    \caption{\justifying In an FDA, each antenna element slightly offset the carrier frequency.}
    \label{fig:placeholder}
\end{figure}
%


Under the unified model given in \eqref{eq:AF_unified}, different array paradigms can be distinguished by how the design-variable space is expanded or constrained. In simple terms, the more adjustable parameters, or degrees of freedom (DoFs), an array has, the greater its flexibility in controlling the beampattern across different domains. In the following, we discuss the beampatterns of several representative antenna array configurations:  
\\
{\textit{a}) \bf Phased-Array: Spatial-Phase Driven.} 
When $\Delta f_m = 0$,  the propagation-induced phase across the transmit aperture reduces to a purely spatial term, namely
\begin{equation}
\phi_m^{\mathrm{PA}}(\theta) = -2\pi f_c\frac{d_m\sin\theta}{c},
\end{equation}
which depends only on the angular variable \cite{Richards2005Fundamentals,johnson1992array}. Therefore, the effective DoFs arise exclusively from spatial phase steering,
%
\\\\
{\textit{b}) \bf FDA: Frequency-gradient driven.} 
When $\Delta f_m\neq 0$, the element-dependent carrier $f_m$ yields an additional time--range phase gradient \cite{secmen2007frequency,Wang2015FrequencyDiverseArray,antonik2009investigation,basit2018development}, i.e.,
\begin{equation}
\label{eq8}
\phi_m^{\mathrm{FDA}}(t,R,\theta)=\phi_m^{\mathrm{PA}}(\theta)
+2\pi\Delta f_m t-2\pi \Delta f_m\frac{R}{c}.
\end{equation}

The second and third terms constitute explicit time and range phase gradients across the aperture, fundamentally altering the interference structure formed during propagation.

As a result, the system DoFs expand to 

\begin{equation}
\mathrm{DoF}_{\mathrm{FDA}} = \mathrm{Space} \oplus \mathrm{Frequency}. \label{eq9}
\end{equation}
\\\\
{\textit{c}) \bf MIMO: Waveform-Orthogonality Driven.}
Consider a MIMO system with $M$ transmit antennas and $N$ receive antennas. The $m$th transmit element radiates an orthogonal baseband waveform $x_m(t)$ at a common carrier frequency $f_c$ (i.e., no frequency gradient is applied). Let $d_n^{\mathrm{R}}$ and $d_m^{\mathrm{T}}$ denote the locations of the $n$th receive and $m$th transmit antenna elements from the reference transmit and receive antennas, respectively. For a colocated Tx/Rx array with element locations $\{d_n^{\mathrm{R}}\}$ and $\{d_m^{\mathrm{T}}\}$, define the {virtual position} $d_{n,m}^{\mathrm{V}} \triangleq d_m^{\mathrm{T}}+d_n^{\mathrm{R}},$
and the corresponding {virtual phase} as the propagation-kernel phase associated with $(m,n)$.

Under the narrow-band far-field approximation, the received signal at the $n$-th receive channel can be written as \cite{liu2020joint}
\begin{equation}
y_n(t)=\sum_{m=0}^{M-1}\alpha\, x_m(t-\tau)\,
e^{-j2\pi f_c\frac{d_{n,m}^{\mathrm{V}}\sin\theta}{c}}+w_n(t),
\label{eq:mimo_rx}
\end{equation}
where $\tau=2R/c$ and $\alpha$ collects the scattering and propagation coefficients and $w_n(t)$ represents noise term.
By matched filtering with $x_m(t)$ (waveform separability), one obtains the $m$-th {virtual channel} \cite{san2007mimo,chen2008mimo}
\begin{equation}
\begin{split}
z_{n,m} &\triangleq \int y_n(t)\,x_m^*(t-\tau)\,dt\\
        &\approx \alpha e^{-j2\pi f_c\frac{d_{n,m}^{\mathrm{V}}\sin\theta}{c}},
\end{split}
\label{eq:mimo_virtual}
\end{equation}
which exhibits the standard {Tx--Rx separable spatial phase} (virtual array) at a single carrier. Hence, the phase term of MIMO can be expressed as
\begin{equation}
\phi_{n,m}^{\mathrm{MIMO}}(\theta) = -2\pi f_c\frac{d_{n,m}^{\mathrm{V}}\sin\theta}{c},
\end{equation}
with $
\mathrm{DoF}_{\mathrm{MIMO}}=\mathrm{Space}\oplus \mathrm{Waveform}.$ Further, we have 
the compact {virtual array factor} \cite{Li2009MIMORadarSignalbook,AkcakayaNehorai2011MIMO}
\begin{equation}
\AF_{\mathrm{MIMO}}(\theta) \triangleq \sum_{n=0}^{N-1}\sum_{m=0}^{M-1}
e^{-j2\pi f_c\frac{d_{n,m}^{\mathrm{V}}\sin\theta}{c}}. \label{eq:AF_mimo_virtual}
\end{equation}
This explicitly shows that MIMO expands DoF via the virtual aperture $\{d_{n,m}^{\mathrm{V}}\}$,
i.e., $\mathrm{Space}_{\mathrm{Tx}}\oplus \mathrm{Space}_{\mathrm{Rx}}$.
\\\\
{\textit{d}) \bf FDA--MIMO: Composite Expansion.} 
\label{sec2d}
If an frequency coding is further imposed across transmit elements, while maintaining waveform orthogonality $\{x_m(t)\}$, then the received signal becomes \cite{Xu2015Deceptivejammingsuppression}
\begin{align}
y_n(t) &=\sum_{m=0}^{M-1} \alpha x_m(t-\tau ) e^{-j2\pi\frac{\Delta f_m}{c}R} \notag \\
       &\times e^{j2\pi \Delta f_mt}e^{-j2\pi f_c \frac{d_{n,m}^{\mathrm{V}}\sin\theta}{c}}  + w_n(t).
\label{eq:fdamimo_rx}
\end{align}
After matched filtering with $x_m(t)$, the $m$-th virtual channel is (up to a constant scaling) \cite{gui2020generalized}
\begin{align}
z_{n,m}  & \approx \alpha e^{j2\pi \Delta f_m t} e^{-j2\pi\frac{\Delta f_m}{c}R} e^{-j2\pi f_c\frac{d_{n,m}^{\mathrm{V}}\sin\theta}{c}}.
\label{eq:fdamimo_virtual}
\end{align}
Therefore, the\! composite\! phase\! structure\! can thus be written as
\begin{equation}
\phi_{n,m}^{\mathrm{FDA\text{-}MIMO}}(t,R,\theta) = 2\pi f_c\frac{d_{n,m}^{\mathrm{V}}\sin\theta}{c}
+2\pi\Delta f_m t -2\pi\frac{\Delta f_m}{c}R, \notag
\end{equation}
where the DoF are determined by the combined contributions of spatial, frequency, and waveform domains:
\begin{equation}
\label{eq17}
\mathrm{DoF}_{\mathrm{FDA\text{-}MIMO}} = \mathrm{Space}\oplus \mathrm{Frequency}\oplus \mathrm{Waveform}.    
\end{equation}
Accordingly, the compact {virtual FDA array factor} is
\begin{align}
\AF_{\mathrm{FDA-}\-\mathrm{MIMO}}(t,R,\theta ) \triangleq \sum_{n=0}^{N-1}\sum_{m=0}^{M-1} e^{j\phi _{n,m}^{\mathrm{FDA-}\-\mathrm{MIMO}}(t,R,\theta)}. \label{eq:AF_fdamimo_virtual}
\end{align}

\begin{tcolorbox}[colback=blue!5!white, colframe=blue!60!black,title= {\bf Remark 1: Alignment with the Unified AF in~\eqref{eq:AF_unified}}]
\begin{itemize}
    \item The AF expressions of MIMO and FDA–MIMO follow a mathematical form similar to that of FDA.
    \item In contrast to PA and FDA, MIMO introduces a \emph{virtual array} that effectively enlarges the aperture, thereby improving spatial (angular) resolution.
    \item FDA--MIMO further exploit $\Delta f_m$ to induce time- and range-dependent phase gradients, in addition to the virtual aperture.
\end{itemize} 
\end{tcolorbox}


This taxonomy underscores that the FDA is neither a special case of MIMO radar nor a simple variant of PA radar. Instead, it represents an independent paradigm that extends the array manifold through frequency gradients.
\begin{figure*}[htp]
	\centering
	    \subfigure[PA]
	{		{\includegraphics[width=0.43\textwidth]{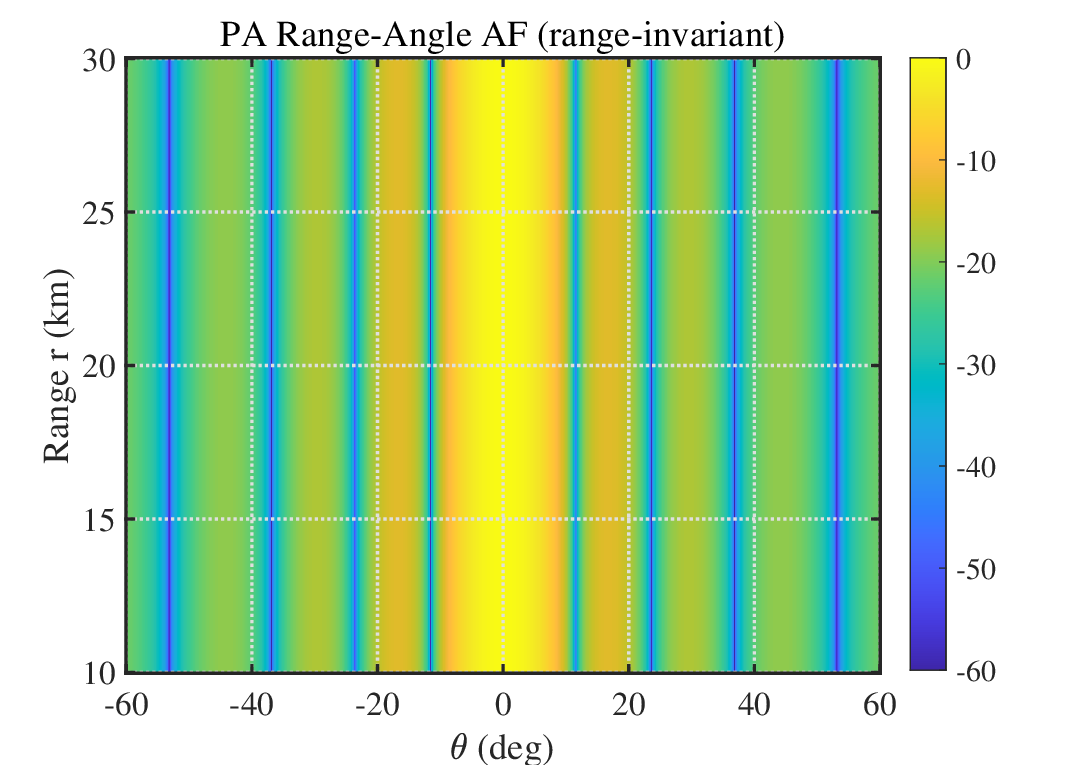}}}   
    	    \subfigure[FDA]
	{		{\includegraphics[width=0.43\textwidth]{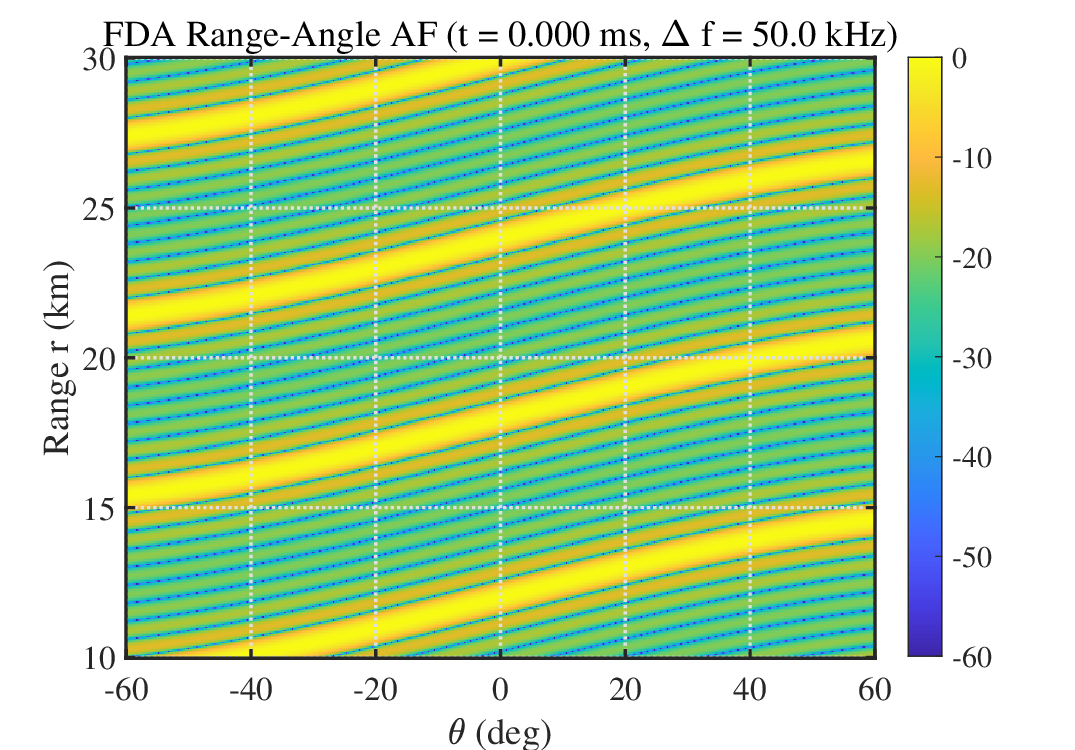}}}   
    	    \subfigure[MIMO]
	{		{\includegraphics[width=0.43\textwidth]{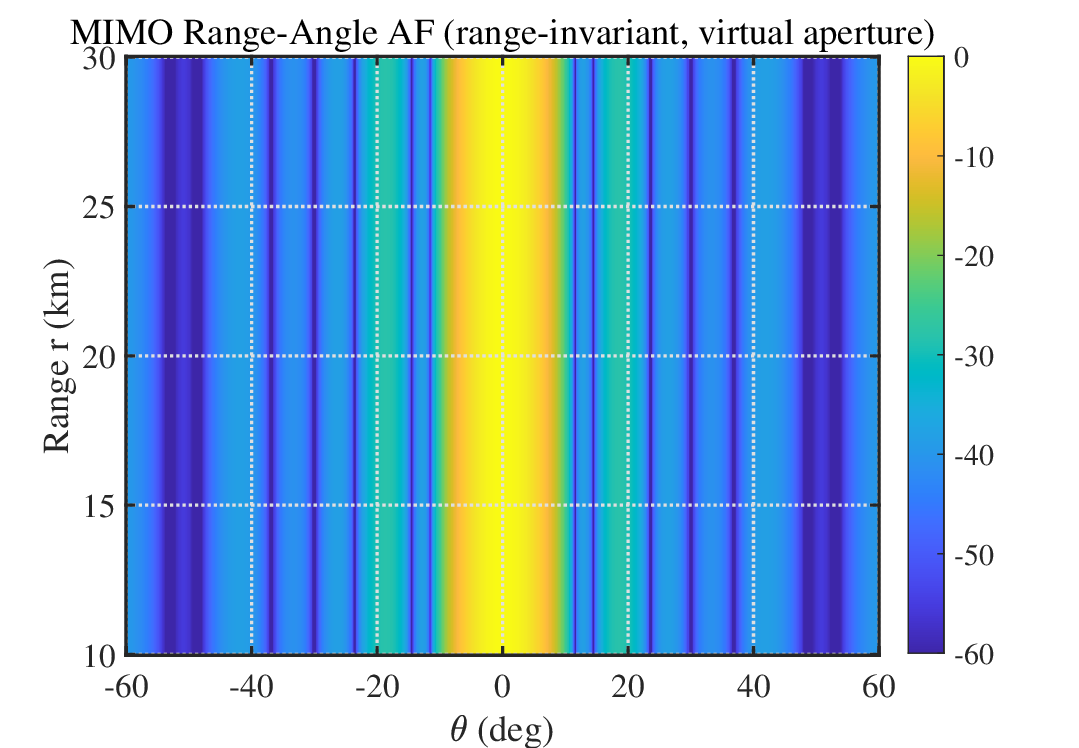}}}   
    	    \subfigure[FDA-MIMO]
	{		{\includegraphics[width=0.43\textwidth]{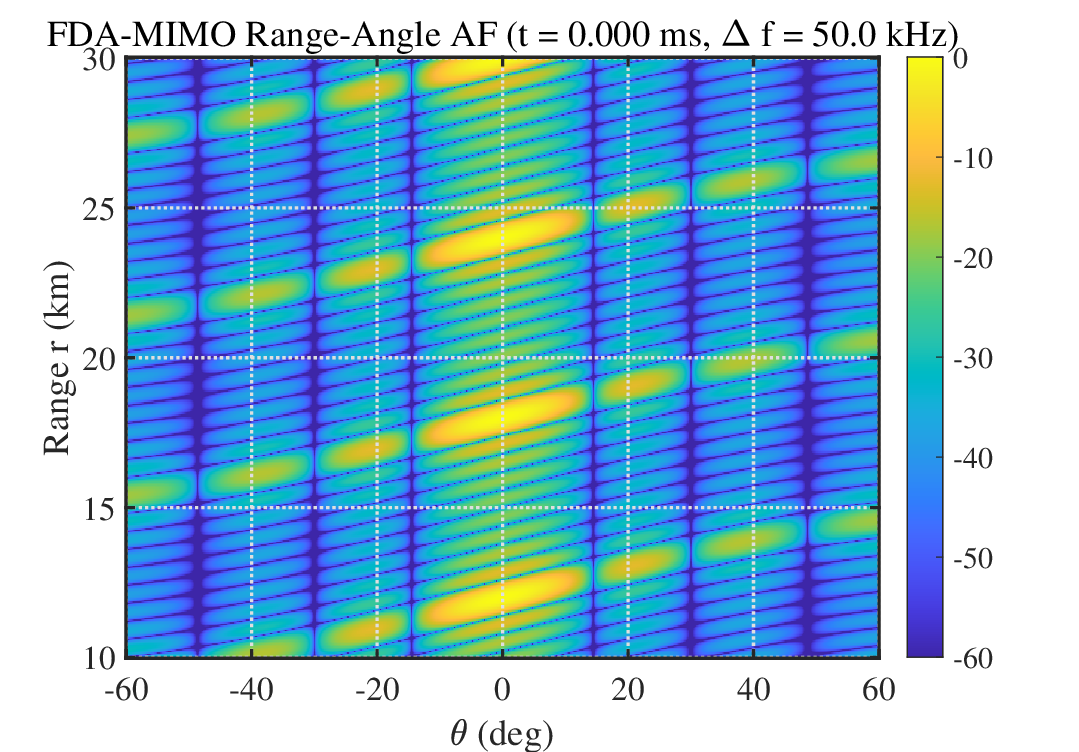}}}   
	\caption{\justifying Range–angle response comparison based on the corresponding array-factor formulations of four array paradigms. PA and MIMO remain range-invariant, whereas FDA introduces range–angle coupling, and FDA–MIMO combines coupling with virtual aperture expansion.}
	\label{fig2}	
\end{figure*}

\begin{figure}[htp]
	\centering
	{\includegraphics[width=0.45\textwidth]{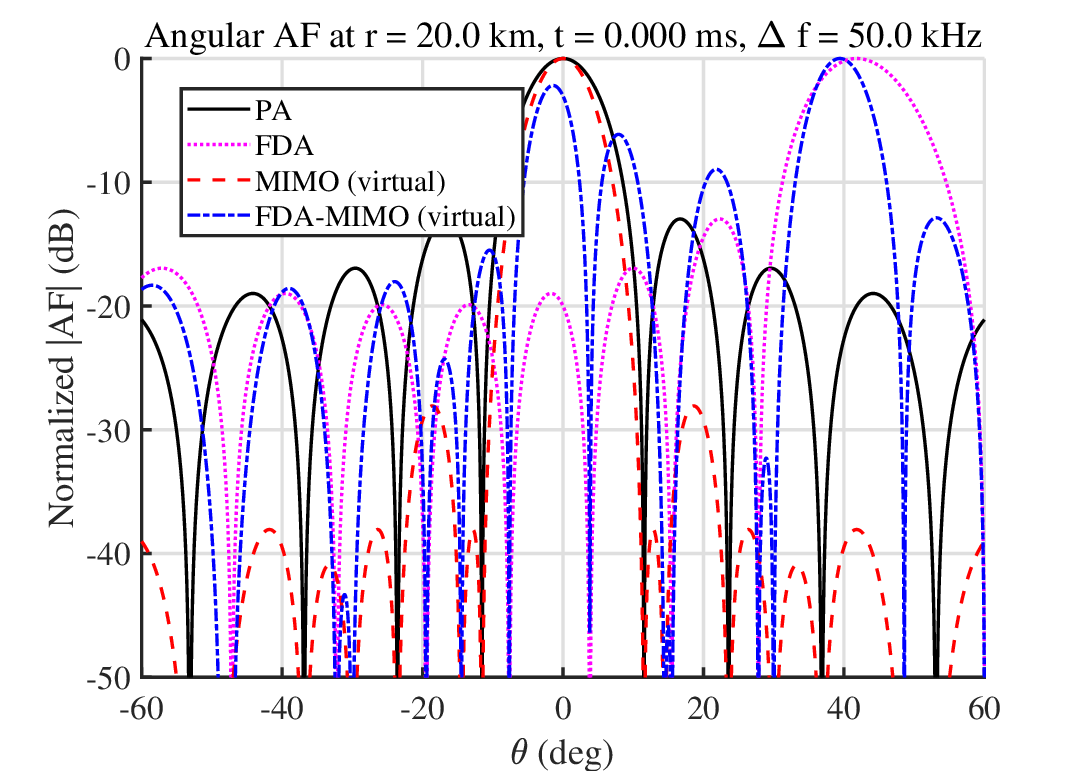}}
	\caption{\justifying Angular AF comparison at a fixed range–time slice.}
	\label{fig3}	
\end{figure}

To illustrate the distinctions among these paradigms, we present a comparative simulation. Fig.~\ref{fig2} depicts the range–angle array factors of PA, FDA, MIMO, and FDA–MIMO systems. In all cases, the spacing between adjacent antenna elements at both the transmitter and receiver is set to $d_T = d_R = \frac{\lambda}{2}$. The results reveal that PA and MIMO exhibit range-invariant behavior, as their phase structures depend exclusively on spatial variables. While MIMO enhances the effective aperture through waveform separation and virtual-array synthesis, it does not introduce transmit-side range dependence, since its phase structure remains purely spatial after matched filtering. In contrast, the FDA and FDA–MIMO configurations inherently incorporate range-dependent phase terms due to their frequency-diverse transmissions, leading to tilted interference contours. 

Fig.~\ref{fig3} further examines the angular array factor at a fixed range–time slice, offering deeper insight into the DoF governing each paradigm. The results demonstrate that waveform-driven DoF predominantly enhance angular resolution, as they exploit spatial diversity to sharpen the beam pattern. In contrast, frequency-driven DoF fundamentally reshape the propagation-induced interference structure, introducing range-dependent phase variations that enable joint range–angle processing. This distinction underscores the complementary roles of waveform and frequency diversity in FDA-based systems, where the interplay between spatial, frequency, and waveform domains unlocks unique operational capabilities. These observations confirm that the fundamental distinction among array paradigms lies in their propagation-kernel structure.
%
%
\subsection{Design Variables and DoF: A Physical-Layer Criterion vs. Equivalent-Layer Phenomena}\label{subsec:dof-criterion}
Before examining whether FDA introduces an additional physical DoFs, 
it is necessary to clarify that the notion of “frequency diversity” 
has been used in conceptually different ways in the literature. 

In some architectures \cite{faruque2018frequency,saadia2020single,li2006orthogonal,weinstein2009history}, frequency serves merely as a signal-separation 
mechanism or an indexing resource, without altering the propagation structure. 
In contrast, FDA embeds element-dependent frequency offsets directly 
into the propagation kernel, thereby modifying the range-dependent phase progression itself. 
Confusing these roles may lead to interpreting equivalent processing-level 
effects as genuine physical-layer DoF expansion. 
Therefore, from a signal-processing perspective, frequency can play 
fundamentally different roles in system design, i.e.,
\begin{itemize}
\item {Frequency gradient (FDA)}: $\Delta f_m$ remains in the propagation-kernel phase and induces inter-channel {range-phase-gradient differences}. The gradient exists before reception and shapes the interference pattern during propagation.
\item {Frequency-domain orthogonality (FDMA-style separation)}: frequency is used solely for channel separation. 
The propagation kernel is effectively shared at $f_c$, and no inter-channel 
range-gradient difference is introduced.
\item {Subcarrier structure (e.g., OFDM)}: frequency acts as an indexing resource for modulation and multiplexing. 
Although multi-frequency components are present, they do not necessarily induce 
element-dependent range-phase gradients.
\end{itemize}

The above distinction highlights that frequency-driven behavior may arise 
either from propagation physics or from signal-level multiplexing, 
and these two mechanisms are fundamentally different in terms of DoF interpretation.

\subsubsection{Physical Range DoF: A Reducibility/Irreducibility Criterion}
The range-dependent component of the propagation kernel for the $m$-th channel is
\begin{equation}
e^{-j2\pi(f_c+\Delta f_m)\frac{R_m}{c}}.
\end{equation}

Define the range-phase gradient of each channel as
\begin{equation}
\label{eq18}
\nabla_{R_m} \phi_m = -2\pi  (f_c+\Delta f_m)/c.
\end{equation}

All channels share the common baseline gradient $-2\pi f_c/c$, similar to PA. 
However, FDA introduces additional inter-channel gradient differences given by
\begin{equation}
\nabla_{R_m} \phi_m - \nabla_{R_n} \phi_n
= -2\pi\frac{\Delta f_m - \Delta f_n}{c}.
\end{equation}

If $\nabla_{R_m} \phi_m \neq \nabla_{R_n} \phi_n$ for $m\neq n$,
the propagation exhibits a multi-gradient structure. This leads to a fundamental question:

\emph{Do these additional gradients represent a genuine physical DoF created during wave propagation, 
or can they be completely removed through independent per-channel linear compensation at the receiver?}

\begin{definition}[Physical Range DoF -- Reducibility Criterion:]
If there exists a set of operators $\{\mathcal{L}_m\}$, one for each channel $m$, such that for all $R_m$,
\begin{equation}
\mathcal{L}_m
\left\{
e^{-j\frac{2\pi}{c}(f_c+\Delta f_m)R_m}
\right\} = 
e^{-j\frac{2\pi}{c}f_c R_m},
\quad \forall m,
\end{equation}
then the system is reducible (or DoF free). Otherwise, if no such operator exist, the system has irreducible physical DoF.  
\end{definition}

For FDA architectures in which the frequency offsets remain embedded in the
propagation kernel and jointly affect the received superposition across channels,
per-channel compensation operators generally do not exist without altering
the multi-channel interference structure. 
This is because the additional range-phase gradients are formed prior to reception
through wave propagation, thereby shaping the interference geometry itself. 
Consequently, such gradients constitute an irreducible physical range DoF.

FDA becomes reducible only under degenerate conditions, 
such as when all frequency offsets are identical, 
or when compensation is tailored to a single known range point.
In general multi-range scenarios, however, the element-dependent
range-phase gradients introduce propagation-structural differences
that cannot be eliminated by independent per-channel linear operations.
In essence, FDA is reducible only when its frequency offsets
do not induce inter-channel range-phase curvature differences
over the scene of interest.

Hence, the reducibility criterion serves as a physics-grounded boundary 
that separates true propagation-level DoF creation from equivalent 
processing-layer reformulation.

\begin{tcolorbox}[colback=blue!5!white,colframe=blue!60!black,title= {\bf Remark 2: Physical vs. Equivalent Processing Layer}]
\begin{itemize}
    \item The reducibility criterion emphasizes that system taxonomy should be determined 
by the structure of the propagation kernel, rather than by post-reception signal manipulation. 
Even if the receiver implements channel-wise compensation, 
the interference structure has already been formed during propagation. 
\item Frequency-driven behavior is fundamentally a property of wave physics, 
not merely a processing-level artifact. The essence of physical DoF lies in whether a dimension is created during propagation, 
not whether it can be algebraically removed afterward.
\item  The reducibility criterion provides a physics-grounded boundary 
between genuine DoF expansion and equivalent signal reparameterization.
\end{itemize}
\end{tcolorbox}



\subsection{The Answers to Three Major Structural Debates}\label{subsec:three-debates}

\subsubsection{Taxonomy of FDA-MIMO -- Orthogonal-Waveforms vs. Orthogonal-Frequency realizations}

From a system-design perspective, FDA--MIMO architectures can be
more naturally categorized according to how channel orthogonality
is achieved, namely through {waveform orthogonality} or
{frequency orthogonality}. This classification is particularly
relevant for FDA--MIMO, since the orthogonalization strategy not
only determines the receiver processing architecture, but also
affects whether the propagation-induced range-phase gradients are
preserved, suppressed, or re-synthesized in the final range--angle
response.
Eq.\eqref{eq8} shows that the range-dependent component
$-2\pi\Delta f_m \frac{R_0}{c}$
represents the FDA-induced range-phase gradient.
Although this gradient is always embedded in the propagation kernel,
its manifestation in the final system response depends strongly on
the adopted orthogonality mechanism.

\paragraph{Type I - Waveform-Orthogonal FDA--MIMO}

In waveform-orthogonal FDA--MIMO, the transmit channels are separated
by orthogonal waveforms while the element-dependent carrier offsets
remain jointly embedded in the propagation field.
As a result, the FDA-induced range-phase gradients coexist during
propagation and physically contribute to the transmitted radiation
pattern.
Therefore, the range--angle coupling is directly preserved in the
common propagation field, and the resulting FDA behavior is
observed without requiring cross-band reconstruction at the receiver. 

\paragraph{Type II - Frequency-Orthogonal FDA--MIMO (FDMA Realization)}

In frequency-orthogonal FDA--MIMO, the frequency offsets are chosen
such that the transmitted spectra occupy non-overlapping subbands,
for example under FDMA-style multiplexing with $\Delta f = B$, where $B$ is the bandwidth.
The transmitted signals are therefore orthogonal in frequency and
can be separated at the receiver.
Importantly, the element-dependent carrier frequencies
$f_m$ remain embedded in the propagation kernel.
Consequently, the radiated field during propagation still exhibits
range–angle coupling and forms the characteristic FDA slanted
interference structure.
However, unlike waveform-orthogonal FDA--MIMO, the final range--angle
response is no longer determined solely by propagation. Instead, it
depends on how the separated subbands are processed at the receiver.
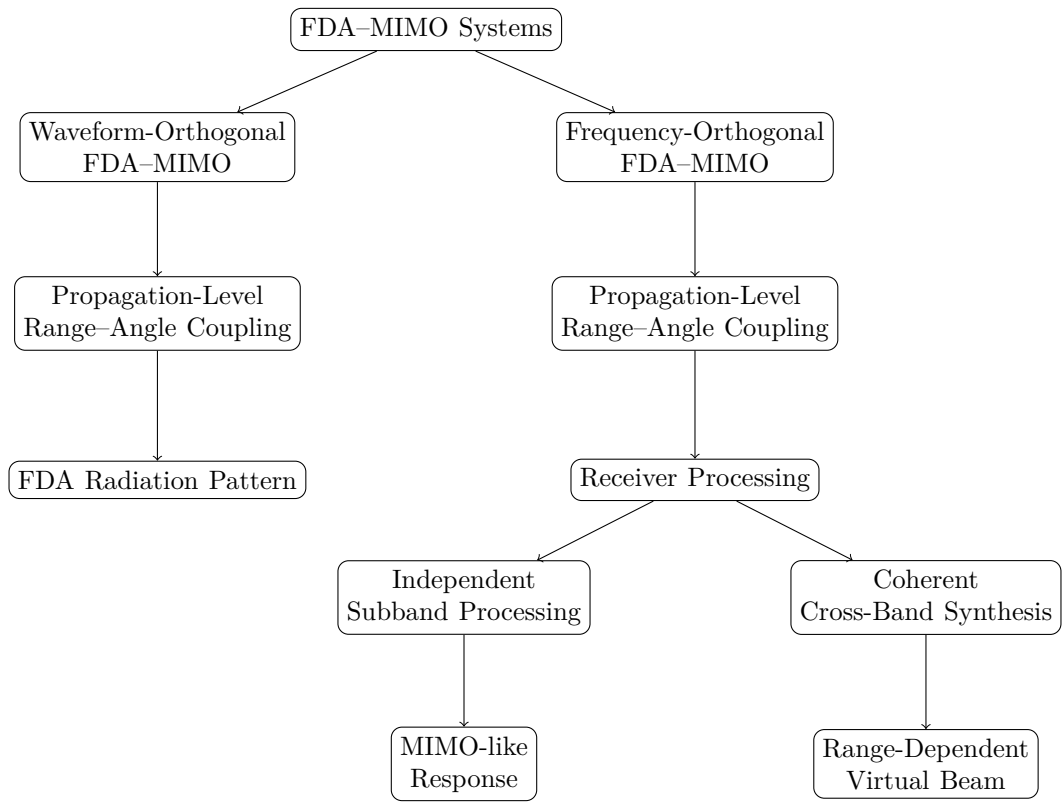
\begin{figure*}[htp]
\centering
\begin{tikzpicture}[
node distance=2.2cm,
every node/.style={draw, rectangle, rounded corners, align=center},
]
\node (root) {FDA--MIMO Systems};

\node (type1) [below left of=root, xshift=-2cm] {Waveform-Orthogonal\\FDA--MIMO};
\node (type2) [below right of=root, xshift=2cm] {Frequency-Orthogonal\\FDA--MIMO};

\node (p1) [below of=type1] {Propagation-Level\\Range--Angle Coupling};
\node (p2) [below of=p1] {FDA Radiation Pattern};

\node (p3) [below of=type2] {Propagation-Level\\Range--Angle Coupling};

\node (f1) [below of=p3] {Receiver Processing};

\node (f2) [below left of=f1, xshift=-1.5cm] {Independent\\Subband Processing};
\node (f3) [below right of=f1, xshift=1.5cm] {Coherent\\Cross-Band Synthesis};

\node (f4) [below of=f2] {MIMO-like\\Response};

\node (f5) [below of=f3] {Range-Dependent\\Virtual Beam};

\draw[->] (root) -- (type1);
\draw[->] (root) -- (type2);

\draw[->] (type1) -- (p1);
\draw[->] (p1) -- (p2);

\draw[->] (type2) -- (p3);
\draw[->] (p3) -- (f1);

\draw[->] (f1) -- (f2);
\draw[->] (f1) -- (f3);

\draw[->] (f2) -- (f4);
\draw[->] (f3) -- (f5);

\end{tikzpicture}

\caption{\justifying Taxonomy of FDA--MIMO architectures based on the orthogonality
mechanism used to separate transmit channels. Both waveform-orthogonal
and frequency-orthogonal FDA--MIMO preserve propagation-level
range--angle coupling due to element-dependent carrier offsets.
However, in frequency-orthogonal FDA--MIMO the final system response
depends on the receiver processing strategy, which may either suppress
or exploit the propagation-induced gradients.}
\label{figu5}
\end{figure*}

\begin{itemize}

\item \textbf{Independent Subband Processing.}
Each frequency channel is matched-filtered (MF) and beamformed
independently. In this case the cross-channel phase gradients
are not coherently combined, and the system behavior reduces
to that of a conventional MIMO radar with virtual spatial
aperture expansion.

\item \textbf{Coherent Multiband Synthesis.}
If the separated subbands are phase-aligned and coherently
combined across frequency, the propagation-induced gradients
can be jointly exploited. The resulting array response forms
a range-dependent virtual beam pattern analogous to that of
propagation-coupled FDA structures.

\end{itemize}

Therefore, frequency-orthogonal FDA--MIMO remains an FDA system at
the propagation level, but its final range--angle behavior depends
on whether the receiver performs independent subband processing or
coherent cross-band synthesis.

The above discussion leads to a conceptual taxonomy of FDA--MIMO
architectures, as illustrated in Fig.~\ref{figu5}. The classification
is based on the orthogonality mechanism used to separate the transmit
channels, namely waveform orthogonality and frequency orthogonality.
In waveform-orthogonal FDA--MIMO, the element-dependent carrier
offsets coexist within a common propagation field, leading to
propagation-coupled range--angle interference patterns.
In contrast, frequency-orthogonal FDA--MIMO separates the channels
spectrally. Although the propagation field still retains the same
range-phase gradients, the final range--angle behavior depends on
the receiver processing strategy.

\begin{figure}[htp]
\centering
\subfigure[]{
\includegraphics[width=0.5\textwidth]{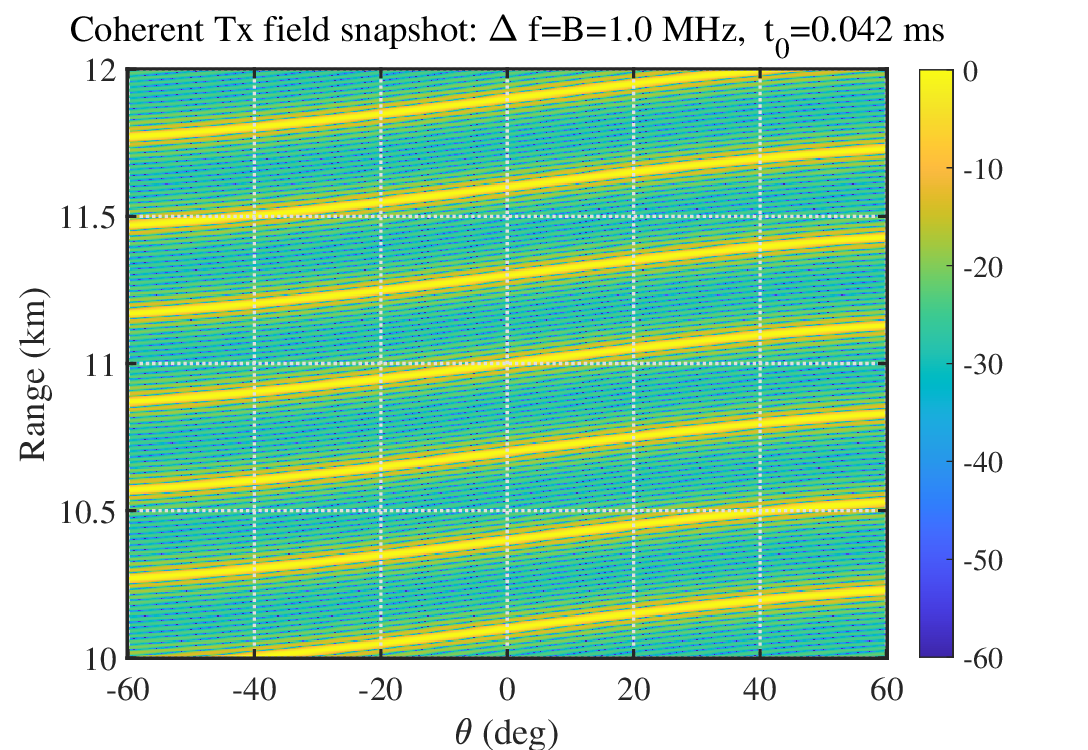}}
\subfigure[]{
\includegraphics[width=0.5\textwidth]{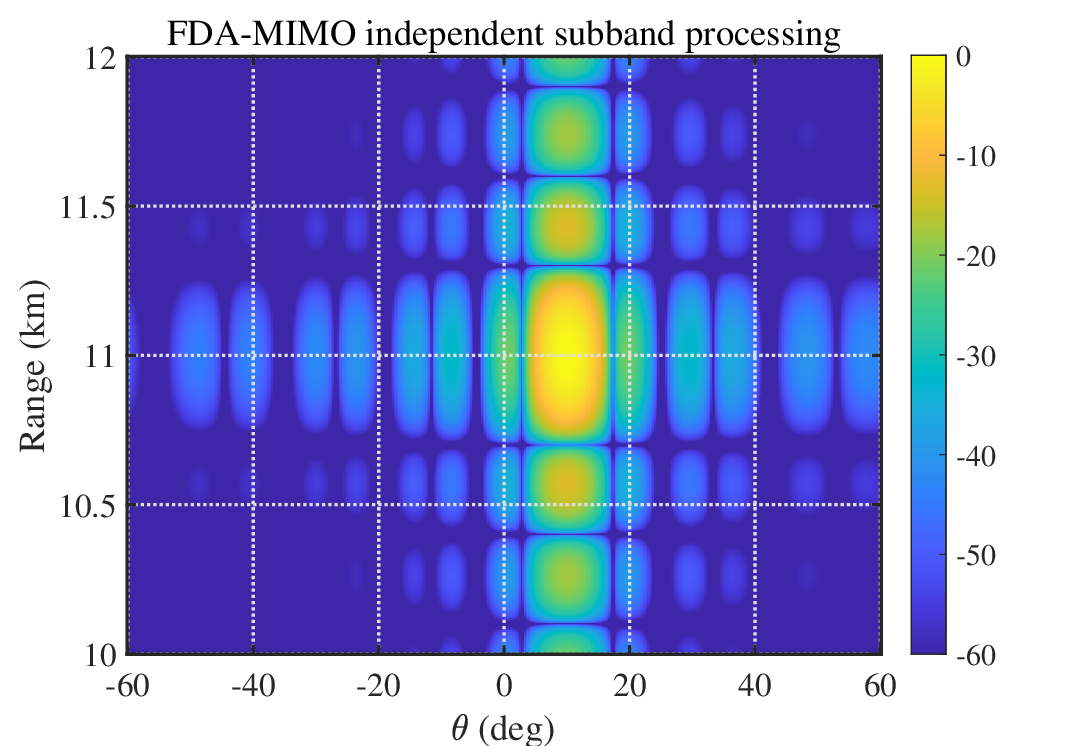}}
\subfigure[]{
\includegraphics[width=0.5\textwidth]{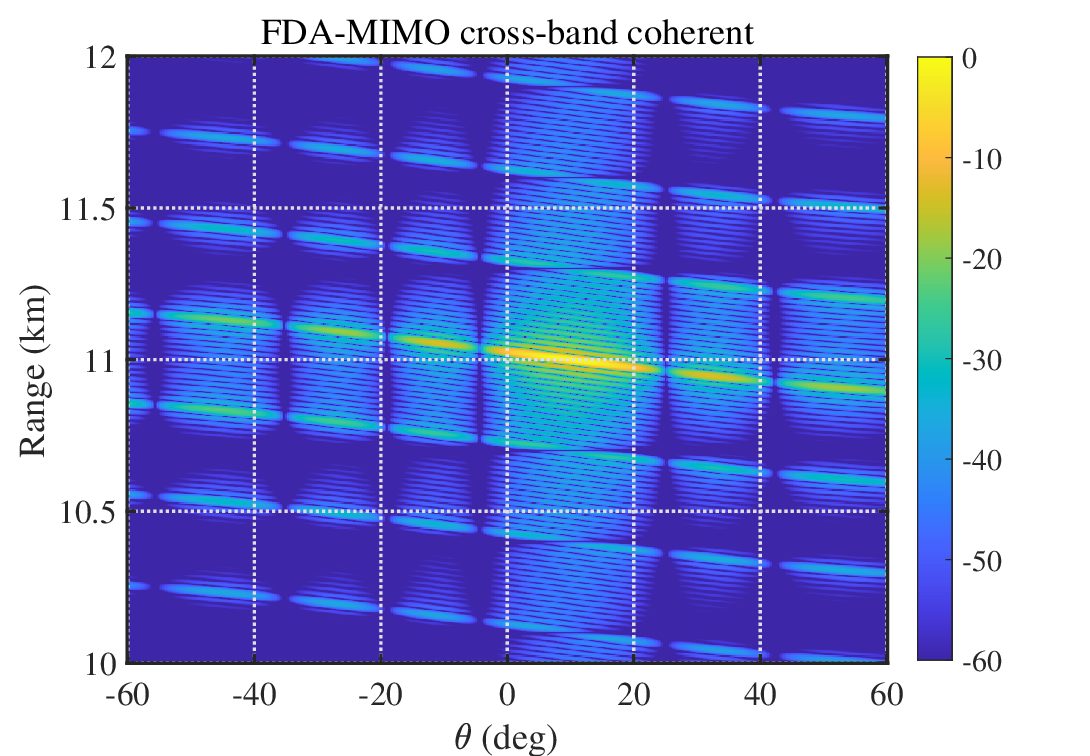}}
\caption{\justifying Mechanism of frequency-orthogonal FDA--MIMO.
(a) During propagation the transmitted field still exhibits
FDA-type range–angle coupling for an LFM waveform.
(b) Independent subband processing leads to a conventional
MIMO-like virtual aperture pattern.
(c) Coherent cross-band synthesis jointly exploits the
propagation-induced gradients and produces a range-dependent
virtual beam.}
\label{fig4}
\end{figure}
To further illustrate the mechanism of frequency-orthogonal
FDA--MIMO, Fig.~\ref{fig4} visualizes the propagation field and
the corresponding receiver responses under different processing
strategies.
During propagation, the transmitted field still exhibits the
characteristic FDA slanted range–angle interference structure,
as shown in Fig.~\ref{fig4}(a).
If each subband is processed independently, the cross-band
gradients are not combined and the resulting response reduces
to a conventional MIMO virtual aperture pattern, as shown in
Fig.~\ref{fig4}(b).
In contrast, when coherent cross-band synthesis is performed,
the propagation-induced gradients are jointly exploited and a
range-dependent virtual beam is formed, as illustrated in
Fig.~\ref{fig4}(c).

\textbf{Key Insight:}
FDA range–angle coupling is fundamentally a propagation phenomenon,
while the orthogonality mechanism determines whether the resulting
gradients are directly observed or reconstructed through receiver
processing.

\subsubsection{Range--Angle coupling: Redundancy or DoF?}

Range--angle coupling \cite{TangJiang2020RangeAngle,Wang2013RangeAngleDependent} arises from the superposition of
frequency-gradient terms across the transmit aperture.
In conventional PA processing frameworks,
where range and angle are typically assumed to be separable
parameters, such coupling is often regarded as undesirable.
Its practical value, however, depends strongly on the
processing framework adopted by the system:

\begin{itemize}

\item \textbf{Redundancy Viewpoint:}
If one insists on PA-style ``separable-parameter'' processing \cite{mailloux2017phased},
range--angle coupling manifests as estimation difficulty,
beam distortion, and defocusing.

\item \textbf{DoF Viewpoint:}
If joint $(R_0,\theta)$ filtering, compensation, or coding
is employed, the same coupling structure can be exploited
to enable range-selective beamforming \cite{XuShi2015RangeAngleDependent,MahmoodMir2018FrequencyDiverseArray,MirAlbasha2025FrequencyDiverseArray}, interference
suppression \cite{LiaoTang2022ALowSidelobe,TanWang2021ANoveleceptive}, and ambiguity-function shaping \cite{LiuWang2024AmbiguityFunction,wang2017fda,wang2014linear}.

\end{itemize}

Therefore, range--angle coupling is neither inherently
beneficial nor detrimental. Rather, it represents a
{structured degree of freedom in the range--angle response} whose value becomes
explicit only under an appropriate joint processing framework.

\subsubsection{ Time-Invariant Spatial Focusing: Physical Infeasibility and ``Equivalent Focusing''}
At the physical-field or transmit-array-factor level, the FDA response
contains an explicit time gradient, i.e.,
\begin{equation}
\nabla_t\phi_m=2\pi\Delta f_m.
\end{equation}
Hence, as long as $\Delta f_m\neq 0$ for some $m$, $AF(t,R_0,\theta)$ is intrinsically time-varying \cite{liao2023estimation,Wang2021LPI}. The ``time-invariant focusing'' reported in much of the literature \cite{GongWang2018TimeInvariantJoint,cheng2017time,liao2023time,hei2024ann,chen2019accurate,zhai2021joint,bilal2026measurement} typically corresponds to two equivalent implementations, namely
\begin{itemize}
\item \textbf{Instantaneous Focusing}: achieving local focusing at a particular time instant within a finite time window (or under pulsed operation);
\item \textbf{Post-Processing Equivalent Focusing}: constructing a stable peak in the output domain via matched filtering/phase compensation at the receiver.
\end{itemize}

Because the range gradient and the time gradient originate from the
same frequency-offset term, any attempt to remove time variation
at the physical-field level would generally suppress the associated
range-gradient effect as well, thereby driving the structure toward
PA-like degeneration. Therefore, strictly time-invariant spatial focusing at the
physical-field level is incompatible with the FDA mechanism. Instead,
the time-varying structure should be regarded as an intrinsic
feature of FDA arrays and exploited to enhance detection,
estimation, and interference suppression.

\subsection{A Forward-Looking Answer: Reinterpreting Coherent FDA Beyond Pulse Compression}

\begin{figure*}[htp]
\centering
\includegraphics[width=0.88\textwidth]{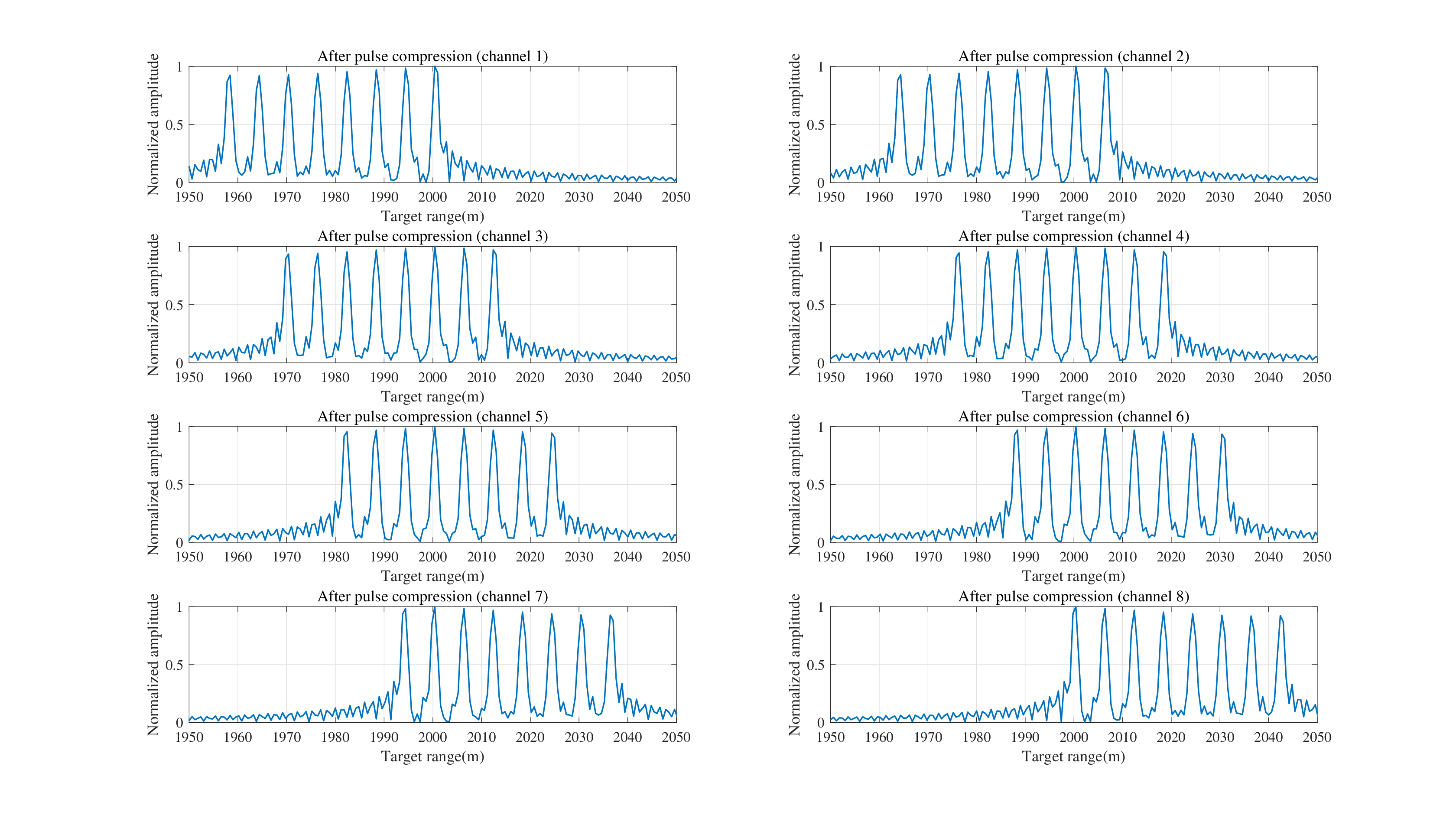}
\caption{\justifying Pulse-compression outputs for different channels in a coherent FDA system. Because each transmit element operates at a different carrier frequency, conventional pulse compression produces multiple peaks and nonuniform range alignment across channels.}
\label{fig_receiver}
\end{figure*}

The above discussions also suggest a potentially important re-evaluation of coherent FDA.
Under the conventional pulse-compression framework, coherent FDA is often regarded as difficult to process, since the element-dependent carrier offsets may produce misaligned compressed responses and multiple nonuniform range peaks across channels, as shown in Fig.~\ref{fig_receiver}.
From this viewpoint, coherent FDA appears incompatible with the standard delay-domain ranging pipeline.

However, this apparent difficulty may stem less from the FDA mechanism itself than from the long-standing assumption that distance must first be recovered through conventional pulse compression.
If range in FDA/FDA--MIMO is no longer interpreted exclusively as a delay-domain quantity, but instead as an irreducible parameter embedded in the propagation manifold, then coherent FDA may become a much more meaningful architecture.
In that case, its propagation-level phase coupling and coherent field gain need not be treated as processing obstacles, but rather as exploitable structural resources.
This perspective opens the possibility of a manifold-based processing framework for coherent FDA.
Instead of forcing the received signals into a conventional pulse-compression pipeline, one may directly exploit the joint time--range--angle response through manifold matching, coherent hypothesis testing, joint parameter estimation, or structure-aligned focusing.
Under such a viewpoint, coherent FDA may offer a form of coherent gain that is not fully captured by traditional delay-domain processing.

Therefore, an important open question is whether the true value of coherent FDA has been partially obscured by the classical pulse-compression paradigm.
Revisiting coherent FDA from a manifold-based ranging and detection perspective may provide a new avenue for unlocking its physical advantages and clarifying its role within future FDA/FDA--MIMO systems.

\section{Fundamental Design Dimensions of FDA: From Mechanism to Structural Abstraction}
\label{sec:fundamental-dimensions}

Having established the propagation-kernel-based taxonomy in the previous section, we now examine the fundamental properties of FDA from an array-manifold perspective. 
The goal of this section is not simply to enumerate several characteristic FDA phenomena, but to show that they originate from a common structural mechanism: the frequency gradient across the transmit aperture reshapes the array manifold and enlarges the set of controllable dimensions of the radiated field. 
Under this viewpoint, time variability, beam scanning, and range--angle coupling are direct manifestations of the same underlying expansion, while the integrated transmit beampattern, the ambiguity function, and the scattering response can be understood as higher-level consequences of this altered manifold structure. 
This mechanism-to-property perspective helps connect the basic physical behavior of FDA with its broader implications for sensing, imaging, and communication applications.

In conventional phased arrays (PAs), the radiated response depends only on the angular variable, which can be represented as
$
\mathcal{A}_{\mathrm{PA}}(\theta).
$
By contrast, FDA introduces element-dependent carrier frequencies, so that the array response becomes a joint function of time, range, and angle,
$
\mathcal{A}_{\mathrm{FDA}}(t,R_0r,\theta).
$
This expansion from an angle-only response to a time--range--angle-dependent manifold is the fundamental structural distinction between FDA and conventional arrays. 
We begin with time variability, which is the most immediate consequence of this additional manifold dimension.

\subsection{Time Variability as Manifold Expansion}
\label{subsec:time-variability}

Introducing a frequency gradient $\Delta f$ fundamentally alters the structure of the array manifold. Unlike conventional PA, whose array responses are independent of time, the FDA manifold explicitly includes time as an intrinsic dimension.
\begin{figure}[htp]
\centering
\subfigure[PA]{
\includegraphics[width=0.23\textwidth]{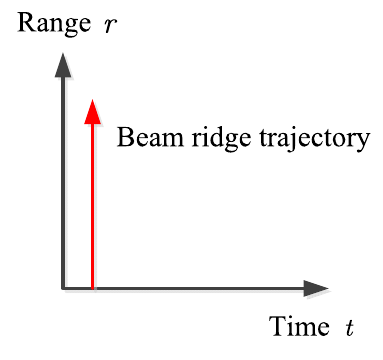}}
\subfigure[FDA]{
\includegraphics[width=0.23\textwidth]{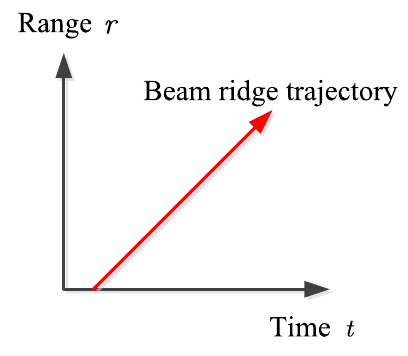}}
\caption{\justifying Comparison of beam behavior in FDA and PA in the range--time plane.
The PA beam remains spatially stationary, whereas the FDA beam forms a
propagating constructive-interference ridge that moves along the range
direction as time increases.}
\label{fig6}
\end{figure}
Fig.~\ref{fig6} illustrates the fundamental difference between PA and
FDA in the range--time plane. For PA, the beam pattern remains
stationary because the array response depends only on angle. In
contrast, the element-dependent frequency offsets in FDA introduce
intrinsic time--range coupling, causing the constructive-interference
ridge to propagate along the range direction as time evolves.

\begin{figure}[htp]
\centering
\subfigure[]{
\includegraphics[width=0.5\textwidth]{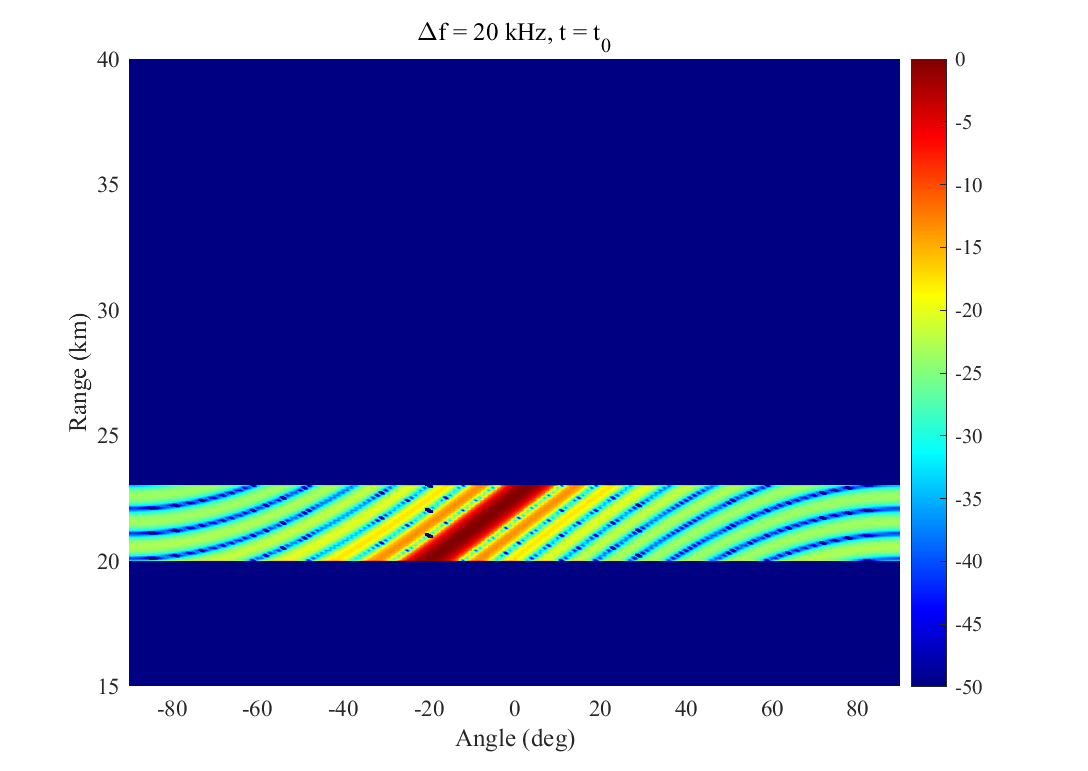}}
\subfigure[]{
\includegraphics[width=0.5\textwidth]{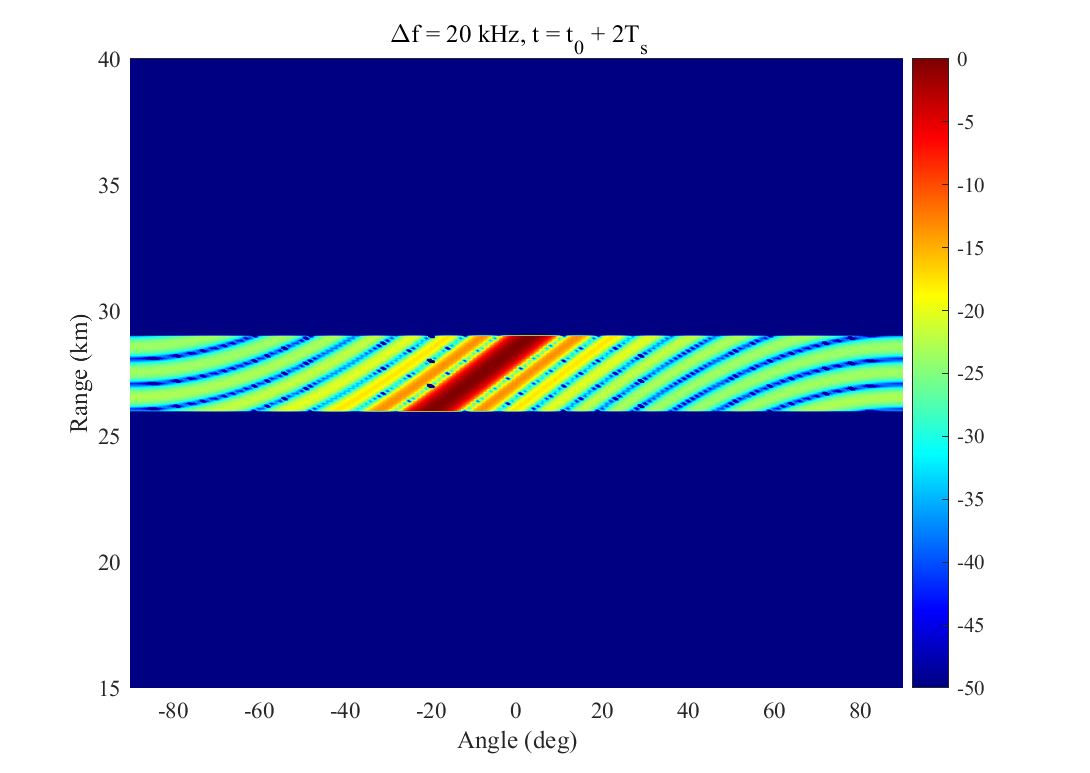}}
\subfigure[]{
\includegraphics[width=0.5\textwidth]{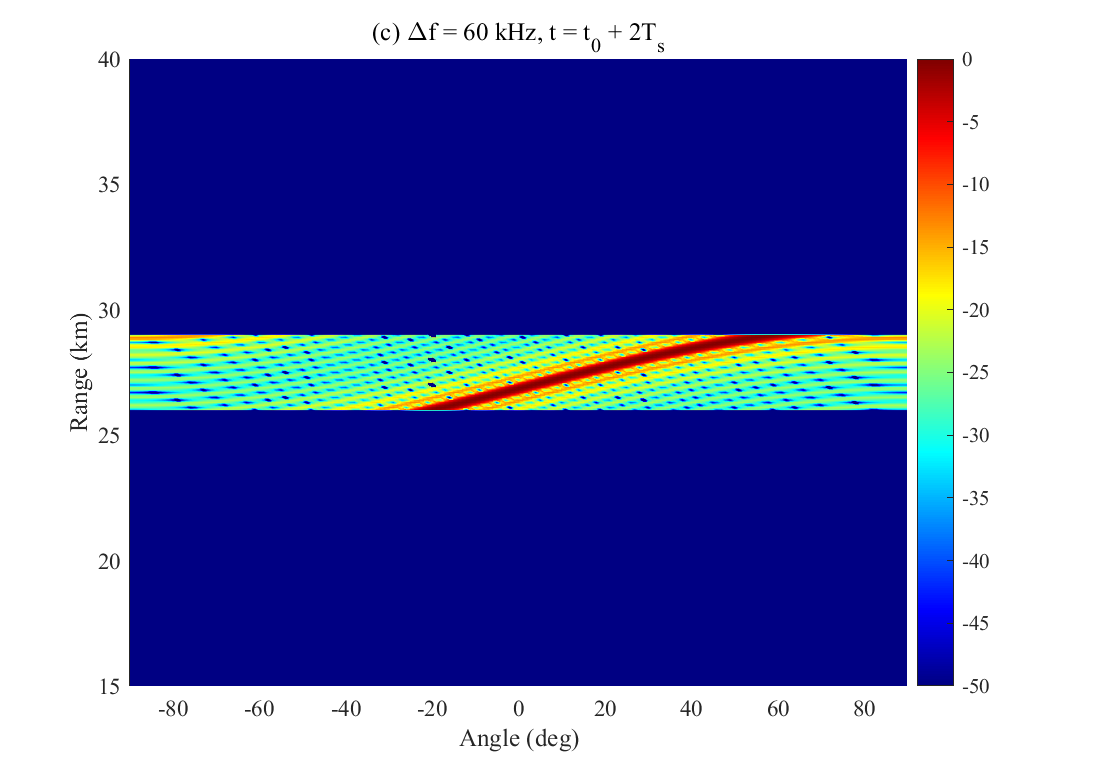}}
\caption{\justifying Time evolution of the FDA transmit beampattern under different
frequency increments. For a fixed $\Delta f$, the mainlobe ridge shifts
toward farther ranges as time increases. A larger $\Delta f$ further
strengthens the range--angle--time coupling and changes the ridge
trajectory.}
\label{fig5}
\end{figure}

Under far-field conditions, the trajectory of the FDA mainlobe can be
approximately described by
\begin{equation}
R_0 = ct - \ell \frac{c}{\Delta f}
+ \left(1+\frac{f_c}{\Delta f}\right)d_T \sin\theta ,
\end{equation}
where $R_0$ is the range, $t$ is time, $c$ is the speed of light, $d_T$
is the transmit element spacing, $\theta$ is the observation angle, and
$\ell$ denotes the beam index associated with different propagation
ridges.
This expression reveals an important physical insight: the FDA beam
ridge propagates approximately at the speed of light along the range
direction. Therefore, the FDA beam pattern is intrinsically
time-varying rather than spatially stationary.
The above analytical result is further illustrated in
Fig.~\ref{fig5}.

Because of this inherent beam evolution, many studies have attempted to
synthesize time-invariant FDA beampatterns. One line of research
introduces time-varying frequency offsets to compensate for the beam
drift. Khan \textit{et al.}~\cite{khan2014frequency} proposed such a
time-varying frequency offset strategy. Another line of work relies on
time-modulated frequency design. Yao \textit{et al.}~\cite{yao2016frequency}
introduced a time-modulated frequency offset scheme for
time-invariant spatial focusing, and further extended it to multi-time
focusing~\cite{yao2015single} and near-range multi-point focusing
scenarios~\cite{yao2016solutions}. In addition, array-structure-based
solutions have also been explored. Cheng \textit{et al.}~\cite{cheng2017time}
proposed an array pointing modulation approach, while
Yang \textit{et al.}~\cite{yang2018optimization} developed a
time-invariant beamforming method based on sparse frequency diverse
arrays~\cite{chen2015sparse}.

Nevertheless, from a physical viewpoint, the time-varying characteristic
of FDA beams originates from the outward propagation of electromagnetic
energy. Hence, such approaches may reshape or compensate the beampattern
at selected instants, but they cannot fundamentally force the beam peak
to remain permanently at a fixed range.
Therefore, the time variability of FDA should not be regarded as a mere
modulation artifact. Rather, it is a structural consequence of manifold
expansion from the spatial response
$\mathcal{A}_{\mathrm{PA}}(\theta)$
to the joint time--range--angle response
$\mathcal{A}_{\mathrm{FDA}}(t,r,\theta)$.

\label{subsec:scanning-mechanism}
\begin{figure*}[htp]
	\centering
	    \subfigure[]
	{		{\includegraphics[width=0.43\textwidth]{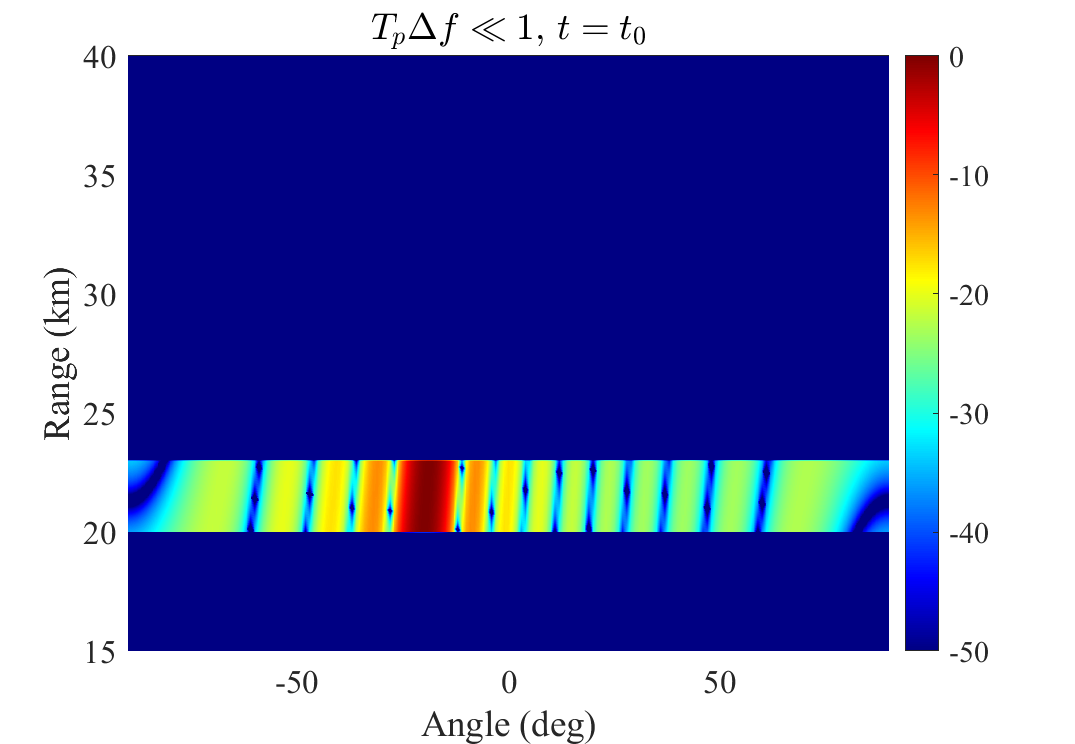}}}   
    	    \subfigure[]
	{		{\includegraphics[width=0.43\textwidth]{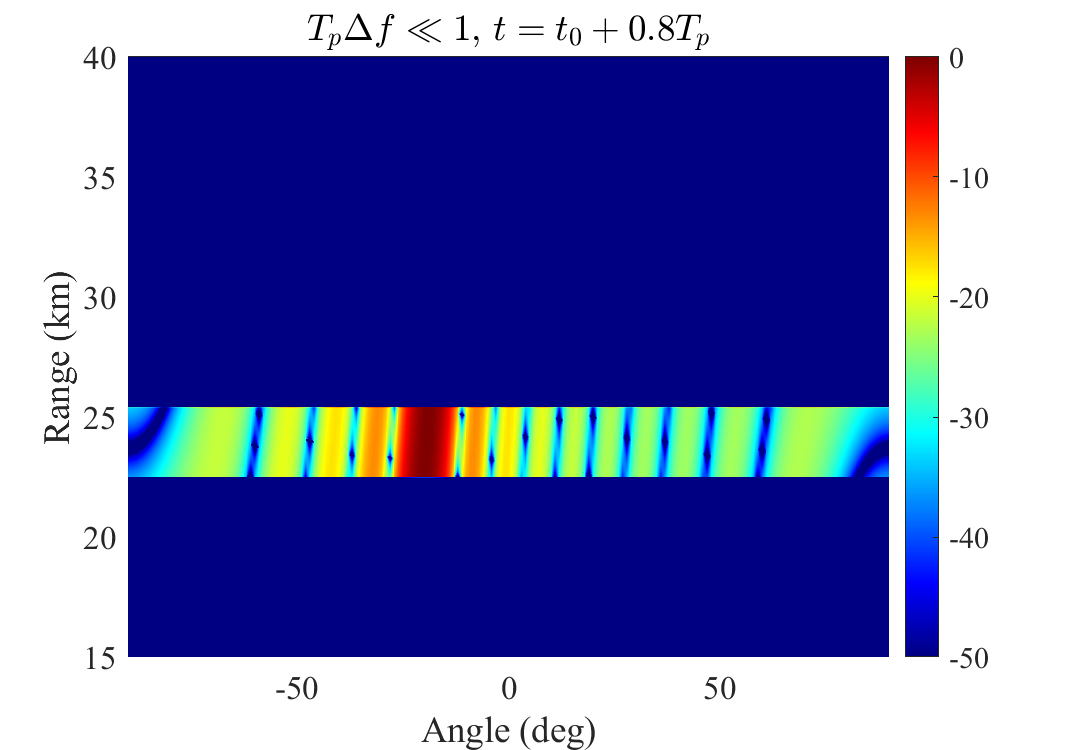}}}   
    	    \subfigure[]
	{		{\includegraphics[width=0.43\textwidth]{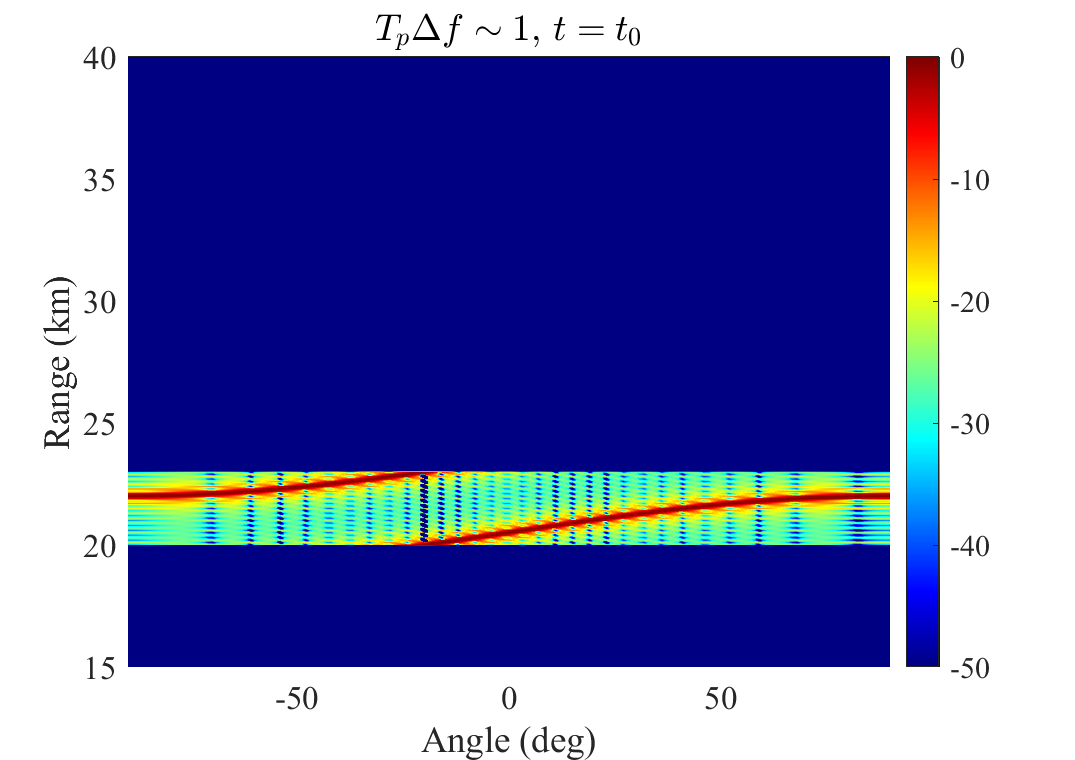}}}   
    	    \subfigure[]
	{		{\includegraphics[width=0.43\textwidth]{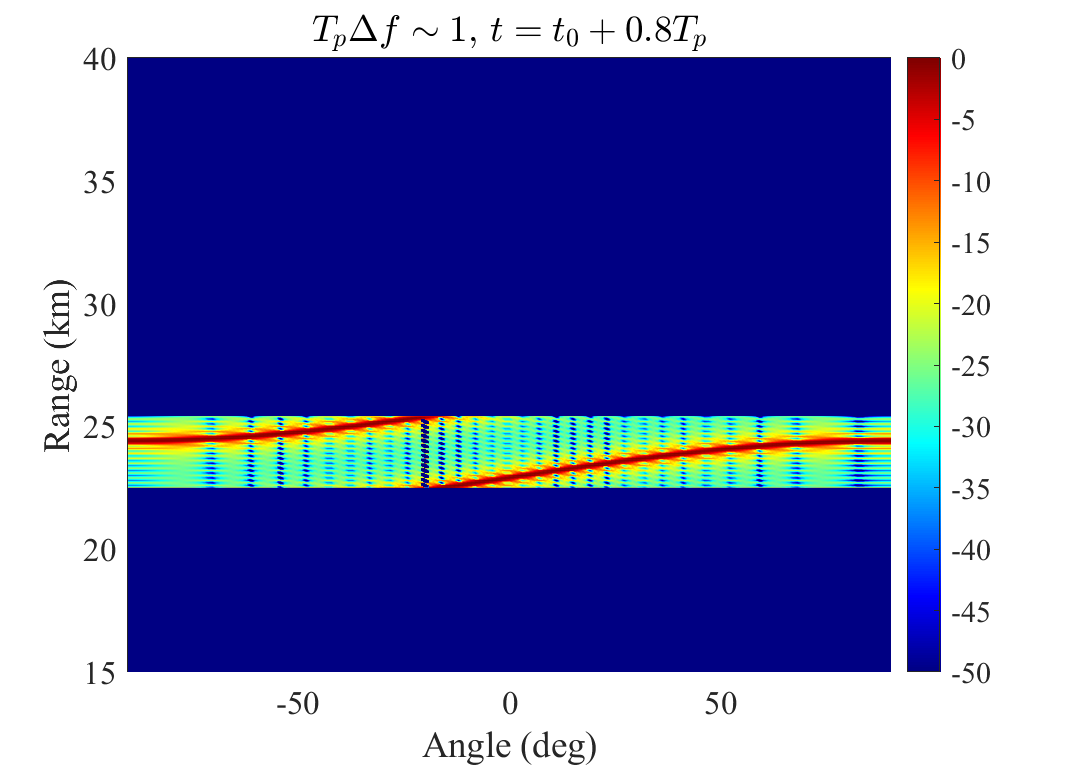}}}     
\caption{\justifying Illustration of two FDA operating regimes determined by the
product $T_p\Delta f$. When $T_p\Delta f \ll 1$, the beampattern remains
approximately unchanged within a pulse. When $T_p\Delta f \sim 1$, the
mainlobe ridge evolves significantly within the same pulse duration,
resulting in noticeable beam scanning.}
	\label{fig7}	
\end{figure*}
\subsection{Frequency Gradient and Scanning Mechanism}
The automatic beam scanning behavior of FDA can be interpreted
from the perspective of array manifold evolution.
As discussed in the previous subsection, the FDA array response
extends from a purely spatial manifold
$\mathcal{A}_{\mathrm{PA}}(\theta)$
to a joint time--range--angle manifold
$\mathcal{A}_{\mathrm{FDA}}(t,r,\theta)$.
Within this expanded manifold, the beam peak corresponds to a
constructive-interference ridge rather than a stationary point.
As time evolves, this ridge propagates across the manifold,
and its projection onto the angular dimension manifests as
an apparent beam scanning phenomenon.

From the ridge trajectory derived previously, the temporal
variation of the beam pointing direction can be approximated as
\begin{equation}
\frac{\partial \theta}{\partial t}
=
\frac{-c\Delta f}
{(f_c+\Delta f)d_T \cos\theta},
\end{equation}
which represents the angular scanning rate of the FDA mainlobe.
This expression indicates that the scanning speed is directly
controlled by the frequency increment $\Delta f$.
Therefore, the frequency gradient determines the scale and
rate of beam scanning, rather than the existence of scanning
itself.

Two distinct operating regimes can be identified depending on
the product $T_p\Delta f$, where $T_p$ denotes the pulse duration.
When $T_p\Delta f \ll 1$, the phase evolution induced by the
frequency gradient is negligible within a single pulse, and the
beam can therefore be regarded as approximately stationary.
In contrast, when $T_p\Delta f \sim 1$, the manifold ridge evolves
significantly within one pulse duration, resulting in noticeable
beam scanning, as illustrated by the representative beampattern
snapshots in Fig.~\ref{fig7}.

\subsection{Range–Angle Coupling in the FDA Manifold}
\label{subsec:ra-coupling}

Another fundamental characteristic introduced by frequency
diversity is the coupling between range and angle.
In conventional PA, the array manifold depends only
on the angular variable, i.e., $\mathbf{a}(\theta)$, and the
range parameter affects only the signal delay rather than the
spatial phase structure.
In FDA systems, however, the element-dependent frequency
offsets introduce an additional range-dependent phase term.
As a result, the equivalent beam pointing direction can be
expressed as
\begin{equation}
\theta_e =
\sin^{-1}
\left[
\left(1+\frac{\Delta f}{f_c}\right)\sin\theta
-
\frac{\Delta f}{f_c d_T} R_0
\right],
\end{equation}
which explicitly depends on both the target angle $\theta$
and range $R_0$.

From the manifold perspective, this implies that the array
response is no longer described by the one-dimensional
manifold $\mathbf{a}(\theta)$, but instead by a
two-dimensional manifold $\mathbf{a}(R_0,\theta)$.
In other words, the parameter space expands from a purely
angular dimension to a joint range--angle space.
Therefore, the range--angle coupling observed in FDA systems
can be interpreted as a parameter-mixing phenomenon caused by
this manifold-dimension expansion. In practice, such coupling
may degrade the focusing capability of the transmit beam and
affect target localization accuracy. Consequently, a number of
studies have attempted to mitigate this effect through frequency
design.
For example,
Khan \textit{et al.} proposed a logarithmic frequency increment
scheme to suppress the range--angle coupling effect and generate
discrete point-like beams in the propagation space
\cite{khan2014frequency}. Subsequent works
\cite{shao2016dot,xiong2016frequency} further explored nonlinear
frequency increment strategies, such as exponential
\cite{gao2016decoupled} and quadratic frequency offsets
\cite{xu2021fda}, which improve beam focusing performance.
In addition, frequency coding strategies, including Costas
frequency coding \cite{wang2016range} and random frequency
assignments \cite{liu2016random}, have also been investigated
to alleviate the coupling effect in the transmit beampattern.

Despite these efforts, such approaches essentially reshape the
frequency distribution rather than eliminating the underlying
physical mechanism. From the manifold perspective, the coupling
originates from the intrinsic expansion of the array manifold
from $\mathbf{a}(\theta)$ to $\mathbf{a}(R_0,\theta)$, and thus
cannot be fundamentally removed.
Instead of treating range--angle coupling purely as an
undesirable artifact, it can also be interpreted as an additional
structural degree of freedom introduced by the expanded manifold.
Although the coupling complicates conventional separable
parameter estimation methods, it can be exploited through
coupling-aware joint processing for enhanced sensing and
beamforming capabilities.

\subsection{Integrated Transmit Beampattern from a Manifold Perspective}
\label{subsec:integrated-beampattern}

From the manifold viewpoint, the instantaneous FDA transmit
beampattern can be interpreted as the response of the
time-varying array manifold to the transmit signal.
When the transmit field is integrated over time, the resulting
beampattern corresponds to the statistical average of this
manifold response.
Specifically, the integrated transmit beampattern can be written as
\begin{equation}
P_{\mathrm{FDA}}(\theta )
=
\mathbb{E}_t
\!\left[
\left|
\mathbf{a}^H(\theta)\mathbf{s}(t)
\right|^2
\right]
=
\left| \mathbf{a}_{}^{H}(\theta )\mathbf{R}_T\mathbf{a}_{}(\theta ) \right|
,
\end{equation}
where $\mathbf{s}(t)$ denotes the transmit signal vector and
$\mathbf{R}_T=\mathbb{E}_t[\mathbf{s}(t)\mathbf{s}^H(t)]$ is the
transmit signal correlation matrix.
From this perspective, the integrated beampattern reflects the
statistical spatial response determined by the correlation
structure of the transmitted signals. This expression resembles
the conventional MIMO radar beampattern, but here it arises
from integrating the time-varying FDA transmit response over time.

\begin{figure}[htp]
\centering
\includegraphics[width=0.41\textwidth]{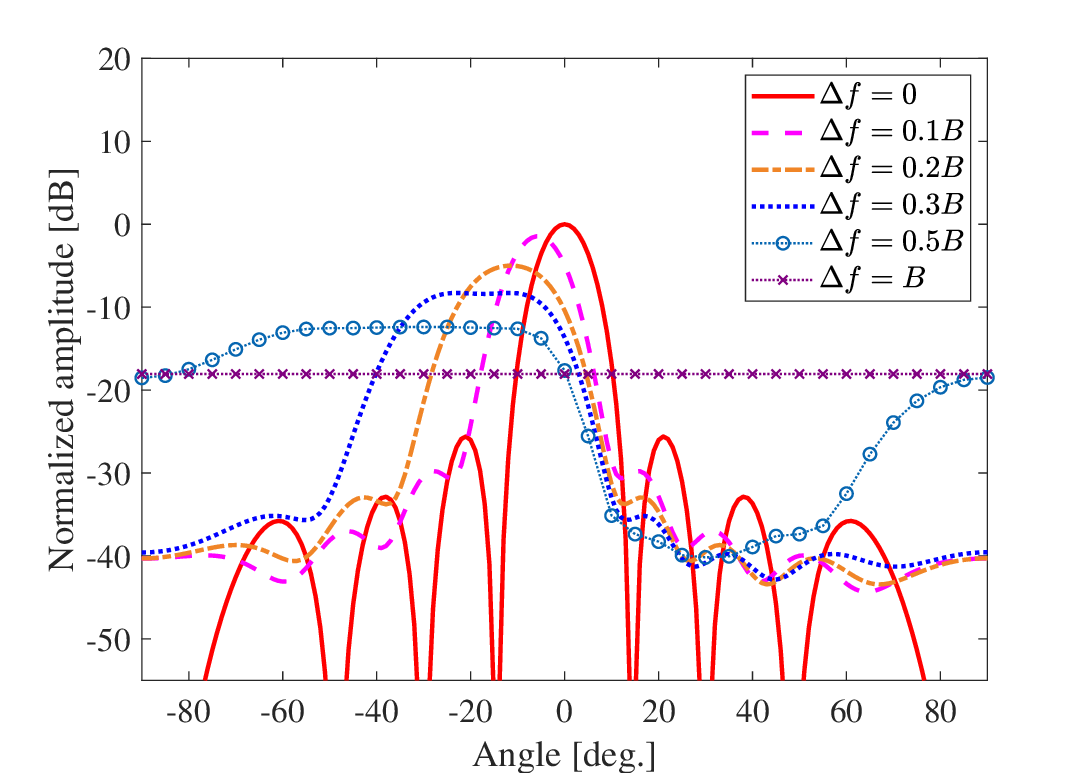}
\caption{\justifying Integrated transmit beampattern of FDA for different
frequency increments \cite{gui2020generalized}.}
\label{fig9}
\end{figure}

An important transition occurs as the frequency increment
$\Delta f$ increases. As illustrated in Fig.~\ref{fig9}, when
$\Delta f$ is much smaller than the signal bandwidth $B$,
the transmit signals across the array elements remain highly
correlated, and the integrated beampattern is approximately
consistent with that of a conventional phased array.
As $\Delta f$ increases, the inter-element correlation gradually
decreases, leading to a progressive broadening of the integrated
beampattern. When $\Delta f \geq B$, the transmit signals become
nearly orthogonal, and the integrated beampattern approaches an
approximately omnidirectional radiation pattern.

These observations highlight that the instantaneous and
integrated FDA beampatterns describe two complementary aspects
of the same system. The instantaneous beampattern captures the
time-varying focusing behavior produced by frequency diversity,
whereas the integrated beampattern characterizes the
correlation-dependent average radiation pattern of the array.
\subsection{Ambiguity-Function Interpretation of the FDA Manifold}
\label{subsec:gaf-shaping}

The ambiguity function provides a useful framework for
characterizing the resolution and parameter separability of
radar systems. From a manifold perspective, the ambiguity function can be
interpreted as the correlation (inner product) between array
responses corresponding to two parameter vectors, i.e.,
\begin{equation}
\label{eq26}
\chi_{\mathrm{FDA}}(\mathbf{\Theta}_1,\mathbf{\Theta}_2)
=
\langle
\mathbf{a}(\mathbf{\Theta}_1),
\mathbf{a}(\mathbf{\Theta}_2)
\rangle,
\end{equation}
where $\mathbf{\Theta}$ denotes the parameter vector that may
include range, angle, and Doppler variables
\cite{gui2020generalized}. In this sense, the ambiguity
function reflects the correlation structure of the array
manifold in the parameter space.

In FDA systems, the element-dependent frequency offsets modify
the phase structure of the array manifold, which consequently
reshapes the multidimensional ambiguity function. Compared
with conventional phased arrays, the introduced frequency
gradient generates additional coupling among range, angle, and
time dimensions, leading to a more intricate ambiguity
surface.
From the ambiguity-function perspective, the most notable
impact of frequency diversity appears in the range dimension.
Unlike conventional phased arrays, the effective range
resolution of FDA systems is jointly determined by the signal
bandwidth and the frequency increment across the transmit
array,
\begin{equation}
[\Delta R]_{\mathrm{res}}
=
\min
\left\{
\frac{c}{2B},
\frac{c}{2M\Delta f}
\right\}.
\end{equation}
This result indicates that the additional phase diversity
introduced by the frequency gradient can provide an extra
range discrimination capability beyond the conventional
bandwidth-limited resolution\footnote{This observation also raises a broader conceptual question regarding the interpretation of range in FDA systems.
If the effective range discrimination is no longer determined solely by the delay--bandwidth relation, but also by the frequency-gradient-induced manifold structure, then the classical pulse-compression-based notion of range resolution may no longer be sufficient to fully characterize the system.
In this sense, the role of pulse compression in FDA may need to be reconsidered:
rather than serving as the unique mechanism for range acquisition, it may become only one component in a broader ranging framework where distance is also reflected in the distinguishability of range-dependent array responses.
Accordingly, whether distance in FDA/FDA--MIMO should still be measured exclusively through propagation delay, or jointly interpreted through delay and manifold discrimination, remains an open but important question.}.

\begin{figure}[htp]
\centering
\includegraphics[width=0.41\textwidth]{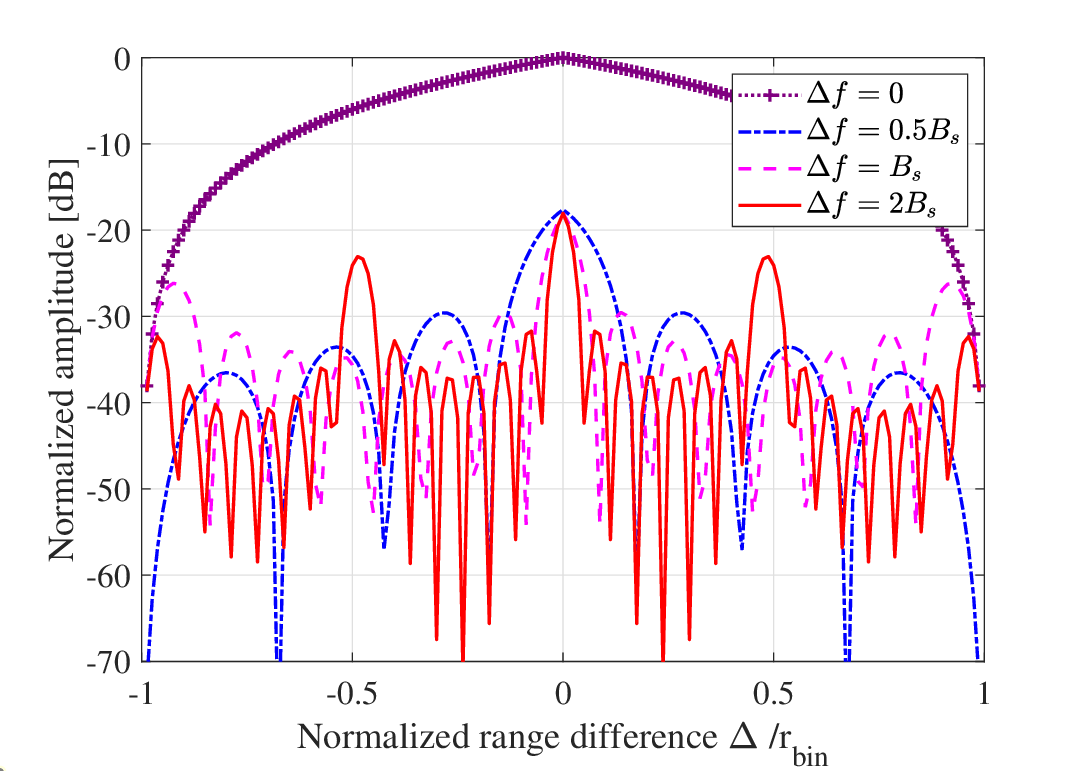}
\caption{\justifying Range-cut of the FDA ambiguity function for different
frequency increments $\Delta f$ \cite{gui2020generalized}.}
\label{fig10}
\end{figure}

Fig.~\ref{fig10} shows the range-cut of the FDA ambiguity
function for different frequency increments $\Delta f$.
The conventional range bin size is given by
$R_{\mathrm{bin}} = c/(2B)$. As $\Delta f$ increases, the
additional phase diversity across the transmit array enhances
the range discrimination capability.
However, when the frequency increment becomes excessively
large, multiple dominant peaks may appear within a single
range bin. This phenomenon can be interpreted as the emergence
of {secondary range cells}, which introduce additional
range ambiguities in the FDA response. Physically, this
behavior originates from the additional range-dependent phase
term induced by the frequency gradient, which creates multiple
constructive-interference regions within a single conventional
range resolution cell.

In addition to modifying the range resolution, frequency
diversity also changes the geometric structure of the
ambiguity surface. Eq.\eqref{eq26} shows that the phase difference between two parameter points
$(r_1,\theta_1)$ and $(r_2,\theta_2)$ becomes

\begin{equation}
\Delta\phi_m
=
2\pi \frac{f_c}{c} m d_T
(\sin\theta_2-\sin\theta_1)
-
2\pi \frac{\Delta f_m}{c}(R_2-R_1).
\end{equation}

Constructive interference occurs when the phase difference
remains approximately constant across the array aperture,
which leads to the approximate relation

\begin{equation}
\Delta R
\approx
\frac{f_c d_T}{\Delta f}\,\Delta\sin\theta .
\end{equation}

This relation indicates that the mainlobe of the ambiguity
function forms a slanted ridge in the
$(\Delta R,\Delta\sin\theta)$ plane rather than a separable
peak at $(0,0)$. The resulting tilted ambiguity surface
directly reflects the intrinsic range--angle coupling induced
by the frequency gradient.
By contrast, the angular resolution under far-field conditions
remains primarily determined by the physical aperture of the
array, while the Doppler resolution depends on the coherent
observation time, as in conventional radar systems. Therefore,
frequency diversity mainly reshapes the ambiguity structure in
the range dimension rather than fundamentally altering the
angular or Doppler resolution limits.

\subsection{Scattering Response from a Manifold Perspective}
\label{subsec:rcs-pointer}

The scattering response of a target can also be interpreted
through the array manifold. Consider a target consisting of
multiple scattering centers $\{(R_k,\theta_k)\}$ with complex
amplitudes $\alpha_k$. The received echo can be expressed as
the coherent superposition of the manifold responses
associated with these scattering centers, namely \cite{cetintepe2014examination}
\begin{equation}
y(t)=\sum_k{\alpha _k\mathbf{1}^T\boldsymbol{a}(t,R_k,\theta _k),}
\end{equation}
where $\mathbf{a}(t,r,\theta)$ denotes the FDA array manifold
and $\mathbf{1}$ denotes the vector whose entries are all
equal to one. This expression indicates that the received
signal results from the coherent projection of the array
manifold responses corresponding to different scattering
points.

Expanding the manifold response reveals that each scattering
contribution contains a phase term determined by the
propagation delay and the element-dependent carrier
frequency. In FDA systems, the introduced frequency offsets
generate an additional time-dependent phase component,
which causes the relative phases among the scattered
contributions to evolve continuously. The received signal can
therefore be written as
$y(t)
=
\sum_k
\alpha_k
e^{-j\phi_k(t)},$
with
\begin{equation}
\phi_k(t)
=
2\pi f_c \tau_k
+
2\pi \Delta f_k t,
\end{equation}
where $\tau_k$ denotes the propagation delay associated with
the $k$th scattering center and $\Delta f_k$ represents the
effective frequency offset contributing to the time-varying
phase evolution.
Besides, the equivalent radar cross section (RCS) is defined as the
squared magnitude of the received echo, i.e., \cite{HuangYan2022RadarCrossSection}

\begin{equation}
\sigma_{\mathrm{FDA}}(t,\Delta f)
=|y(t)|^2 .
\end{equation}

\begin{figure}[htp]
\centering
\includegraphics[width=0.41\textwidth]{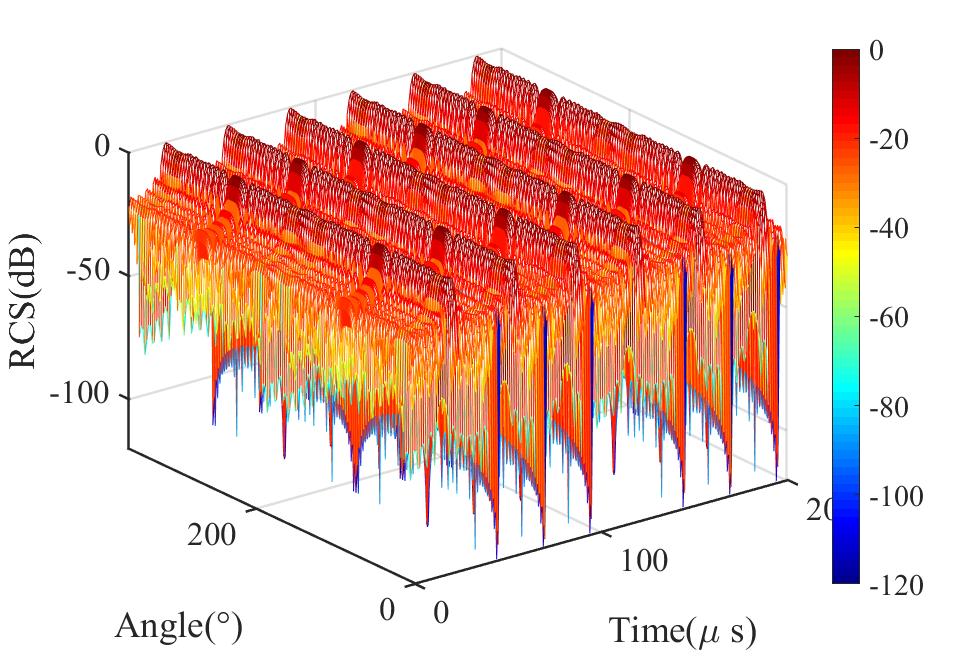}
\caption{\justifying Time-varying RCS of an FDA radar target as a function
of time and observation angle. The periodic fluctuations
demonstrate that, due to the element-dependent frequency
offsets, the coherent superposition of scattering
contributions evolves over time, resulting in a time-varying
equivalent RCS even for a static target \cite{HuangYan2022RadarCrossSection}}
\label{fig12}
\end{figure}

Unlike conventional phased arrays, where the relative phases
among scattering contributions remain constant for a fixed
target geometry, the FDA-induced phase evolution causes the
coherent superposition to vary with time, like Fig.\ref{fig12}. 

However, the aforementioned time-varying RCS behavior relies
on the coherence of the scattered echoes across the array
elements. If the frequency increment becomes excessively
large, the echoes corresponding to different transmit
elements may become decorrelated within a single range
resolution cell.
To preserve the coherent superposition among array elements,
the frequency increment must satisfy the approximate
constraint \cite{gui2020target}
\begin{equation}
\Delta f
\le
\frac{c}{4(M-1)\Delta r},
\end{equation}
which ensures that the phase variation across the array
remains sufficiently small within one range cell.
Since the range resolution is given by $\Delta r=c/(2B)$,
this condition can be equivalently written as
\cite{gui2020phdResearch,Huang2023phdResearch}
\begin{equation}
(M-1)\Delta f \le B.
\end{equation}

When this condition is violated, the scattered echoes from
different transmit elements lose their mutual coherence,
and the FDA echo model approaches that of a frequency-agile
or MIMO radar. In such cases, the coherent scattering
mechanism described above no longer holds, and the RCS
statistics follow the conventional models used for
decorrelated radar returns.

\begin{figure}[htp]
\centering
\includegraphics[width=0.41\textwidth]{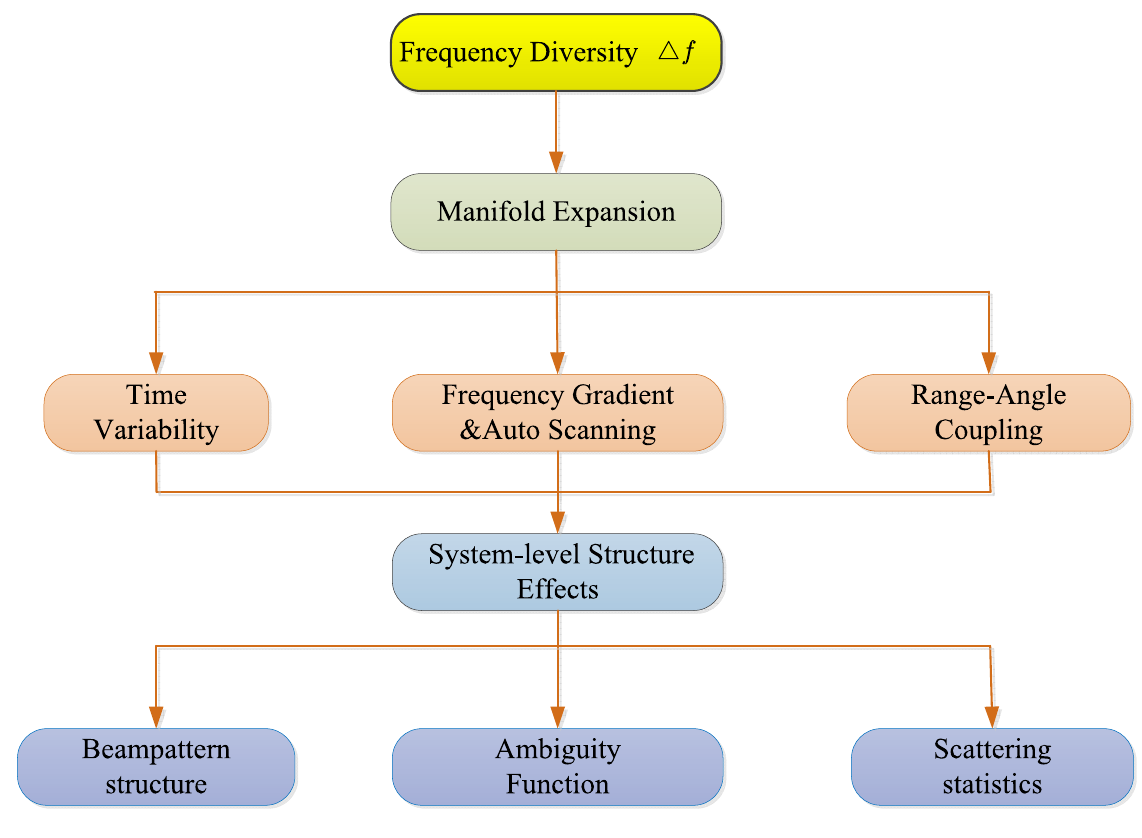}
\caption{\justifying Unified structural interpretation of FDA from the array
manifold perspective. Frequency diversity induces manifold
expansion, which gives rise to time variability, automatic
scanning, and range–angle coupling, and subsequently
manifests in the beampattern structure, ambiguity function,
and scattering statistics.}
\label{fig11}
\end{figure}

From a statistical viewpoint, this phenomenon indicates that
frequency diversity fundamentally reshapes the geometry of
the FDA array manifold. Consequently, the transmit
beampattern, ambiguity structure, and coherent scattering
statistics are all determined by this manifold expansion,
as summarized in Fig.~\ref{fig11}.

\section{Frequency-Gradient and Time-Coding Array Paradigms
}
\label{sec:paradigm-dof-comparison}
In the literature, FDA and FDA--MIMO are often discussed together with other array paradigms capable of producing range--angle dependent responses, such as element--pulse coding (EPC) arrays \cite{xu2020resolving,liu2023resolving,liu2025range,lan2024mainlobe,zhu2024simultaneous,lan2020mainlobe} and space--time coding arrays (STCA) \cite{wang2024sidelobe,wang2025range,wang2025mainlobe,wang2025monopulse,lan2018subarray,liu2023signal,wang2021transmit}. Because these architectures may produce similar output behaviors, they are sometimes interpreted as belonging to the same class of range--angle dependent arrays. However, the mechanisms that generate these behaviors are fundamentally different.
To clarify these relationships, this section compares FDA/FDA--MIMO and representative time-coding paradigms from a structural perspective. The comparison focuses on the design variables involved in signal generation, the resulting propagation-phase structures, and the corresponding system DoF. In particular, we distinguish between {physical range dependence} created directly by carrier-frequency gradients and {processing-induced range selectivity} obtained through temporal coding and receiver-side reconstruction.
\subsection{Propagation Phase and Design Dimensions}
\label{subsec:unified-phase-structure}

To reveal the fundamental differences among array paradigms, it is instructive to examine how the design variables affect the propagation phase of the transmitted signal.
Consider the signal transmitted by the $m$-th array element and observed at location $(r,\theta)$. Under far-field narrowband propagation, its phase can be written as

\begin{equation}
\phi_m(t,R_0,\theta)
=
2\pi f_m t
-
\frac{2\pi f_m}{c}R_0
+
\frac{2\pi f_m d_m \sin\theta}{c}.
\end{equation}

This expression shows that the received phase is determined by three physical quantities: the carrier frequency $f_m$, the element position $x_m$, and the propagation delay $R_0/c$. Consequently, different array paradigms can be interpreted according to which of these variables they manipulate during signal generation.
From this viewpoint, several fundamental design dimensions can be identified:

\begin{itemize}

\item \textbf{Spatial dimension:}  
Element positions $x_m$ determine the spatial phase distribution across the aperture and are responsible for conventional angular selectivity.

\item \textbf{Frequency dimension:}  
Element-dependent carrier frequencies $f_m$ introduce inter-channel frequency gradients, which directly modify the propagation phase and can produce range-dependent interference structures.

\item \textbf{Temporal coding dimension:}  
Time-domain modulation across pulses or symbols provides structured signal diversity and may generate equivalent range selectivity after matched filtering and receiver-side reconstruction.

\item \textbf{Waveform diversity dimension:}  
Orthogonal waveforms enable channel separability and facilitate virtual-array formation in MIMO-type systems.

\end{itemize}

Different array paradigms activate different subsets of these dimensions. As will be discussed in the following subsections, this distinction is particularly important for understanding the fundamental difference between frequency-gradient arrays (e.g., FDA) and time-coding-based arrays including EPC and STCA, since only the former directly modify the propagation phase associated with range.
Among the design dimensions discussed above, the frequency dimension plays a unique role because it directly modifies the carrier-dependent propagation phase. This property gives rise to the class of frequency-gradient arrays, represented by FDA and FDA--MIMO.
\subsection{Frequency-Driven Paradigm: FDA and FDA--MIMO}
\label{subsec:freq-driven-fda}
Frequency diverse arrays introduce element-dependent carrier frequencies $f_m$ across the transmit aperture, which fundamentally modifies the propagation-phase structure of the transmitted field. 
Substituting $f_m=f_c+\Delta f_m$ reveals that the element-dependent frequency offsets introduce additional phase terms proportional to both time and range. In particular, the range-phase gradient becomes \eqref{eq18}.
Since $\Delta f_m$ varies across array elements, different transmit channels exhibit different range-phase gradients,
i.e.,
\begin{equation}
\Delta(\nabla_R\phi_m)
=
-\frac{2\pi \Delta f_m}{c}.
\end{equation}

This inter-channel gradient difference creates range-dependent interference directly during electromagnetic propagation. As a result, FDA is able to generate transmit-side range–angle dependent responses without relying on receiver-side temporal reconstruction.
From a system-design perspective, the introduction of element-dependent carrier frequencies therefore expands the design space of the array manifold Eq.\eqref{eq9}.
When orthogonal waveforms are further transmitted across array elements (FDA--MIMO), waveform diversity provides an additional signal-space dimension, leading to Eq.\eqref{eq17}.

The key characteristic of this paradigm is that range-dependent behavior is embedded directly in the propagation phase through carrier-frequency gradients, rather than being reconstructed through temporal coding or receiver-side processing.

\subsection{Time-Coding-Driven Paradigm $\mathrm{I}$: EPC}
\label{subsec:epc}

EPC arrays \cite{qiu2023range,kan2025non,yu2023mainbeam,wang2025fast} represent a class of transmit architectures that synthesize structured transmit diversity by applying coding weights jointly across array elements and pulses. Instead of transmitting identical pulses from all elements, EPC introduces element-dependent coding coefficients across successive pulse repetition intervals, thereby coupling the spatial dimension (array aperture) with the slow-time dimension (pulse index).

\begin{figure}[htp]
\centering
\includegraphics[width=0.41\textwidth]{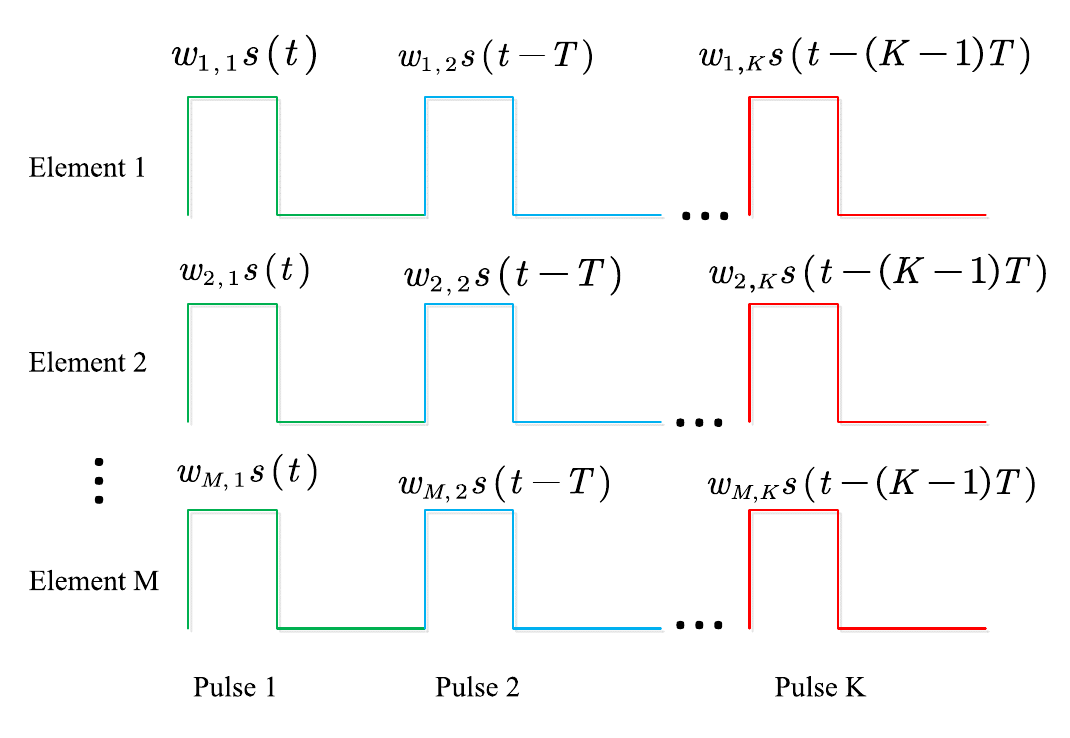}
\caption{\justifying Illustration of the EPC structure. 
Each array element transmits a sequence of coded pulses across successive pulse repetition intervals. 
The coding weights $w_{m,p}$ form an element–pulse coding matrix that couples the spatial and slow-time dimensions.}
\label{fig13}
\end{figure}
As shown in Fig.\ref{fig13}, in EPC architectures, the signal transmitted by the $m$-th element can be written as
\begin{equation}
s_m(t)
=
\sum_{p} w_{m,p} \, s(t-pT),
\end{equation}
where $p$ denotes the pulse index, $T$ is the pulse repetition interval, and $w_{m,p}$ represents the coding weight applied to the $p$-th pulse at the $m$-th element.
The coefficients $\{w_{m,p}\}$ form an element–pulse coding matrix that jointly controls the spatial and temporal structure of the transmitted signals. As a result, different array elements transmit differently weighted pulse sequences over time. This structure effectively expands the design space of the transmit array beyond the purely spatial aperture.

After propagation, the echoes corresponding to different pulses arrive at the receiver with different time delays. When matched filtering is applied with respect to the transmitted waveform $s(t)$, the echoes associated with different pulses are separated in the range domain. By properly designing the coding matrix $\{w_{m,p}\}$, the receiver can combine these delayed echoes to synthesize transmit responses that depend on both angle and range.
From a system-design viewpoint, EPC introduces an additional temporal degree of freedom beyond the spatial dimension, leading to the effective system DoF, namely

\begin{equation}
\mathrm{DoF}_{\mathrm{EPC}}
=
\mathrm{Space}
\oplus
\mathrm{Time}.
\end{equation}

It is important to emphasize that all transmit elements share the same carrier frequency. Consequently, the propagation kernel
$e^{-j2\pi f_c r/c}$
remains identical across array channels. No inter-element range-phase gradient difference is introduced during propagation. Therefore, the range-dependent behavior observed in EPC arrays arises primarily from pulse-domain processing and temporal reconstruction at the receiver rather than from transmit-side propagation-phase gradients.

\subsection{Time-Coding-Driven Paradigm $\mathrm{II}$: STCA}
\label{subsec:stca}

STCA represent another class of transmit architectures capable of producing range–angle dependent responses. Unlike FDA, STCA introduces element-dependent time delays across the transmit aperture, thereby coupling the spatial and temporal dimensions of the transmitted signals.

\begin{figure}[htp]
\centering
\includegraphics[width=0.41\textwidth]{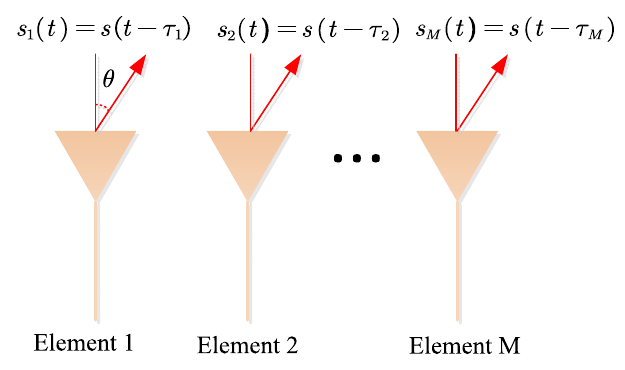}
\caption{\justifying Illustration of the STCA structure.
Each transmit element radiates a delayed replica of the same waveform,
$s_m(t)=s(t-\tau_m)$, where $\tau_m$ denotes the element-dependent time
delay. Unlike FDA that employ carrier-frequency
offsets, STCA introduces coupling between space and time through
relative time delays across the transmit aperture.}
\label{fig14}
\end{figure}

As illustrated in Fig.~\ref{fig14}, the signal transmitted by the $m$-th element can be expressed as
\begin{equation}
s_m(t)=s(t-\tau_m),
\end{equation}
where $\tau_m$ denotes the relative time delay applied to the $m$-th array element. Through these element-dependent delays, the transmitted waveform across the array aperture becomes jointly structured in the spatial and temporal domains.
More generally, STCA can also be described using a space--time coding formulation
\begin{equation}
s_m(t)=\sum_{p} a_{m,p}s_p(t),
\end{equation}
where $a_{m,p}$ denotes the space--time coding coefficient applied at the $p$-th time slot and $s_p(t)$ represents the transmitted waveform during that slot. This formulation highlights that STCA jointly exploits spatial weighting, temporal coding, and waveform diversity within a unified transmit framework.

Due to the element-dependent time delays, the transmitted steering vector becomes dependent on both range and angle. Consequently, STCA can produce range--angle coupled responses in the transmitted field.
From a system-design perspective, STCA introduces additional temporal and waveform dimensions beyond the spatial aperture, leading to the effective degrees of freedom, namely

\begin{equation}
\mathrm{DoF}_{\mathrm{STCA}}
=
\mathrm{Space}
\oplus
\mathrm{Time}
\oplus
\mathrm{Waveform}.
\end{equation}

It is worth noting that all transmit elements typically share the same carrier frequency. Therefore, the propagation kernel $e^{-j2\pi f_c r/c}$ remains identical across array channels. As a result, the range-dependent behavior observed in STCA systems primarily originates from space--time signal processing rather than from propagation-phase gradients across the transmit aperture.

\begin{table*}[htp]
\centering
\caption{\justifying Structural comparison of representative array paradigms in terms of their activated design variables.}
\begin{tabular}{c|c|c|c|c}
\hline
Paradigm & Frequency gradient & Temporal modulation & Waveform orthogonal & Physical range dependence \\
\hline
FDA & $\checkmark$ & $\times$ & $\times$ & $\checkmark$ \\
FDA--MIMO & $\checkmark$ & $\times$ & $\checkmark$ & $\checkmark$ \\
EPC & $\times$ & $\checkmark$ & $\times$ & Processing-induced \\
STCA & $\times$ & $\checkmark$ & $\checkmark$ & Processing-induced \\
\hline
\end{tabular}
\label{tab1}
\end{table*}

\subsection{Structural  Comparison of Array Paradigms}
\label{subsec:unified-table}

Although FDA, EPC, and STCA may produce similar range--angle dependent responses in certain signal-processing frameworks, the mechanisms that generate these behaviors originate from fundamentally different design dimensions. A structural comparison can therefore be made in terms of the design variables that shape the transmitted field.

Table~\ref{tab1} summarizes the representative paradigms and the corresponding design dimensions they activate. FDA introduces element-dependent carrier offsets, which create range-dependent phase gradients directly during electromagnetic propagation. As a result, the transmitted steering vector inherently depends on both range and angle.
In contrast, time-coding-based architectures such as EPC and STCA rely on temporal modulation across the transmit array. EPC achieves this through coding weights applied to successive pulses, while STCA introduces relative time delays or more general space--time coding structures across the array aperture. These mechanisms enable structured transmit diversity and can synthesize range--angle dependent responses after signal processing.
However, since all transmit elements typically share the same carrier frequency in these architectures, the propagation kernel remains identical across channels. Consequently, the observed range selectivity arises primarily from temporal coding and receiver-side reconstruction rather than from transmit-side propagation-phase gradients.

Therefore, among the paradigms summarized above, only frequency-gradient arrays (e.g., FDA and FDA--MIMO) introduce range-dependent structure directly in the physical propagation phase.

\section{FDA System Capability Mapping Framework}
\label{sec:capability-mapping}
While the previous sections focus on the structural properties of FDA, this section interprets these properties in terms of system capabilities.
Building upon the unified signal model and the manifold-expansion theory developed in the previous sections, this section establishes a {capability mapping framework} for FDA systems. The objective is to relate the underlying design variables of the transmit architecture to the system-level capabilities enabled by the expanded array manifold.
In particular, we interpret FDA capabilities through the structural relationship between design variables, the resulting propagation-phase structure, the physical degrees of freedom introduced in the array manifold, and the system capabilities that these degrees of freedom enable.

Under the unified signal representation, the phase of the $m$-th transmit channel can be written as
\begin{equation}
\phi_m(t,r,\theta)
=
\phi_m^{\mathrm{space}}
+
\phi_m^{\mathrm{freq}}
+
\phi_m^{\mathrm{time}}
+
\phi_m^{\mathrm{waveform}},
\end{equation}
where each component corresponds to a different design dimension of the transmit architecture. Different array paradigms activate different phase-gradient components, which determine how the transmit manifold expands in the joint space of range, angle, and time.

For FDA systems, the introduction of element-dependent carrier frequencies creates additional propagation-phase gradients that expand the array manifold along new physical dimensions. As a result, several distinctive capabilities emerge beyond those of conventional phased arrays.
In the following subsections, we analyze three representative capabilities enabled by this manifold expansion: \emph{range--angle selectivity}, \emph{time-evolving beam control}, and \emph{frequency-domain diversity}.

\subsection{Range--Angle Selectivity}
\label{subsec:cap_range_angle}

A fundamental capability introduced by FDA is the ability to generate transmit responses that depend jointly on range and angle. 
In conventional PA, the transmit steering vector depends only on the spatial direction, i.e., $\mathbf{a}(\theta)$, which leads to a one-dimensional angular response.
In contrast, FDA introduces element-dependent carrier frequencies, which expand the transmit manifold into a joint range--angle domain,
$\mathbf{a}_{\mathrm{FDA}}(r,\theta)$.
As a result, the array response becomes inherently two-dimensional in $(r,\theta)$, enabling selective focusing or interference control across both dimensions.

The origin of this capability lies in the range-dependent propagation phase associated with element-dependent carrier frequencies. 
When frequency offsets $\Delta f_m$ are introduced across array elements, the propagation phase exhibits different range-phase gradients across channels. 
These gradient differences produce controllable interference structures along the range dimension during electromagnetic propagation. 
Consequently, the transmitted field no longer forms a purely angular beampattern, but instead generates a range--angle dependent response whose structure can be shaped through the frequency-gradient design.

An additional consequence of this frequency-gradient structure is the emergence of a finer discrimination scale along the range dimension. 
When multiple transmit elements operate with slightly different carrier frequencies, the superposition of their radiated fields forms a periodic interference pattern in range. 
Under coherent conditions, this structure can provide a secondary range discrimination capability beyond the conventional bandwidth-limited resolution \cite{li2024knowledge,huang2023adaptive}.
For example, when a uniform frequency increment $\Delta f$ is applied across $M$ transmit elements, the effective sub-range discrimination scale can be approximated as \cite{huang2022adaptive,gui2020generalized,huang2022bayesian}
\begin{equation}
\Delta r_{\mathrm{sub}} \approx \frac{c}{2M\Delta f},
\end{equation}
which corresponds to the spacing between adjacent constructive interference regions generated by the multi-frequency transmit aperture.

In practice, the effectiveness of this range--angle selectivity depends on several factors. 
First, excessively large frequency offsets may introduce inter-element decorrelation or violate the narrowband assumption. 
Second, coherent combination of the multi-frequency components is required in order to preserve the interference structure. 
Finally, propagation environments with strong multipath or heterogeneous clutter may distort the range--angle coupling pattern.

Overall, range--angle selectivity represents one of the most fundamental capabilities enabled by the frequency-gradient architecture of FDA, reflecting the expansion of the array manifold into the joint range--angle domain.

\subsection{Time-Evolving Beam Control}
\label{subsec:cap_time_beam}

Another distinctive capability of FDA is the ability to generate transmit responses that evolve over time. 
In conventional PA, the transmit beampattern is typically time-invariant once the spatial weights are fixed. 
In contrast, FDA introduces element-dependent carrier frequencies, which cause the relative phases across the array aperture to change continuously with time. 
As a result, the resulting beampattern becomes inherently time-varying, producing dynamic beam trajectories in the joint range--angle domain.
The time evolution of the FDA beampattern originates from the temporal phase gradient introduced by the frequency offsets across array elements. 
Because each element transmits with a slightly different carrier frequency, the inter-element phase differences vary continuously with time. 
This temporal phase evolution modifies the interference structure of the transmitted field, leading to dynamic redistribution of radiated energy in space.

Consequently, time becomes an intrinsic dimension of the transmit manifold, and the resulting beam structure can evolve along both the range and angular directions.
The apparent beam dynamics depend on the relationship between the pulse duration and the frequency offsets applied across the array.

\begin{itemize}

\item \textbf{Small Frequency Offsets:}  
When the product of pulse duration and frequency offset is small, the beampattern can be regarded as approximately static within a single pulse.

\item \textbf{Larger Frequency Offsets:}  
When the temporal phase evolution becomes significant over the pulse duration, the beam structure may vary noticeably within a pulse, leading to continuous scanning or drifting effects in the range--angle domain.

\end{itemize}

The time-varying nature of FDA beams also introduces several practical considerations:

\begin{itemize}

\item Perfect time-invariant focusing cannot be maintained when nonzero frequency offsets are present.
\item The apparent beam trajectory depends on both waveform duration and frequency-gradient design.
\item Receiver-side processing is often required to reconstruct stationary responses for parameter estimation.

\end{itemize}

Overall, the time-evolving beam behavior reflects the intrinsic temporal phase gradients introduced by frequency diversity, distinguishing FDA from conventional arrays with static spatial beampatterns.
\subsection{Frequency-Domain Diversity}
\label{subsec:cap_freq_diversity}

Another important capability introduced by frequency diverse arrays is frequency-domain diversity. 
Because different transmit elements operate at slightly different carrier frequencies, the radiated field can be interpreted as a superposition of multiple narrowband components distributed across the frequency domain.
Compared with conventional phased arrays, where all elements share a common carrier frequency, FDA effectively creates a distributed multi-frequency transmit aperture. 
This structure introduces additional diversity in the frequency domain and expands the available signal space for sensing and communication tasks.
The origin of this capability lies in the element-dependent carrier frequencies
$f_m$
which produce distinct spectral components across the transmit aperture. 
As a result, the transmitted field contains multiple frequency-dependent responses that propagate simultaneously through the environment.

These frequency components experience different propagation phases and scattering responses, providing additional diversity beyond the purely spatial domain.
The frequency diversity introduced by FDA enables several useful signal-processing mechanisms:

\begin{itemize}

\item \textbf{Multi-Frequency Observation:}  
Different carrier frequencies probe the scene with slightly different propagation characteristics, providing additional independent observations.

\item \textbf{Frequency-Dependent Interference Shaping:}  
The superposition of multiple frequency components allows the spatial–spectral structure of the transmitted field to be adjusted through the design of frequency offsets.

\item \textbf{Enhanced Parameter Discrimination:}  
Frequency diversity improves the separability of targets or channels that may be indistinguishable in purely spatial processing.

\end{itemize}

The additional diversity provided by multiple carrier frequencies can be exploited in several system-level tasks, such as enhanced parameter estimation, interference mitigation, and joint sensing–communication waveform design. 
However, the achievable benefits depend on maintaining coherence across the array elements and on the appropriate selection of frequency offsets relative to the system bandwidth and propagation conditions.

Overall, frequency-domain diversity represents another fundamental capability enabled by the frequency-gradient architecture of FDA, complementing the range--angle selectivity and time-evolving beam behaviors discussed in the previous subsections. This capability plays an important role in modern sensing and integrated sensing–communication (ISAC) systems where multi-frequency observations provide additional information for joint environment perception and data transmission.

\subsection{Capability Mapping Summary}
\label{subsec:cap_summary}

The above discussions reveal that the distinctive capabilities of FDA systems originate from the additional phase gradients introduced by element-dependent carrier frequencies. 
These gradients expand the array manifold beyond the purely spatial dimension, enabling new forms of signal diversity and beam control. 
The capability mapping discussed above can be summarized in Table~\ref{tab:capability_mapping}.
\begin{table*}[htbp]
\centering
\caption{\justifying Capability mapping of FDA systems based on propagation phase gradients and manifold expansion.}
\begin{tabular}{c|c|c|c}
\hline
Capability & Design Variable & Phase-Gradient Origin & Manifold Expansion \\
\hline
Range--Angle Selectivity 
& Frequency offsets $\Delta f_m$ 
& Range phase gradient 
& Range--angle manifold \\

Time-Evolving Beam Control 
& Frequency offsets $\Delta f_m$ 
& Temporal phase gradient 
& Time--range manifold \\

Frequency-Domain Diversity 
& Multi-frequency transmit aperture 
& Spectral diversity 
& Spectral manifold \\
\hline
\end{tabular}
\label{tab:capability_mapping}
\end{table*}
The key contribution of FDA is not merely the introduction of frequency offsets, but the expansion of the array manifold through additional propagation-phase gradients. 
These gradients create new physical degrees of freedom that enable capabilities beyond those of conventional phased arrays.

In practice, these capabilities are realized through concrete signal and array design strategies, including frequency-offset optimization, subarray architectures, and array-geometry design. 
A large portion of the FDA literature has approached the problem from the perspective of transmit beampattern synthesis, with the goal of shaping the range--angle-dependent radiation pattern induced by the frequency-gradient structure. 
From this viewpoint, such design methods form the mechanism-level bridge between the physical capabilities summarized above and the radar task families discussed next.

\section{FDA Applications Across Radar Task Families}
\label{sec:task-family}

Building on the capability mapping developed in the previous section, we now examine how the structural properties of FDA translate into performance gains across representative radar task families. Rather than reviewing the literature chronologically, we adopt a task-oriented perspective and focus on the relationship among the task objective, the required physical degrees of freedom, the structural mechanisms introduced by FDA, and the practical conditions under which these mechanisms can be effectively exploited.

The applications discussed in this section can be interpreted through three core FDA capabilities established earlier: range--angle manifold expansion, time-evolving beam behavior, and multi-frequency transmit diversity. Different radar tasks draw on one or more of these capabilities in different ways. This perspective helps clarify not only where FDA can provide genuine advantages, but also why such advantages may disappear when the underlying structural effects cannot be coherently preserved or effectively utilized.

\subsection{Parameter Estimation and Target Localization}
\label{subsec:task_estimation}

Target parameter estimation and localization represent one of the most
fundamental tasks in radar systems. In conventional PA
radar, the transmit steering vector depends only on the angular
direction, and range estimation is typically performed independently
through matched filtering. In contrast, FDA introduces element-dependent carrier frequencies,
which expand the transmit manifold into a joint range--angle domain.
This capability corresponds to the range--angle manifold expansion
identified in the capability mapping framework, and provides a
structural basis for joint parameter estimation and target localization. As a result, the received signal model can be written as
\begin{equation}
\mathbf{y} = \alpha\, \mathbf{a}_{\mathrm{FDA}}(R_0,\theta) + \mathbf{n}.
\end{equation}
This additional manifold dimension provides
new opportunities for joint parameter estimation and improved target
localization performance
\cite{wang2013range,xu2015joint,chen2018space,kan2025joint}.
However, the range--angle coupling introduced by FDA also complicates parameter estimation, since the conventional separation between range processing and spatial processing no longer strictly holds. 
To address this challenge, a variety of signal processing strategies have been developed to either decouple the parameters or directly perform joint estimation \cite{zhong2024multiparameter,xin2025joint}.

One representative class of approaches exploits cooperative or hybrid architectures that combine conventional PA signals with FDA signals. 
For example, dual-pulse schemes have been proposed in which one pulse operates in the conventional PA mode to estimate the target angle, while the other pulse employs FDA signals to estimate range using the known angular information \cite{wang2013range}. 
This idea has been further extended to FDA--MIMO radar configurations to improve parameter identifiability \cite{khan2015double}. 
Related cooperative schemes have also been proposed where different portions of the transmit array operate in PA and FDA modes, enabling sequential estimation of angle and range parameters \cite{zhu2021cooperative}.

Another widely studied class of approaches relies on subarray-based FDA structures. 
In such architectures, coherent signals are transmitted within each subarray while different subarrays employ distinct frequency offsets or orthogonal waveforms. 
By exploiting the resulting spatial–frequency diversity, joint range--angle estimation can be performed using beamforming or subspace-based signal processing techniques \cite{wang2014transmit,wang2014subarray,wang2018subarray,xu2015joint}. 
Analytical studies have also investigated the theoretical estimation performance of FDA--MIMO systems in terms of the Cramér–Rao lower bound (CRLB) and mean-square error (MSE) properties \cite{Xiong2018FDAMIMO}.

Despite the large number of algorithms proposed in the literature, several challenges remain. 
First, the parameter coupling introduced by frequency offsets increases computational complexity and may degrade the performance of classical subspace estimators when the frequency gradient becomes large. 
Second, many existing algorithms rely on multiple snapshots or high signal-to-noise ratios, which may not always be available in practical scenarios. 
To address these limitations, recent research has explored more robust estimation frameworks, including compressive sensing and tensor-decomposition-based approaches, which can reduce the required number of snapshots and improve robustness to noise and model mismatch \cite{fu2020CS,Kolda2009TensorReview,Guo2011Parafac}.
Furthermore, extending FDA parameter estimation frameworks to more
complex array configurations, such as planar arrays, coprime arrays,
conformal arrays, multistatic radar systems, or polarization-diverse
systems, introduces additional parameters and coupling relationships
that remain active research topics \cite{Ni2021RangeDependent,qin2016frequency,cui2018search,li2018successive}.

Overall, although a large number of algorithms have been developed for
FDA parameter estimation \cite{ma2026multiparameter,wang2025frequency,kan2025joint}, most existing studies focus on linear array
configurations and idealized signal models. Extending these techniques
to more practical scenarios therefore remains an important direction
for future research.

\subsection{Target Detection in Complex Environments}
\label{subsec:task_detection}

\subsubsection{ Detection Model and FDA Structural Effect}

Target detection is one of the most fundamental functions of radar
systems. The detection problem is commonly formulated as a binary
hypothesis test, namely
\begin{equation}
\begin{cases}
\mathcal{H}_0: \mathbf{y} = \mathbf{n},\\
\mathcal{H}_1: \mathbf{y} = \alpha \mathbf{a}(R_0,\theta) + \mathbf{n},
\end{cases}
\end{equation}
where $\mathbf{y}$ denotes the received signal vector, $\alpha$ is the
target reflection coefficient, and $\mathbf{n}$ denotes disturbance
including thermal noise, clutter, and interference.

For multichannel radar systems, target detection is commonly addressed
within the framework of binary hypothesis testing, where classical
detectors such as the generalized likelihood ratio test (GLRT),
adaptive matched filter (AMF), Rao test, and Wald test have been
extensively studied
\cite{Kay1993Fundamentals2,Kelly1986AnAdaptiveDetection,Robey1992ACFARadaptive,
DeMaio2004Anewderivation,DeMaio2007RaoTestforAdaptive,Lan2020GLRTbasedAdaptive,LanRosamilia2022AdaptiveTargetDetection}.
In conventional adaptive radar, many detectors rely on secondary or
training data to estimate the disturbance covariance matrix and achieve
constant false alarm rate (CFAR) performance in unknown environments
\cite{chong2010mimo}. At the same time, training-free or reduced-training
detection strategies \cite{Liu2015Adaptivedetectionwithout,Liuliu2016Performancepredictionofsubspacebased,liu2018distributed} have also been developed for scenarios where
homogeneous secondary data are unavailable or insufficient.
Overall, detection performance depends not only on the detector
structure itself, but also on the availability of prior information \cite{Liu2018BayesianDetection,LiYang2017TwostepBayesian},
training support, and the accuracy of the assumed signal model \cite{ConteDeMaio2003Distributedtargetdetection,XuLi2008Targetdetection,Liu2015Persymmetricadaptivetarget}.

When FDA is introduced, the received signal structure becomes
explicitly range-dependent due to the element-dependent carrier
frequencies. This property is consistent with the
range--angle manifold expansion discussed in the previous section and
fundamentally modifies the statistical structure of the received data.
Meanwhile, the use of multiple carrier frequencies across the transmit
aperture introduces an inherent {multi-frequency transmit diversity},
which provides additional spectral degrees of freedom beyond conventional
spatial processing.

As a result, the clutter and interference characteristics observed by
FDA radar can differ from those of conventional arrays, creating new
opportunities for target detection in complex environments such as
range-ambiguous clutter, deceptive jamming, and high-speed target
scenarios where Doppler--frequency coupling can be effectively exploited.
\vspace{0.5em}
\subsubsection{ Detection in Range-Ambiguous Clutter}

In high-PRF airborne radar systems, clutter echoes from different
range intervals may fold into the same observation cell, resulting in
range-ambiguous clutter. Conventional PA radar mainly relies on angular
discrimination or space--time adaptive processing (STAP) to suppress
such clutter \cite{sun2025nonlinear,xu2016space,liu2023range}. However,
these approaches operate primarily in the spatial or space--time domain
and do not explicitly exploit range-dependent structure.
In contrast, FDA enables joint range--angle beampattern shaping due to
its element-dependent carrier frequencies, allowing the radiated field
to exhibit controllable gain variations across range. This additional
degree of freedom can be exploited to suppress ambiguous clutter
originating from undesired range regions by placing nulls or low-gain
zones at specific ranges \cite{LiuZhu2023RangeAmbiguous,WangZhu2021RangeAmbiguousClutter}.

Building upon this property, FDA--MIMO systems further extend the
processing capability by employing orthogonal transmit waveforms,
which enable reconstruction of the transmit aperture at the receiver.
This facilitates joint space--time--range adaptive processing for
clutter suppression and target detection \cite{xu2015range,xu2016space,
Xu2017AnAdaptive,qiu2024range,wang2021range}. 
Moreover, several studies have demonstrated that FDA-based processing
can significantly improve the detection performance of moving targets
in range-ambiguous clutter environments \cite{GuiWang2021FDAradar,LIU2023103942,liuCooperatedMovingTarget2023},
highlighting the effectiveness of exploiting range-dependent diversity
in challenging scenarios.

\vspace{0.5em}
\subsubsection{Detection Under Deceptive Jamming}

FDA radar exhibits inherent advantages in countering deceptive jamming,
particularly in mainlobe scenarios where conventional angle-domain
discrimination becomes ineffective. In classical PA radar, deceptive
signals generated by repeaters can closely match the spatial steering
vector of the true target, making them difficult to distinguish and
suppress using spatial processing alone.
In contrast, due to the element-dependent carrier frequencies, FDA
introduces a range-dependent phase structure, leading to a
range--angle coupled manifold. As a result, false targets generated by
repeaters generally cannot simultaneously match both the range and
angle characteristics of the true FDA echo, unless precise knowledge
of the frequency offsets and target parameters is available. This
inherent mismatch provides an additional discrimination mechanism
beyond conventional spatial processing.

Moreover, the multi-frequency transmit diversity of FDA further
enhances robustness against deceptive jamming. Specifically, the
frequency-dependent echo structure introduces additional constraints
on the jammer, making it difficult to coherently replicate the full
signal model across the aperture, especially under dynamic or
partially known conditions \cite{huang2024fda}. As such, FDA transforms
the deceptive jamming suppression problem from a purely spatial
discrimination task into a joint range--angle--frequency inference
problem.

\begin{figure}[htp]
\centering
\subfigure[MIMO ($\Delta f = 0$)]{
\includegraphics[width=0.41\textwidth]{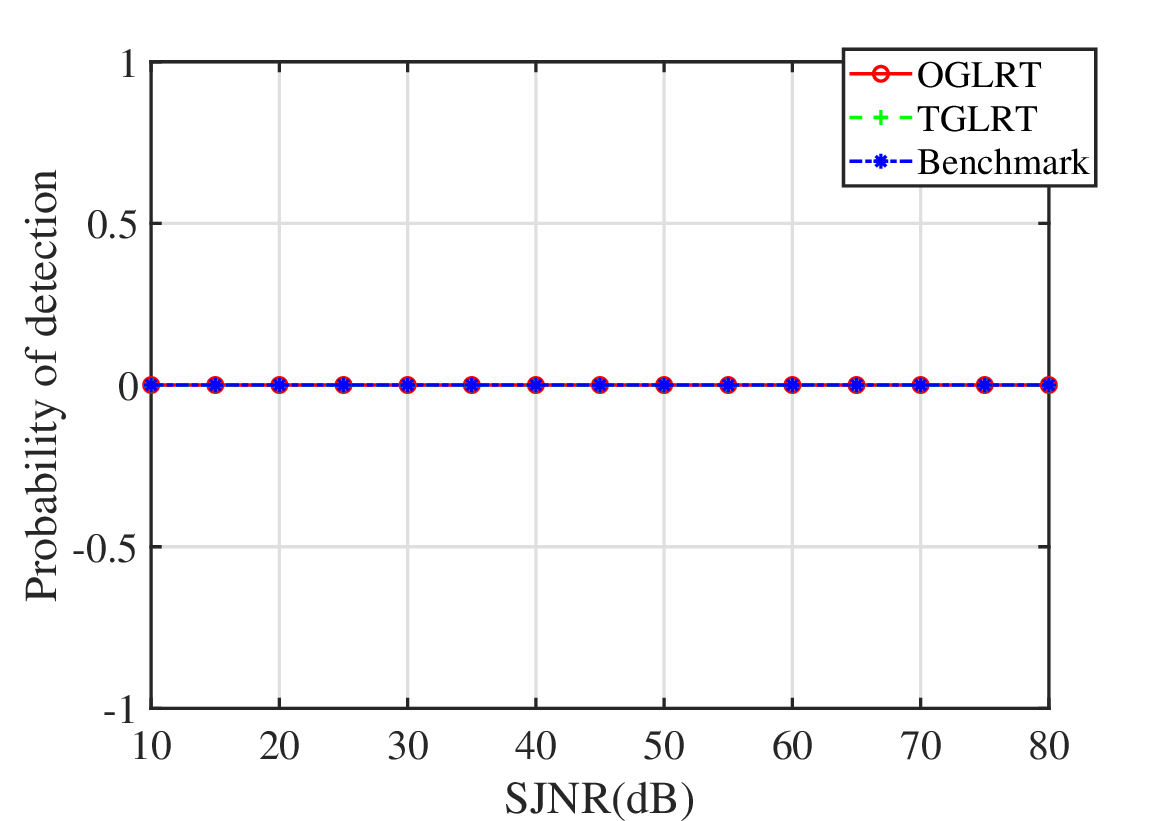}}
\subfigure[FDA-MIMO ($\Delta f \neq 0$)]{
\includegraphics[width=0.41\textwidth]{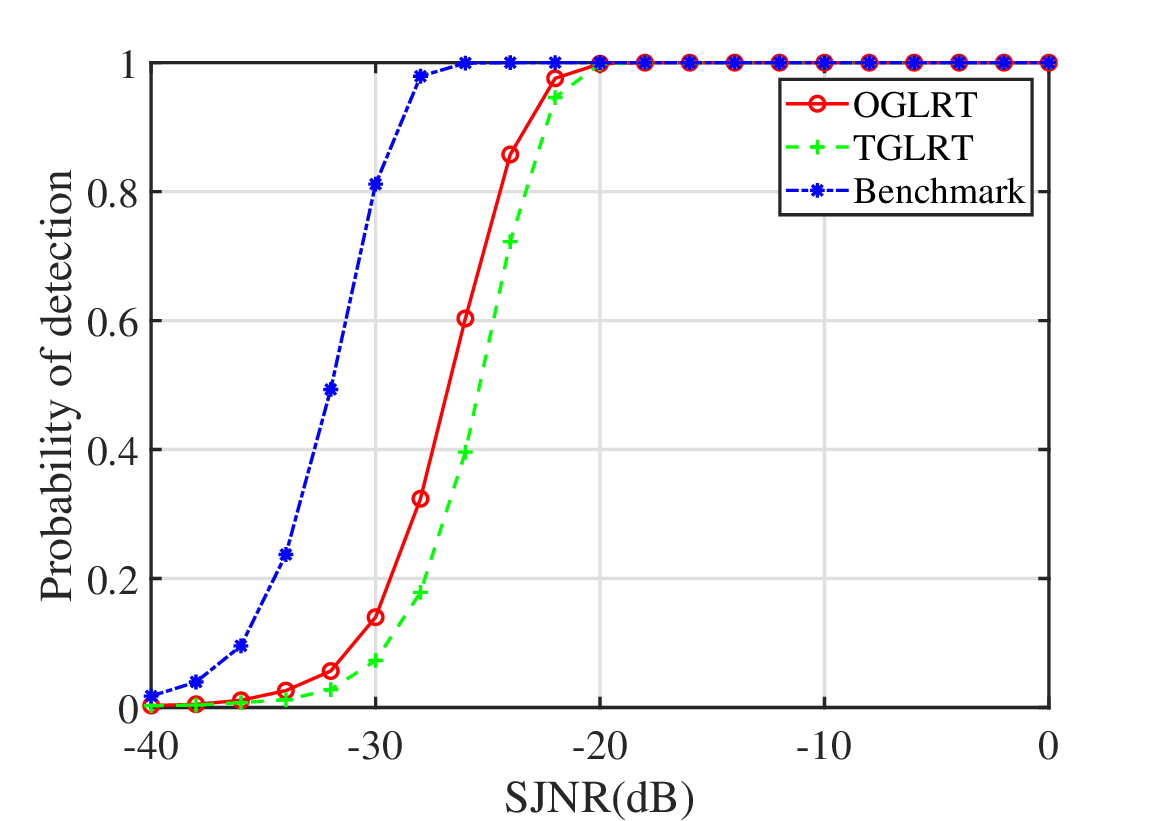}}
\caption{\justifying Detection performance under mainlobe deceptive jamming \cite{huang2025glrt}.
When $\Delta f = 0$, corresponding to conventional MIMO radar,
all detectors fail to achieve reliable detection.
In contrast, when $\Delta f \neq 0$, FDA enables effective target
detection by exploiting range-dependent diversity, even in the
presence of strong mainlobe deceptive jamming.
}
\label{fig15}
\end{figure}

Building upon these properties, adaptive detection schemes based on
GLRT, Rao, and Wald tests have been extended to FDA--MIMO systems under
deceptive and suppression jamming scenarios
\cite{Gui2020Low,HuangBasit2022AdaptiveMoving,hehuang2026parametricGLRT}. In particular,
GLRT-based adaptive detectors have been developed for FDA--MIMO radar
operating in mainlobe deceptive jamming environments, where the jammer
and target share similar angular signatures but differ in their
range-dependent structures \cite{huang2025glrt,guihuang2025robustdetector}. Extensions to
partially homogeneous noise environments have also been investigated,
demonstrating robustness under practical conditions
\cite{huang2024adaptive}.
To further highlight the fundamental role of frequency diversity, we
consider the limiting case where $\Delta f = 0$, under which waveform-orthogonality FDA-MIMO
degenerates to conventional MIMO radar. The corresponding detection
performance under mainlobe deceptive jamming is illustrated in
Fig.~\ref{fig15}. It is observed that, when $\Delta f = 0$, all
detectors fail to achieve reliable detection, whereas when
$\Delta f \neq 0$, FDA enables effective target detection by
exploiting range-dependent diversity.

This result clearly indicates that FDA does not merely improve
detection performance, but fundamentally enables reliable target
detection in scenarios where conventional MIMO radar becomes
ineffective.
Furthermore, joint target--jammer detection frameworks have been
proposed to simultaneously identify desired targets and deceptive
signals within the mainlobe \cite{Zhu2023SimultaneousDetection}. These
methods exploit the additional degrees of freedom introduced by FDA to
enable more effective separation of target and jammer components,
further improving detection performance in challenging electronic
countermeasure scenarios.

\vspace{0.5em}
\subsubsection{Detection for High-Speed Target }

High-speed or maneuvering targets introduce significant challenges for
radar detection due to Doppler ambiguity, Doppler spreading, and range
migration effects. In conventional radar systems, target motion induces
a Doppler shift $f_d = \frac{2v}{c}f_c$, which can be effectively
handled when the velocity is moderate. However, for high-speed targets,
Doppler ambiguity and spectral spreading degrade coherent integration
performance and reduce detection sensitivity.
In FDA radar, the presence of element-dependent carrier frequencies
introduces an additional Doppler-related component. Specifically, the
Doppler shift becomes
\begin{equation}
f_{\mathrm{obs}} = \frac{2v}{c}(f_c + \Delta f_m),
\end{equation}
which can be decomposed as a conventional Doppler term plus an
FDA-induced Doppler expansion term. Although $\Delta f_m \ll f_c$,
these additional Doppler components introduce phase accumulation
errors across pulses and array elements. As pointed out in
\cite{jia2025long}, neglecting these effects leads to defocusing
of the target response in the range--angle--Doppler domain, thereby
degrading detection performance.

Unlike conventional systems where Doppler effects are treated purely
as impairments, FDA provides an additional degree of freedom to exploit
these frequency-dependent Doppler variations. The induced Doppler
expansion and phase diversity can be leveraged to enhance target
identifiability and improve detection performance, especially for
high-speed or maneuvering targets.
Building upon this property, several FDA-based detection frameworks
have been developed. For example, Doppler-spread-aware detection
methods have been proposed to account for motion-induced spectral
broadening \cite{jia2025fda}. Long-time coherent integration
and joint range--angle beamforming techniques have been designed to
compensate for motion-induced phase errors and improve detection
sensitivity \cite{jia2025long}. FDA has also been applied to
resolve Doppler ambiguity of high-speed targets by exploiting its
multi-frequency structure \cite{wan2021resolving}.

\begin{figure}[htp]
\centering
\subfigure[SCNR loss versus normalized Doppler.]{
\includegraphics[width=0.41\textwidth]{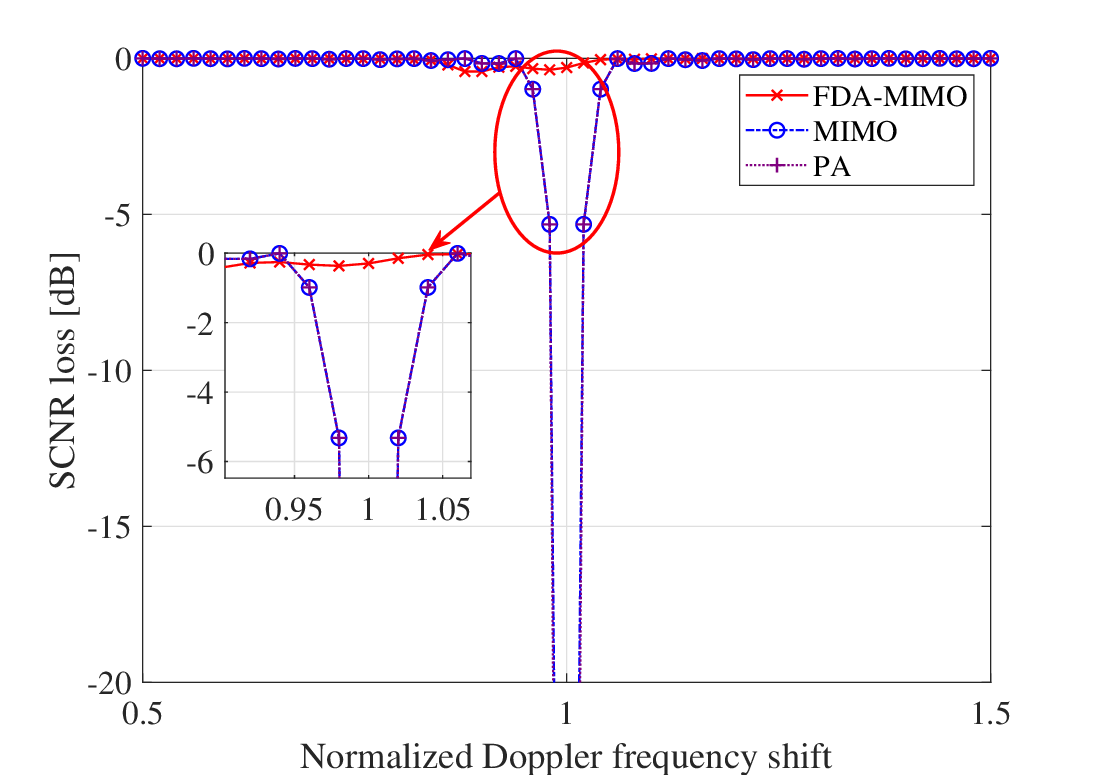}}
\subfigure[Detection probability versus input SNR]{
\includegraphics[width=0.41\textwidth]{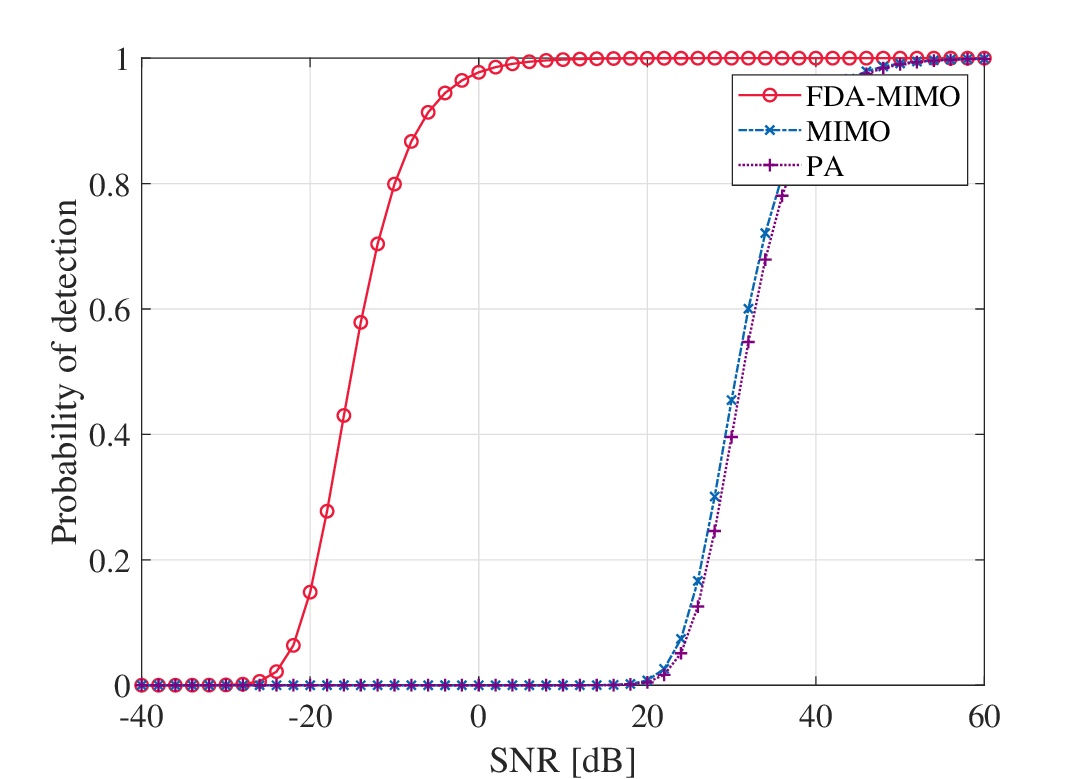}}
\caption{\justifying  High-speed target detection performance under Doppler spreading \cite{gui2021fda}.
}
\label{fig16}
\end{figure}

To illustrate the impact of Doppler ambiguity in high-speed target detection,
Fig.~\ref{fig16}(a) shows the SCNR loss versus normalized Doppler frequency.
It can be observed that conventional MIMO and PA radar suffer from severe SCNR degradation around the Doppler-ambiguous region (i.e., $f_d \approx \mathrm{PRF}$), where deep notches appear due to blind Doppler.
In contrast, FDA--MIMO effectively mitigates this degradation and maintains a relatively stable SCNR level, demonstrating its inherent robustness against Doppler ambiguity.
This advantage directly translates into improved detection performance.
As shown in Fig.~\ref{fig16}(b), FDA--MIMO achieves a significantly higher detection probability than conventional MIMO and PA, especially in low-SNR regimes.
Furthermore, FDA enables effective detection of maneuvering targets
in complex environments. Detection frameworks incorporating Doppler
spread and range migration have been developed for FDA--MIMO radar
systems \cite{liu2022detecting}. In the presence of mainlobe
clutter or deceptive jamming, FDA-based methods can jointly exploit
frequency diversity and motion-induced signatures to improve
detection robustness \cite{gui2021fda}.

Overall, FDA transforms high-speed target detection from a conventional
Doppler processing problem into a joint range--angle--frequency--Doppler
inference problem. This additional structural diversity provides new
opportunities for detecting high-speed and maneuvering targets in
challenging scenarios.

\vspace{0.5em}
\subsubsection{Detection in Secondary Range Cells}

A fundamental yet often overlooked phenomenon in FDA systems
stems from the inherent range--frequency coupling introduced by
element-dependent frequency offsets. From the perspective of the
ambiguity function, FDA does not produce a single dominant
range mainlobe as in conventional phased-array radar. Instead,
it generates a set of periodic range responses, with peaks located at
\begin{equation}
\Delta R_{\mathrm{peak}} = \ell \frac{c}{2\Delta f},
\end{equation}
where $\ell$ is an integer index.
This periodic structure reflects a transmit-induced
{range super-resolution} capability, whereby multiple
distinguishable responses can be formed across different range
locations. As a consequence, the effective degrees of freedom
in the range domain are no longer solely limited by the signal
bandwidth, but are further enriched by the frequency diversity
across the transmit aperture.

From a detection perspective, however, this structural property
fundamentally alters the conventional interpretation of a range cell.
In classical radar systems, each range cell is typically modeled as
containing a single dominant scatterer. In contrast, due to the
presence of secondary peaks, an FDA range cell may correspond to
multiple effective scattering components, which can be viewed as a
{distributed target structure} induced by the waveform \cite{ZhuZhu2022AdaptiveMultiTarget}.
This shift in the signal model has two important implications.
First, the coexistence of multiple peaks may introduce ambiguities,
especially when secondary responses overlap with clutter or
interference. To address this issue, various ambiguity mitigation
strategies have been developed, including the use of nonlinear
frequency offsets, multi-pulse diversity, and adaptive processing
techniques, which aim to decorrelate or suppress undesired
secondary peaks
\cite{GuiWang2018Generalreceiver,Lang2022LambWave,liu2025jointdoa}.
Second, and more importantly, these multiple responses can be
actively exploited to enhance detection performance.
By interpreting the secondary range peaks\footnote{A broader open issue concerns whether the conventional range-cell model should continue to serve as the primary interpretive framework for FDA/FDA--MIMO systems. If secondary range peaks are directly incorporated into target detection rather than regarded merely as byproducts of range compression, it remains unclear whether they may support a different form of range discrimination beyond the classical resolution limit.} as structured
multi-component signals, recent works have developed
distributed detection frameworks that leverage the additional
degrees of freedom provided by FDA. Such approaches enable the
detection of multiple targets within a single conventional range cell,
or equivalently, improve detection performance in scenarios where
the target exhibits extended or multi-scattering characteristics
\cite{HuangJian2022AdaptiveDistributed,Huang2023Adaptivemultiple,
huang2022bayesian,li2024knowledge,lihuang2025adaptivedetection}.

\begin{figure}[htp]
\centering
\includegraphics[width=0.41\textwidth]{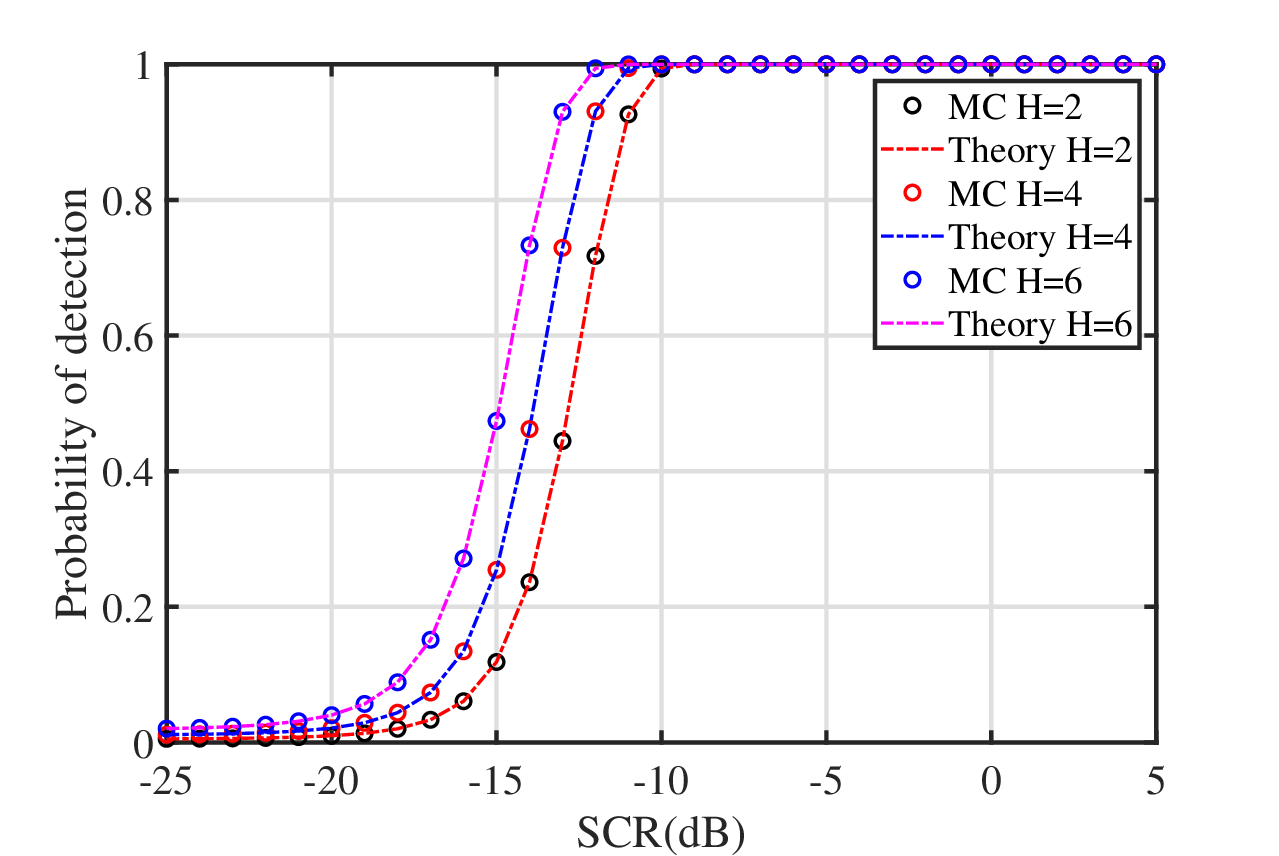}
\caption{\justifying 
Probability of detection versus SCR for different numbers of
scattering components within a single range cell \cite{HuangJian2022AdaptiveDistributed}.
Here, $H$ denotes the number of scatterers (i.e., effective peaks)
within the cell, and ``MC'' represents Monte Carlo simulations.
The results show that the detection performance is strongly affected
by the intra-cell scattering structure, highlighting the impact of
distributed targets in FDA secondary range cell detection.
}
\label{fig17}
\end{figure}

To illustrate this effect, Fig.~\ref{fig17} shows the detection
probability as a function of the signal-to-clutter ratio (SCR) for
different numbers of scattering components within a single range cell.
Here, $H$ represents the number of effective scatterers, which can
be interpreted as the number of FDA-induced peaks contributing to
the received signal. It can be observed that increasing $H$ leads
to improved detection performance, as multiple components
contribute constructively to the received energy.
This observation highlights a key distinction between FDA and
conventional radar systems: while secondary range responses are
traditionally regarded as ambiguity, in FDA they constitute a
structured source of diversity that enables a transition from
single-target detection to distributed detection. As such, the
secondary range cells in FDA should not be viewed solely as a
limitation, but rather as a fundamental structural feature that can
be either mitigated or exploited depending on the system design
objective.

\vspace{0.5em}
\subsubsection{Challenges and Opportunities}

Despite its significant potential, FDA-based detection also
faces several fundamental challenges that stem from its
range--angle--time coupled signal structure.

\emph{1) Non-stationary and range-dependent environments:}
The range-dependent and non-stationary nature of clutter in FDA
systems complicates the construction of homogeneous training
data, which is critical for conventional adaptive detection.
The violation of the stationarity assumption poses a major
challenge for covariance estimation and limits the effectiveness
of classical STAP-like methods.

\emph{2) High-dimensional signal processing:}
The expanded signal dimensionality introduced by FDA, while
beneficial for detection performance, leads to increased
computational complexity. Joint processing across range, angle,
and Doppler dimensions requires high-dimensional matrix
operations, which may hinder real-time implementation.

\emph{3) Hardware and synchronization constraints:}
FDA relies on precise control of element-dependent frequency
offsets and strict phase coherence across transmit channels.
Hardware impairments and synchronization mismatches may distort
the intended interference structure, degrading performance.

\emph{4) Ambiguity--diversity trade-off:}
The secondary range responses in FDA introduce an inherent
trade-off between ambiguity suppression and diversity
exploitation, making system design more challenging.

\vspace{0.5em}
\noindent\textbf{Opportunities:}

While these challenges present practical limitations, they also
reveal several promising research opportunities.

\emph{1) Exploiting Doppler--frequency coupling for high-speed targets:}
FDA inherently couples Doppler and frequency, providing a new
mechanism to mitigate Doppler ambiguity and Doppler spreading.
Fully exploiting this property for maneuvering and high-speed
targets, especially under range migration, remains an open
direction.

\emph{2) Detection in heterogeneous and non-Gaussian environments:}
The non-stationary nature of FDA signals suggests that
conventional homogeneous Gaussian assumptions are insufficient.
Developing detection frameworks for non-homogeneous clutter,
range-dependent interference, and non-Gaussian disturbances is
a critical research direction.

\emph{3) Distributed detection and structured signal modeling:}
The secondary range responses in FDA can be interpreted as
structured multi-component signals within a single range cell.
This enables a transition from single-target detection to
distributed detection, opening new possibilities for exploiting
intra-cell diversity.

\emph{4) Emerging applications in low-altitude scenarios:}
FDA provides additional degrees of freedom for distinguishing
targets in complex environments, which is particularly promising
for low-altitude economy (LAE) scenarios involving low-slow-small (LSS)
targets. Leveraging FDA for reliable detection in such highly
dynamic and cluttered environments is an important future
direction.

Overall, these observations suggest that FDA not only introduces
new challenges, but also fundamentally expands the design space
of radar detection, calling for a unified co-design of waveform,
hardware, and signal processing tailored to application-specific
requirements.
\vspace{0.5em}
\subsubsection{ Summary of FDA Detection Advantages}

In summary, FDA introduces a fundamentally different detection
paradigm by extending the conventional spatial processing
framework to a joint range--angle--time domain.
Compared with traditional phased-array and MIMO radar systems,
FDA offers several distinctive advantages:

\begin{itemize}
\item \emph{Expanded Degrees of Freedom:} The introduction of
frequency diversity enables additional controllable dimensions
in the transmit signal, providing enhanced capability for
clutter and interference suppression.

\item \emph{Intrinsic discrimination capability:} The
range-dependent phase structure allows FDA to distinguish
targets from deceptive jamming and interference that share
similar angular signatures but differ in range characteristics.

\item \emph{Robustness in ambiguous environments:} FDA can
effectively mitigate performance degradation in range-ambiguous
or Doppler-spread scenarios by exploiting its structured
signal diversity.

\item \emph{Distributed detection capability:} The presence of
secondary range responses enables a transition from
single-target detection to distributed detection, allowing
multiple scattering components to be resolved and exploited
within a single conventional range cell.
\end{itemize}

Overall, FDA shifts the role of ambiguity from a limiting factor
to a structured source of diversity, thereby opening new
opportunities for target detection in complex and contested
environments. This paradigm shift positions FDA as a promising
candidate for next-generation radar systems.

\subsection{Anti-Deception and Anti-Jamming}
\label{subsec:task_antijam}

Electronic counter-countermeasure (ECCM) is one of the most
compelling application domains of FDA and FDA--MIMO radar.
In particular, FDA exhibits a fundamental advantage in
countering {mainlobe deceptive jamming}, a scenario
where conventional phased-array and MIMO radars are known
to fail.

\textbf{Mainlobe Deceptive Jamming:}
Moreover, the robustness of FDA against mainlobe deceptive
jamming can be understood from the perspective of
{signal separability at the receiver}. Unlike conventional
radar systems, where target and jammer signals share the same
angle-dependent steering vector within the mainlobe, FDA
introduces an additional range-dependent dimension into the
signal manifold.
As a result, even when a jammer lies within the mainlobe and
matches the angular signature of the target, the corresponding
received signals remain distinguishable due to their different
ranges. This range-dependent structure enables joint
range--angle processing, allowing the receiver to separate
or suppress deceptive signals that would otherwise be
indistinguishable in conventional systems \cite{lan2020mainlobe,wang2025mainlobe,LanXu2022MainlobeDeceptive,Lan2020SuppressionofMainbeam}.

From a structural viewpoint, this separability originates from
the range-dependent phase gradient induced by the element-dependent
carrier frequencies. Specifically, the propagation-induced phase
exhibits a range derivative that varies across the array, and any
mismatch between the target and jammer can be characterized as
\begin{equation}
\Delta(\nabla_R \phi_m)_{\mathrm{jam}}
\neq
\Delta(\nabla_R \phi_m)_{\mathrm{radar}},
\end{equation}
which prevents the jammer from aligning with the target in the
joint range--angle domain \cite{tangong2023range,houwang2025mainlobe,wangjia2025ucnn,yuanhe2024suppress,liuwang2022discrimination,zhongtao2023mainlobedeceptive,lanzhang2024suppressingmainlobe,liuwang2025mainlobeJamming}.

\begin{figure}[htp]
\centering
\subfigure[MIMO]{
\includegraphics[width=0.41\textwidth]{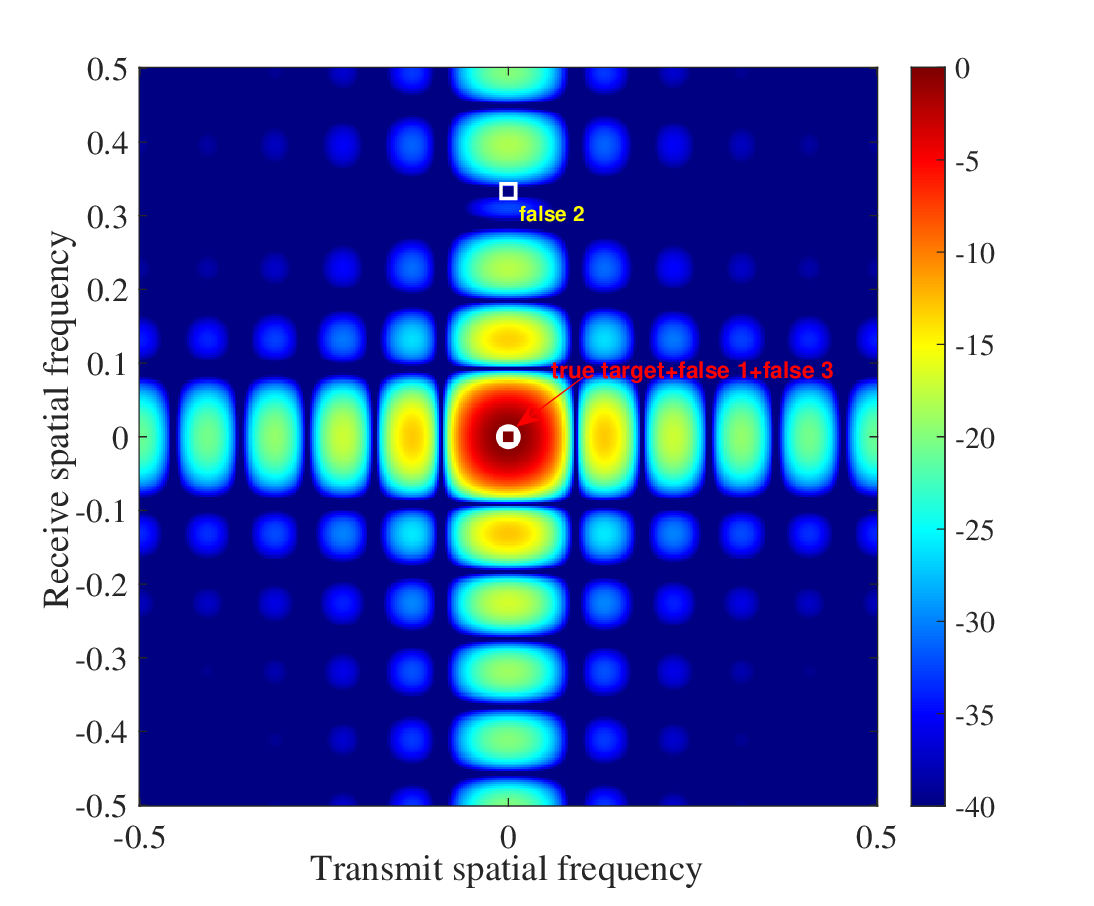}}
\subfigure[FDA-MIMO]{
\includegraphics[width=0.41\textwidth]{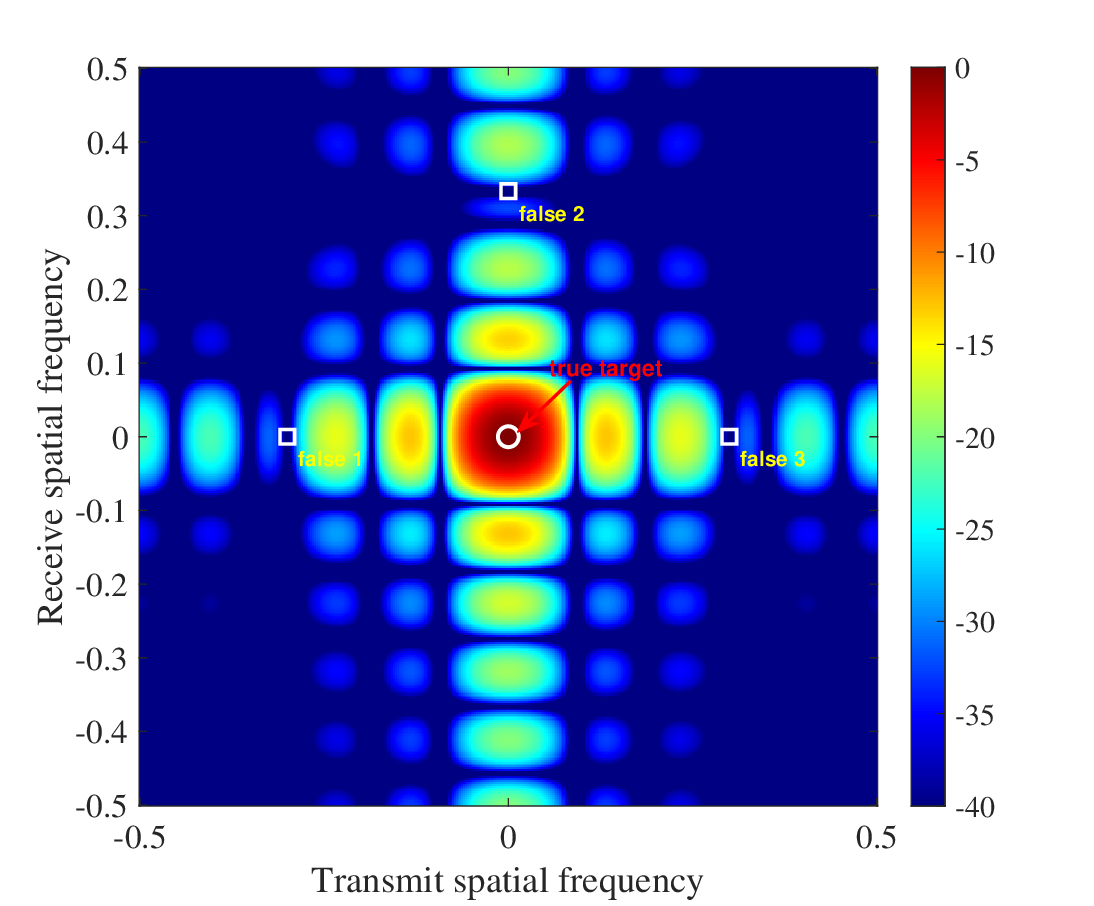}}
\caption{\justifying 
Comparison of the joint transmit--receive spatial-frequency responses
for MIMO and FDA-MIMO. The true target and three deceptive false targets
are indicated. In the MIMO case, false target 2 is separable from the true
target in the receive spatial-frequency domain and can therefore be suppressed,
whereas false targets 1 and 3 lie within the same mainlobe region as the true
target and are indistinguishable using receive-domain processing alone.
In contrast, FDA-MIMO introduces range-dependent transmit spatial frequencies,
which enable additional separation in the transmit dimension, allowing effective
suppression of mainlobe deceptive interference.
}
\label{fig18}
\end{figure}

Fig.~\ref{fig18} illustrates the fundamental difference between
conventional MIMO and FDA-MIMO in handling deceptive targets.
In Fig.~\ref{fig18}(a), target separability relies solely on the
receive spatial-frequency dimension. While false target 2 can be
distinguished from the true target, false targets 1 and 3 fall
within the same mainlobe region and remain inseparable, leading
to strong ambiguity.
In contrast, Fig.~\ref{fig18}(b) shows that FDA-MIMO introduces an
additional degree of freedom through range-dependent transmit
spatial frequencies. Although false targets 1 and 3 still overlap
with the true target in the receive domain, they become separable
in the transmit dimension. This joint-domain separability enables
effective suppression of deceptive targets located within the
mainlobe.
These results highlight that the key advantage of FDA-MIMO lies in
its ability to break the mainlobe ambiguity inherent in conventional
MIMO systems by exploiting transmit-side range dependence.
In practical terms, FDA does not prevent the generation of deceptive
signals, but instead enables the receiver to discriminate them by
leveraging the intrinsic range-dependent structure.

Furthermore, the multi-frequency transmit mechanism provides
additional diversity across the array, which enhances robustness
against model mismatch and improves reliability in challenging
deceptive jamming scenarios.

\vspace{0.5em}
\textbf{Range-Ambiguous Clutter and General ECCM:}
Beyond deceptive jamming, FDA also provides additional
degrees of freedom for suppressing range-ambiguous clutter,
which is a critical issue in high-PRF airborne radar systems.
In such scenarios, clutter from different range intervals may
fold into the same observation cell, severely degrading
detection performance.

FDA--MIMO radar enables joint range--angle beampattern
shaping, allowing clutter from different ambiguous regions
to be separated or suppressed. Early studies demonstrated
that the range-dependent transmit beampattern can be used to
mitigate range ambiguities \cite{baizert2006forward}, and
subsequent works extended this concept to adaptive
space--time--range processing frameworks for improved clutter
suppression and target detection
\cite{xu2015range,xu2016space,Xu2017AnAdaptive}.
Existing ECCM approaches can be broadly categorized into
three groups. The first exploits range-dependent transmit
beampattern shaping to place nulls at interference locations
\cite{xu2015deceptive,Wang2020MainBeamRange}. The second
combines FDA--MIMO with adaptive space--time--range
processing to separate target, clutter, and jammer
components in a higher-dimensional domain
\cite{Xu2015SpaceTimeRange,Xu2017RobustAdaptive}. The third
focuses on robust processing strategies, including
interference-aware training data selection and
optimization-based beamforming
\cite{Lan2018Suppression,Wen2018Enhanced,wen2019clutter}.
Recent works have also explored hybrid coding schemes, such
as quadratic phase coding (QPC) and EPC-assisted FDA, to further enhance robustness
in severe interference environments
\cite{Liu2023RangeAmbiguous,Qiu2023RangeAmbiguous,Lan2020SuppressionofMainbeam,wanxu2025clutterSuppression}.

\vspace{0.5em}
\textbf{Discussion:}
Overall, the key advantage of FDA in ECCM does not merely lie
in improved suppression capability, but in its ability to
{fundamentally reshape the signal manifold}. In
particular, the range-dependent phase structure introduces a
new discrimination dimension that makes coherent deceptive
reconstruction significantly more difficult. This capability
is especially critical in mainlobe jamming scenarios, where
conventional radars lack effective countermeasures.
Nevertheless, practical deployment remains challenging.
Reliable target--jammer discrimination, robust covariance
estimation under range-dependent clutter, and the high
computational complexity of space--time--range processing
remain open problems. These challenges indicate that the
full potential of FDA-based ECCM can only be realized through
joint waveform, hardware, and signal processing design.

\subsection{Low-Probability-of-Intercept Radar}
\label{subsec:task_lpi}

FDA architectures provide a fundamentally new mechanism for 
low-probability-of-intercept (LPI) radar design by introducing 
spatio-temporal energy dispersion at the waveform propagation level. 
Unlike conventional PA, whose radiation pattern is 
stationary once the steering direction is fixed, FDA produces a 
time-varying and range-dependent field distribution. As a result, 
the instantaneous spatial power density
\begin{equation}
P(t,R_0,\theta) = |\mathcal{A}_{\mathrm{FDA}}(t,R_0,\theta)|^2,
\end{equation}
becomes inherently non-stationary, exhibiting dynamic evolution over 
the coupled range--angle--time domain.
From a structural perspective, this behavior can be interpreted as a 
{spatio-temporal energy dispersion} mechanism, where transmit 
energy is redistributed across multiple dimensions rather than 
concentrated at a fixed spatial location. This fundamentally alters 
the interceptability of the emitted signal, as passive receivers can 
no longer rely on persistent high-power observations.

To quantitatively characterize this effect, we define the 
{detection persistence ratio} as
\begin{equation}
\eta = \frac{1}{T} \int_0^T 
\mathbb{I}\big(P(t,R_0,\theta_0) > P_{\mathrm{th}}\big)\, dt,
\end{equation}
where $\mathbb{I}(\cdot)$ is the indicator function, 
$P_{\mathrm{th}}$ denotes the interception threshold, and 
$(R_0,\theta_0)$ represents the location of a passive interceptor. 
This metric captures the fraction of time during which the signal 
remains detectable and directly reflects the feasibility of 
interception and coherent integration.

\begin{figure}[htp]
\centering
\includegraphics[width=0.41\textwidth]{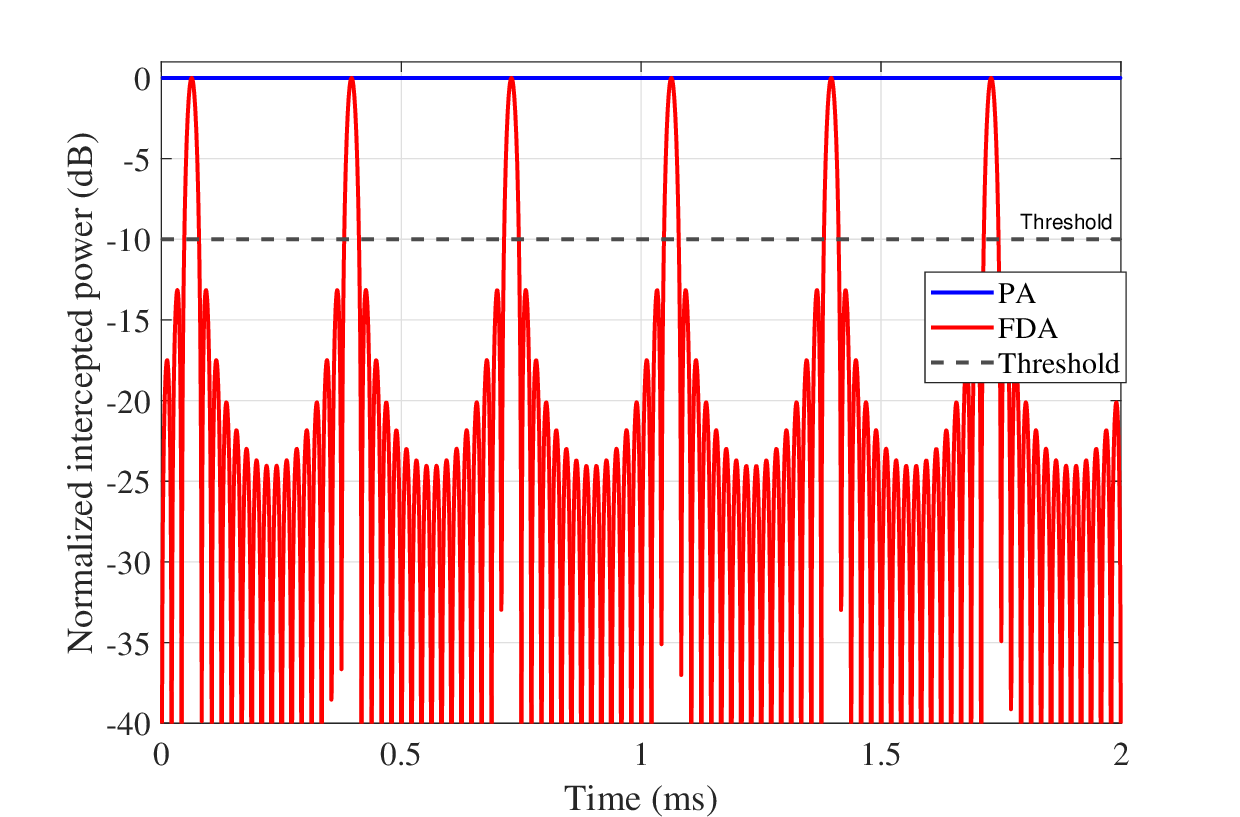}
\caption{\justifying Intercepted power trajectories at a passive receiver for PA and FDA.
A detection threshold is introduced to characterize interceptability.
While PA maintains a nearly constant received power above the threshold,
FDA exhibits strong temporal fluctuations due to its range-dependent and
time-varying radiation structure. Consequently, the FDA signal achieves
a significantly reduced detection persistence ratio (approximately $9.25\%$),
compared to $100\%$ for PA, indicating substantially lower interceptability.}
\label{fig19}
\end{figure}

Fig.~\ref{fig19} illustrates the intercepted power trajectories at a 
passive receiver for both PA and FDA. For PA, the received power remains 
almost constant and continuously exceeds the detection threshold, leading 
to $\eta \approx 1$. This enables stable interception and facilitates 
long-time coherent integration at the receiver.
In contrast, FDA produces a highly time-varying received power due to 
its intrinsic range--angle--time coupling. As shown in Fig.~\ref{fig19}, 
the received power intermittently crosses the detection threshold, 
resulting in a much smaller detection persistence ratio 
(e.g., $\eta \approx 9.25\%$ in this example). 

This observation reveals that FDA-enabled LPI is fundamentally achieved 
not by reducing the instantaneous transmit power, but by suppressing the 
{temporal persistence of detectability}. The intermittent nature of 
the received signal prevents reliable energy accumulation and disrupts 
coherent processing at passive interceptors.
From a physical standpoint, the reduction of $\eta$ can be directly 
attributed to the additional time and range phase gradients induced by 
frequency offsets across the array elements. These gradients continuously 
reshape the interference pattern in space, causing the high-energy regions 
to drift over time and preventing sustained illumination of any fixed 
location.

Existing studies have explored several design strategies to exploit this 
property. One line of work leverages FDA-induced range--angle coupling 
and temporal variability to introduce ambiguity or deception in passive 
localization \cite{guan2021Passive,wangwang2021lpiproperty}. Another 
direction focuses on joint transmit--receive optimization, where adaptive 
beampattern shaping is employed to balance detection performance and LPI 
requirements \cite{Gong2022JointDesign}. This concept has also been 
extended to integrated radar--communication systems, where LPI-oriented 
waveform design is jointly optimized with communication signaling to 
ensure coexistence and interference mitigation \cite{Gong2023Optimizationof}. 
In addition, cognitive FDA frameworks have been proposed, in which 
frequency offsets and array phases are dynamically adapted to enhance 
radio-frequency stealth in complex environments \cite{Wang2016MovingTarget}.

Overall, FDA provides a promising pathway for LPI radar by transforming 
interceptability from a static power-level problem into a dynamic 
spatio-temporal behavior. The introduced detection persistence ratio 
serves as a unifying metric that links transmit-side waveform design 
to receiver-side detectability. However, the actual LPI performance 
depends not only on the transmit strategy but also on the interception 
capability of adversarial receivers, suggesting that receiver-aware 
and scenario-dependent design remains an important open research direction.

\subsection{Imaging and Wide-Area Surveillance}
\label{subsec:task_imaging}

Building upon the capability-mapping framework established in the previous section, 
FDA-enabled imaging can be interpreted as a direct consequence of 
range--angle selectivity and frequency-domain diversity. 
These capabilities fundamentally reshape the signal formation process in radar imaging, 
providing new mechanisms for resolution enhancement, ambiguity suppression, 
and scene reconstruction.
In contrast to conventional PA-based imaging systems, where the transmit 
response depends primarily on angle, FDA introduces explicit range-dependent 
propagation characteristics. As a result, the imaging process is no longer 
governed solely by receiver-side aperture synthesis, but also by the 
transmit-side manifold expansion.

\textbf{Range-Domain Resolution and Multi-Frequency Synthesis}:
From the perspective of frequency-domain diversity, FDA can be viewed as a 
distributed multi-frequency transmit aperture. Each array element radiates 
at a slightly different carrier frequency, leading to a composite spectrum 
that spans a wider frequency range. Under ideal coherent synthesis across frequency channels, the composite 
spectrum spans a frequency range approximately given by
\begin{equation}
B_{\mathrm{eff}} \approx B + 
\left( M-1 \right) 
 \Delta f.
\end{equation}
However, it should be noted that this extended bandwidth can only 
translate into improved range resolution if coherent processing 
across different frequency components is properly achieved.

More fundamentally, this effect originates from the range-phase gradients 
introduced by frequency offsets, which create additional interference 
structures along the range dimension. Compared with conventional SAR, 
where range resolution is determined solely by signal bandwidth, FDA 
introduces a secondary range discrimination mechanism associated with 
the multi-frequency transmit aperture \cite{chenZhang2020elevatedfrequency,zhouWang2020highresolution,wang2024generalbandwidth}.

\textbf{Ambiguity Suppression via Range--Angle Coupling}:
Beyond resolution enhancement, FDA also provides new mechanisms for 
ambiguity suppression through its intrinsic range--angle selectivity. 
Because the transmit response depends jointly on range and angle, echoes 
originating from different ambiguous regions may exhibit distinguishable 
signatures even when they overlap in conventional SAR processing.

\begin{figure}[htp]
\centering
\subfigure[The part of airport runway from GF3]{
\includegraphics[width=0.41\textwidth]{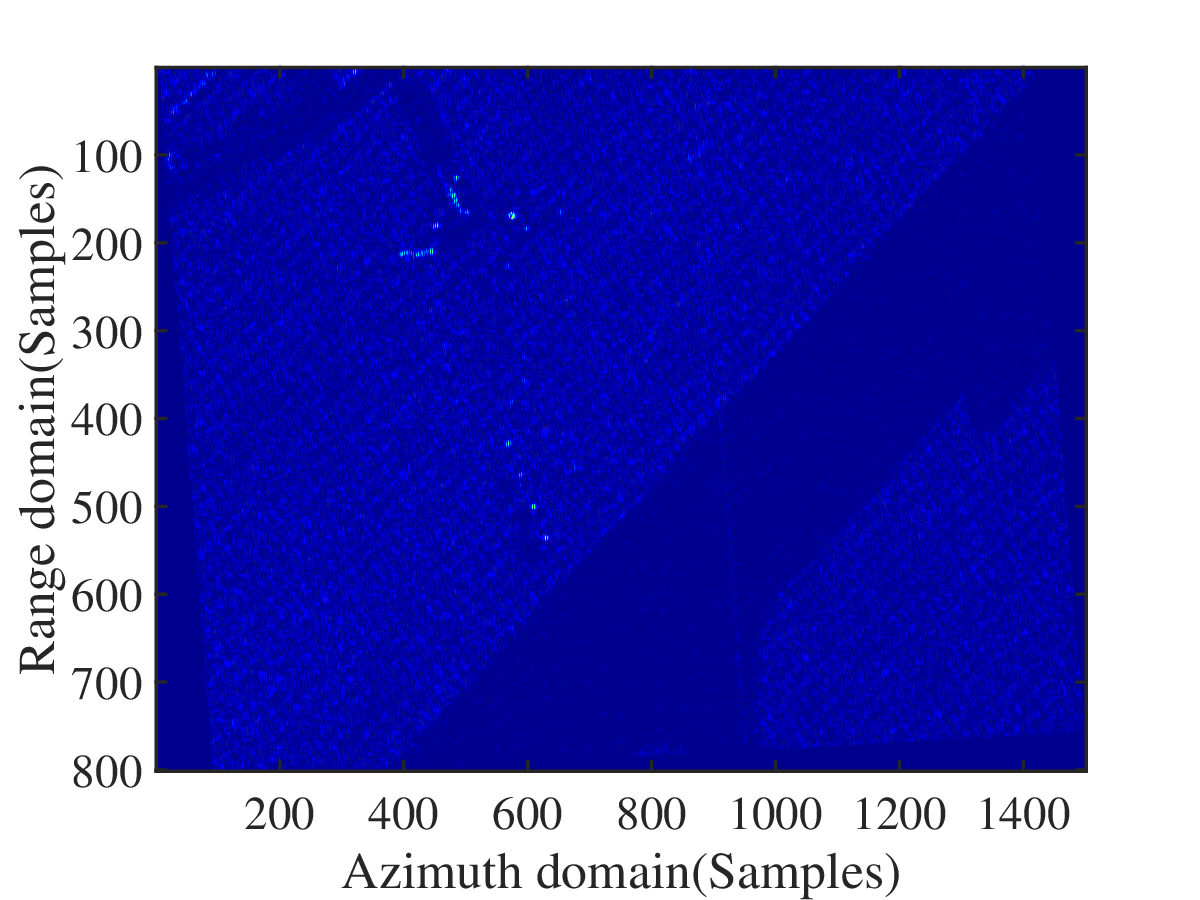}}
\subfigure[Jamming scene template]{
\includegraphics[width=0.41\textwidth]{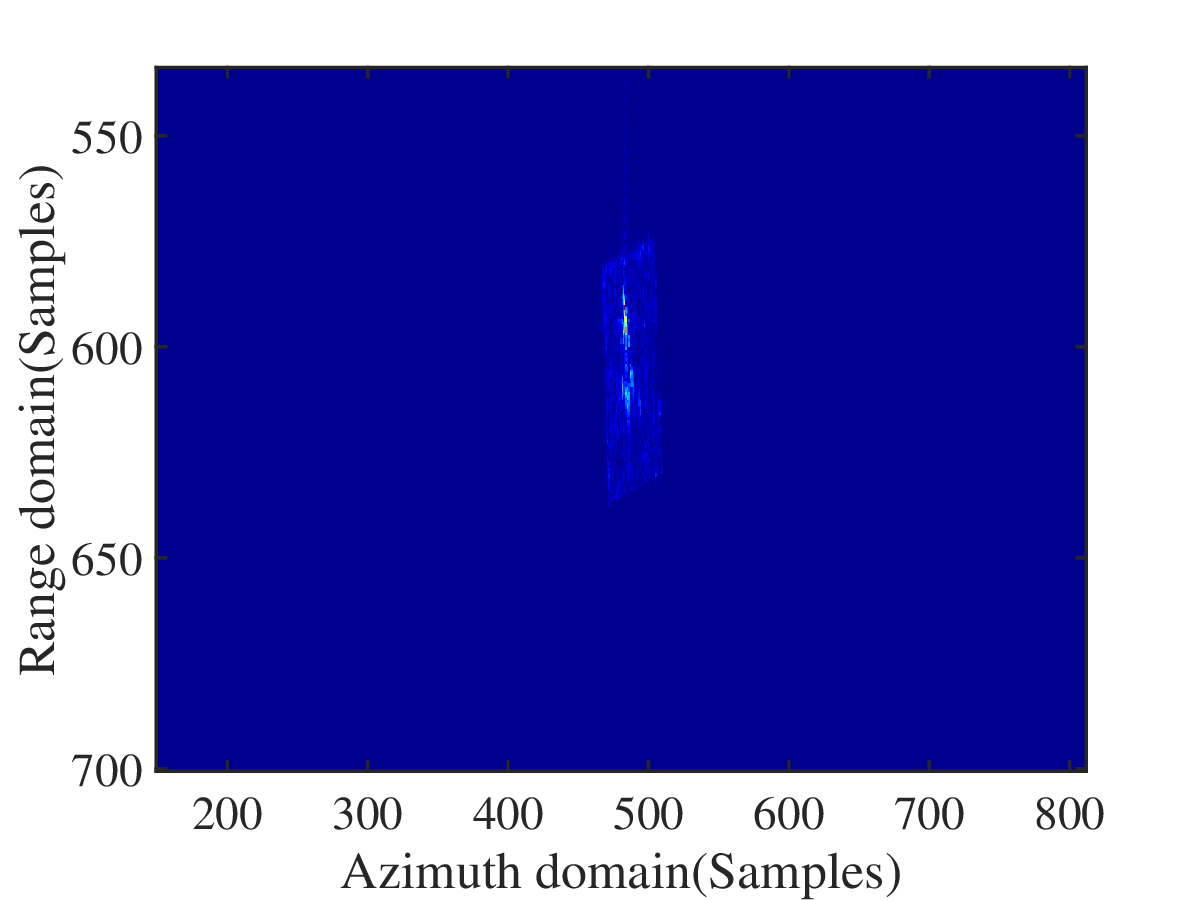}}
\subfigure[SAR jamming imaging results]{
\includegraphics[width=0.41\textwidth]{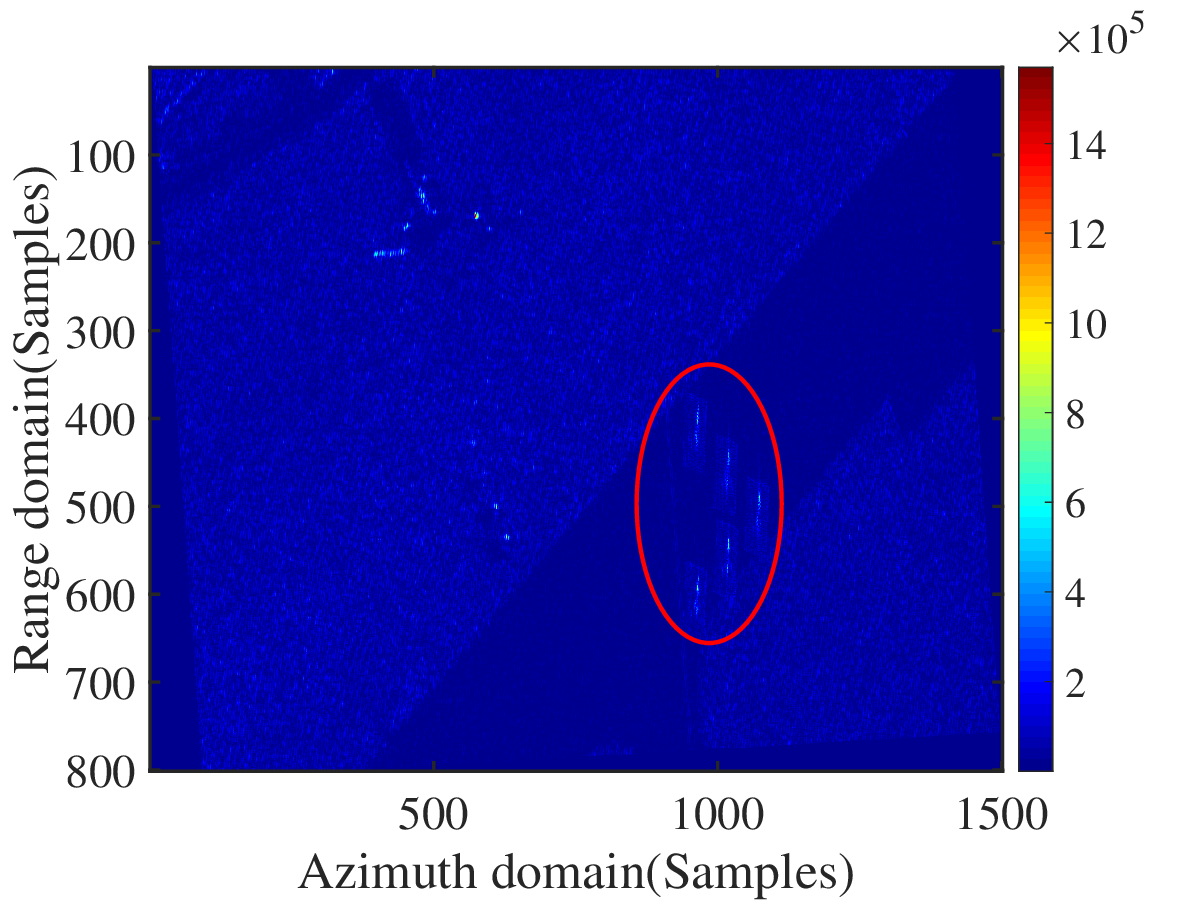}}
\caption{\justifying 
Example of scene-template-based deceptive jamming in SAR using FDA \cite{HuangWang2022FDABased}.
This example illustrates that FDA can be exploited not only for sensing and imaging enhancement, but also for image-domain manipulation and electronic countermeasures through controllable range--angle-dependent signal synthesis.}
\label{fig:sar_deception}
\end{figure}

This property is particularly beneficial for high-resolution wide-swath (HRWS) 
SAR systems \cite{wangXu2016rangeambiguity,zhouWang2021novelHigh,wenZhang2023frequencydiverse,linHuang2017unambiguoussignal,zhuYu2019applicationfrequency}, 
where range and azimuth ambiguities pose fundamental limitations. By exploiting 
the range-dependent transmit field, FDA enables additional separation 
mechanisms that can be leveraged through adaptive beamforming or spectral 
reconstruction techniques.
Similarly, in forward-looking SAR configurations, where Doppler diversity 
is inherently limited, the range-dependent transmit structure provides an 
alternative source of diversity, thereby improving imaging performance in 
scenarios where conventional approaches are constrained \cite{Wang2016Forwardlooking,baizert2006forward,shen2021front,zhang2020unambiguous}.

\textbf{Imaging Manipulation and Electronic Countermeasures}:
In addition to performance enhancement, the same multi-dimensional 
characteristics of FDA can also be exploited in electronic countermeasure (ECM) 
scenarios.
One representative line of work employs scene-template-based deceptive jamming, in which a desired false target template is first constructed and then embedded into the SAR echo formation process to manipulate the final reconstructed image. 
An example is shown in Fig.~\ref{fig:sar_deception}, where an aircraft template is injected into a real GF-3 runway scene, resulting in structured false targets appearing in the final SAR image.
From a system perspective, FDA-induced range--angle coupling 
enables more flexible control over the spatial and spectral distribution 
of transmitted signals.
In SAR deception scenarios, this flexibility can be used to generate 
artificial scatterers or manipulate reconstructed images by injecting 
multi-frequency and spatially distributed interference 
\cite{HuangWang2022FDABased,Wang2020MultiSceneDeception,Huang2019ANovelApproach,Zhu2018DeceptiveJamming,huang2019deceptive,zhangJin2025fastrepeater,yuNie2020scatteredwave,lou2022joint,ji2023template}. 
Compared with conventional jamming techniques, FDA-based approaches provide 
additional degrees of freedom for shaping false targets and distributed 
interference patterns in the image domain.

Overall, despite these potential advantages, the practical realization of 
FDA-based imaging remains challenging. On the one hand, multi-frequency 
synthesis requires strict phase coherence across transmit channels, 
imposing demanding hardware synchronization and calibration requirements. 
On the other hand, the range--angle coupling introduced by FDA breaks the 
separability assumptions underlying conventional SAR processing, thereby 
necessitating new reconstruction algorithms that explicitly account for 
the expanded transmit manifold.

Beyond these fundamental issues, most existing FDA imaging studies remain 
largely theoretical. Practical challenges such as bandwidth constraints, 
computational complexity, and robustness in heterogeneous environments 
have yet to be fully addressed. While the time-evolving beam property may 
offer additional diversity in dynamic scenarios, its role in imaging is 
not yet well established and remains an open topic for future investigation. A further open issue arises if FDA-based SAR imaging moves beyond the conventional pulse-compression-based treatment of the range dimension. In that case, classical SAR imaging algorithms and processing chains may no longer be directly applicable, and it remains an open question how HRWS SAR imaging should be reformulated under an FDA-oriented imaging framework.

\subsection{Target Tracking and Adaptive Sensing}
\label{subsec:task_tracking}

Target tracking requires radar systems to continuously estimate the kinematic state of a target while maintaining sufficient illumination and measurement quality over time. 
In conventional phased-array systems, beam steering is achieved through spatial phase control, resulting in angle-focused but range-invariant transmit beampatterns that remain essentially static once configured.
By contrast, the element-dependent carrier frequencies of FDA introduce explicit time--range phase gradients across the transmit aperture, leading to inherently time-varying and range--angle-coupled transmit responses. 
This structural property provides additional flexibility for shaping the spatiotemporal distribution of radiated energy, and has motivated a series of studies on FDA-based target tracking.

A central line of work focuses on exploiting the range--angle-dependent beampattern of FDA for tracking purposes. 
However, the intrinsic time-variant and range-periodic characteristics of conventional FDA beampatterns significantly limit their direct applicability, as targets are only intermittently illuminated with sufficient energy. 
To overcome this limitation, several studies have proposed the synthesis of {time-invariant} and {range--angle-focused} beampatterns through nonlinear or time-dependent frequency offset designs, as well as multicarrier FDA architectures \cite{wang2022fdafast,basit2020adaptivetransmit,basit2019cognitive,wang2016moving,sun2020doaestimation}. 
These approaches enable the formation of pencil-shaped beams that remain concentrated at a desired range--angle location over time, thereby improving energy accumulation and facilitating more reliable tracking. 
Extensions to multi-target scenarios have also been investigated, 
primarily through time-division-based beam synthesis approaches, 
where multiple focused beams are sequentially formed within a 
pulse duration to enable multi-target tracking \cite{wang2017timeinvariant}.

Beyond transmit beampattern synthesis, FDA has also been explored from the perspective of adaptive sensing. 
By dynamically adjusting the frequency offsets based on prior observations, the transmit field can be reconfigured to emphasize specific range regions where targets are likely to appear. 
This capability is closely related to the concept of cognitive radar \cite{gui2018cognitive,wangWang2016cognitive,yang2023cognitivefdamimo,gui2021cognitivefda,yanTao2024cognitivefdamimo}, in which waveform parameters are adaptively updated in a closed-loop manner according to the estimated target state. 
In this sense, FDA provides a form of range-aware adaptive illumination that is not directly available in conventional phased-array systems. 
Nevertheless, existing works in this direction are still primarily focused on transmit-side design, and the integration with complete tracking frameworks, such as Kalman or particle filtering with FDA, specific measurement models—remains relatively limited.

Despite these developments, the practical use of FDA in target tracking still faces several fundamental challenges. 
First, tracking-oriented FDA designs usually rely on carefully engineered nonlinear or time-dependent frequency offsets, which increase waveform design complexity and make real-time implementation more difficult. 
Second, even when approximately time-invariant focusing is synthesized, residual range--angle coupling and inherent range periodicity may still undermine stable target illumination, especially in cluttered or multi-target scenarios. 
Third, because FDA signals are intrinsically time-varying and structurally coupled, it remains difficult to formulate measurement models that are both physically consistent and convenient for recursive tracking algorithms. 
Finally, most existing studies remain focused on transmit-side beampattern synthesis, whereas the end-to-end advantages of FDA in complete tracking pipelines, including detection, association, filtering, and robustness analysis, are still far from fully established.

Overall, FDA introduces a distinctive mechanism for range--angle-dependent beam control and adaptive illumination, which may provide additional flexibility for target tracking. 
However, compared with its more mature applications in interference suppression and range-aware sensing, the role of FDA in target tracking is still in an early stage and requires further systematic investigation.
\subsection{Summary and Key Insight}

The above analysis reveals that the performance gains of FDA radar are not determined by individual signal processing techniques, but by whether the underlying phase-gradient structure induced at the propagation level can be effectively aligned with the requirements of a given task.
This relationship is summarized in Table~\ref{tab:fda_task_mapping}, which provides a mapping between FDA physical degrees of freedom, their induced structural mechanisms, and the corresponding radar task capabilities.
From this perspective, different radar tasks exploit fundamentally different manifestations of FDA-induced diversity. Estimation and localization primarily benefit from range--angle manifold expansion, detection in complex environments relies on range diversity for clutter suppression, while anti-deception capability arises from joint range--angle separability that prevents coherent matching by deceptive signals. LPI performance is enabled by spatio-temporal energy dispersion, whereas imaging and tracking tasks depend on virtual aperture extension and dynamic range-aware beam control, respectively.

These observations indicate that FDA does not provide a uniform improvement over conventional radar systems. Instead, its effectiveness is fundamentally governed by a {structure--task matching principle}: performance gains can only be realized when the FDA-induced structural properties remain physically irreducible and can be coherently exploited under given processing and environmental conditions. Otherwise, the additional degrees of freedom collapse into equivalent conventional representations, yielding limited or no practical advantage.

\begin{table*}[t]
\centering
\caption{\justifying Mapping between FDA physical degrees of freedom, underlying mechanisms, and radar task capabilities.}
\begin{tabular}{c|c|c}
\hline
Task & Primary FDA DoF & Underlying Mechanism \\
\hline
Parameter Estimation \& Localization 
& Frequency 
& Range--angle manifold expansion \\

Detection in Complex Environments 
& Frequency 
& Range diversity for clutter suppression \\

Anti-Deception \& Anti-Jamming 
& Frequency + Space 
& Joint range--angle separability \\

Low-Probability-of-Intercept (LPI) 
& Time + Frequency 
& Spatio-temporal energy dispersion \\

Imaging \& Wide-Area Surveillance 
& Frequency + Space 
& Virtual bandwidth and aperture extension \\

Target Tracking \& Adaptive Sensing 
& Time + Frequency 
& Dynamic range-aware beam control \\
\hline
\end{tabular}
\label{tab:fda_task_mapping}
\end{table*}





\section{Extensions Toward Communications and Integrated Sensing and Communications (ISAC)}
\label{sec_communication}
Beyond conventional radar applications, FDA has also attracted increasing attention in communication and integrated sensing and communication (ISAC) systems. 
This extension naturally raises a fundamental question: whether the frequency-gradient-induced range dependence introduces genuinely new communication-theoretic capabilities, or simply reshapes existing spatial resources in a different form.
To address this question, FDA-enabled communication and ISAC systems can be interpreted from a structural perspective. 
Specifically, the frequency offsets at the transmit side modify the propagation kernel, which in turn reshapes the channel structure and influences the resulting information-theoretic performance metrics, ultimately leading to new sensing–communication trade-offs.
From this viewpoint, the role of FDA in communication is not merely to provide an additional design parameter, but to alter the geometry of the underlying channel representation. 
This perspective enables a interpretation of FDA-based secure transmission, index modulation (IM), and ISAC waveform design within a common analytical framework.

\subsection{Physical-Layer Secure Communications}

Physical-layer security is one of the earliest and most extensively investigated application directions of FDA in communication systems. 
By introducing element-dependent frequency offsets, FDA generates range-dependent beampatterns, enabling selective signal focusing in both angle and range domains. 
This allows FDA to distinguish users that are closely aligned in angle but separated in range, thereby providing a fundamental mechanism for enhancing secrecy performance beyond conventional angle-only beamforming.

At a more fundamental level, such security capability originates from the irreducible range-phase gradient introduced by FDA across the transmit aperture, given by
\begin{equation}
\nabla_R \phi_m = -\frac{2\pi(f_c+\Delta f_m)}{c}.
\end{equation}
This gradient induces intrinsic differences in the propagation responses toward users located at different ranges, resulting in a range–angle-dependent channel structure that cannot be replicated by conventional PA systems. 
Consequently, FDA enables physical-layer security through propagation-layer differentiation, rather than relying solely on signal-domain processing.

From a communication-theoretic perspective, consider a legitimate user located at $(R_L,\theta_L)$ and an eavesdropper at $(R_E,\theta_E)$. 
Under FDA transmission, the effective channels can be expressed as
\begin{equation}
h_L = \mathbf{w}^H \mathbf{a}(R_L,\theta_L), 
\qquad
h_E = \mathbf{w}^H \mathbf{a}(R_E,\theta_E),
\end{equation}
where $\mathbf{a}(R_0,\theta)$ denotes the FDA array manifold. 
Unlike conventional phased-array systems with $h(\theta)$, the FDA channel explicitly depends on both range and angle.
The resulting secrecy rate is given by
\begin{equation}
C_s = \left[ \log_2(1+\gamma_L) - \log_2(1+\gamma_E) \right]^+,
\end{equation}
which is fundamentally governed by the separability of $\mathbf{a}(r,\theta)$ in the joint range–angle domain. 
In particular, secure transmission becomes feasible when
\begin{equation}
\mathbf{a}(R_L,\theta_L) \neq \mathbf{a}(R_E,\theta_E),
\end{equation}
even if $\theta_L \approx \theta_E$ but $R_L \neq R_E$, highlighting the role of range-dependent channel discrimination.

Motivated by this structural property, existing FDA-based physical-layer security techniques can be broadly categorized into two main paradigms: artificial noise (AN) injection and directional modulation (DM).
The AN-based approach enhances secrecy by projecting interference into the orthogonal range–angle subspace of the legitimate user, thereby degrading the reception quality of potential eavesdroppers~\cite{Ji2018SecrecyCapacityAnalysis,Qiu2018ArtificialNoiseAided,Hu2017ArtificialNoiseAided}. 
In contrast, DM-based schemes directly manipulate the transmitted constellation in a location-dependent manner, such that only the intended receiver can correctly demodulate the signal, while distorted constellations are observed at other locations~\cite{ChengWang2021PhysicalLayerSecurity,WangYan2021SecrecyZone,Qiu2019MultiBeamDirectional,wang2017dm,xiong2017directional}. 
These two paradigms exploit the range–angle-dependent transmission property of FDA from complementary perspectives and have been further extended to more complex scenarios. 
For example, Jian \textit{et al.}~\cite{Jian2023PhysicalLayer} investigated multi-hop FDA systems and jointly exploited AN and DM to enhance secrecy performance across multiple transmission stages.
Beyond these canonical paradigms, FDA has also been integrated with advanced modulation schemes to further improve both security and spectral efficiency. 
For instance, Basit \textit{et al.}~\cite{BasitWang2021FDABased} combined FDA with quadrature spatial modulation (QSM), enabling information embedding through frequency offsets. 
This line of work further suggests that FDA extends secure transmission from angle-only to joint range–angle domains.

\begin{figure}[htp]
\centering
\includegraphics[width=0.41\textwidth]{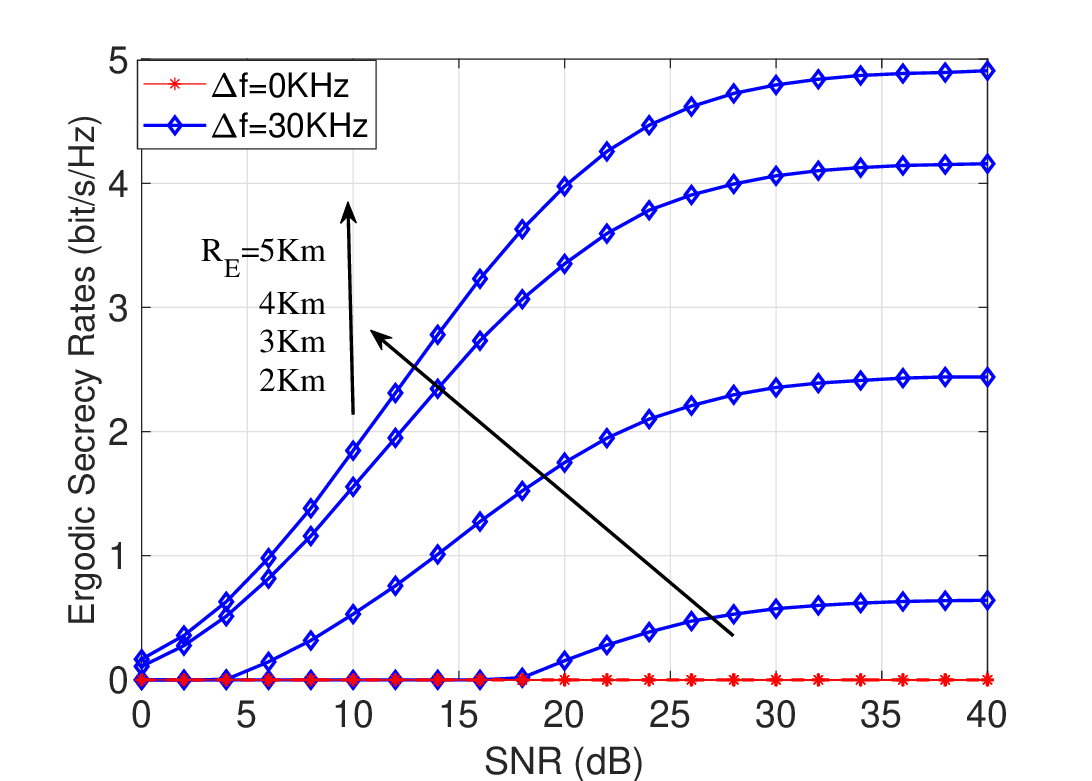}
\caption{\justifying 
Ergodic secrecy rate versus SNR for PA ($\Delta f=0$) and FDA ($\Delta f=30$ kHz) under different eavesdropper distances $R_E$ \cite{Jian2023PhysicalLayer} 
While PA yields nearly zero secrecy rate, FDA achieves significant secrecy gains that increase with SNR and improve as the eavesdropper moves farther away, demonstrating its inherent range-dependent security capability.
}
\label{fig20}
\end{figure}

Fig.~\ref{fig20} demonstrates the secrecy performance difference between PA and FDA systems.
Due to the lack of range selectivity, the PA scheme fails to provide positive secrecy rate, as both the legitimate user and the eavesdropper experience similar channel gains. In contrast, FDA introduces a frequency-induced range–angle coupling, which enables spatially selective energy focusing not only in angle but also in range.
As a result, FDA effectively suppresses signal leakage toward the eavesdropper, leading to a significant secrecy rate improvement. Furthermore, as the eavesdropper moves away from the intended range, the mismatch in the FDA beampattern becomes more pronounced, yielding higher secrecy gains.

Despite these advantages, such gains are not universally guaranteed and depend on several practical conditions. 
First, the achievable secrecy gain strongly relies on accurate range information, and the advantage diminishes when the eavesdropper is located close to the legitimate user. 
Second, if the frequency offsets are known and can be compensated, the range-dependent structure may be weakened, reducing the security benefit. 
Finally, practical issues such as time-varying beampatterns, synchronization requirements, and hardware complexity further constrain real-world deployment.
\subsection{Index Modulation and Hybrid Transmission Schemes}
\label{subsec:isac_im}

Index modulation (IM) improves spectral efficiency (SE) by encoding information into transmission resource indices, antennas, subcarriers, time slots, carrier frequencies, or channel states \cite{ZhangZou2024RISAidedIndex,xu2024polar,aydinCogen2019codeindex,huang2026movableantennaindexmodulationmaim}. 
In FDA systems, element-dependent frequency offsets introduce a fundamentally new indexing mechanism, where information can be embedded directly into the frequency-gradient domain.
From a physical perspective, this capability originates from the frequency-gradient structure of FDA. 
By assigning
$f_m = f_c + \Delta f_m(b),$
the transmitted waveform becomes dependent on the selected frequency-offset pattern indexed by $b$. 
Unlike conventional IM schemes that rely on discrete antenna or subcarrier selection, FDA enables indexing through structured frequency variations, thereby introducing a propagation-level signaling dimension.

From a signal representation viewpoint, classical IM can be expressed as
\begin{equation}
\mathbf{s} = \mathbf{e}_k x,
\end{equation}
where $\mathbf{e}_k \in \mathbb{C}^{N_t}$ denotes the $k$-th canonical basis vector (i.e., the $k$-th transmit antenna or subcarrier is activated), and $x \in \mathbb{C}$ is the modulated symbol drawn from a given constellation.
FDA-based IM extends this principle by mapping information bits to frequency-offset configurations.
Consequently, the effective signaling space can be characterized as
\begin{equation}
\mathrm{DoF}_{\mathrm{IM-FDA}}
=
\mathrm{Space} \oplus \mathrm{Frequency},
\end{equation}
indicating that FDA expands the index domain beyond conventional spatial or subcarrier-based schemes. 
This additional degree of freedom enables more flexible information embedding and provides new opportunities for SE enhancement.

From a design perspective, existing FDA-based IM schemes can be interpreted as progressively exploiting this frequency-gradient-induced dimension. 
Early works focus on frequency-offset-only indexing, where information is conveyed by selecting or rearranging frequency offsets~\cite{Huang2020IndexModulation,jian2023mimo,jian2023fda}. 
Subsequent approaches introduce joint space–frequency indexing, combining antenna selection with frequency offsets to improve SE~\cite{Nusenu2020SpaceFrequency}. 
More recent studies further integrate FDA-based IM with system-level functionalities, such as receiver-efficient designs and multi-user security-oriented transmission~\cite{Basit2021FDABasedQSM,Qiu2020MultiBeam}. 
Overall, this evolution reflects a transition from single-domain indexing to multi-domain joint modulation across space, frequency, and system dimensions.
Motivated by this expanded signaling space, FDA has also been combined with various communication frameworks to form hybrid transmission schemes. 
Typical examples include integration with OFDM for joint subcarrier–frequency indexing, space-time coding for diversity enhancement, artificial-noise-aided transmission for secure IM \cite{huang2018fda,wachowiak2025frequency,nusenu2018range,nusenu2022powerallocation,nusenuShao2020spacefrequency}. 
These designs exploit multiple domains jointly to further enhance flexibility and performance \cite{huang2025generalized}.

\begin{figure}[htp]
\centering
\includegraphics[width=0.41\textwidth]{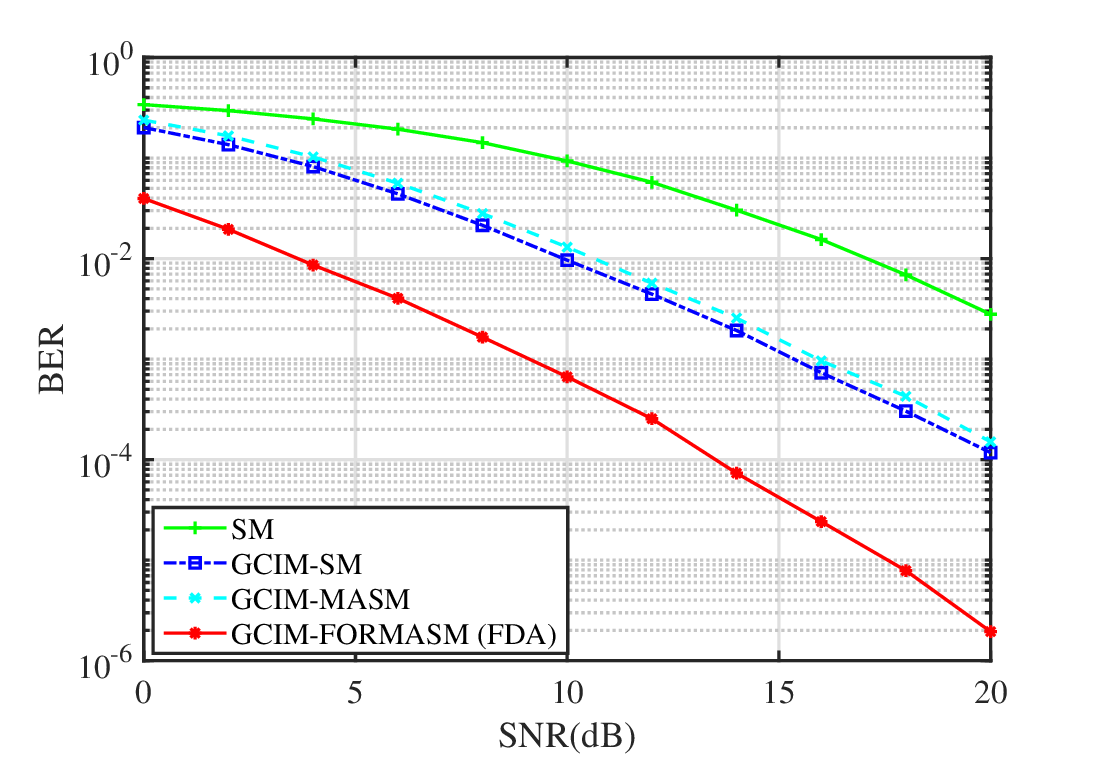}
\caption{\justifying 
BER performance of index modulation schemes with different indexing domains \cite{huang2025generalized}. 
}
\label{fig211}
\end{figure}

Fig.~\ref{fig211} shows the BER performance of different IM schemes with increasing indexing dimensions. 
Conventional schemes such as spatial modulation (SM) \cite{gandotra2017survey}, generalized code index modulation-SM (GCIM-SM) \cite{cogen2020generalized}, and generalized CM modulation-aided multiple-antenna SM (GCIM-MASM) \cite{huang2025generalized} exploit spatial and coding domains for information embedding.
In contrast, the proposed FDA-based scheme introduces the frequency-offset dimension as an additional index domain. 
The inclusion of this frequency index increases the overall indexing dimensionality, resulting in a larger set of distinguishable transmission states. 
Consequently, the receiver can better differentiate between index combinations, leading to improved detection reliability and reduced BER.

From a broader perspective, a fundamental question remains: whether FDA-based IM introduces genuinely new physical degrees of freedom, or merely redistributes existing signaling resources across spatial and frequency domains. 
Answering this question is essential for understanding the ultimate performance limits of FDA-enabled communication systems.

\subsection{FDA-Enabled ISAC Systems}

Motivated by the unique physical-layer characteristics of FDA-MIMO, recent studies have begun to explore its integration into ISAC systems. 
Although this research direction is still in its early stage, accumulating works suggest that FDA-MIMO introduces fundamentally new design opportunities beyond conventional ISAC paradigms.
Unlike conventional PA or MIMO-based ISAC systems, where sensing and communication primarily share spatial (angle-domain) resources, FDA-MIMO introduces a frequency-gradient-induced propagation structure, which gives rise to an additional range-dependent degree of freedom. 
Specifically, by assigning element-dependent carrier frequencies, the resulting transmit field becomes inherently dependent on time, range, and angle. 
This leads to a range-selective propagation kernel, enabling the transmit energy distribution to be jointly controlled over both angle and range dimensions. 
As a result, range is no longer merely an observable parameter for sensing, but becomes a controllable resource that can be actively exploited for joint sensing and communication design.

From this perspective, existing FDA-MIMO-based ISAC schemes can be reinterpreted under a unified framework, where both sensing and communication functionalities are realized through range-dependent signal shaping.

\textit{1) Communication-oriented designs:}
Several works embed communication information into FDA-MIMO radar waveforms by exploiting the frequency-offset dimension. 
For instance, information bits can be conveyed through the sign or configuration of frequency offsets~\cite{Ji2018FDAISAC}, or embedded via time-modulated spreading sequences~\cite{Nusenu2018TimeModFDA}. 
More advanced designs further improve communication rate by dividing pulses into multiple chips carrying PSK symbols~\cite{Zhou2021FDMIW}, or by introducing index modulation over waveform selection and frequency-offset patterns~\cite{Li2023FIM,Jian2024FOPIM}. 
These approaches effectively utilize the range-dependent channel variations induced by FDA to realize communication signaling.

\textit{2) Sensing-oriented enhancements:}
On the sensing side, FDA-MIMO enables improved robustness against interference and ambiguity. 
For example, frequency-offset design can mitigate range ambiguities and facilitate multi-target parameter estimation~\cite{Jian2024CCIE}. 
Furthermore, FDA's range-dependent beampattern provides inherent capability to suppress mainlobe deceptive jamming and clutter~\cite{Xu2025TransceiverDesign}, which are difficult to handle using conventional angle-only beamforming.

\textit{3) Joint ISAC optimization:}
Beyond isolated designs, several works explicitly exploit FDA-MIMO for joint sensing–communication optimization. 
These include transmit waveform and receive filter co-design for LPI communication~\cite{Gong2023Optimizationof}, region-focused sensing with multi-user communication via joint beamforming~\cite{Jian2025DualFunction}, and RIS-assisted FDA-MIMO ISAC systems for clutter suppression and rate maximization~\cite{Yang2025RISFDA}. 
These studies demonstrate the flexibility of FDA in balancing sensing and communication performance.

\begin{figure}[htp]
\centering
\subfigure[MIMO]{
\includegraphics[width=0.41\textwidth]{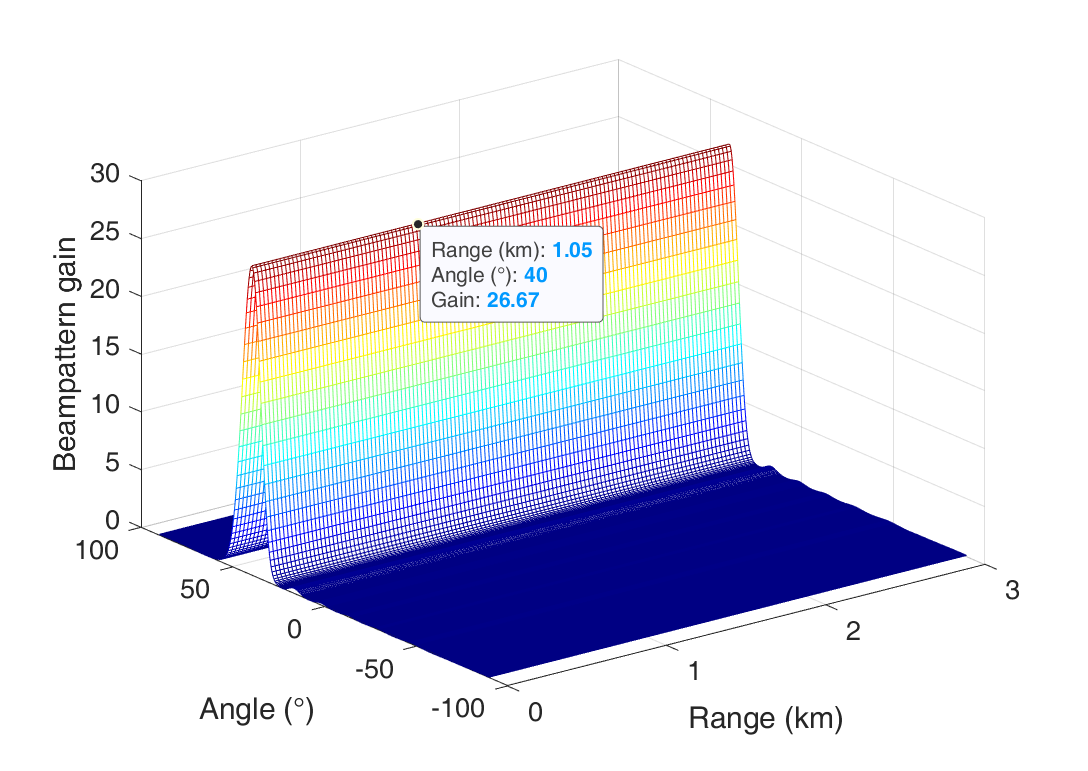}}
\subfigure[FDA-MIMO]{
\includegraphics[width=0.41\textwidth]{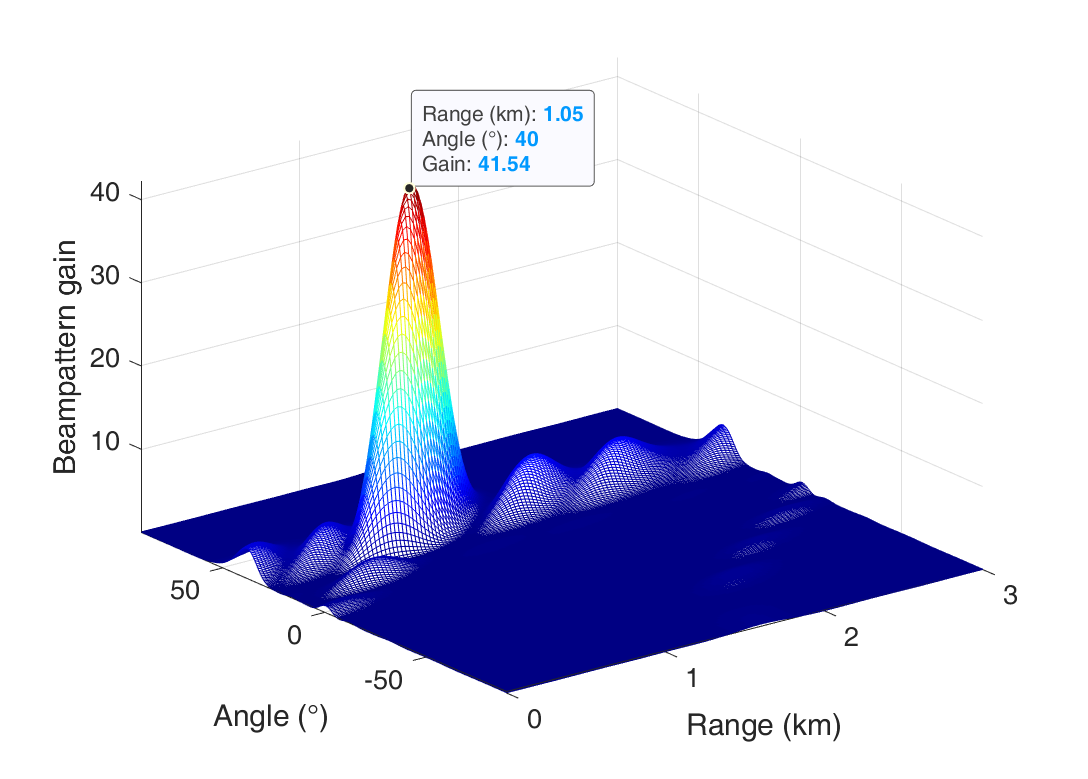}}
\caption{\justifying 
 Comparison of area surveillance performance between MIMO \cite{Hassanien2010PhasedMIMORadar} and FDA-MIMO \cite{Jian2025DualFunction}.  MIMO provides angle-only beamforming, while FDA-MIMO achieves joint angle–range focusing for area-selective surveillance. 
}
\label{fig21}
\end{figure}

\begin{figure}[htp]
\centering
\includegraphics[width=0.41\textwidth]{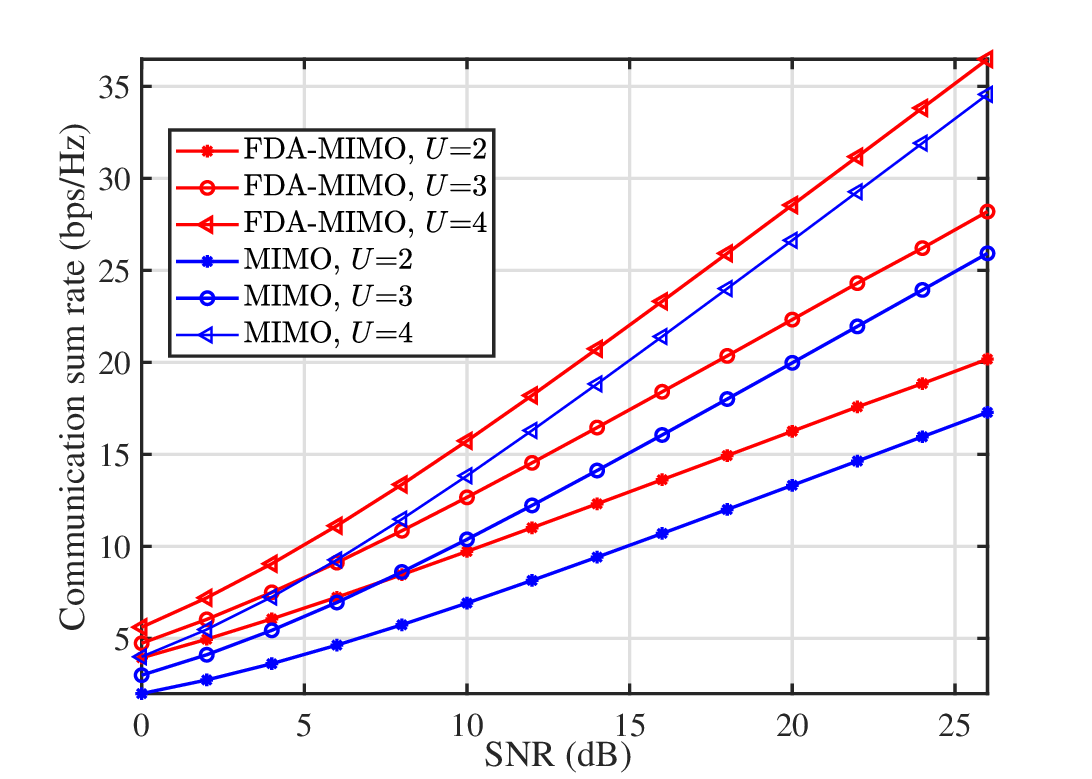}
\caption{\justifying 
Communication sum rate versus SNR, showing that FDA-MIMO outperforms MIMO under multi-user transmission.
}
\label{fig22}
\end{figure}

Figs.~\ref{fig21}-\ref{fig22} illustrate the fundamental difference between conventional MIMO and FDA-MIMO in ISAC systems. 
As shown in Fig.~\ref{fig21}, conventional MIMO exhibits angle-only beamforming, where the transmit energy is distributed uniformly along the range dimension and thus cannot selectively focus on a specific spatial region. 
In contrast, FDA-MIMO enables joint angle–range focusing due to its frequency-gradient-induced propagation structure, allowing energy to be concentrated within a desired range–angle region for area-selective surveillance.
This structural capability directly impacts the communication performance. 
As shown in Fig.~\ref{fig22}, FDA-MIMO achieves higher communication sum rates compared to MIMO under multi-user scenarios. 
By properly designing the frequency offsets and transmit strategy, FDA-MIMO can exploit the range-dependent channel variations to improve user separability and enhance communication efficiency.
More importantly, these gains are not independent improvements in sensing and communication. 
Instead, both the area-selective sensing capability and the communication rate enhancement originate from the same underlying mechanism, namely the range-dependent signal shaping enabled by the frequency gradient. 
This demonstrates that FDA-MIMO does not merely provide additional functionality, but fundamentally enables a new ISAC operating paradigm in which sensing and communication are jointly controlled through the propagation structure.

Despite their diverse implementations, the aforementioned works share a common underlying principle: both communication and sensing functionalities are enabled by the same range-dependent propagation mechanism introduced by FDA. 
From a communication perspective, techniques such as IM and waveform embedding exploit the distinguishability of range-dependent channel responses, while from a sensing perspective, interference suppression and ambiguity mitigation arise from the ability to shape the transmit energy distribution across range. 
Therefore, these seemingly different functionalities can be interpreted as different manifestations of a unified concept, namely {range-selective signal shaping}.

This unified mechanism also leads to a fundamental sensing–communication trade-off governed by the frequency gradient. 
On one hand, increasing the frequency offset enhances the range-dependent phase variation across the array, improving the sensitivity of the signal with respect to range and angle parameters and thus boosting sensing performance. 
On the other hand, the same mechanism introduces channel decorrelation and phase dispersion, which reduces the coherence of the effective communication channel and degrades communication reliability. 
As a result, the frequency offset acts as a shared structural variable that couples sensing and communication, giving rise to a Pareto frontier between sensing accuracy and communication efficiency.

From a higher-level perspective, the key contribution of FDA to ISAC is not merely the introduction of an additional range dimension, but the transformation of how sensing and communication are fundamentally coupled.
In conventional ISAC systems, sensing and communication primarily share spatial resources, leading to an angle-centric design paradigm in which the two functionalities inevitably compete for the same degrees of freedom. 
As a result, the sensing–communication trade-off is largely determined by resource allocation strategies.
In contrast, FDA introduces a frequency-gradient-induced propagation structure that jointly governs both the sensing signal manifold and the communication channel. 
Under this mechanism, sensing and communication are no longer independent functionalities sharing common resources, but are intrinsically coupled through the same range-dependent propagation process.

Consequently, the sensing–communication trade-off in FDA-enabled ISAC systems is not resource-driven but structure-induced. 
The frequency gradient simultaneously controls the sensitivity of the signal with respect to environmental parameters and the coherence properties of the communication channel, thereby shaping both sensing and communication performance in a unified manner.
This reveals a fundamental paradigm shift: FDA transforms ISAC design from resource sharing to structure coupling, enabling sensing and communication to be coordinated through the same physical mechanism rather than balanced through resource partitioning.

\section{Limitations, Misconceptions, and Future Research Directions}
\label{sec:limitations_future}

While FDA introduces a structurally enriched propagation model with expanded manifold degrees of freedom, its advantages are neither universal nor unconditional. 
Many reported gains rely on idealized assumptions regarding coherence, linear phase gradients, and hardware realizability, which may not hold in practical scenarios. 
More importantly, several commonly cited properties of FDA, such as time-invariant focusing or arbitrary range–angle decoupling, are often misunderstood or overstated. 
A rigorous examination reveals that these behaviors are intrinsically constrained by the underlying frequency-gradient mechanism and its interaction with propagation physics.
In this section, we revisit these issues from a unified perspective, aiming to (i) clarify prevalent misconceptions, (ii) identify the fundamental conditions under which FDA provides meaningful gains, and (iii) outline key research directions for bridging the gap between theoretical potential and practical deployment.

\subsection{On the Feasibility of Time-Invariant Focusing}
\label{subsec:limit_time_invariant}

A recurring claim in parts of the early FDA literature is that time-invariant spatial focusing can be achieved through appropriate frequency design. 
However, as discussed earlier, the frequency-gradient mechanism of FDA inherently introduces temporal phase evolution into the transmit manifold. 
As a result, strict time-invariant focusing is not physically compatible with the basic FDA operating principle: once nonzero frequency offsets are used to create range-dependent behavior, the corresponding array response necessarily evolves with time.

From this perspective, the notion of time-invariant focusing should be interpreted with caution. 
What many existing studies actually realize is not a truly time-invariant FDA field, but rather one of several practical approximations, such as instantaneous focusing at selected snapshots, receive-side compensation, or multi-pulse synthesis that stabilizes the apparent focusing effect over a longer observation interval.

The key structural point is that time invariance and frequency-gradient-induced range selectivity arise from conflicting requirements. 
The former requires suppressing temporal phase evolution, whereas the latter fundamentally depends on it. 
Therefore, strict time-invariant focusing and genuine FDA range selectivity cannot be simultaneously achieved within the same single-transmit physical mechanism. 
This distinction is important for avoiding overstatement in the interpretation of FDA beampattern design results.

\subsection{Near-Field and Far-Field Applicability Boundary}
\label{subsec:limit_field_boundary}

Most existing FDA analyses are established under the far-field approximation, where the propagation distance of the $m$-th element is linearized as
\begin{equation}
R_m \approx R_0 - d_m \sin\theta.
\end{equation}
Under this assumption, the FDA-induced phase variation can be interpreted as a linear gradient over range and angle, forming the basis of the commonly adopted range–angle coupling model.
However, in near-field regimes, the exact propagation distance follows
\begin{equation}
R_m = \sqrt{R_0^2 + d_m^2 - 2 R_0 d_m \sin\theta},
\end{equation}
which introduces intrinsic spherical-wave effects that fundamentally alter the signal structure.

In the near-field, range and angle are inherently coupled by geometry, independent of any frequency offset. 
As a result, the coupling induced by FDA is no longer the sole source of range–angle dependence, but instead interacts with the natural spherical-wave coupling.
This leads to two key consequences:

\begin{itemize}
\item The classical interpretation of FDA as a {frequency-gradient-induced linear coupling} becomes invalid,
\item The overall signal manifold is governed by a superposition of geometric coupling and frequency-induced effects.
\end{itemize}

The far-field FDA model remains valid only when the linear approximation holds, i.e.,
\begin{equation}
\frac{D^2}{\lambda_0} \ll R_0,
\end{equation}
where $D$ denotes the array aperture.
When this condition is violated, near-field effects dominate, and the FDA signal model must be reformulated based on spherical-wave propagation.
FDA does not introduce range–angle coupling in isolation; rather, it modifies an existing propagation structure. 
In the far-field, this structure is approximately separable and can be shaped by frequency gradients, whereas in the near-field, the intrinsic geometric coupling becomes dominant.
Consequently, the role of FDA shifts from {creating} range–angle coupling to {modulating} an already coupled propagation manifold.

\subsection{Frequency-Offset Constraints under Wideband Conditions}
\label{subsec:limit_wideband}

Most FDA analyses are derived under the narrowband assumption
\begin{equation}
B \ll f_c,
\end{equation}
which ensures that the propagation-induced phase can be approximated as a linear function of the frequency offset. 
Under this condition, the frequency gradient $\Delta f$ induces a well-defined and coherent phase structure across array elements.
However, practical radar and communication systems often operate in wideband regimes, where this assumption no longer holds.
To preserve the linear phase-gradient interpretation, the frequency offset must satisfy
\begin{equation}
\Delta f \ll B.
\end{equation}
This condition ensures that the frequency diversity across array elements remains small relative to the signal bandwidth, thereby maintaining inter-element coherence.

When
$\Delta f \sim B,$
the FDA structure undergoes a qualitative change:

\begin{itemize}
\item Inter-element signals become partially decorrelated,
\item The linear phase-gradient model breaks down,
\item The resulting ambiguity function becomes distorted and irregular.
\end{itemize}

The frequency offset is further constrained by hardware limitations, typically requiring
\begin{equation}
\Delta f \leq \min(B, f_c/Q),
\end{equation}
where $Q$ characterizes oscillator stability and frequency accuracy.
These constraints reveal an inherent trade-off: increasing $\Delta f$ enhances range-dependent discrimination, but simultaneously weakens coherence and violates the assumptions underlying FDA modeling.

Wideband signaling and frequency-gradient design impose conflicting structural requirements. 
The former relies on frequency diversity across the signal spectrum, while the latter requires coherence across array elements with controlled frequency offsets.
Consequently, the benefits of FDA cannot be arbitrarily amplified by increasing either bandwidth or frequency gradient alone. 
Effective system design must balance these two factors to preserve both gradient interpretability and signal coherence.

\subsection{Experimental Validation and Dataset Challenges}
\label{subsec:limit_experiment}

A major limitation of current FDA research lies in the lack of large-scale, reproducible, and systematically calibrated experimental validation. 
Although a substantial body of work has reported promising analytical and simulation-based results, the extent to which these gains can be sustained under realistic operating conditions remains largely unclear. 
At present, most FDA studies still rely on numerical simulations or small-scale laboratory prototypes, while publicly available benchmark datasets and broadly accessible experimental platforms remain extremely limited.

This gap is particularly critical because FDA is more sensitive to implementation imperfections than many conventional array architectures. 
Its expected behavior depends on the accurate preservation of frequency-gradient-induced phase structures across the transmit aperture. 
Consequently, practical FDA systems impose stringent requirements on inter-channel phase calibration, frequency synchronization, and environmental stability. 
Even small mismatches in phase, timing, oscillator stability, or channel responses may distort the intended range-dependent interference structure, thereby weakening or even obscuring the very effects that FDA is expected to provide.

From this perspective, the experimental challenge in FDA is not merely one of hardware scale or engineering complexity. 
More fundamentally, it concerns whether the frequency-gradient structure can be generated, maintained, and measured with sufficient fidelity for the claimed physical mechanisms to remain observable. 
This also explains why direct comparison across different FDA designs is currently difficult: without common datasets, standardized calibration procedures, and shared evaluation protocols, it remains hard to determine whether performance differences are caused by algorithmic improvements or by differences in hardware assumptions and implementation conditions.

Dataset construction is equally challenging. 
Unlike conventional radar benchmarks, FDA-oriented datasets must preserve not only target and clutter information, but also detailed calibration metadata associated with frequency offsets, phase alignment, synchronization quality, and scene stability. 
Without such information, many FDA-specific phenomena cannot be reliably reproduced or fairly evaluated. 
As a result, the absence of open and standardized datasets has become a significant bottleneck for reproducibility, cross-comparison, and data-driven FDA research.

Future progress therefore requires more than isolated hardware demonstrations. 
It calls for a broader validation ecosystem, including scalable multi-channel FDA testbeds, open benchmark datasets with explicit calibration and synchronization records, and standardized evaluation protocols that enable fair comparison across different array architectures, signal designs, and processing methods. 
Only through such efforts can the field move from proof-of-concept demonstrations toward robust and reproducible experimental evidence.

Overall, the challenge of experimental validation in FDA should be viewed as a structural issue rather than a purely engineering one. 
Because the core FDA mechanism is highly sensitive to implementation fidelity, robustness, repeatability, and calibration transparency are central to translating FDA from theoretical promise into practically credible systems.

\subsection{Future Research Directions}
\label{subsec:future_directions}

Building upon the aforementioned limitations, future FDA research should move beyond incremental algorithm design toward a deeper understanding of its structural properties and practical realizability. 
Several key research directions are outlined as follows.

\subsubsection*{(1) Coherent FDA Modeling and Processing}

A fundamental challenge lies in maintaining and exploiting coherence in FDA systems. 
Future work should investigate coherent FDA frameworks, including phase-consistent waveform design, coherent accumulation strategies, and the impact of frequency offsets on coherent integration.
In particular, understanding how coherence can be preserved or reconstructed across time, frequency, and array elements remains an open problem.

\subsubsection*{(2) Hybrid FDA--PA and Multi-Mode Architectures}

Rather than viewing FDA as a standalone paradigm, hybrid architectures that combine FDA and conventional PA offer a promising direction for balancing flexibility and stability. 
Such designs enable controlled cooperation between angle-dominant (PA-like) and range–angle coupled (FDA-like) transmission modes.
In hybrid FDA--PA architectures, only a subset of array elements employ frequency offsets, while the remaining elements operate at a common carrier frequency. 
By properly selecting the number and placement of frequency-offset elements, the strength of range--angle coupling can be regulated, allowing the system to retain part of the additional degrees of freedom introduced by FDA while mitigating excessive coupling and instability.

Beyond spatial hybridization, multi-mode transmission strategies can also be realized in the temporal domain. 
In particular, multi-pulse FDA schemes introduce diversity across pulses by allowing the frequency offsets to vary over time:
\begin{equation}
\Delta f^{(p)}, \quad p = 1, \dots, P.
\end{equation}
Such frequency scheduling mechanisms effectively reshape the transmit response across pulses, providing additional control over the resulting signal structure.
These approaches have been shown to improve ambiguity characteristics, mitigate secondary range ambiguities, and enhance robustness in detection and estimation tasks.

Hybrid FDA--PA architectures and multi-pulse frequency scheduling can be viewed as two complementary mechanisms for controlling FDA-induced coupling. 
The former regulates coupling in the spatial domain by partially introducing frequency diversity, while the latter redistributes coupling in the temporal domain through pulse-to-pulse variation.
Together, they provide a flexible framework for trading off range selectivity, stability, and robustness in practical FDA systems.

\subsubsection*{(3) Space--Time--Range Joint Design}

FDA inherently extends the signal model to a joint space–time–range domain. 
Future research should systematically investigate joint design across these dimensions, including waveform synthesis, beamforming, and resource allocation.
This direction also calls for new analytical tools to characterize the expanded signal manifold and its associated degrees of freedom.

\subsubsection*{(4) Fundamental Channel Modeling and Information Theory}

The introduction of frequency gradients fundamentally alters the communication channel structure. 
Key open questions include:

\begin{itemize}
\item What is the canonical form of the FDA communication channel matrix?
\item How does range-dependent coupling affect channel rank, capacity, and identifiability?
\item What are the fundamental limits of FDA-enabled secure communication?
\end{itemize}

Establishing rigorous channel models and information-theoretic limits is essential for moving beyond heuristic designs.

\subsubsection*{(5) FDA-Enabled ISAC and Secure Transmission}

While FDA has shown potential in ISAC, its unique role in joint sensing–communication remains underexplored. 
Future research should clarify how range-dependent propagation can be leveraged for:

\begin{itemize}
\item Interference-aware communication,
\item Physical-layer security via spatial–range selectivity,
\item Joint sensing–communication resource allocation.
\end{itemize}

In particular, experimentally validating FDA-based secure communication schemes remains an open challenge.

\subsubsection*{(6) Retrofitting Existing Radar Platforms for FDA Realization}

A practically important open question is how FDA functionality can be realized by modifying existing radar hardware rather than developing entirely new platforms from scratch. 
In many scenarios, it may be more realistic to retrofit current phased-array, MIMO radar, or software-defined radio infrastructures through partial frequency-offset control, subarray-level frequency scheduling, multi-pulse transmission strategies, or digital compensation frameworks. 
Such an approach could significantly lower the barrier to experimental validation and practical deployment.

This direction raises several fundamental questions. 
First, which FDA capabilities genuinely require element-level frequency offsets at the hardware level, and which can be approximated through hybrid architectures or signal processing? 
Second, what are the performance limits of FDA-like operation when implemented on legacy platforms with restricted frequency agility and synchronization accuracy? 
Third, how should existing transmit--receive chains, local oscillators, and calibration procedures be modified to preserve the intended frequency-gradient structure without compromising coherence? 
Addressing these questions is essential for translating FDA from a conceptually attractive array model into a practically deployable architecture.

\subsubsection*{(7) Large-Scale Experimental Platforms and Benchmarking}

Next, bridging theory and practice requires the development of scalable FDA testbeds, open datasets, and standardized evaluation protocols. 
This includes multi-channel calibration, synchronization frameworks, and reproducible experimental setups for fair comparison across studies.
\subsubsection*{(8) FDA and Near-Field / Holographic MIMO Integration}

Recent advances in near-field and holographic MIMO systems have highlighted the importance of spherical-wave propagation, where range-dependent effects arise naturally from geometry. 
This raises a fundamental question regarding the role of FDA in such regimes.
In conventional near-field MIMO, range–angle coupling is inherently induced by spherical-wave propagation, without requiring frequency diversity. 
In contrast, FDA introduces range dependence through frequency-gradient-induced phase modulation, even under far-field conditions.

This distinction suggests that FDA and near-field MIMO provide two different mechanisms for achieving range-dependent channel structures:

\begin{itemize}
\item \textbf{Geometric Coupling:} Induced by spherical-wave propagation in near-field systems,
\item \textbf{Frequency-Induced Coupling:} Introduced by element-dependent frequency offsets in FDA.
\end{itemize}

Understanding the interaction between these two mechanisms is an open research problem. 
In particular, key questions include:

\begin{itemize}
\item Whether FDA provides additional degrees of freedom in near-field regimes where coupling already exists,
\item How frequency-gradient design interacts with spherical-wave channel models,
\item Whether hybrid designs can jointly exploit geometric and frequency-induced coupling for enhanced sensing and communication.
\end{itemize}

This direction is closely related to emerging 6G concepts such as holographic MIMO and extremely large-scale antenna arrays, where spatial and range-domain effects become tightly intertwined.
\subsection{Concluding Perspective}

FDA is neither a universal replacement for conventional phased arrays nor merely a theoretical construct. 
Its true significance lies in introducing a new form of propagation-embedded design, where frequency gradients reshape the array manifold beyond purely spatial dimensions.

This structural expansion, however, is fundamentally conditional. 
Its effectiveness depends critically on the ability to preserve a coherent and irreducible frequency-gradient structure under practical constraints. 
When this structure is distorted, decorrelated, or averaged out, the system behavior collapses toward conventional array paradigms.
In this sense, FDA should not be interpreted as simply adding an additional degree of freedom, but as redefining how degrees of freedom are embedded within the propagation process itself.
The practical value of FDA therefore hinges on three intertwined factors:

\begin{itemize}
\item Preserving gradient irreducibility in the presence of temporal, spectral, and geometric constraints,
\item Maintaining coherence under hardware imperfections and synchronization limitations,
\item Aligning the resulting structural degrees of freedom with task-specific objectives in sensing and communication.
\end{itemize}

Future progress will likely emerge from a deeper integration of propagation theory, hardware-aware design, and information-theoretic analysis. 
More broadly, FDA highlights a paradigm shift from spatial-domain beamforming to structure-aware signal design, where waveform, array, and channel are no longer separable components but jointly shape system capability.

\section{Conclusion}
\label{sec:conclusion}

This paper presented a unified structural perspective on FDA, connecting design variables, manifold expansion, physical degrees of freedom, and system-level capabilities within a coherent analytical framework.
Rather than treating FDA as a collection of application-specific techniques, we reformulated it through a generalized phase-gradient viewpoint. 
From this perspective, the defining feature of FDA is not merely frequency diversity, but the introduction of inter-element frequency gradients that embed range dependence directly into the propagation process. 
This mechanism extends the array manifold from $\mathbf{a}(\theta)$ to $\mathbf{a}(t,R_0,\theta)$, thereby enabling new forms of structural degrees of freedom.

A key insight established in this work is that frequency diversity alone does not guarantee a new physical degree of freedom. 
The decisive factor is irreducibility: only when the gradient-induced range-phase structure cannot be removed through linear compensation does a genuine range-domain DoF emerge. 
Otherwise, the system effectively collapses to conventional phased-array or waveform-equivalent models.
Within this unified framework, we linked FDA-induced structural DoF to a range of capabilities, including range–angle selectivity, time-varying beam evolution, anti-deception performance, and sensing–communication coupling. 
At the same time, we emphasized that these gains are inherently conditional. 
Their realization depends on maintaining coherence, preserving gradient structure, and operating within the validity bounds of underlying modeling assumptions.

We further clarified several common misconceptions. 
In particular, strict time-invariant focusing is physically infeasible under nonzero frequency gradients, while near-field and wideband regimes fundamentally alter the validity of classical FDA interpretations. 
These observations highlight that FDA performance must be evaluated in conjunction with propagation conditions and hardware constraints.
Ultimately, FDA should be understood not as a universal replacement for phased arrays, but as a propagation-embedded design paradigm that reshapes how degrees of freedom are created and utilized. 
Its significance lies in redefining the interaction between signal design and propagation physics, rather than simply increasing system dimensionality.

Looking forward, the evolution of FDA will depend on advancing coherent and robust gradient design, extending manifold theory to near-field and wideband regimes, establishing rigorous channel and information-theoretic models, and developing large-scale experimental platforms. 
More broadly, FDA points toward a shift from spatial-domain beamforming to structure-aware system design, where waveform, array, and channel are jointly optimized to meet task-specific objectives in radar and ISAC systems.

\section{Acknowledgment}
The authors would like to sincerely thank all the reviewers for their valuable comments and constructive suggestions, which have helped improve the quality of this manuscript. The authors also gratefully acknowledge Dr. Jiangwei Jian from the University of Electronic Science and Technology of China for his valuable suggestions on the secure communication and ISAC applications covered in this paper.

\bibliographystyle{IEEEtran}
\bibliography{referencesv1}

@article{gui2020generalized,
	title={Generalized ambiguity function for {FDA} radar joint range, angle and doppler resolution evaluation},
	author={Gui, Ronghua and Huang, Bang and Wang, Wen-Qin and Sun, Yan},
	journal={IEEE Geoscience and Remote Sensing Letters},
	volume={19},
	pages={1--5},
	year={2020},
	publisher={IEEE}
}

@book{lutkepohl1997handbook,
	title={Handbook of matrices},
	author={L{\"u}tkepohl, Helmut},
	year={1997},
	publisher={John Wiley \& Sons}
}

@ARTICLE{HuangYan2022RadarCrossSection,
	author={Huang, Bang and Yan, Yisheng and Basit, Abdul and Wang, Wen-Qin and Cheng, Jie},
	journal={IEEE Transactions on Aerospace and Electronic Systems}, 
	title={Radar Cross Section Characterization of Frequency Diverse Array Radar}, 
	year={2022},
	volume={},
	number={},
	pages={1-11},
	doi={10.1109/TAES.2022.3185023}}

@inproceedings{cetintepe2014examination,
  title={Examination of target {RCS} characteristics in {FDA} radars},
  author={Cetintepe, Cagri and Demir, Simsek},
  booktitle={2014 IEEE Antennas and Propagation Society International Symposium (APSURSI)},
  pages={1754--1755},
  year={2014},
  organization={IEEE}
}

@inproceedings{wang2024sidelobe,
  title={A Sidelobe Optimization Method using Spatial Code Based on Space-Time Coding Array},
  author={Wang, Shenjing and He, Feng and Dong, Zhen},
  booktitle={2024 IEEE International Conference on Signal, Information and Data Processing (ICSIDP)},
  pages={1--5},
  year={2024},
  organization={IEEE}
}

@article{wang2025range,
  title={Range-Ambiguous Clutter Modeling, Separation, and Reduced-Dimensional STAP for {MIMO-STCA} Radar},
  author={Wang, Huake and Li, Sijia and Liao, Guisheng and Quan, Yinghui},
  journal={IEEE Transactions on Aerospace and Electronic Systems},
  year={2025},
  publisher={IEEE}
}

@article{wang2025mainlobe,
  title={Mainlobe Jamming Suppression Using {MIMO-STCA} Radar},
  author={Wang, Huake and Cai, Bairui and Liao, Guisheng},
  journal={arXiv preprint arXiv:2505.09112},
  year={2025}
}

@article{wang2025monopulse,
  title={Monopulse Parameter Estimation based on {MIMO-STCA} Radar in the Presence of Multiple Mainlobe Jammings},
  author={Wang, Huake and Zhang, Dongchang and Liao, Guisheng and Quan, Yinghui},
  journal={IEEE Transactions on Aerospace and Electronic Systems},
  year={2025},
  publisher={IEEE}
}

@article{lan2018subarray,
  title={Subarray-based time-delay low sidelobes methods for space-time coding array},
  author={Lan, Lan and Liao, Guisheng and Xu, Jingwei and Zhu, Shengqi and Wang, Zhirui},
  journal={IET Radar, Sonar \& Navigation},
  volume={12},
  number={8},
  pages={807--814},
  year={2018},
  publisher={Wiley Online Library}
}

@article{liu2023signal,
  title={A signal model based on the space--time coding array and a novel imaging method based on the hybrid correlation algorithm for {F-SCAN SAR}},
  author={Liu, Yuqing and Wang, Pengbo and Men, Zhirong and Guo, Yanan and He, Tao and Bao, Rui and Cui, Lei},
  journal={Remote Sensing},
  volume={15},
  number={17},
  pages={4276},
  year={2023},
  publisher={MDPI}
}

@article{wang2021transmit,
  title={Transmit beampattern synthesis for chirp space-time coding array by time delay design},
  author={Wang, Huake and Liao, Guisheng and Zhang, Yuhong and Xu, Jingwei and Zhu, Shengqi and Huang, Lei},
  journal={Digital Signal Processing},
  volume={110},
  pages={102901},
  year={2021},
  publisher={Elsevier}
}

@article{wang2025fast,
  title={Fast-Moving Target Detection with Frequency Diverse Element-Pulse Coding Radar},
  author={Wang, Yanxing and Lan, Lan and Zhu, Shengqi and Li, Ximin and Liao, Guisheng},
  journal={IEEE Transactions on Aerospace and Electronic Systems},
  year={2025},
  publisher={IEEE}
}

@article{yu2023mainbeam,
  title={Mainbeam deceptive jammer suppression with joint element-pulse phase coding},
  author={Yu, Kun and Zhu, Shengqi and Lan, Lan and Zhu, Jingjing and Li, Ximin},
  journal={IEEE Transactions on Vehicular Technology},
  volume={73},
  number={2},
  pages={2332--2344},
  year={2023},
  publisher={IEEE}
}

@article{kan2025non,
  title={Non-cooperative Bistatic Denial with {EPC-MIMO} Radar Transmitter},
  author={Kan, Qingyun and Xu, Jingwei and Zhang, Yuhong and Wang, Weiwei and Xu, Yanhong},
  journal={IEEE Transactions on Aerospace and Electronic Systems},
  year={2025},
  publisher={IEEE}
}

@article{qiu2023range,
  title={Range-ambiguous clutter suppression for space-based early warning radar using vertical {FDA and horizontal EPC}},
  author={Qiu, Zizhou and Liao, Zhipeng and Xu, Jingwei and Duan, Keqing},
  journal={IEEE Geoscience and Remote Sensing Letters},
  volume={20},
  pages={1--5},
  year={2023},
  publisher={IEEE}
}

@inproceedings{lan2020mainlobe,
  title={Mainlobe deceptive jammer suppression with MIMO radar using element-pulse coding},
  author={Lan, Lan and Liao, Guisheng and Xu, Jingwei and Zhang, Yuhong},
  booktitle={2020 IEEE Radar Conference (RadarConf20)},
  pages={1--6},
  year={2020},
  organization={IEEE}
}

@inproceedings{zhu2024simultaneous,
  title={Simultaneous Detection and Estimation with Distributed {FDA-MIMO} and {EPC-MIMO} Radars},
  author={Zhu, Jiayun and Zhang, Xiang and Lan, Lan and Ma, Runlong and Cui, Lei and Chen, Guozhong and Xu, Jingwei and Liao, Guisheng},
  booktitle={2024 IEEE International Conference on Signal, Information and Data Processing (ICSIDP)},
  pages={1--6},
  year={2024},
  organization={IEEE}
}

@article{lan2024mainlobe,
  title={Mainlobe deceptive jammer mitigation with multipath in {EPC-MIMO} radar exploiting matrix decomposition},
  author={Lan, Lan and Zhang, Yitao and Liao, Ruiqian and Liao, Guisheng and Xu, Jingwei and So, Hing Cheung},
  journal={IEEE Transactions on Vehicular Technology},
  volume={73},
  number={10},
  pages={14674--14688},
  year={2024},
  publisher={IEEE}
}

@article{liu2025range,
  title={Range ambiguity mitigation and range-angle estimation with {EPC-FDA-MIMO} radar},
  author={Liu, Feilong and Zhu, Shengqi and Xu, Jingwei and Li, Ximin and Lan, Lan and Liao, Guisheng},
  journal={IEEE Transactions on Aerospace and Electronic Systems},
  volume={61},
  number={3},
  pages={6280--6294},
  year={2025},
  publisher={IEEE}
}

@inproceedings{liu2023resolving,
  title={Resolving range ambiguity in {EPC SAR} based on beampattern precise control},
  author={Liu, Yangyang and Lan, Lan and Xu, Jingwei and Liao, Guisheng and Zhang, Yongwei},
  booktitle={IET Conference Proceedings CP874},
  volume={2023},
  number={47},
  pages={3324--3330},
  year={2023},
  organization={IET}
}

@article{xu2020resolving,
  title={Resolving range ambiguity via multiple-input multiple-output radar with element-pulse coding},
  author={Xu, Jingwei and Zhang, Yuhong and Liao, Guisheng and So, Hing Cheung},
  journal={IEEE Transactions on Signal Processing},
  volume={68},
  pages={2770--2783},
  year={2020},
  publisher={IEEE}
}

@article{Jian2022physical,
	author = {Jiangwei Jian and Bang Huang and Wen-Qin Wang},
	title = {Physical-layer security with frequency diverse array for {DF} multi-antenna relaying {SWIPT} system},
	journal = {International Journal of Electronics Letters},
	volume = {},
	number = {},
	pages = {1-9},
	year  = {2022},
	publisher = {Taylor & Francis},
	doi = {10.1080/21681724.2022.2087911}, 
}

@INBOOK{Richards2005Fundamentals,  
	author={M. Richards},  
	booktitle={ New York: McGraw-Hil},   
	title={Fundamentals of Radar Signal
	Processing},   
	year={2005},  volume={},  number={},  
	pages={}, 
	doi={}}

@book{johnson1992array,
  title={Array signal processing: concepts and techniques},
  author={Johnson, Don H and Dudgeon, Dan E},
  year={1992},
  publisher={Simon \& Schuster, Inc.}
}

@inproceedings{secmen2007frequency,
  title={Frequency diverse array antenna with periodic time modulated pattern in range and angle},
  author={Secmen, Mustafa and Demir, Simsek and Hizal, Altunkan and Eker, Taylan},
  booktitle={2007 IEEE Radar Conference},
  pages={427--430},
  year={2007},
  organization={IEEE}
}

@ARTICLE{Wang2017ARangeAmbiguity,
	author={C. {Wang} and J. {Xu} and G. {Liao} and X. {Xu} and Y. {Zhang}},
	journal={IEEE Journal of Selected Topics in Signal Processing}, 
	title={A Range Ambiguity Resolution Approach for High-Resolution and Wide-Swath {SAR} Imaging Using Frequency Diverse Array}, 
	year={2017},
	volume={11},
	number={2},
	pages={336-346},
	doi={10.1109/JSTSP.2016.2605064}}

@ARTICLE{Wang2020MultiSceneDeception,
	author={H. {Wang} and S. {Zhang} and W. {Wang} and B. {Huang} and Z. {Zheng} and Z. {Lu}},
	journal={IEEE Access}, 
	title={Multi-Scene Deception Jamming on {SAR} Imaging With {FDA} Antenna}, 
	year={2020},
	volume={8},
	number={},
	pages={7058-7069},
	doi={10.1109/ACCESS.2019.2963042}}

@ARTICLE{Ji2018SecrecyCapacityAnalysis,
	author={S. {Ji} and W. {Wang} and H. {Chen} and Z. {Zheng}},
	journal={IEEE Wireless Communications Letters}, 
	title={Secrecy Capacity Analysis of AN-Aided {FDA} Communication Over Nakagami- ${m}$  Fading}, 
	year={2018},
	volume={7},
	number={6},
	pages={1034-1037},
	doi={10.1109/LWC.2018.2850896}}

@ARTICLE{Qiu2019MultiBeamDirectional,
	author={B. {Qiu} and M. {Tao} and L. {Wang} and J. {Xie} and Y. {Wang}},
	journal={IEEE Transactions on Information Forensics and Security}, 
	title={Multi-Beam Directional Modulation Synthesis Scheme Based on Frequency Diverse Array}, 
	year={2019},
	volume={14},
	number={10},
	pages={2593-2606},
	doi={10.1109/TIFS.2019.2900942}}

@ARTICLE{Wang2013RangeAngleDependent,
	author={W. {Wang}},
	journal={IEEE Transactions on Antennas and Propagation}, 
	title={Range-Angle Dependent Transmit Beampattern Synthesis for Linear Frequency Diverse Arrays}, 
	year={2013},
	volume={61},
	number={8},
	pages={4073-4081},
	doi={10.1109/TAP.2013.2260515}}

@ARTICLE{Lan2020SuppressionofMainbeam,
	author={L. {Lan} and J. {Xu} and G. {Liao} and Y. {Zhang} and F. {Fioranelli} and H. C. {So}},
	journal={IEEE Transactions on Vehicular Technology}, 
	title={Suppression of Mainbeam Deceptive Jammer With {FDA-MIMO} Radar}, 
	year={2020},
	volume={69},
	number={10},
	pages={11584-11598},
	doi={10.1109/TVT.2020.3014689}}

@book{Skolnik2001IntroductionRadarbook,
	title={Introduction to Radar Systems},
	author={M. Skolnik},
	year={2001},
	publisher={New York, NY: McGrowHill}
}

@book{Li2009MIMORadarSignalbook,
	title={{MIMO} Radar Signal Processing},
	author={J. Li and P. Stoica},
	year={2009},
	publisher={New York, USA: John Wiley \& Sons, Inc}
}

@article{Gui2020Low,
	title={Low-complexity {GLRT} for {FDA} radar without training data},
	author={Gui, Ronghua and Wang, {W.-Q} and Zheng, Zhi},
	journal={Digital Signal Processing},
	volume={107},
	year={2020},
}

@article{Xu2015Deceptivejammingsuppression,
  title={Deceptive jamming suppression with frequency diverse MIMO radar},
  author={Xu, Jingwei and Liao, Guisheng and Zhu, Shengqi and So, Hing Cheung},
  journal={Signal Processing},
  volume={113},
  pages={9--17},
  year={2015},
  publisher={Elsevier}
}

@ARTICLE{Liu2015Adaptivedetectionwithout,
	author={W. {Liu} and Y. {Wang} and J. {Liu} and W. {Xie} and H. {Chen} and W. {Gu}},
	journal={IEEE Transactions on Aerospace and Electronic Systems}, 
	title={Adaptive detection without training data in colocated {MIMO} radar}, 
	year={2015},
	volume={51},
	number={3},
	pages={2469-2479},
	doi={10.1109/TAES.2015.130754}}

@ARTICLE{Lan2020GLRTbasedAdaptive,
	author={L. {Lan} and A. {Marino} and A. {Aubry} and A. {De Maio} and G. {Liao} and J. {Xu} and Y. {Zhang}},
	journal={IEEE Transactions on Aerospace and Electronic Systems}, 
	title={{GLRT}-based Adaptive Target Detection in {FDA-MIMO} Radar}, 
	year={2020},
	volume={},
	number={},
	pages={1-1},
	doi={10.1109/TAES.2020.3028485}}

@ARTICLE{DeMaio2007RaoTestforAdaptive,
	author={A. {De Maio}},
	journal={IEEE Transactions on Signal Processing}, 
	title={Rao Test for Adaptive Detection in Gaussian Interference With Unknown Covariance Matrix}, 
	year={2007},
	volume={55},
	number={7},
	pages={3577-3584},
	doi={10.1109/TSP.2007.894238}}

@ARTICLE{Robey1992ACFARadaptive,
	author={F. C. {Robey} and D. R. {Fuhrmann} and E. J. {Kelly} and R. {Nitzberg}},
	journal={IEEE Transactions on Aerospace and Electronic Systems}, 
	title={A {CFAR} adaptive matched filter detector}, 
	year={1992},
	volume={28},
	number={1},
	pages={208-216},
	doi={10.1109/7.135446}}

@ARTICLE{Liu2018BayesianDetection,
	author={J. {Liu} and J. {Han} and Z. {Zhang} and J. {Li}},
	journal={IEEE Transactions on Signal Processing}, 
	title={Bayesian Detection for {MIMO} Radar in {Gaussian} Clutter}, 
	year={2018},
	volume={66},
	number={24},
	pages={6549-6559},
	doi={10.1109/TSP.2018.2879038}}

@ARTICLE{Liu2015Persymmetricadaptivetarget,
	author={J. {Liu} and H. {Li} and B. {Himed}},
	journal={IEEE Transactions on Aerospace and Electronic Systems}, 
	title={Persymmetric adaptive target detection with distributed {MIMO} radar}, 
	year={2015},
	volume={51},
	number={1},
	pages={372-382},
	doi={10.1109/TAES.2014.130652}}

@book{Kay1993Fundamentals2,
	title={Fundamentals of Statistical Signal Processing, Volume II},
	author={Steven Kay},
	volume={},
	year={1993},
	publisher={Prentice Hall}
}

@ARTICLE{XuLiao2018Anovervoew,
	author={J. {Xu} and S. {Zhu} and G. {Liao} and Y. {Zhang}},
	journal={Journal of Radars}, 
	title={An overview of frequency diverse array radar technology}, 
	year={2018. (Chinese)},
	volume={7},
	number={2},
	pages={167-182},
	doi={10.1200/JR18023}}

@ARTICLE{AkcakayaNehorai2011MIMO,
	author={M. {Akcakaya} and A. {Nehorai}},
	journal={IEEE Transactions on Signal Processing}, 
	title={{MIMO} Radar Sensitivity Analysis for Target Detection}, 
	year={2011},
	volume={59},
	number={7},
	pages={3241-3250},
	doi={10.1109/TSP.2011.2141665}}

@ARTICLE{ChengWang2021PhysicalLayerSecurity,
	author={Q. {Cheng} and S. {Wang} and V. {Fusco} and F. {Wang} and J. {Zhu} and C. {Gu}},
	journal={IEEE Transactions on Wireless Communications}, 
	title={Physical-Layer Security for Frequency Diverse Array-Based Directional Modulation in Fluctuating Two-Ray Fading Channels}, 
	year={2021},
	volume={},
	number={},
	pages={1-1},
	doi={10.1109/TWC.2021.3056521}}

@ARTICLE{TanWang2021CorrectionAnalysis,
	author={Tan, Ming and Wang, Chunyang and Li, Zhihui},
	journal={IEEE Transactions on Antennas and Propagation}, 
	title={Correction Analysis of Frequency Diverse Array Radar About Time}, 
	year={2021},
	volume={69},
	number={2},
	pages={834-847},
	doi={10.1109/TAP.2020.3016508}}

@article{GuiWang2021FDAradar,
	title = {{FDA} radar with {Doppler}-spreading consideration: Mainlobe clutter suppression for blind-{Doppler} target detection},
	journal = {Signal Processing},
	volume = {179},
	pages = {107773},
	year = {2021},
	issn = {0165-1684},
	author = {Ronghua Gui and Wen-Qin Wang and Alfonso Farina and Hing Cheung So}
}

@ARTICLE{WangZhu2021RangeAmbiguousClutter,
	author={Wang, Yuzhuo and Zhu, Shengqi},
	journal={IEEE Transactions on Vehicular Technology}, 
	title={Range Ambiguous Clutter Suppression for {FDA-MIMO} Forward Looking Airborne Radar Based on Main Lobe Correction}, 
	year={2021},
	volume={70},
	number={3},
	pages={2032-2046},
	doi={10.1109/TVT.2021.3057436}}

@ARTICLE{TangJiang2020RangeAngle,
	author={Tang, Wen-Gen and Jiang, Hong and Zhang, Qi},
	journal={IEEE Signal Processing Letters}, 
	title={Range-Angle Decoupling and Estimation for {FDA-MIMO} Radar via Atomic Norm Minimization and Accelerated Proximal Gradient}, 
	year={2020},
	volume={27},
	number={},
	pages={366-370},
	doi={10.1109/LSP.2020.2972470}}

@ARTICLE{HuangBasit2022AdaptiveDetection,
	author={Huang, Bang and Basit, Abdul and Wang, Wen-Qin and Zhang, Shunsheng},
	journal={IEEE Geoscience and Remote Sensing Letters}, 
	title={Adaptive Detection With {Bayesian} Framework for {FDA-MIMO} Radar}, 
	year={2022},
	volume={19},
	number={},
	pages={1-5},
	doi={10.1109/LGRS.2021.3123654}}

@INPROCEEDINGS{Antonik2006Frequencydiverse,
	author={Antonik, P. and Wicks, M.C. and Griffiths, H.D. and Baker, C.J.},
	booktitle={2006 IEEE Conference on Radar}, 
	title={Frequency diverse array radars}, 
	year={2006},
	volume={},
	number={},
	pages={3 pp.-},
	doi={10.1109/RADAR.2006.1631800},
	address={Verona, NY, USA}}

@patent{wicks2008frequency,
		title={Frequency diverse array with independent modulation of frequency, amplitude, and phase},
		author={Wicks, Michael C and Antonik, Paul},
		publisher={Google Patents},
		country ={US Patent 7,319,427}
	}

@patent{wicks2009method,
		title={Method and apparatus for a frequency diverse array},
		author={Wicks, Michael C and Antonik, Paul},
		publisher={Google Patents},
		country={US Patent 7,511,665}
	}

@phdthesis{antonik2009investigation,
		title={An investigation of a frequency diverse array},
		author={Antonik, Paul},
		year={2009},
		institution={University College London},
		address={London},
		school={UCL (University College London)}
	}

@phdthesis{aytun2010frequency,
	title={Frequency diverse array radar},
	author={Aytun, Alper},
	year={2010},
	institution={Naval Postgraduate School},
	address={Monterey, California}
	}

@article{liu2020joint,
  title={Joint transmit beamforming for multiuser {MIMO} communications and MIMO radar},
  author={Liu, Xiang and Huang, Tianyao and Shlezinger, Nir and Liu, Yimin and Zhou, Jie and Eldar, Yonina C},
  journal={IEEE Transactions on Signal Processing},
  volume={68},
  pages={3929--3944},
  year={2020},
  publisher={IEEE}
}

@article{huang2022bayesian,
  title={Bayesian detection of distributed targets for {FDA-MIMO} radar in Gaussian interference},
  author={Huang, Bang and Wang, Wen-Qin and Orlando, Danilo and Basit, Abdul and Liu, Jun},
  journal={IEEE Signal Processing Letters},
  volume={29},
  pages={2168--2172},
  year={2022},
  publisher={IEEE}
}

@article{huang2022adaptive,
  title={Adaptive distributed target detection for {FDA-MIMO} radar in Gaussian clutter without training data},
  author={Huang, Bang and Jian, Jiangwei and Basit, Abdul and Gui, Ronghua and Wang, Wen-Qin},
  journal={IEEE Transactions on Aerospace and Electronic Systems},
  volume={58},
  number={4},
  pages={2961--2972},
  year={2022},
  publisher={IEEE}
}

@article{huang2023adaptive,
  title={Adaptive multiple targets detection for {FDA-MIMO} radar with Gaussian clutter},
  author={Huang, Bang and Orlando, Danilo and Wang, Wen-Qin and Liu, Weijian and Lan, Lan},
  journal={Signal Processing},
  volume={205},
  pages={108893},
  year={2023},
  publisher={Elsevier}
}

@article{li2024knowledge,
  title={Knowledge-aided Bayesian detection of distributed target for {FDA-MIMO} radar in Gaussian clutter},
  author={Li, Ping and Huang, Bang and Wang, Wen-Qin},
  journal={IEEE Transactions on Radar Systems},
  volume={2},
  pages={344--354},
  year={2024},
  publisher={IEEE}
}

@phdthesis{Huang2023phdResearch,
  author  = {Huang, Bang},
  title   = {Research on Target Detection Algorithm for
{FDA-MIMO} Radar},
  school  = {University of Electronic Science and Technology of China},
  address = {Chengdu, China},
  year    = {2023,(Chinese)}
}

@article{guan2021Passive,
  title={Passive Localization Countermeasure Based on Frequency Diverse Array},
  author={GUAN Haoliang and ZHANG Shunsheng and WANG Wenqin},
  journal={Journal of Radar},
  year={2021},
     number={6},
volume={10}
}

@article{wanxu2025clutterSuppression,
  title={Clutter Suppression for Spaceborne {FDA-MIMO} Radar with {QPC}},
  author={Wan, Fuhai and Xu, Jingwei and Xu, Yanhong and Wang, Weiwei and Liao, Guisheng and Zhang, Yuhong},
  journal={IEEE Transactions on Aerospace and Electronic Systems},
  year={2025},
  publisher={IEEE}
}

@article{ji2023template,
  title={A template-modulation jamming against SAR based on frequency-diverse array},
  author={Ji, Penghui and Xing, Shiqi and Dai, Dahai and Pang, Bo and Feng, Dejun},
  journal={IEEE Geoscience and Remote Sensing Letters},
  volume={20},
  pages={1--5},
  year={2023},
  publisher={IEEE}
}

@article{lou2022joint,
  title={Joint optimal and adaptive 2-D spatial filtering technique for {FDA-MIMO SAR} deception jamming separation and suppression},
  author={Lou, Mingyue and Yang, Jianyu and Li, Zhongyu and Ren, Hang and An, Hongyang and Wu, Junjie},
  journal={IEEE Transactions on Geoscience and Remote Sensing},
  volume={60},
  pages={1--14},
  year={2022},
  publisher={IEEE}
}

@inproceedings{yuNie2020scatteredwave,
  title={Scattered wave deception jamming against squint SAR using frequency diverse array},
  author={Yu, Jianfei and Nie, Wei and Zhou, Mu and Tian, Zengshan and Huang, Bang},
  booktitle={2020 IEEE Asia-Pacific Microwave Conference (APMC)},
  pages={979--981},
  year={2020},
  organization={IEEE}
}

@article{zhangJin2025fastrepeater,
  title={A fast repeater mainlobe deceptive jamming suppression method for {FDA-MIMO SAR} under complex motion condition},
  author={Zhang, Hanqing and Jin, Guodong and Zhang, Hongbo and Wang, Yu and Cheng, Yuan and Guo, Zhenyu and Ye, Shaohua and Zhu, Daiyin},
  journal={IEEE Transactions on Aerospace and Electronic Systems},
  year={2025},
  publisher={IEEE}
}

@inproceedings{huang2019deceptive,
  title={A deceptive jamming against high and low orbit bistatic SAR using frequency diversity array},
  author={Huang, Bang and Nusenu, Shaddrack Yaw and Zhang, Shunsheng and Wang, Wen-Qin and Liao, Yi and Wang, Zhibin},
  booktitle={2019 6th Asia-Pacific Conference on Synthetic Aperture Radar (APSAR)},
  pages={1--5},
  year={2019},
  organization={IEEE}
}

@article{wang2024generalbandwidth,
  title={General bandwidth synthesis approach for multiresolution {SAR} imaging with frequency diverse array},
  author={Wang, Keyi and Liao, Guisheng and Xu, Jingwei and Xu, Yanhong and Zhang, Yuhong},
  journal={IEEE Transactions on Geoscience and Remote Sensing},
  volume={62},
  pages={1--15},
  year={2024},
  publisher={IEEE}
}

@inproceedings{zhuYu2019applicationfrequency,
  title={Application of frequency diverse array to resolve range ambiguity for {SAR} imaging},
  author={Zhu, Jingjing and Yu, Kun and Zhu, Shengqi and Wang, Bo and Wang, Rujie and Wang, Lei},
  booktitle={2019 6th Asia-Pacific Conference on Synthetic Aperture Radar (APSAR)},
  pages={1--5},
  year={2019},
  organization={IEEE}
}

@article{linHuang2017unambiguoussignal,
  title={Unambiguous signal reconstruction approach for SAR imaging using frequency diverse array},
  author={Lin, Chenchen and Huang, Puming and Wang, Weiwei and Li, Yu and Xu, Jingwei},
  journal={IEEE Geoscience and Remote Sensing Letters},
  volume={14},
  number={9},
  pages={1628--1632},
  year={2017},
  publisher={IEEE}
}

@article{wenZhang2023frequencydiverse,
  title={A frequency diverse array {SAR} processing framework based on the segmented phase code waveform for {HRWS} imaging},
  author={Wen, Yuhao and Zhang, Zhimin and Chen, Zhen and Zhao, Haonan and Zhang, Yongwei and Fan, Huaitao},
  journal={IEEE Geoscience and Remote Sensing Letters},
  volume={20},
  pages={1--5},
  year={2023},
  publisher={IEEE}
}

@inproceedings{zhouWang2021novelHigh,
  title={A Novel High-Resolution and Wide-Swath {SAR} Imaging Mode Using Frequency Diverse Planar Array},
  author={Zhou, Yashi and Wang, Wei and Chen, Zhen and Zhao, Qingchao and Deng, Yunkai and Wang, Robert},
  booktitle={EUSAR 2021; 13th European Conference on Synthetic Aperture Radar},
  pages={1--5},
  year={2021},
  organization={VDE}
}

@article{wangXu2016rangeambiguity,
  title={A range ambiguity resolution approach for high-resolution and wide-swath {SAR} imaging using frequency diverse array},
  author={Wang, Chenghao and Xu, Jingwei and Liao, Guisheng and Xu, Xuefei and Zhang, Yuhong},
  journal={IEEE Journal of Selected Topics in Signal Processing},
  volume={11},
  number={2},
  pages={336--346},
  year={2016},
  publisher={IEEE}
}

@article{zhouWang2020highresolution,
  title={High-resolution and wide-swath {SAR} imaging mode using frequency diverse planar array},
  author={Zhou, Yashi and Wang, Wei and Chen, Zhen and Zhao, Qingchao and Zhang, Heng and Deng, Yunkai and Wang, Robert},
  journal={IEEE Geoscience and Remote Sensing Letters},
  volume={18},
  number={2},
  pages={321--325},
  year={2020},
  publisher={IEEE}
}

@article{chenZhang2020elevatedfrequency,
  title={Elevated frequency diversity array: A novel approach to high resolution and wide swath imaging for synthetic aperture radar},
  author={Chen, Zhen and Zhang, Zhimin and Zhou, Yashi and Zhao, Qingchao and Wang, Wei},
  journal={IEEE Geoscience and Remote Sensing Letters},
  volume={19},
  pages={1--5},
  year={2020},
  publisher={IEEE}
}

@article{wangwang2021lpiproperty,
  title={{LPI} property of {FDA} transmitted signal},
  author={Wang, Liu and Wang, Wen-Qin and Guan, Haoliang and Zhang, Shunsheng},
  journal={IEEE Transactions on Aerospace and Electronic Systems},
  volume={57},
  number={6},
  pages={3905--3915},
  year={2021},
  publisher={IEEE}
}

@article{liuwang2025mainlobeJamming,
  title={A Mainlobe Jamming Suppression Method for an FDA--MIMO Radar System with an Ultralarge Sparse Aperture},
  author={Liu, Quanhua and Wang, Changjie and Tian, Dezhi and Pu, Weiming and Zhou, Xiangrong and Liang, Zhennan},
  journal={IEEE Transactions on Aerospace and Electronic Systems},
  year={2025},
  publisher={IEEE}
}

@article{lanzhang2024suppressingmainlobe,
  title={Suppressing mainlobe deceptive jammers via two-low-rank matrix decomposition in {FDA-MIMO} radar},
  author={Lan, Lan and Zhang, Yitao and Xu, Jingwei and Liao, Guisheng and So, Hing Cheung},
  journal={IEEE Transactions on Aerospace and Electronic Systems},
  volume={61},
  number={2},
  pages={2885--2898},
  year={2024},
  publisher={IEEE}
}

@inproceedings{zhongtao2023mainlobedeceptive,
  title={Mainlobe deceptive jamming suppression with polarimetric characteristic {FDA-MIMO} radar},
  author={Zhong, Tiantian and Tao, Haihong and Cao, Han and Liao, Haiyun},
  booktitle={IET Conference Proceedings CP874},
  volume={2023},
  number={47},
  pages={2380--2383},
  year={2023},
  organization={IET}
}

@article{liuwang2022discrimination,
  title={Discrimination of mainlobe deceptive target with meter-wave {FDA-MIMO} radar},
  author={Liu, Yibin and Wang, Chunyang and Gong, Jian and Chen, Geng},
  journal={IEEE Communications Letters},
  volume={26},
  number={5},
  pages={1131--1135},
  year={2022},
  publisher={IEEE}
}

@article{yuanhe2024suppress,
  title={Suppress mainlobe deceptive jamming target under unambiguous range compensation based on {FDA-MIMO} radar},
  author={Yuan, Tao and He, Feng and Dong, Zhen and Su, Yi and Fan, Zhuoya and Yu, Lei},
  journal={IEEE Transactions on Aerospace and Electronic Systems},
  volume={60},
  number={5},
  pages={6853--6868},
  year={2024},
  publisher={IEEE}
}

@inproceedings{wangjia2025ucnn,
  title={{UC-CNN}: An U-Shaped Complex {CNN} for Mainlobe Deceptive Jamming Suppression with {FDA-MIMO} Radar},
  author={Wang, Ruida and Jia, Yizhen and Wang, Wenbiao and Wang, Wenqin},
  booktitle={2025 IEEE Radar Conference (RadarConf25)},
  pages={682--686},
  year={2025},
  organization={IEEE}
}

@article{houwang2025mainlobe,
  title={Mainlobe Deceptive Jamming Discrimination Using Distributed {MIMO-FDA} Radar},
  author={Hou, Yudian and Wang, Wen-Qin},
  journal={IEEE Signal Processing Letters},
  year={2025},
  publisher={IEEE}
}

@article{tangong2023range,
  title={Range dimensional monopulse approach with {FDA-MIMO } radar for mainlobe deceptive jamming suppression},
  author={Tan, Ming and Gong, Jian and Wang, Chunyang},
  journal={IEEE Antennas and Wireless Propagation Letters},
  volume={23},
  number={2},
  pages={643--647},
  year={2023},
  publisher={IEEE}
}

@article{guihuang2025robustdetector,
  title={Robust detector designing for {FDA-MIMO} radar with training data in gaussian noise},
  author={Gui, Limin and Huang, Bang and Wang, Wen-Qin},
  journal={IEEE Transactions on Aerospace and Electronic Systems},
  year={2025},
  publisher={IEEE}
}

@article{hehuang2026parametricGLRT,
  title={Parametric {GLRT-based} Adaptive Target Detection for {FDA-MIMO} Radar in Gaussian Clutter},
  author={He, Changshan and Huang, Bang and Wang, Jianping and Liu, Lei and Zhang, Running},
  journal={IEEE Transactions on Aerospace and Electronic Systems},
  year={2026},
  publisher={IEEE}
}

@article{lihuang2025adaptivedetection,
  title={Adaptive Detection of Distributed Target for {FDA-MIMO} Radar in Compound-Gaussian Clutter},
  author={Li, Ping and Huang, Bang and Jia, Wenkai and Wang, Wen-Qin},
  journal={IEEE Transactions on Aerospace and Electronic Systems},
  year={2025},
  publisher={IEEE}
}

@phdthesis{gui2020phdResearch,
  author  = {Gui, Ronghua},
  title   = {Research on Adaptive Processing Technology for
Frequency Diverse Array Radar},
  school  = {University of Electronic Science and Technology of China},
  address = {Chengdu, China},
  year    = {2020,(Chinese)}
}

@inproceedings{gui2020target,
  title={Target reflectivity characterization for FDA radar},
  author={Gui, Ronghua and Wang, Wen-Qin and So, Hing-Cheung and Cui, Can},
  booktitle={2020 IEEE 11th Sensor Array and Multichannel Signal Processing Workshop (SAM)},
  pages={1--5},
  year={2020},
  organization={IEEE}
}

@article{cheng2017time,
  title={Time-invariant angle-range dependent directional modulation based on time-modulated frequency diverse arrays},
  author={Cheng, Qian and Zhu, Jiang and Xie, Tao and Luo, Junshan and Xu, Zuohong},
  journal={IEEE access},
  volume={5},
  pages={26279--26290},
  year={2017},
  publisher={IEEE}
}

@article{xiong2016frequency,
  title={Frequency diverse array transmit beampattern optimization with genetic algorithm},
  author={Xiong, Jie and Wang, Wen-Qin and Shao, Huaizong and Chen, Hui},
  journal={IEEE Antennas and Wireless Propagation Letters},
  volume={16},
  pages={469--472},
  year={2016},
  publisher={IEEE}
}

@article{shao2016dot,
  title={Dot-shaped range-angle beampattern synthesis for frequency diverse array},
  author={Shao, Huaizong and Dai, Jun and Xiong, Jie and Chen, Hui and Wang, Wen-Qin},
  journal={IEEE antennas and wireless propagation letters},
  volume={15},
  pages={1703--1706},
  year={2016},
  publisher={IEEE}
}

@article{liu2016random,
  title={The random frequency diverse array: A new antenna structure for uncoupled direction-range indication in active sensing},
  author={Liu, Yimin and Ruan, Hang and Wang, Lei and Nehorai, Arye},
  journal={IEEE Journal of Selected Topics in Signal Processing},
  volume={11},
  number={2},
  pages={295--308},
  year={2016},
  publisher={IEEE}
}

@article{wang2016range,
  title={Range-azimuth decouple beamforming for frequency diverse array with Costas-sequence modulated frequency offsets},
  author={Wang, Zhe and Wang, Wen-Qin and Shao, Huaizong},
  journal={EURASIP Journal on Advances in Signal Processing},
  volume={2016},
  number={1},
  pages={124},
  year={2016},
  publisher={Springer}
}

@article{xu2021fda,
  title={{FDA} beampattern synthesis with both nonuniform frequency offset and array spacing},
  author={Xu, Wei and Zhang, Lihua and Bi, Hui and Huang, Pingping and Tan, Weixian},
  journal={IEEE Antennas and Wireless Propagation Letters},
  volume={20},
  number={12},
  pages={2354--2358},
  year={2021},
  publisher={IEEE}
}

@article{gao2016decoupled,
  title={Decoupled frequency diverse array range--angle-dependent beampattern synthesis using non-linearly increasing frequency offsets},
  author={Gao, Kuandong and Wang, Wen-Qin and Cai, Jingye and Xiong, Jie},
  journal={IET Microwaves, Antennas \& Propagation},
  volume={10},
  number={8},
  pages={880--884},
  year={2016},
  publisher={Wiley Online Library}
}

@article{khan2014frequency,
  title={Frequency diverse array radar with logarithmically increasing frequency offset},
  author={Khan, Waseem and Qureshi, Ijaz Mansoor and Saeed, Sarah},
  journal={IEEE antennas and wireless propagation letters},
  volume={14},
  pages={499--502},
  year={2014},
  publisher={IEEE}
}

@inproceedings{chen2015sparse,
  title={Sparse reconstruction based frequency diverse array transmit beampattern synthesis},
  author={Chen, Hui and Shao, Huaizong and Wang, Wenqin},
  booktitle={2015 3rd International Workshop on Compressed Sensing Theory and its Applications to Radar, Sonar and Remote Sensing (CoSeRa)},
  pages={253--257},
  year={2015},
  organization={IEEE}
}

@article{yao2015single,
  title={Single-sideband time-modulated phased array},
  author={Yao, A-Min and Wu, Wen and Fang, Da-Gang},
  journal={IEEE Transactions on Antennas and Propagation},
  volume={63},
  number={5},
  pages={1957--1968},
  year={2015},
  publisher={IEEE}
}

@article{yang2018optimization,
  title={Optimization of sparse frequency diverse array with time-invariant spatial-focusing beampattern},
  author={Yang, Yu-Qian and Wang, Hao and Wang, Hai-Qing and Gu, Si-Qi and Xu, Da-Long and Quan, Shuang-Long},
  journal={IEEE Antennas and Wireless Propagation Letters},
  volume={17},
  number={2},
  pages={351--354},
  year={2018},
  publisher={IEEE}
}

@article{yao2016solutions,
  title={Solutions of time-invariant spatial focusing for multi-targets using time modulated frequency diverse antenna arrays},
  author={Yao, A-Min and Wu, Wen and Fang, Da-Gang},
  journal={IEEE Transactions on Antennas and Propagation},
  volume={65},
  number={2},
  pages={552--566},
  year={2016},
  publisher={IEEE}
}

@article{yao2016frequency,
  title={Frequency diverse array antenna using time-modulated optimized frequency offset to obtain time-invariant spatial fine focusing beampattern},
  author={Yao, A-Min and Wu, Wen and Fang, Da-Gang},
  journal={IEEE Transactions on Antennas and Propagation},
  volume={64},
  number={10},
  pages={4434--4446},
  year={2016},
  publisher={IEEE}
}

@article{zhai2021joint,
  title={Joint optimization of sparse {FDAs} for time invariant transmit beampattern synthesis},
  author={Zhai, Weitong and Wang, Xiangrong and Greco, Maria S and Gini, Fulvio},
  journal={IEEE Signal Processing Letters},
  volume={29},
  pages={110--114},
  year={2021},
  publisher={IEEE}
}

@article{chen2019accurate,
  title={Accurate models of time-invariant beampatterns for frequency diverse arrays},
  author={Chen, Kejin and Yang, Shiwen and Chen, Yikai and Qu, Shi-Wei},
  journal={IEEE Transactions on Antennas and Propagation},
  volume={67},
  number={5},
  pages={3022--3029},
  year={2019},
  publisher={IEEE}
}

@article{hei2024ann,
  title={{ANN}-Assisted Quasi-Time-Invariant Beamforming for Retrodirective Frequency Diverse Array},
  author={Hei, Yong Qiang and Ju, Xiao Yan and Ma, Long Yuan and Li, Wen Tao and Shi, Xiao Wei},
  journal={IEEE Transactions on Antennas and Propagation},
  volume={72},
  number={5},
  pages={4271--4282},
  year={2024},
  publisher={IEEE}
}

@article{liao2023time,
  title={Time-variance analysis for frequency-diverse array beampatterns},
  author={Liao, Yi and Zeng, Guanghui and Luo, Zhibang and Liu, Qing Huo},
  journal={IEEE Transactions on Antennas and Propagation},
  volume={71},
  number={8},
  pages={6558--6567},
  year={2023},
  publisher={IEEE}
}

@article{wang2021lpi,
  title={{LPI} property of {FDA} transmitted signal},
  author={Wang, Liu and Wang, Wen-Qin and Guan, Haoliang and Zhang, Shunsheng},
  journal={IEEE Transactions on Aerospace and Electronic Systems},
  volume={57},
  number={6},
  pages={3905--3915},
  year={2021},
  publisher={IEEE}
}

@article{liao2023estimation,
  title={Estimation of time-varying channels in virtual angular domain for massive MIMO systems},
  author={Liao, Mingduo and Zakharov, Yuriy},
  journal={IEEE Access},
  volume={11},
  pages={1923--1933},
  year={2023},
  publisher={IEEE}
}

@article{wang2014linear,
  title={Linear frequency diverse array manifold geometry and ambiguity analysis},
  author={Wang, Yongbing and Wang, Wen-Qin and Chen, Hui},
  journal={IEEE Sensors Journal},
  volume={15},
  number={2},
  pages={984--993},
  year={2014},
  publisher={IEEE}
}

@article{wang2017fda,
  title={{FDA} radar ambiguity function characteristics analysis and optimization},
  author={Wang, Wen-Qin and Dai, Miaomiao and Zheng, Zhi},
  journal={IEEE Transactions on Aerospace and Electronic Systems},
  volume={54},
  number={3},
  pages={1368--1380},
  year={2017},
  publisher={IEEE}
}

@ARTICLE{LiuWang2024AmbiguityFunction,
  author={Liu, Mingjie and Wang, Chunyang and Gong, Jian and Tan, Ming and Bao, Lei and Zhou, Changlin},
  journal={IEEE Geoscience and Remote Sensing Letters}, 
  title={Ambiguity Function Analysis and Optimization of Coherent FDA Radar}, 
  year={2024},
  volume={21},
  number={},
  pages={1-5},
  doi={10.1109/LGRS.2024.3406435}}

@ARTICLE{TanWang2021ANoveleceptive,
  author={Tan, Ming and Wang, Chunyang and Xue, Bin and Xu, Jingwei},
  journal={IEEE Sensors Journal}, 
  title={A Novel Deceptive Jamming Approach Against Frequency Diverse Array Radar}, 
  year={2021},
  volume={21},
  number={6},
  pages={8323-8332},
  doi={10.1109/JSEN.2020.3045757}}

@ARTICLE{MirAlbasha2025FrequencyDiverseArray,
  author={Mir, Hasan Saeed and Albasha, Lutfi and Wong, Kainam Thomas},
  journal={IEEE Wireless Communications Letters}, 
  title={Frequency Diverse Array Using Equivalent Transmit Beamforming: Array Factor for Multiple Receive-Antennas}, 
  year={2025},
  volume={14},
  number={10},
  pages={3259-3263},
  doi={10.1109/LWC.2025.3591496}}

@ARTICLE{MahmoodMir2018FrequencyDiverseArray,
  author={Mahmood, Mobeen and Mir, Hasan},
  journal={IEEE Antennas and Wireless Propagation Letters}, 
  title={Frequency Diverse Array Beamforming Using Nonuniform Logarithmic Frequency Increments}, 
  year={2018},
  volume={17},
  number={10},
  pages={1817-1821},
  doi={10.1109/LAWP.2018.2867085}}

@ARTICLE{LiaoTang2022ALowSidelobe,
  author={Liao, Yi and Tang, Hu and Wang, Wen-Qin and Xing, Mengdao},
  journal={IEEE Transactions on Antennas and Propagation}, 
  title={A Low Sidelobe Deceptive Jamming Suppression Beamforming Method With a Frequency Diverse Array}, 
  year={2022},
  volume={70},
  number={6},
  pages={4884-4889},
  doi={10.1109/TAP.2021.3138529}}

@ARTICLE{XuShi2015RangeAngleDependent,
  author={Xu, Yanhong and Shi, Xiaowei and Xu, Jingwei and Li, Ping},
  journal={IEEE Transactions on Antennas and Propagation}, 
  title={Range-Angle-Dependent Beamforming of Pulsed Frequency Diverse Array}, 
  year={2015},
  volume={63},
  number={7},
  pages={3262-3267},
  doi={10.1109/TAP.2015.2423698}}

@book{mailloux2017phased,
  title={Phased array antenna handbook},
  author={Mailloux, Robert J},
  year={2017},
  publisher={Artech house}
}

@article{chen2008mimo,
  title={{MIMO} radar ambiguity properties and optimization using frequency-hopping waveforms},
  author={Chen, Chun-Yang and Vaidyanathan, PP},
  journal={IEEE Transactions on signal processing},
  volume={56},
  number={12},
  pages={5926--5936},
  year={2008},
  publisher={IEEE}
}

@article{san2007mimo,
  title={{MIMO} radar ambiguity functions},
  author={San Antonio, Geoffrey and Fuhrmann, Daniel R and Robey, Frank C},
  journal={IEEE Journal of Selected Topics in Signal Processing},
  volume={1},
  number={1},
  pages={167--177},
  year={2007},
  publisher={IEEE}
}

@article{basit2018development,
  title={Development of frequency diverse array radar technology: a review},
  author={Basit, Abdul and Khan, Wasim and Khan, Shafqatullah and Qureshi, Ijaz Mansoor},
  journal={IET Radar, Sonar \& Navigation},
  volume={12},
  number={2},
  pages={165--175},
  year={2018},
  publisher={Wiley Online Library}
}

@ARTICLE{LanRosamilia2022AdaptiveTargetDetection,
		author={Lan, Lan and Rosamilia, Massimo and Aubry, Augusto and Maio, Antonio De and Liao, Guisheng and Xu, Jingwei},
		journal={IEEE Transactions on Aerospace and Electronic Systems}, 
		title={Adaptive Target Detection with Polarimetric {FDA-MIMO} Radar}, 
		year={2022},
		volume={},
		number={},
		pages={1-16},
		doi={10.1109/TAES.2022.3210887}}

@ARTICLE{HuangJian2022AdaptiveDistributed,
	author={Huang, Bang and Jian, Jiangwei and Basit, Abdul and Gui, Ronghua and Wang, Wen-Qin},
	journal={IEEE Transactions on Aerospace and Electronic Systems}, 
	title={Adaptive Distributed Target Detection for
	{FDA-MIMO} Radar in {Gaussian} Clutter
	without Training Data}, 
year={2022},
volume={58},
number={4},
pages={2961-2972},
doi={10.1109/TAES.2022.3145781}}

@ARTICLE{HuangBasit2022AdaptiveMoving,
	author={Huang, Bang and Basit, Abdul and Gui, Ronghua and Wang, Wen-Qin},
	journal={IEEE Transactions on Vehicular Technology}, 
	title={Adaptive Moving Target Detection Without Training Data for {FDA-MIMO} Radar}, 
	year={2022},
	volume={71},
	number={1},
	pages={220-232},
	doi={10.1109/TVT.2021.3126781}}

@article{Huang2023Adaptivemultiple,
		title = {Adaptive multiple targets detection for {FDA-MIMO} radar with Gaussian clutter},
		journal = {Signal Processing},
		volume = {205},
		pages = {108893},
		year = {2023},
		issn = {0165-1684},
		doi = {https://doi.org/10.1016/j.sigpro.2022.108893},
		url = {https://www.sciencedirect.com/science/article/pii/S0165168422004327},
		author = {Bang Huang and Danilo Orlando and Wen-Qin Wang and Weijian Liu and Lan Lan}
	}

@INPROCEEDINGS{ZhuZhu2022AdaptiveMultiTarget,
		author={Zhu, Jingjing and Zhu, Shengqi and Lan, Lan and Xu, Jingwei},
		booktitle={2022 IEEE 12th Sensor Array and Multichannel Signal Processing Workshop (SAM)}, 
		title={Adaptive Multi-Target Detection with {FDA-MIMO} Radar}, 
		year={2022},
		volume={},
		number={},
		pages={370-374},
		doi={10.1109/SAM53842.2022.9827880},
		address={Trondheim, Norway}}

@ARTICLE{Zhu2023SimultaneousDetection,
			author={Zhu, Jingjing and Zhu, Shengqi and Xu, Jingwei and Lan, Lan and Liao, Guisheng},
			journal={IEEE Transactions on Aerospace and Electronic Systems}, 
			title={Simultaneous Detection and Discrimination of Mainlobe Deceptive Jammers in {FDA-MIMO} Radar}, 
			year={2023},
			volume={},
			number={},
			pages={1-15},
			doi={10.1109/TAES.2023.3291683}}

@ARTICLE{Kelly1986AnAdaptiveDetection,
	author={ E.J. {Kelly}},
	journal={IEEE Transactions on Aerospace and Electronic Systems}, 
	title={An Adaptive Detection Algorithm}, 
	year={1986},
	volume={AES-22},
	number={2},
	pages={115-127},
	doi={10.1109/TAES.1986.310745}}

@ARTICLE{ConteDeMaio2003Distributedtargetdetection,
	author={{Conte}, E. and {De Maio}, A.},
	journal={IEEE Transactions on Aerospace and Electronic Systems}, 
	title={Distributed target detection in compound-Gaussian noise with {Rao} and {Wald} tests}, 
	year={2003},
	volume={39},
	number={2},
	pages={568-582},
	doi={10.1109/TAES.2003.1207267}}

@ARTICLE{GongWang2018TimeInvariantJoint,
	author={{Gong}, S.Q. and {Wang}, S. and {Chen}, S. and {Xing}, C.W. and {Wei}, X.},
	journal={IEEE Transactions on Signal Processing}, 
	title={Time-Invariant Joint Transmit and Receive Beampattern Optimization for Polarization-Subarray Based Frequency Diverse Array Radar}, 
	year={2018},
	volume={66},
	number={20},
	pages={5364-5379},
	doi={10.1109/TSP.2018.2868041}}

@ARTICLE{WangSo2014TransmitSubaperturing,
	author={{Wang}, W.-Q. and {So}, H. C.},
	journal={IEEE Transactions on Signal Processing}, 
	title={Transmit Subaperturing for Range and Angle Estimation in Frequency Diverse Array Radar}, 
	year={2014},
	volume={62},
	number={8},
	pages={2000-2011},
	doi={10.1109/TSP.2014.2305638}}

@article{liu2025jointdoa,
  title={Joint {DOA-Range} Estimation for Coherent Signals Exploiting Moving Time-Modulated Frequency Diverse Coprime Array},
  author={Liu, Zhihui and Ma, Biyun and Liu, Jiaojiao and Yang, Kun and Wang, Yide},
  journal={IEEE Signal Processing Letters},
  year={2025},
  publisher={IEEE}
}

@ARTICLE{GuiWang2018CoherentPulsedFDA,
	author={{Gui}, R.H. and {Wang}, W.-Q. and {Cui}, C. and {So}, H. C.},
	journal={IEEE Transactions on Signal Processing}, 
	title={Coherent Pulsed-{FDA} Radar Receiver Design With Time-Variance Consideration: {SINR} and {CRB} Analysis}, 
	year={2018},
	volume={66},
	number={1},
	pages={200-214},
	doi={10.1109/TSP.2017.2764860}}

@ARTICLE{BasitWang2021FDABased,
	author={{Basit}, A. and {Wang}, W.Q. and {Nusenu}, S. Y. and {Wali}, S.},
	journal={IEEE Transactions on Wireless Communications}, 
	title={{FDA} Based {QSM} for mmWave Wireless Communications: Frequency Diverse Transmitter and Reduced Complexity Receiver}, 
	year={2021},
	volume={},
	number={},
	pages={1-1},
	doi={10.1109/TWC.2021.3060512}}

@ARTICLE{WangYan2021SecrecyZone,
	author={{Wang}, S.Y. and {Yan}, S.H. and {Zhang}, J. and {Yang}, N. and {Chen}, R.Q. and {Shu}, F.},
	journal={IEEE Transactions on Vehicular Technology}, 
	title={Secrecy Zone Achieved by Directional Modulation With Random Frequency Diverse Array}, 
	year={2021},
	volume={70},
	number={2},
	pages={2001-2006},
	doi={10.1109/TVT.2021.3054803}}

@article{Liuliu2016Performancepredictionofsubspacebased,
	title = {Performance prediction of subspace-based adaptive detectors with signal mismatch},
	journal = {Signal Processing},
	volume = {123},
	pages = {122-126},
	year = {2016},
	issn = {0165-1684},
	author = {Weijian Liu and Jun Liu and Chen Zhang and Hongli Li and Xueke Wang}
}

@ARTICLE{DeMaio2004Anewderivation,
	author={Antonio De Maio},
	journal={IEEE Signal Processing Letters}, 
	title={A new derivation of the adaptive matched filter}, 
	year={2004},
	volume={11},
	number={10},
	pages={792-793},
	doi={10.1109/LSP.2004.835464}}

@ARTICLE{Choi2023Analysisof,
	author={Choi, Woohyeok and Georgiadis, Apostolos and Tentzeris, Manos M. and Kim, Sangkil},
	journal={IEEE Transactions on Antennas and Propagation}, 
	title={Analysis of Exponential Frequency-Diverse Array for Short-Range Beam-Focusing Technology}, 
	year={2023},
	volume={71},
	number={2},
	pages={1437-1447},
	doi={10.1109/TAP.2022.3226155}}

@ARTICLE{LiuZhu2023RangeAmbiguous,
	author={Liu, Zhixin and Zhu, Shengqi and Xu, Jingwei and He, Xiongpeng and Duan, Keqing and Lan, Lan},
	journal={IEEE Transactions on Geoscience and Remote Sensing}, 
	title={Range-Ambiguous Clutter Suppression for {STAP}-Based Radar With Vertical Coherent Frequency Diverse Array}, 
	year={2023},
	volume={61},
	number={},
	pages={1-17},
	doi={10.1109/TGRS.2023.3291738}}

@ARTICLE{LiaoWang2019FrequencyDiverseArray,
  author={Liao, Yi and Wang, Wen-Qin and Zheng, Zhi},
  journal={IEEE Transactions on Antennas and Propagation}, 
  title={Frequency Diverse Array Beampattern Synthesis Using Symmetrical Logarithmic Frequency Offsets for Target Indication}, 
  year={2019},
  volume={67},
  number={5},
  pages={3505-3509},
  doi={10.1109/TAP.2019.2900353}}

@ARTICLE{Zhou2021HighResolutionandWideSwath,
 	author={Zhou, Yashi and Wang, Wei and Chen, Zhen and Zhao, Qingchao and Zhang, Heng and Deng, Yunkai and Wang, Robert},
 	journal={IEEE Geoscience and Remote Sensing Letters}, 
 	title={High-Resolution and Wide-Swath {SAR} Imaging Mode Using Frequency Diverse Planar Array}, 
 	year={2021},
 	volume={18},
 	number={2},
 	pages={321-325},
 	doi={10.1109/LGRS.2020.2974041}}

@ARTICLE{Zhang2023HighResolution,
 	author={Zhang, Mengdi and Liao, Guisheng and Xu, Jingwei and He, Xiongpeng and Liu, Qi and Lan, Lan and Li, Shiyin},
 	journal={IEEE Transactions on Aerospace and Electronic Systems}, 
 	title={High-Resolution and Wide-Swath {SAR} Imaging With Sub-Band Frequency Diverse Array}, 
 	year={2023},
 	volume={59},
 	number={1},
 	pages={172-183},
 	doi={10.1109/TAES.2022.3187386}}

@INPROCEEDINGS{LiYang2017TwostepBayesian,
	author={Li, Na and Yang, Haining and Cui, Guolong and Kong, Lingjiang and Liu, Qing Huo},
	booktitle={2017 IEEE Radar Conference (RadarConf)}, 
	title={Two-step Bayesian detection for {MIMO} radar in compound-Gaussian clutter with Gamma texture}, 
	year={2017},
	volume={},
	number={},
	pages={0146-0151},
	doi={10.1109/RADAR.2017.7944187},
	address={ Seattle, WA, USA}}

@article{chong2010mimo,
	title={{MIMO} radar detection in non-Gaussian and heterogeneous clutter},
	author={Chong, Chin Yuan and Pascal, Fr{\'e}d{\'e}ric and Ovarlez, Jean-Philippe and Lesturgie, Marc},
	journal={IEEE Journal of selected topics in signal processing},
	volume={4},
	number={1},
	pages={115--126},
	year={2010},
	publisher={IEEE}
}

@ARTICLE{XuLi2008Targetdetection,
	author={L. {Xu} and J. {Li} and P. {Stoica}},
	journal={IEEE Transactions on Aerospace and Electronic Systems}, 
	title={Target detection and parameter estimation for {MIMO} radar systems}, 
	year={2008},
	volume={44},
	number={3},
	pages={927-939},
	doi={10.1109/TAES.2008.4655353}}

@ARTICLE{HuangWang2022FDABased,
	author={Huang, Bang and Wang, Wen-Qin and Zhang, Shunsheng and Liao, Yi},
	journal={IEEE Transactions on Aerospace and Electronic Systems}, 
	title={{FDA}-Based Space–Time–Frequency Deceptive Jamming Against {SAR} Imaging}, 
	year={2022},
	volume={58},
	number={3},
	pages={2127-2140},
	doi={10.1109/TAES.2021.3130212}}

@book{stoica2005spectral,
	title={Spectral analysis of signals},
	author={Stoica, Petre and Moses, Randolph L and others},
	volume={452},
	year={2005},
	publisher={Pearson Prentice Hall Upper Saddle River, NJ}
}

@ARTICLE{Hassanien2010PhasedMIMORadar,
	author={Hassanien, Aboulnasr and Vorobyov, Sergiy A.},
	journal={IEEE Transactions on Signal Processing}, 
	title={Phased-{MIMO} Radar: A Tradeoff Between {Phased-Array} and {MIMO} Radars}, 
	year={2010},
	volume={58},
	number={6},
	pages={3137-3151},
	doi={10.1109/TSP.2010.2043976}}

@ARTICLE{Huang2019ANovelApproach, 
	author={B. {Huang} and W. {Wang} and S. {Zhang} and H. {Wang} and R. {Gui} and Z. {Lu}}, 
	journal={IEEE Geoscience and Remote Sensing Letters}, 
	title={A Novel Approach for Spaceborne SAR Scattered-Wave Deception Jamming Using Frequency Diverse Array}, 
	year={2019}, 
	volume={}, 
	number={}, 
	pages={1-5}, 
	doi={10.1109/LGRS.2019.2950454}, 
	ISSN={1558-0571}, 
	month={},}

@INPROCEEDINGS{Zhu2018DeceptiveJamming, 
	author={Y. {Zhu} and H. {Wang} and S. {Zhang} and Z. {Zheng} and W. {Wang}}, 
	booktitle={IGARSS 2018 - 2018 IEEE International Geoscience and Remote Sensing Symposium}, 
	title={Deceptive Jamming on Space-Borne Sar Using Frequency Diverse Array}, 
	year={2018}, 
	volume={}, 
	number={}, 
	pages={605-608}, 
	doi={10.1109/IGARSS.2018.8518451}, 
	ISSN={2153-6996}, 
	month={July},}

@ARTICLE{Lang2022LambWave,
	author={Lang, Yanfeng and Yang, Zhibo and Yang, Laihao and Chen, Xuefeng},
	journal={IEEE Transactions on Ultrasonics, Ferroelectrics, and Frequency Control}, 
	title={Lamb Wave Frequency Diverse Array}, 
	year={2022},
	volume={69},
	number={8},
	pages={2526-2539},
	doi={10.1109/TUFFC.2022.3182419}}

@ARTICLE{Qiu2018ArtificialNoiseAided,
	author={Qiu, Bin and Xie, Jian and Wang, Ling and Wang, Yuexian},
	journal={IEEE Access}, 
	title={Artificial-Noise-Aided Secure Transmission for Proximal Legitimate User and Eavesdropper Based on Frequency Diverse Arrays}, 
	year={2018},
	volume={6},
	number={},
	pages={52531-52543},
	doi={10.1109/ACCESS.2018.2869529}}

@article{wang2017dm,
	title={{DM} using {FDA} antenna for secure transmission},
	author={Wang, Wen-Qin},
	journal={IET Microwaves, Antennas \& Propagation},
	volume={11},
	number={3},
	pages={336--345},
	year={2017},
	publisher={Wiley Online Library}
}

@article{xiong2017directional,
	title={Directional modulation using frequency diverse array for secure communications},
	author={Xiong, Jie and Nusenu, Shaddrack Yaw and Wang, Wen-Qin},
	journal={Wireless Personal Communications},
	volume={95},
	pages={2679--2689},
	year={2017},
	publisher={Springer}
}

@ARTICLE{Jian2023PhysicalLayer,
	author={Jian, Jiangwei and Wang, Wen-Qin and Basit, Abdul and Huang, Bang},
	journal={IEEE Transactions on Wireless Communications}, 
	title={Physical Layer Security for Frequency Diverse Array-based Dual-hop Spatial Modulation}, 
	year={2023},
	volume={},
	number={},
	pages={1-1},
	doi={10.1109/TWC.2023.3253214}}

@ARTICLE{Hu2017ArtificialNoiseAided,
	author={Hu, Jinsong and Yan, Shihao and Shu, Feng and Wang, Jiangzhou and Li, Jun and Zhang, Yijin},
	journal={IEEE Access}, 
	title={Artificial-Noise-Aided Secure Transmission With Directional Modulation Based on Random Frequency Diverse Arrays}, 
	year={2017},
	volume={5},
	number={},
	pages={1658-1667},
	doi={10.1109/ACCESS.2017.2653182}}

@INPROCEEDINGS{GuiWang2018Generalreceiver,
	author={R.H. {Gui} and W.-Q {Wang} and H.Z. {Shao} },
	booktitle={2018 IEEE Radar Conference (RadarConf18)}, 
	title={General receiver design for {FDA} radar}, 
	year={2018},
	volume={},
	number={},
	pages={1-6},
	address={Oklahoma},
	doi={10.1109/RadarConf2043947.2020.9266509}}

@ARTICLE{LanXu2022MainlobeDeceptive,
	author={Lan, Lan and Xu, Jingwei and Liao, Guisheng and Zhu, Shengqi and So, Hing Cheung},
	journal={IEEE Transactions on Vehicular Technology}, 
	title={Mainlobe Deceptive Jammer Suppression in {SF-RDA} Radar: Joint Transmit-Receive Processing}, 
	year={2022},
	volume={71},
	number={12},
	pages={12602-12616},
	doi={10.1109/TVT.2022.3195408}}

@ARTICLE{Liao2020FrequencyDiverseArray,
	author={Y. {Liao} and H. {Tang} and X. {Chen} and W. -Q. {Wang}},
	journal={IEEE Antennas and Wireless Propagation Letters}, 
	title={Frequency Diverse Array Beampattern Synthesis With Taylor Windowed Frequency Offsets}, 
	year={2020},
	volume={19},
	number={11},
	pages={1901-1905},
	doi={10.1109/LAWP.2020.3024710}}

@article{liu2018distributed,
	title={Distributed target detection in partially homogeneous environment when signal mismatch occurs},
	author={Liu, Weijian and Liu, Jun and Du, Qinglei and Wang, Yong-Liang},
	journal={IEEE Transactions on Signal Processing},
	volume={66},
	number={14},
	pages={3918--3928},
	year={2018},
	publisher={IEEE}
}

@ARTICLE{Liu2023RangeAmbiguous,
	author={Liu, Zhixin and Zhu, Shengqi and Xu, Jingwei and He, Xiongpeng and Duan, Keqing and Lan, Lan},
	journal={IEEE Transactions on Geoscience and Remote Sensing}, 
	title={Range-Ambiguous Clutter Suppression for {STAP}-Based Radar With Vertical Coherent Frequency Diverse Array}, 
	year={2023},
	volume={61},
	number={},
	pages={1-17},
	doi={10.1109/TGRS.2023.3291738}}

@ARTICLE{Li2023JointRadarCommunication,
	author={Li, Mengjiao and Wang, Wen-Qin},
	journal={IEEE Access}, 
	title={Joint Radar-Communication System Design Based on {FDA-MIMO } via Frequency Index Modulation}, 
	year={2023},
	volume={11},
	number={},
	pages={67722-67736},
	doi={10.1109/ACCESS.2023.3291462}}

@ARTICLE{Wang2016MovingTarget,
	author={Wang, Wen-Qin},
	journal={IEEE Transactions on Geoscience and Remote Sensing}, 
	title={Moving-Target Tracking by Cognitive RF Stealth Radar Using Frequency Diverse Array Antenna}, 
	year={2016},
	volume={54},
	number={7},
	pages={3764-3773},
	doi={10.1109/TGRS.2016.2527057}}

@INPROCEEDINGS{Nusenu2018Dualfunction,
	author={Nusenu, Shaddrack Yaw and Wang, Wen-Qin},
	booktitle={2018 IEEE Radar Conference (RadarConf18)}, 
	title={Dual-function {FDA MIMO} radar-communications system employing costas signal waveforms}, 
	year={2018},
	volume={},
	number={},
	pages={0033-0038},
	doi={10.1109/RADAR.2018.8378526},
	address={Oklahoma City, OK, USA}}

@ARTICLE{Gong2023Optimizationof,
	author={Gong, Pengcheng and Xu, Kaiyan and Wu, Yuntao and Zhang, Jing and So, Hing Cheung},
	journal={IEEE Wireless Communications Letters}, 
	title={Optimization of {LPI-FDA-MIMO} Radar and {MIMO} Communication for Spectrum Coexistence}, 
	year={2023},
	volume={12},
	number={6},
	pages={1076-1080},
	doi={10.1109/LWC.2023.3261419}}

@ARTICLE{Gong2022JointDesign,
	author={Gong, Pengcheng and Zhang, Zhuoyu and Wu, Yuntao and Wang, Wen-Qin},
	journal={IEEE Signal Processing Letters}, 
	title={Joint Design of Transmit Waveform and Receive Beamforming for {LPI FDA-MIMO} Radar}, 
	year={2022},
	volume={29},
	number={},
	pages={1938-1942},
	doi={10.1109/LSP.2022.3205206}}

@article{aydinCogen2019codeindex,
  title={Code-index modulation aided quadrature spatial modulation for high-rate {MIMO} systems},
  author={Aydin, Erdogan and Cogen, Fatih and Basar, Ertugrul},
  journal={IEEE Transactions on Vehicular Technology},
  volume={68},
  number={10},
  pages={10257--10261},
  year={2019},
  publisher={IEEE}
}

@article{xu2024polar,
  title={Polar-coded modulation schemes for multiple-mode {OFDM} with index modulation system},
  author={Xu, Yun and Ma, Dingfei and Fang, Yi and Lv, Liang and Zhou, Yu},
  journal={IEEE Wireless Communications Letters},
  volume={13},
  number={12},
  pages={3668--3672},
  year={2024},
  publisher={IEEE}
}

@article{yanTao2024cognitivefdamimo,
  title={Cognitive {FDA-MIMO} Radar Network’s Transmit Element Selection Algorithm for Target Tracking in a Complex Interference Scenario},
  author={Yan, Yingfei and Tao, Haihong and Guo, Jingjing and Yang, Biao},
  journal={Remote Sensing},
  volume={17},
  number={1},
  pages={59},
  year={2024},
  publisher={MDPI}
}

@article{gui2021cognitivefda,
  title={Cognitive {FDA} radar transmit power allocation for target tracking in spectrally dense scenario},
  author={Gui, Ronghua and Zheng, Zhi and Wang, Wen-Qin},
  journal={Signal Processing},
  volume={183},
  pages={108006},
  year={2021},
  publisher={Elsevier}
}

@article{sun2020doaestimation,
  title={{DOA} estimation and tracking for {FDA-MIMO} radar signal},
  author={Sun, Yan and Zheng, Zhi and Wang, Wen-Qin and Liao, Tian-xing},
  journal={Digital Signal Processing},
  volume={106},
  pages={102858},
  year={2020},
  publisher={Elsevier}
}

@article{wang2016moving,
  title={Moving-target tracking by cognitive {RF} stealth radar using frequency diverse array antenna},
  author={Wang, Wen-Qin},
  journal={IEEE Transactions on Geoscience and Remote Sensing},
  volume={54},
  number={7},
  pages={3764--3773},
  year={2016},
  publisher={IEEE}
}

@article{wang2017timeinvariant,
  title={Time-invariant range-angle-dependent beampattern synthesis for FDA radar targets tracking},
  author={Wang, Yuxi and Li, Wei and Huang, Guoce and Li, Jinliang},
  journal={IEEE Antennas and Wireless Propagation Letters},
  volume={16},
  pages={2375--2379},
  year={2017},
  publisher={IEEE}
}

@article{yang2023cognitivefdamimo,
  title={Cognitive {FDA-MIMO} radar network for target discrimination and tracking with main-lobe deceptive trajectory interference},
  author={Yang, Biao and Zhu, Shengqi and He, Xiongpeng and Lan, Lan and Li, Ximin},
  journal={IEEE Transactions on Aerospace and Electronic Systems},
  volume={59},
  number={4},
  pages={4207--4222},
  year={2023},
  publisher={IEEE}
}

@inproceedings{wangWang2016cognitive,
  title={Cognitive target tracking using {FDA} radar for increased {SINR} performance},
  author={Wang, Zhe and Wang, Wen-Qin and Xiong, Jie},
  booktitle={2016 IEEE Radar Conference (RadarConf)},
  pages={1--4},
  year={2016},
  organization={IEEE}
}

@article{gui2018cognitive,
  title={Cognitive target tracking via angle-range-Doppler estimation with transmit subaperturing {FDA} radar},
  author={Gui, Ronghua and Wang, Wen-Qin and Pan, Ye and Xu, Jian},
  journal={IEEE Journal of Selected Topics in Signal Processing},
  volume={12},
  number={1},
  pages={76--89},
  year={2018},
  publisher={IEEE}
}

@article{basit2019cognitive,
  title={Cognitive {FDA-MIMO} with channel uncertainty information for target tracking},
  author={Basit, Abdul and Wang, Wen-Qin and Nusenu, Shaddrack Yaw and Zheng, Zhi},
  journal={IEEE Transactions on Cognitive Communications and Networking},
  volume={5},
  number={4},
  pages={963--975},
  year={2019},
  publisher={IEEE}
}

@article{basit2020adaptivetransmit,
  title={Adaptive transmit array sidelobe control using {FDA-MIMO} for tracking in joint radar-communications},
  author={Basit, Abdul and Wang, Wen-Qin and Nusenu, Shaddrack Yaw},
  journal={Digital Signal Processing},
  volume={97},
  pages={102619},
  year={2020},
  publisher={Elsevier}
}

@article{wang2022fdafast,
  title={{FDA-SSD}: fast depth-assisted single-shot multibox detector for 3D tracking based on monocular vision},
  author={Wang, Zihao and Yang, Sen and Shi, Mengji and Qin, Kaiyu},
  journal={Applied Sciences},
  volume={12},
  number={3},
  pages={1164},
  year={2022},
  publisher={MDPI}
}

@article{zhang2020unambiguous,
  title={Unambiguous forward-looking {SAR} imaging on {HSV-R} using frequency diverse array},
  author={Zhang, Mengdi and Liao, Guisheng and He, Xiongpeng and Zhu, Shengqi},
  journal={Sensors},
  volume={20},
  number={4},
  pages={1169},
  year={2020},
  publisher={MDPI}
}

@inproceedings{shen2021front,
  title={Front-Downward-Looking {3D SAR} Imaging Using Frequency Diversity Array},
  author={Shen, Jifa and Liao, Kefei and Ouyang, Shan and Wang, Haitao and Yu, Qiaoying},
  booktitle={2021 IEEE International Geoscience and Remote Sensing Symposium IGARSS},
  pages={3967--3970},
  year={2021},
  organization={IEEE}
}

@article{baizert2006forward,
	title={Forward-looking radar {GMTI} benefits using a linear frequency diverse array},
	author={Baizert, Piotr and Hale, Todd B and Temple, Michael A and Wicks, Michael C},
	journal={Electronics Letters},
	volume={42},
	number={22},
	pages={1311--1312},
	year={2006},
	publisher={IET}
}

@article{xu2015range,
	title={Range ambiguous clutter suppression for airborne {FDA-STAP} radar},
	author={Xu, Jingwei and Zhu, Shengqi and Liao, Guisheng},
	journal={IEEE Journal of Selected Topics in Signal Processing},
	volume={9},
	number={8},
	pages={1620--1631},
	year={2015},
	publisher={IEEE}
}

@article{xu2016space,
	title={Space-time adaptive processing with vertical frequency diverse array for range-ambiguous clutter suppression},
	author={Xu, Jingwei and Liao, Guisheng and So, Hing Cheung},
	journal={IEEE Transactions on Geoscience and Remote Sensing},
	volume={54},
	number={9},
	pages={5352--5364},
	year={2016},
	publisher={IEEE}
}

@ARTICLE{Xu2017AnAdaptive,
	author={Xu, Jingwei and Liao, Guisheng and Zhang, Yuhong and Ji, Hongbing and Huang, Lei},
	journal={IEEE Journal of Selected Topics in Signal Processing}, 
	title={An Adaptive Range-Angle-Doppler Processing Approach for {FDA-MIMO} Radar Using Three-Dimensional Localization}, 
	year={2017},
	volume={11},
	number={2},
	pages={309-320},
	doi={10.1109/JSTSP.2016.2615269}}

@ARTICLE{Xu2017RobustAdaptive,
	author={Xu, Jingwei and Liao, Guisheng and Huang, Lei and So, Hing Cheung},
	journal={IEEE Transactions on Signal Processing}, 
	title={Robust Adaptive Beamforming for Fast-Moving Target Detection With {FDA-STAP} Radar}, 
	year={2017},
	volume={65},
	number={4},
	pages={973-984},
	doi={10.1109/TSP.2016.2628340}}

@ARTICLE{Wang2021RangeAmbiguous,
	author={Wang, Yuzhuo and Zhu, Shengqi},
	journal={IEEE Transactions on Vehicular Technology}, 
	title={Range Ambiguous Clutter Suppression for {FDA-MIMO} Forward Looking Airborne Radar Based on Main Lobe Correction}, 
	year={2021},
	volume={70},
	number={3},
	pages={2032-2046},
	doi={10.1109/TVT.2021.3057436}}

@ARTICLE{Qiu2023RangeAmbiguous,
	author={Qiu, Zizhou and Liao, Zhipeng and Xu, Jingwei and Duan, Keqing},
	journal={IEEE Geoscience and Remote Sensing Letters}, 
	title={Range-Ambiguous Clutter Suppression for Space-Based Early Warning Radar Using Vertical {FDA} and Horizontal {EPC}}, 
	year={2023},
	volume={20},
	number={},
	pages={1-5},
	doi={10.1109/LGRS.2023.3260996}}

@article{wen2019clutter,
	title={Clutter suppression for airborne {FDA-MIMO} radar using multi-waveform adaptive processing and auxiliary channel {STAP}},
	author={Wen, Cai and Tao, Mingliang and Peng, Jinye and Wu, Jianxin and Wang, Tong},
	journal={Signal Processing},
	volume={154},
	pages={280--293},
	year={2019},
	publisher={Elsevier}
}

@article{xu2015deceptive,
	title={Deceptive jamming suppression with frequency diverse {MIMO} radar},
	author={Xu, Jingwei and Liao, Guisheng and Zhu, Shengqi and So, Hing Cheung},
	journal={Signal Processing},
	volume={113},
	pages={9--17},
	year={2015},
	publisher={Elsevier}
}

@ARTICLE{Wang2020MainBeamRange,
	author={Wang, Yuzhuo and Zhu, Shengqi},
	journal={IEEE Sensors Journal}, 
	title={Main-Beam Range Deceptive Jamming Suppression With Simulated Annealing {FDA-MIMO} Radar}, 
	year={2020},
	volume={20},
	number={16},
	pages={9056-9070},
	doi={10.1109/JSEN.2020.2982194}}

@ARTICLE{Xu2015SpaceTimeRange,
	author={Xu, Jingwei and Zhu, Shengqi and Liao, Guisheng},
	journal={IEEE Sensors Journal}, 
	title={Space-Time-Range Adaptive Processing for Airborne Radar Systems}, 
	year={2015},
	volume={15},
	number={3},
	pages={1602-1610},
	doi={10.1109/JSEN.2014.2364594}}

@ARTICLE{Wen2018Enhanced,
	author={Wen, Cai and Peng, Jinye and Zhou, Yan and Wu, Jianxin},
	journal={IEEE Sensors Journal}, 
	title={Enhanced Three-Dimensional Joint Domain Localized {STAP} for Airborne {FDA-MIMO} Radar Under Dense False-Target Jamming Scenario}, 
	year={2018},
	volume={18},
	number={10},
	pages={4154-4166},
	doi={10.1109/JSEN.2018.2820905}}

@ARTICLE{Lan2018Suppression,
	author={Lan, Lan and Liao, Guisheng and Xu, Jingwei and Zhang, Yuhong and Fioranelli, Francesco},
	journal={IEEE Access}, 
	title={Suppression Approach to Main-Beam Deceptive Jamming in {FDA-MIMO} Radar Using Nonhomogeneous Sample Detection}, 
	year={2018},
	volume={6},
	number={},
	pages={34582-34597},
	doi={10.1109/ACCESS.2018.2850816}}

@article{wang2013range,
	title={Range-angle localization of targets by a double-pulse frequency diverse array radar},
	author={Wang, Wen-Qin and Shao, Huaizong},
	journal={IEEE Journal of Selected Topics in Signal Processing},
	volume={8},
	number={1},
	pages={106--114},
	year={2013},
	publisher={IEEE}
}

@article{zhu2021cooperative,
	title={Cooperative range and angle estimation with PA and FDA radars},
	author={Zhu, Jingjing and Zhu, Shengqi and Xu, Jingwei and Lan, Lan and He, Xiongpeng},
	journal={IEEE Transactions on Aerospace and Electronic Systems},
	volume={58},
	number={2},
	pages={907--921},
	year={2021},
	publisher={IEEE}
}

@article{wang2014transmit,
	title={Transmit subaperturing for range and angle estimation in frequency diverse array radar},
	author={Wang, Wen-Qin and So, Hing-Cheung},
	journal={IEEE Transactions on Signal Processing},
	volume={62},
	number={8},
	pages={2000--2011},
	year={2014},
	publisher={IEEE}
}

@article{wang2018subarray,
	title={Subarray-based frequency diverse array for target range-angle localization with monopulse processing},
	author={Wang, Shengbin Luo and Xu, Zhen-Hai and Liu, Xinghua and Dong, Wei and Wang, Guoyu},
	journal={IEEE Sensors Journal},
	volume={18},
	number={14},
	pages={5937--5947},
	year={2018},
	publisher={IEEE}
}

@article{khan2015double,
	title={A double pulse {MIMO} frequency diverse array radar for improved range-angle localization of target},
	author={Khan, Wasim and Qureshi, Ijaz Mansoor and Basit, Abdul and Zubair, Muhammad},
	journal={Wireless Personal Communications},
	volume={82},
	pages={2199--2213},
	year={2015},
	publisher={Springer}
}

@article{xu2015joint,
	title={Joint range and angle estimation using {MIMO} radar with frequency diverse array},
	author={Xu, Jingwei and Liao, Guisheng and Zhu, Shengqi and Huang, Lei and So, Hing Cheung},
	journal={IEEE Transactions on Signal Processing},
	volume={63},
	number={13},
	pages={3396--3410},
	year={2015},
	publisher={IEEE}
}

@ARTICLE{Xiong2018FDAMIMO,
	author={Xiong, Jie and Wang, Wen-Qin and Gao, Kuandong},
	journal={IEEE Transactions on Aerospace and Electronic Systems}, 
	title={{FDA-MIMO} Radar Range–Angle Estimation: {CRLB, MSE}, and Resolution Analysis}, 
	year={2018},
	volume={54},
	number={1},
	pages={284-294},
	doi={10.1109/TAES.2017.2756498}}

@article{wang2014subarray,
	title={Subarray-based frequency diverse array radar for target range-angle estimation},
	author={Wang, Wen-Qin},
	journal={IEEE Transactions on Aerospace and Electronic Systems},
	volume={50},
	number={4},
	pages={3057--3067},
	year={2014},
	publisher={IEEE}
}

@article{chen2018space,
title={Space-range-Doppler focus-based low-observable moving target detection using frequency diverse array MIMO radar},
author={Chen, Xiaolong and Chen, Baoxin and Guan, Jian and Huang, Yong and He, You},
journal={IEEE Access},
volume={6},
pages={43892--43904},
year={2018},
publisher={IEEE}
}

@article{chen2023monopulse,
	title={Monopulse Parameter Estimation for {FDA-MIMO} Radar under Mainlobe Deception Jamming},
	author={Chen, Hao and Li, Rongfeng and Chen, Hui and Qu, Qizhe and Zhou, Bilei and Li, Binbin and Wang, Yongliang},
	journal={Remote Sensing},
	volume={15},
	number={16},
	pages={3947},
	year={2023},
	publisher={MDPI}
}

@article{fu2020CS,
	title={{2-D DOA} Estimation for Nested Conformal Arrays via Sparse Reconstruction},
	author={Fu, Mingcheng and Zheng, Zhi and Wang, Wen-Qin},
	journal={IEEE Commun. Lett.},
	volume={25},
	number={3},
	pages={980--984},
	year={2020},
	month = {Mar.}
}

@article{Kolda2009TensorReview,
	author = {T. G. Kolda and B. W. Bader},
	title = {Tensor Decompositions and Applications},
	journal = {SIAM Rev.},
	volume = {51},
	number = {3},
	pages = {455-500},
	year = {2009},
	month = {Jul.}
}

@ARTICLE{Guo2011Parafac,
	author={X. Guo and S. Miron and D. Brie and S. Zhu and X. Liao},
	journal={IEEE Trans. on Signal Process.},
	title={A CANDECOMP/PARAFAC Perspective on Uniqueness of DOA Estimation Using a Vector Sensor Array},
	volume={59},
	number={7},
	pages={3475-3481},
	year={2011},
	month = {Jul.}
}

@ARTICLE{Wang2015FrequencyDiverseArray,
	author={W.-Q {Wang}},
	journal={IEEE Antennas and Propagation Magazine}, 
	title={Frequency Diverse Array Antenna: New Opportunities}, 
	year={2015},
	volume={57},
	number={2},
	pages={145-152},
	doi={10.1109/MAP.2015.2414692}}

@article{wang2016overview,
	title={An overview on time/frequency modulated array processing},
	author={Wang, Wen-Qin and So, Hing Cheung and Farina, Alfonso},
	journal={IEEE Journal of Selected Topics in Signal Processing},
	volume={11},
	number={2},
	pages={228--246},
	year={2016},
	publisher={IEEE}
}

@ARTICLE{Khan2015FrequencyDiverse,
	author={Khan, Waseem and Qureshi, Ijaz Mansoor and Saeed, Sarah},
	journal={IEEE Antennas and Wireless Propagation Letters}, 
	title={Frequency Diverse Array Radar With Logarithmically Increasing Frequency Offset}, 
	year={2015},
	volume={14},
	number={},
	pages={499-502},
	doi={10.1109/LAWP.2014.2368977}}

@article{wang2020dot,
	title={Dot-shaped beamforming analysis based on OSB log-{FDA}},
	author={Wang, Bo and Xie, Junwei and Zhang, Jing and Zhang, Haowei},
	journal={Journal of Systems Engineering and Electronics},
	volume={31},
	number={2},
	pages={312--320},
	year={2020},
	publisher={BIAI}
}

@article{basit2017beam,
	title={Beam pattern synthesis for an {FDA} radar with Hamming window-based nonuniform frequency offset},
	author={Basit, Abdul and Qureshi, Ijaz Mansoor and Khan, Wasim and ur Rehman, Shuja and Khan, Muhammad Mohsin},
	journal={IEEE Antennas and Wireless Propagation Letters},
	volume={16},
	pages={2283--2286},
	year={2017},
	publisher={IEEE}
}

@ARTICLE{Liao2020FrequencyDiverse,
	author={Liao, Yi and Tang, Hu and Chen, Xiaolong and Wang, Wen-Qin},
	journal={IEEE Antennas and Wireless Propagation Letters}, 
	title={Frequency Diverse Array Beampattern Synthesis With Taylor Windowed Frequency Offsets}, 
	year={2020},
	volume={19},
	number={11},
	pages={1901-1905},
	doi={10.1109/LAWP.2020.3024710}}

@ARTICLE{Liao202AntennaBeampattern,
	author={Liao, Yi and Tang, Hu and Chen, Xiaolong and Wang, Wen-Qin and Xing, Mengdao and Zheng, Zhi and Wang, Jian and Liu, Qing Huo},
	journal={IEEE Access}, 
	title={Antenna Beampattern With Range Null Control Using Weighted Frequency Diverse Array}, 
	year={2020},
	volume={8},
	number={},
	pages={50107-50117},
	doi={10.1109/ACCESS.2020.2979942}}

@article{saeed2016tangent,
	title={Tangent hyperbolic circular frequency diverse array radars},
	author={Saeed, Sarah and Qureshi, Ijaz Mansoor and Khan, Waseem and Salman, Ayesha},
	journal={the Journal of Engineering},
	volume={2016},
	number={3},
	pages={23--28},
	year={2016},
	publisher={Wiley Online Library}
}

@article{eker2013exploitation,
	title={Exploitation of linear frequency modulated continuous waveform (LFMCW) for frequency diverse arrays},
	author={Eker, T and Demir, S and Hizal, A},
	journal={IEEE Transactions on Antennas and Propagation},
	volume={61},
	number={7},
	pages={3546--3553},
	year={2013},
	publisher={IEEE}
}

@ARTICLE{Xiong2017FrequencyDiverse,
	author={Xiong, Jie and Wang, Wen-Qin and Shao, Huaizong and Chen, Hui},
	journal={IEEE Antennas and Wireless Propagation Letters}, 
	title={Frequency Diverse Array Transmit Beampattern Optimization With Genetic Algorithm}, 
	year={2017},
	volume={16},
	number={},
	pages={469-472},
	doi={10.1109/LAWP.2016.2584078}}

@article{qin2016frequency,
	title={Frequency diverse coprime arrays with coprime frequency offsets for multitarget localization},
	author={Qin, Si and Zhang, Yimin D and Amin, Moeness G and Gini, Fulvio},
	journal={IEEE Journal of Selected Topics in Signal Processing},
	volume={11},
	number={2},
	pages={321--335},
	year={2016},
	publisher={IEEE}
}

@article{liao2018generalized,
	title={Generalized linear frequency diverse array manifold curve analysis},
	author={Liao, Tianxing and Pan, Ye and Wang, Wen-Qin},
	journal={IEEE Signal Processing Letters},
	volume={25},
	number={6},
	pages={768--772},
	year={2018},
	publisher={IEEE}
}

@article{sammartino2013frequency,
	title={Frequency diverse {MIMO} techniques for radar},
	author={Sammartino, Pier Francesco and Baker, Christopher J and Griffiths, Hugh D},
	journal={IEEE Transactions on Aerospace and Electronic Systems},
	volume={49},
	number={1},
	pages={201--222},
	year={2013},
	publisher={IEEE}
}

@article{mahmood2018frequency,
	title={Frequency diverse array beamforming using nonuniform logarithmic frequency increments},
	author={Mahmood, Mobeen and Mir, Hasan},
	journal={IEEE Antennas and Wireless Propagation Letters},
	volume={17},
	number={10},
	pages={1817--1821},
	year={2018},
	publisher={IEEE}
}

@ARTICLE{Xu2022RangeAngle,
	author={Xu, Yanhong and Wang, Anyi and Xu, Jingwei},
	journal={IEEE Transactions on Aerospace and Electronic Systems}, 
	title={Range–Angle Transceiver Beamforming Based on Semicircular-{FDA} Scheme}, 
	year={2022},
	volume={58},
	number={2},
	pages={834-843},
	doi={10.1109/TAES.2021.3111792}}

@ARTICLE{Shao2021FrequencyDiverse,
	author={Shao, Xiaolang and Hu, Taiyang and Xiao, Zelong and Zhang, Jinyu},
	journal={IEEE Antennas and Wireless Propagation Letters}, 
	title={Frequency Diverse Array Beampattern Synthesis With Modified Sinusoidal Frequency Offset}, 
	year={2021},
	volume={20},
	number={9},
	pages={1784-1788},
	doi={10.1109/LAWP.2021.3096980}}

@article{yaw2020frequency,
	title={Frequency-modulated diverse array transmit beamforming with bat metaheuristic optimisation},
	author={Yaw Nusenu, Shaddrack and Basit, Abdul},
	journal={IET Radar, Sonar \& Navigation},
	volume={14},
	number={9},
	pages={1338--1342},
	year={2020},
	publisher={Wiley Online Library}
}

@INPROCEEDINGS{Ni2021RangeDependent,
	author={Ni, Tianheng and Liu, Shengheng and Mao, Zihuan and Huang, Yongming},
	booktitle={2021 IEEE Radar Conference (RadarConf21)}, 
	title={Range-Dependent Beamforming Using Space-Frequency Virtual Difference Coarray}, 
	year={2021},
	volume={},
	number={},
	pages={1-5},
	address={Atlanta, GA, USA},
	doi={10.1109/RadarConf2147009.2021.9455207}}

@article{cui2018search,
	title={Search-free {DOD}, {DOA} and range estimation for bistatic {FDA-MIMO} radar},
	author={Cui, Can and Xu, Jian and Gui, Ronghua and Wang, Wen-Qin and Wu, Wen},
	journal={IEEE Access},
	volume={6},
	pages={15431--15445},
	year={2018},
	publisher={IEEE}
}

@article{li2018successive,
	title={Successive ESPRIT algorithm for joint DOA-range-polarization estimation with polarization sensitive FDA-MIMO radar},
	author={Li, Binbin and Bai, Weixiong and Zheng, Guimei},
	journal={IEEE Access},
	volume={6},
	pages={36376--36382},
	year={2018},
	publisher={IEEE}
}

@inproceedings{farooq2007application,
	title={Application of frequency diverse arrays to synthetic aperture radar imaging},
	author={Farooq, Jawad and Temple, Michael A and Saville, Michael A},
	booktitle={2007 International Conference on Electromagnetics in Advanced Applications},
	pages={447--449},
	year={2007},
	organization={IEEE},
	address={Turin, Italy}
}

@inproceedings{farooq2008exploiting,
	title={Exploiting frequency diverse array processing to improve SAR image resolution},
	author={Farooq, Jawad and Temple, Michael A and Saville, Michael A},
	booktitle={2008 IEEE Radar Conference},
	pages={1--5},
	year={2008},
	organization={IEEE},
	address={Rome, Italy}
}

@article{lin2017unambiguous,
	title={Unambiguous signal reconstruction approach for SAR imaging using frequency diverse array},
	author={Lin, Chenchen and Huang, Puming and Wang, Weiwei and Li, Yu and Xu, Jingwei},
	journal={IEEE Geoscience and Remote Sensing Letters},
	volume={14},
	number={9},
	pages={1628--1632},
	year={2017},
	publisher={IEEE}
}

@INPROCEEDINGS{Wang2016Forwardlooking,
	author={Wang, Wen-Qin},
	booktitle={2016 IEEE International Geoscience and Remote Sensing Symposium (IGARSS)}, 
	title={Forward-looking {SAR} imaging with frequency diverse array antenna}, 
	year={2016},
	volume={},
	number={},
	pages={4191-4194},
	doi={10.1109/IGARSS.2016.7730092},
	address={Beijing, China}}

@article{huang2020frequency,
	title={Frequency diverse array with random logarithmically increasing frequency offset},
	author={Huang, Gaojian and Ding, Yuan and Ouyang, Shan and Fusco, Vincent},
	journal={Microwave and Optical Technology Letters},
	volume={62},
	number={7},
	pages={2554--2561},
	year={2020},
	publisher={Wiley Online Library}
}

@ARTICLE{Ge2022FuzzyEntropy,
	author={Ge, Jiaang and Xie, Junwei and Wang, Bo and Chen, Chushu},
	journal={IEEE Transactions on Antennas and Propagation}, 
	title={Fuzzy Entropy for Frequency Diverse Array Beampattern Synthesis}, 
	year={2022},
	volume={70},
	number={11},
	pages={11172-11176},
	doi={10.1109/TAP.2022.3195523}}

@article{ulrich2018wavelength,
	title={Wavelength-diverse MIMO radar: parameter-coupling, array-carrier optimization and direction-of-arrival estimation},
	author={Ulrich, Michael and Yang, Bin},
	journal={IEEE Transactions on Aerospace and Electronic Systems},
	volume={55},
	number={4},
	pages={1920--1932},
	year={2018},
	publisher={IEEE}
}

@article{xu2015flat,
	title={Flat-top beampattern synthesis in range and angle domains for frequency diverse array via second-order cone programming},
	author={Xu, Yanhong and Shi, Xiaowei and Li, Wentao and Xu, Jingwei},
	journal={IEEE Antennas and Wireless Propagation Letters},
	volume={15},
	pages={1479--1482},
	year={2015},
	publisher={IEEE}
}

@article{xu2017range,
	title={Range--angle-decoupled beampattern synthesis with subarray-based frequency diverse array},
	author={Xu, Yanhong and Shi, Xiaowei and Xu, Jingwei and Huang, Lei and Li, Wentao},
	journal={Digital Signal Processing},
	volume={64},
	pages={49--59},
	year={2017},
	publisher={Elsevier}
}

@article{chen2017transmit,
	title={Transmit beampattern synthesis for the {FDA} radar},
	author={Chen, Baoxin and Chen, Xiaolong and Huang, Yong and Guan, Jian},
	journal={IEEE Antennas and Wireless Propagation Letters},
	volume={17},
	number={1},
	pages={98--101},
	year={2017},
	publisher={IEEE}
}

@article{gong2018time,
	title={Time-invariant joint transmit and receive beampattern optimization for polarization-subarray based frequency diverse array radar},
	author={Gong, Shiqi and Wang, Shuai and Chen, Sheng and Xing, Chengwen and Wei, Xing},
	journal={IEEE Transactions on Signal Processing},
	volume={66},
	number={20},
	pages={5364--5379},
	year={2018},
	publisher={IEEE}
}

@article{liao2019frequency,
	title={Frequency diverse array beampattern synthesis using symmetrical logarithmic frequency offsets for target indication},
	author={Liao, Yi and Wang, Wen-Qin and Zheng, Zhi},
	journal={IEEE Transactions on Antennas and Propagation},
	volume={67},
	number={5},
	pages={3505--3509},
	year={2019},
	publisher={IEEE}
}

@article{kan2025joint,
  title={Joint {DOA, DOD} and Range Parameter Estimation for Bistatic {FDA-MIMO} Radar via Triangular Geometry and Maximum Likelihood Criterion},
  author={Kan, Qingyun and Xu, Jingwei and Wang, Weiwei and Lan, Lan and Liao, Guisheng and So, Hing Cheung},
  journal={IEEE Transactions on Aerospace and Electronic Systems},
  year={2025},
  publisher={IEEE}
}

@article{wang2025frequency,
  title={Frequency Diverse Array With Discrete Fourier Transform for Single Target Estimation},
  author={Wang, Kai and Jin, Zhiyuan and Yu, Zichuan and Zhong, Feiyang and Tang, Lu and Tang, Xusheng},
  journal={IEEE Transactions on Aerospace and Electronic Systems},
  year={2025},
  publisher={IEEE}
}

@article{LIU2023103942,
  title = {Moving Target Detection in Range-Ambiguous Clutter Scenario with {{PA-FDA}} Dual-Mode Radar},
  author = {Liu, Zhixin and Zhu, Shengqi and Xu, Jingwei and He, Xiongpeng and Liu, Qi},
  year = 2023,
  journal = {Digital Signal Processing},
  volume = {135},
  pages = {103942},
  issn = {1051-2004},
  doi = {10.1016/j.dsp.2023.103942},
}

@article{liuCooperatedMovingTarget2023,
  title = {Cooperated {{Moving Target Detection Approach}} for {{PA-FDA Dual-Mode Radar}} in {{Range-Ambiguous Clutter}}},
  author = {Liu, Zhixin and Zhu, Shengqi and Xu, Jingwei and Lan, Lan and He, Xiongpeng and Li, Ximin},
  year = 2023,
  month = jan,
  journal = {Remote Sensing},
  volume = {15},
  number = {3},
  pages = {692},
  publisher = {Multidisciplinary Digital Publishing Institute},
  issn = {2072-4292},
  doi = {10.3390/rs15030692},
  urldate = {2026-03-20},
}

@inproceedings{liu2022detecting,
  title={Detecting moving target with Doppler spread and range migration for {FDA-MIMO} radar},
  author={Liu, Meihui and Zhang, Shunsheng and Wang, Wenqin},
  booktitle={IGARSS 2022-2022 IEEE International Geoscience and Remote Sensing Symposium},
  pages={2809--2812},
  year={2022},
  organization={IEEE}
}

@article{wan2021resolving,
  title={Resolving Doppler ambiguity of high-speed moving targets via {FDA-MIMO} radar},
  author={Wan, Weitao and Zhang, Shunsheng and Wang, Wen-Qin},
  journal={IEEE Geoscience and Remote Sensing Letters},
  volume={19},
  pages={1--5},
  year={2021},
  publisher={IEEE}
}

@article{jia2025long,
  title={Long-Time Coherent Integration and Range-Angle Beamforming for Detecting High-Speed Maneuvering Targets Using {FDA} Radar},
  author={Jia, Wenkai and Jian, Jiangwei and Li, Ping and Fu, Mingcheng and Huang, Bang and Wang, Wen-Qin},
  journal={IEEE Transactions on Aerospace and Electronic Systems},
  year={2025},
  publisher={IEEE}
}

@article{jia2025fda,
  title={{FDA}-based maneuvering target detection with Doppler-spread consideration},
  author={Jia, Mingjie and Huang, Bang and Basit, Abdul and Wang, Wen-Qin},
  journal={Digital Signal Processing},
  volume={159},
  pages={104990},
  year={2025},
  publisher={Elsevier}
}

@article{gui2021fda,
  title={{FDA} radar with doppler-spreading consideration: Mainlobe clutter suppression for blind-doppler target detection},
  author={Gui, Ronghua and Wang, Wen-Qin and Farina, Alfonso and So, Hing Cheung},
  journal={Signal Processing},
  volume={179},
  pages={107773},
  year={2021},
  publisher={Elsevier}
}

@article{huang2024adaptive,
  title={Adaptive target detection for an {FDA-MIMO} radar in a mainlobe deceptive jamming and a partially homogeneous noise},
  author={Huang, Bang and Wang, Wen-Qin and Li, Ping and Jian, Jiangwei and Jia, Yizhen and Jia, Wenkai and Liao, Tianxing},
  journal={Digital Signal Processing},
  volume={152},
  pages={104583},
  year={2024},
  publisher={Elsevier}
}

@article{huang2025glrt,
  title={{GLRT-based} adaptive target detection for {FDA-MIMO} radar in mainlobe deceptive jamming},
  author={Huang, Bang and Orlando, Danilo and Wang, Wen-Qin and Jian, Jiangwei and Jia, Yizhen and Jia, Wenkai and Liu, Weijian},
  journal={IEEE Sensors Journal},
  year={2025},
  publisher={IEEE}
}

@article{huang2024fda,
  title={{FDA-MIMO }radar target detection with limited test samples},
  author={Huang, Bang and Li, Ping and Jian, Jiangwei and Wang, Wen-Qin and Basit, Abdul},
  journal={IEEE Transactions on Aerospace and Electronic Systems},
  volume={60},
  number={4},
  pages={4390--4406},
  year={2024},
  publisher={IEEE}
}

@article{wang2021range,
  title={Range ambiguous clutter suppression for {FDA-MIMO} forward looking airborne radar based on main lobe correction},
  author={Wang, Yuzhuo and Zhu, Shengqi},
  journal={IEEE Transactions on Vehicular Technology},
  volume={70},
  number={3},
  pages={2032--2046},
  year={2021},
  publisher={IEEE}
}

@article{qiu2024range,
  title={Range-Ambiguous Clutter Suppression for Space-Based Early Warning Radar via {EPC--FDA--MIMO} With Nonorthogonal Waveforms},
  author={Qiu, Zizhou and Duan, Keqing and Yang, Xingjia and Wang, Yongliang},
  journal={IEEE Transactions on Aerospace and Electronic Systems},
  volume={60},
  number={5},
  pages={7106--7124},
  year={2024},
  publisher={IEEE}
}

@article{liu2023range,
  title={Range-ambiguous clutter suppression for {STAP-based} radar with vertical coherent frequency diverse array},
  author={Liu, Zhixin and Zhu, Shengqi and Xu, Jingwei and He, Xiongpeng and Duan, Keqing and Lan, Lan},
  journal={IEEE Transactions on Geoscience and Remote Sensing},
  volume={61},
  pages={1--17},
  year={2023},
  publisher={IEEE}
}

@article{sun2025nonlinear,
  title={Nonlinear frequency offset optimization strategy for solving secondary range ambiguity in planar {FDA-STAP} radar},
  author={Sun, Yan and Shao, Shuai and Wang, Wen-Qin and Greco, Maria Sabrina and Gini, Fulvio and Zhang, Shunsheng},
  journal={IEEE Transactions on Aerospace and Electronic Systems},
  year={2025},
  publisher={IEEE}
}

@article{ma2026multiparameter,
  title={Multiparameter Estimation for Bistatic {EMVS-FDA-MIMO} Radar with Arbitrarily Configured Arrays},
  author={Ma, Huihui and Tao, Haihong and Yue, Yaxing and Zhong, Tiantian and Fang, Yunfei and Wang, Le},
  journal={Digital Signal Processing},
  pages={105928},
  year={2026},
  publisher={Elsevier}
}

@article{bilal2026measurement,
  title={Measurement of Time-Invariant Range-Angle Maps in Frequency Diverse Array Radar Using {USRP}},
  author={Bilal, Ahmad and Shah, Yash H and Hadee, Abdul and Bhattacharjee, Sohom and Srihari, Pathipati and Cho, Choon Sik},
  journal={IEEE Sensors Journal},
  year={2026},
  publisher={IEEE}
}

@inproceedings{xin2025joint,
  title={Joint Range-Angle Estimation Method for Sparse Terahertz Frequency Diverse Arrays Based on the MUSIC Algorithm},
  author={Xin, Chongyang and Yang, Qi and Zhang, Lepeng and Wang, Hongqiang and Xu, Kaiyan},
  booktitle={2025 17th International Conference on Signal Processing Systems (ICSPS)},
  pages={229--233},
  year={2025},
  organization={IEEE}
}

@article{zhong2024multiparameter,
  title={Multiparameter estimation for monostatic {FDA-MIMO} radar with polarimetric antenna},
  author={Zhong, Tiantian and Tao, Haihong and Cao, Han and Liao, Haiyun},
  journal={IEEE Transactions on Antennas and Propagation},
  volume={72},
  number={3},
  pages={2524--2539},
  year={2024},
  publisher={IEEE}
}

@ARTICLE{ZhangZou2024RISAidedIndex,
	author={Zhang, Tong and Zou, Yikun and Wang, Gang and Chaaban, Anas and Liu, Gongliang and Cheng, Julian},
	journal={IEEE Transactions on Communications}, 
	title={{RIS-Aided} Index Modulation for {OFDM} Systems: Analysis and Code Design for Flat-Fading Channels}, 
	year={2024},
	volume={72},
	number={10},
	pages={6192-6208},
	doi={10.1109/TCOMM.2024.3400920}}

@article{jian2023fda,
  title={{FDA-MIMO}-based Integrated Sensing and Communication System with Frequency Offsets Permutation Index Modulation},
author={Jian, Jiangwei and Huang, Qimao and Huang, Bang and Wang, Wen-Qin},
journal={IEEE Transactions on Communications},
year={2024},
publisher={IEEE}
}

@article{gandotra2017survey,
	title={A survey on green communication and security challenges in 5G wireless communication networks},
	author={Gandotra, Pimmy and Jha, Rakesh Kumar},
	journal={Journal of Network and Computer Applications},
	volume={96},
	pages={39--61},
	year={2017},
	publisher={Elsevier}
}

@article{jian2023mimo,
	title={{MIMO-FDA} communications with frequency offsets index modulation},
	author={Jian, Jiangwei and Wang, Wen-Qin and Huang, Bang and Zhang, Lei and Imran, Muhammad Ali and Huang, Qimao},
	journal={IEEE Transactions on Wireless Communications},
	year={2023},
	publisher={IEEE}
}

@article{cogen2020generalized,
	title={Generalized code index modulation and spatial modulation for high rate and energy-efficient MIMO systems on rayleigh block-fading channel},
	author={Cogen, Fatih and Aydin, Erdogan and Kabaoglu, Nihat and Basar, Ertugrul and Ilhan, Haci},
	journal={IEEE Systems Journal},
	volume={15},
	number={1},
	pages={538--545},
	year={2020},
	publisher={IEEE}
}

@ARTICLE{Huang2020IndexModulation,
	author={Huang, Gaojian and Ouyang, Shan and Ding, Yuan and Fusco, Vincent},
	journal={IEEE Antennas and Wireless Propagation Letters}, 
	title={Index Modulation for Frequency Diverse Array}, 
	year={2020},
	volume={19},
	number={1},
	pages={49-53},
	doi={10.1109/LAWP.2019.2952576}}

@ARTICLE{Nusenu2020SpaceFrequency,
	author={Nusenu, Shaddrack Yaw and Huaizong, Shao and Pan, Ye and Basit, Abdul},
	journal={IEEE Transactions on Vehicular Technology}, 
	title={Space-Frequency Increment Index Modulation Approach for Fifth Generation and Beyond Wireless Communication Systems}, 
	year={2020},
	volume={69},
	number={6},
	pages={6286-6298},
	doi={10.1109/TVT.2020.2984912}}

@ARTICLE{Basit2021FDABasedQSM,
		author={Basit, Abdul and Wang, Wen-Qin and Nusenu, Shaddrack Yaw and Wali, Samad},
		journal={IEEE Transactions on Wireless Communications}, 
		title={{FDA} Based {QSM} for mmWave Wireless Communications: Frequency Diverse Transmitter and Reduced Complexity Receiver}, 
		year={2021},
		volume={20},
		number={7},
		pages={4571-4584},
		doi={10.1109/TWC.2021.3060512}}

@ARTICLE{Qiu2020MultiBeam,
	author={Qiu, Bin and Wang, Ling and Xie, Jian and Zhang, Zhaolin and Wang, Yuexian and Tao, Mingliang},
	journal={IEEE Transactions on Vehicular Technology}, 
	title={Multi-Beam Index Modulation With Cooperative Legitimate Users Schemes Based on Frequency Diverse Array}, 
	year={2020},
	volume={69},
	number={10},
	pages={11028-11041},
	doi={10.1109/TVT.2020.3007003}}

@article{huang2018fda,
  title={{FDA-OFDM} for integrated navigation, sensing, and communication systems},
  author={Huang, He and Wang, Wen-Qin},
  journal={IEEE Aerospace and Electronic Systems Magazine},
  volume={33},
  number={5-6},
  pages={34--42},
  year={2018},
  publisher={IEEE}
}

@article{nusenuShao2020spacefrequency,
  title={Space-frequency increment index modulation approach for fifth generation and beyond wireless communication systems},
  author={Nusenu, Shaddrack Yaw and Huaizong, Shao and Pan, Ye and Basit, Abdul},
  journal={IEEE Transactions on Vehicular Technology},
  volume={69},
  number={6},
  pages={6286--6298},
  year={2020},
  publisher={IEEE}
}

@article{nusenu2022powerallocation,
  title={Power allocation and equivalent transmit fda beamspace for 5G mmwave noma networks: Meta-heuristic optimization approach},
  author={Nusenu, Shaddrack Yaw and Huaizong, Shao and Ye, Pan},
  journal={IEEE Transactions on Vehicular Technology},
  volume={71},
  number={9},
  pages={9635--9646},
  year={2022},
  publisher={IEEE}
}

@article{nusenu2018range,
  title={Range-dependent spatial modulation using frequency diverse array for {OFDM} wireless communications},
  author={Nusenu, Shaddrack Yaw and Wang, Wen-Qin},
  journal={IEEE Transactions on Vehicular Technology},
  volume={67},
  number={11},
  pages={10886--10895},
  year={2018},
  publisher={IEEE}
}

@inproceedings{wachowiak2025frequency,
  title={Frequency diverse array OFDM system for joint communication and sensing},
  author={Wachowiak, Marcin and Bourdoux, Andr{\'e} and Pollin, Sofie},
  booktitle={2025 IEEE 5th International Symposium on Joint Communications \& Sensing (JC\&S)},
  pages={1--6},
  year={2025},
  organization={IEEE}
}

@misc{huang2026movableantennaindexmodulationmaim,
      title={Movable-Antenna Index Modulation {(MA-IM)}: System Framework and Performance Analysis}, 
      author={Bang Huang and Shunyuan Shang and Mohamed-Slim Alouini},
      year={2026},
      eprint={2603.26153},
      archivePrefix={arXiv},
      primaryClass={eess.SP},
      url={https://arxiv.org/abs/2603.26153}, 
}

@article{huang2025generalized,
  title={Generalized code-frequency-space index modulation: A next-generation green communication solution},
  author={Huang, Bang and Xu, Jiajie and Alouini, Mohamed-Slim},
  journal={IEEE Transactions on Wireless Communications},
  year={2025},
  publisher={IEEE}
}

@inproceedings{Ji2018FDAISAC,
  author    = {S. Ji and H. Chen and Q. Hu and others},
  title     = {A Dual-Function Radar-Communication System Using Frequency Diverse Array},
  booktitle = {Proc. IEEE Radar Conf. (RadarConf)},
  pages     = {224--229},
  year      = {2018}
}

@article{Nusenu2018TimeModFDA,
  author  = {S. Y. Nusenu and W.-Q. Wang and A. Basit},
  title   = {Time-Modulated FD-MIMO Array for Integrated Radar and Communication Systems},
  journal = {IEEE Antennas Wireless Propag. Lett.},
  volume  = {17},
  number  = {6},
  pages   = {1015--1019},
  year    = {2018}
}

@article{Zhou2021FDMIW,
  author  = {X. Zhou and L. Tang and Y. Bai and others},
  title   = {Performance Analysis and Waveform Optimization of Integrated {FD-MIMO} Radar-Communication Systems},
  journal = {IEEE Trans. Wireless Commun.},
  volume  = {20},
  number  = {11},
  pages   = {7490--7502},
  year    = {2021}
}

@article{Li2023FIM,
  author  = {M. Li and W.-Q. Wang},
  title   = {Joint Radar-Communication System Design Based on {FDA-MIMO} via Frequency Index Modulation},
  journal = {IEEE Access},
  volume  = {11},
  pages   = {67722--67736},
  year    = {2023}
}

@article{Jian2024FOPIM,
  author  = {J. Jian and Q. Huang and B. Huang and others},
  title   = {{FDA-MIMO-Based} Integrated Sensing and Communication System With Frequency Offsets Permutation Index Modulation},
  journal = {IEEE Trans. Commun.},
  volume  = {72},
  number  = {11},
  pages   = {6707--6721},
  year    = {2024}
}

@article{Jian2024CCIE,
  author  = {J. Jian and B. Huang and W. Jia and others},
  title   = {{FDA-MIMO-Based} Integrated Multi-Target Sensing and Communication System With Complex Coefficients Information Embedding},
  journal = {arXiv preprint arXiv:2409.02447},
  year    = {2024}
}

@article{Xu2025TransceiverDesign,
  author  = {Qihang Xu and Lan Lan and Guisheng Liao and Kaiwen Wang and Tongxing Zheng},
  title   = {Transceiver Design for an {FDA-MIMO Radar and MIMO } Communication Spectral Coexistence System},
  journal = {Journal of Radars},
 vol={14},
no={4},
  year    = {2025}
}

@book{buehrer2022code,
  title={Code division multiple access (CDMA)},
  author={Buehrer, R Michael},
  year={2022},
  publisher={Springer Nature}
}

@article{barton1988modern,
  title={Modern radar system analysis},
  author={Barton, David K},
  journal={Norwood},
  year={1988}
}

@book{visser2006array,
  title={Array and phased array antenna basics},
  author={Visser, Hubregt J},
  year={2006},
  publisher={John Wiley \& Sons}
}

@article{fulton2016digital,
  title={Digital phased arrays: Challenges and opportunities},
  author={Fulton, Caleb and Yeary, Mark and Thompson, Daniel and Lake, John and Mitchell, Adam},
  journal={Proceedings of the IEEE},
  volume={104},
  number={3},
  pages={487--503},
  year={2016},
  publisher={IEEE}
}

@article{saadia2020single,
  title={Single carrier-frequency division multiple access radar: Waveform design and analysis},
  author={Saadia, Rahat and Khan, Noor M},
  journal={IEEE Access},
  volume={8},
  pages={35742--35751},
  year={2020},
  publisher={IEEE}
}

@incollection{faruque2018frequency,
  title={Frequency division multiple access ({FDMA})},
  author={Faruque, Saleh},
  booktitle={radio frequency multiple access techniques made easy},
  pages={21--33},
  year={2018},
  publisher={Springer}
}

@article{rappaport2002wireless,
  title={Wireless Communications--Principles and Practice, (The Book End).},
  author={Rappaport, Theodore S},
  journal={Microwave Journal},
  volume={45},
  number={12},
  pages={128--129},
  year={2002},
  publisher={Horizon House Publications, Inc.}
}

@book{rahman2019fundamental,
  title={Fundamental principles of radar},
  author={Rahman, Habibur},
  year={2019},
  publisher={CRC Press}
}

@book{eaves2012principles,
  title={Principles of modern radar},
  author={Eaves, Jerry and Reedy, Edward},
  year={2012},
  publisher={Springer Science \& Business Media}
}

@book{de2008multiantenna,
  title={Multiantenna systems for {MIMO} communications},
  author={De Flaviis, Franco},
  volume={7},
  year={2008},
  publisher={Morgan \& Claypool Publishers}
}

@book{duman2008coding,
  title={Coding for {MIMO} communication systems},
  author={Duman, Tolga M and Ghrayeb, Ali},
  year={2008},
  publisher={John Wiley \& Sons}
}

@book{gershman2005space,
  title={Space-time processing for {MIMO} communications},
  author={Gershman, Alex B and Sidiropoulos, Nikos D},
  year={2005},
  publisher={Wiley Online Library}
}

@book{hampton2013introduction,
  title={Introduction to {MIMO} communications},
  author={Hampton, Jerry R},
  year={2013},
  publisher={Cambridge university press}
}

@article{weinstein2009history,
  title={The history of orthogonal frequency-division multiplexing [History of Communications]},
  author={Weinstein, Stephen B},
  journal={IEEE Communications Magazine},
  volume={47},
  number={11},
  pages={26--35},
  year={2009},
  publisher={IEEE}
}

@book{li2006orthogonal,
  title={Orthogonal frequency division multiplexing for wireless communications},
  author={Li, Ye Geoffrey and Stuber, Gordon L},
  year={2006},
  publisher={Springer Science \& Business Media}
}

@book{zigangirov2004theory,
  title={Theory of code division multiple access communication},
  author={Zigangirov, Kamil Sh},
  year={2004},
  publisher={John Wiley \& Sons}
}

@article{Jian2025DualFunction,
  author  = {J. Jian and W. Jia and W.-Q. Wang and others},
  title   = {Dual-Function System for Area Surveillance and Multi-User Communications With {FDA-MIMO}},
  journal = {IEEE Trans. Veh. Technol.},
  year    = {2025},
  note    = {Early access}
}

@article{Yang2025RISFDA,
  author  = {H. Yang and S. Gong and H. Liu and others},
  title   = {Frequency Diverse Array-Enabled {RIS}-Aided Integrated Sensing and Communication},
  journal = {IEEE Trans. Wireless Commun.},
  year    = {2025},
  note    = {Early access}
}




\end{document}